%% file: TheAtmosphere.tex
\documentclass[12pt]{article}
\usepackage[margin=1in]{geometry}
\usepackage[labelfont=bf]{caption}

\usepackage{hyperref}
\hypersetup{
    colorlinks=true,
    linkcolor=blue,
    filecolor=blue,      
    urlcolor=blue,
}
\usepackage{amsmath}    
\usepackage{amssymb}
\usepackage{graphicx, color}
\usepackage{caption, subcaption}
\usepackage{mathabx}
\usepackage[capbesideposition={bottom},facing=yes,capbesidesep=quad]{floatrow}

\usepackage{booktabs}
\usepackage{multirow}
\usepackage{tabularx}
\usepackage[final]{pdfpages} 

\usepackage{mathptmx}       
\usepackage[scaled=.90]{helvet}
\usepackage{courier}

\usepackage{xcolor}
\usepackage{enumitem}

\usepackage{titlesec}

\titleformat*{\section}{\Large\bfseries}
\titleformat*{\subsection}{\large\bfseries}
\titleformat*{\subsubsection}{\bfseries}

\begin{document}

\parindent 0.2in

\vspace*{0.5in}

\begin{center}
\begin{LARGE}
\textbf{The Atmosphere}
\end{LARGE}

\bigskip

\begin{large}
A flipped course

\vspace{1in}

\begin{figure}[h!]
\begin{center}
  \includegraphics[width=0.9\textwidth]{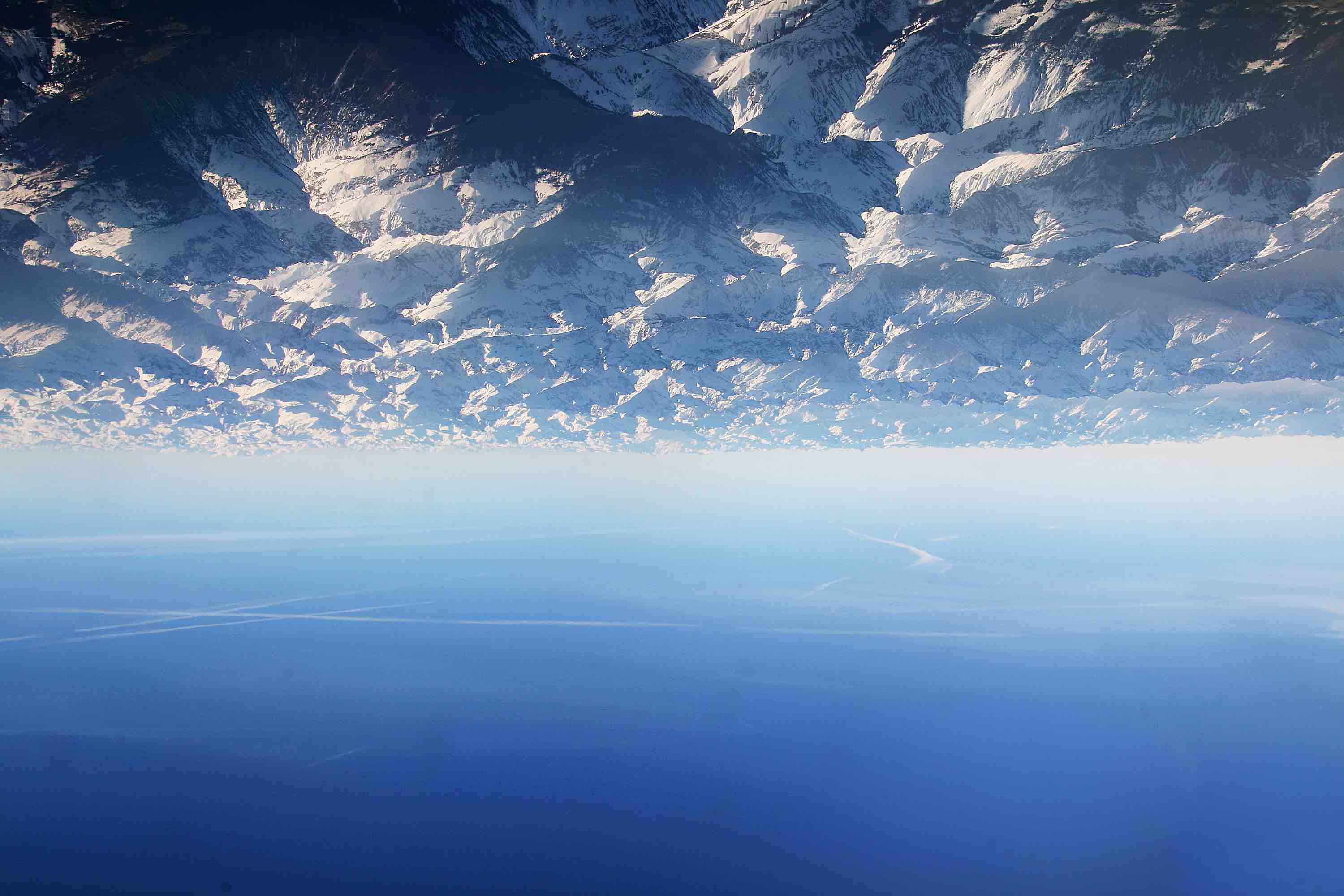}
\end{center}
\end{figure}

\vspace{1in}

Dorian S. Abbot

\end{large}
\end{center}

\thispagestyle{empty}

\clearpage

\tableofcontents

\thispagestyle{empty}

\newpage

\setcounter{page}{1}

\section{Introduction}

\clearpage

\input{Introduction/origin.tex}

\clearpage

\input{Introduction/howto.tex}

\clearpage

\input{Introduction/acknowledgements.tex}

\clearpage

\section{Syllabus}

\clearpage

\input{Syllabus/syllabus.tex}

\clearpage

\section{Thermodynamics}

\clearpage

\input{Flipped/orient_estimation.tex}

\clearpage

\input{Flipped/orientation.tex}

\clearpage

\input{Flipped/pressure_cyclones.tex}

\clearpage

\input{Flipped/composition.tex}

\clearpage

\input{Flipped/virtual_temperature.tex}

\clearpage

\input{Flipped/hydrostatic_scale_height.tex}

\clearpage

\input{Flipped/lapse_rate_theta_1.tex}

\clearpage

\input{Flipped/stability.tex}

\clearpage

\input{Flipped/water_vapor.tex}

\clearpage

\input{Flipped/moist_convection.tex}

\clearpage

\input{Flipped/carnot.tex}

\clearpage

\section{Radiation}

\clearpage

\input{Flipped/electromagnetic.tex}

\clearpage

\input{Flipped/blackbody.tex}

\clearpage

\input{Flipped/greenhouse.tex}

\clearpage

\input{Flipped/abs_scattering.tex}

\clearpage

\input{Flipped/rad_vertical.tex}

\clearpage

\input{Flipped/energy_balance.tex}

\clearpage

\section{Dynamics}

\clearpage

\input{Flipped/partial_derivative_streamfunction.tex}

\clearpage

\input{Flipped/divergence.tex}

\clearpage

\input{Flipped/curl.tex}

\clearpage

\input{Flipped/coriolis.tex}

\clearpage

\input{Flipped/pressure_grad.tex}

\clearpage

\input{Flipped/geostrophic.tex}

\clearpage

\input{Flipped/geostrophic2.tex}

\clearpage

\input{Flipped/potential_vorticity.tex}

\clearpage

\input{Flipped/tornadoes.tex}

\clearpage

\section{Bonus}

\clearpage

\input{Flipped/clouds.tex}

\clearpage

\section{Problem Sets}

\clearpage

\input{PSets/pset01.tex}

\clearpage

\input{PSets/pset02.tex}

\clearpage

\input{PSets/pset03.tex}

\clearpage

\input{PSets/pset04.tex}

\clearpage

\input{PSets/pset05.tex}

\clearpage

\input{PSets/pset06.tex}

\clearpage

\input{PSets/pset07.tex}

\clearpage

\input{PSets/pset08.tex}

\clearpage

\input{PSets/pset09.tex}

\clearpage

\input{PSets/pset10.tex}

\clearpage

\section{Midterms}

\clearpage

\input{Midterms/midterm1_2017.tex}

\clearpage

\begin{figure}[h!]
\begin{center}
  \includegraphics[width=\textwidth]{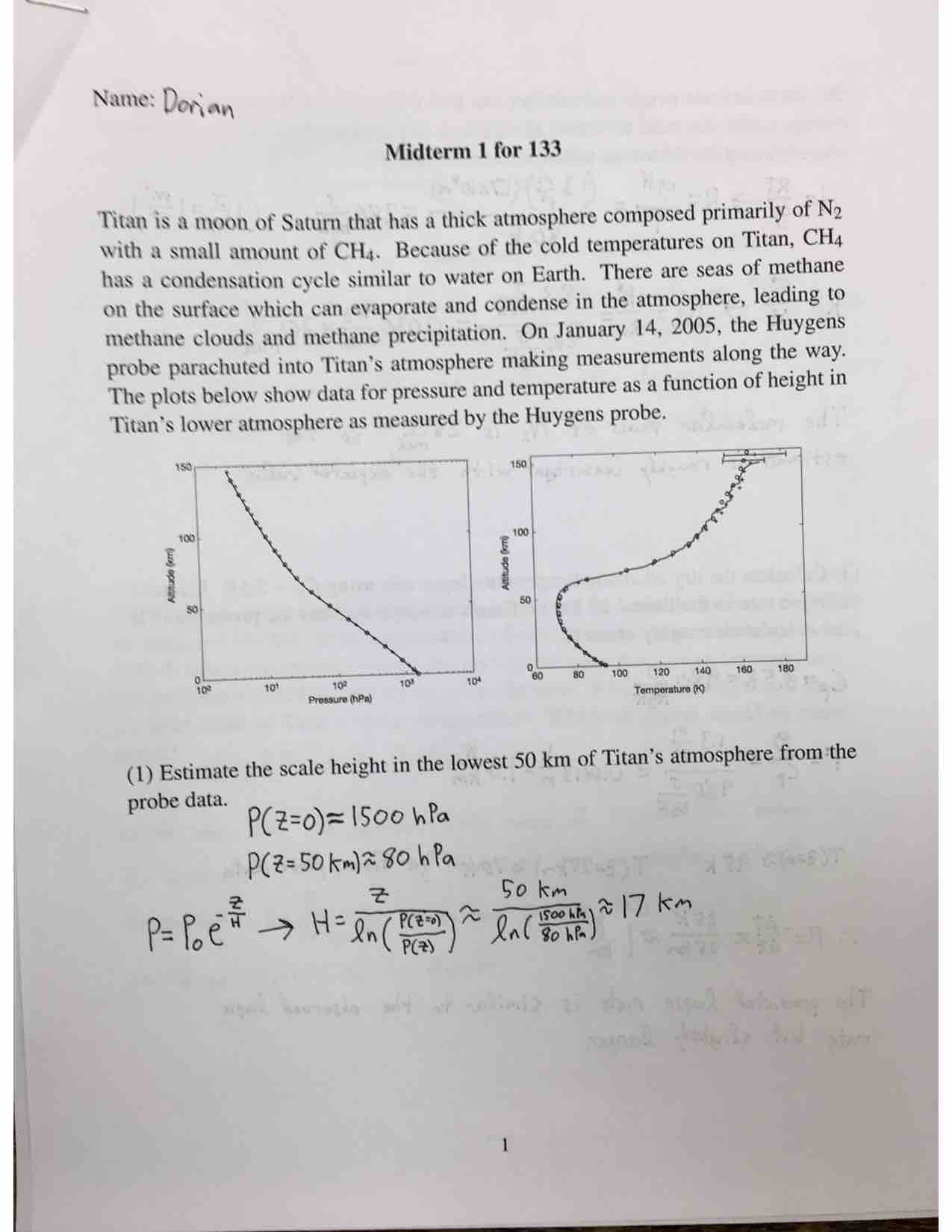}
\end{center}
\end{figure}

\begin{figure}[h!]
\begin{center}
  \includegraphics[width=\textwidth]{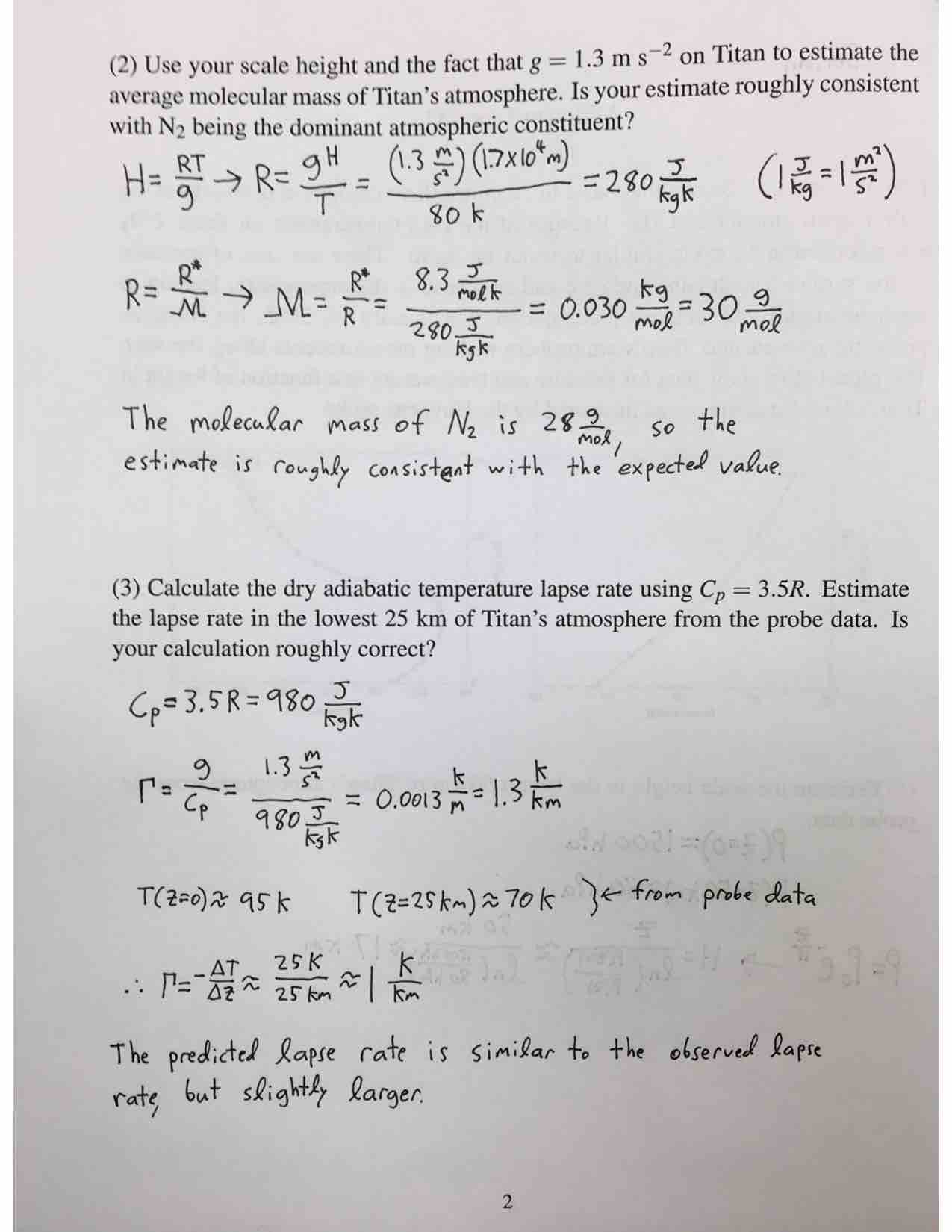}
\end{center}
\end{figure}

\begin{figure}[h!]
\begin{center}
  \includegraphics[width=\textwidth]{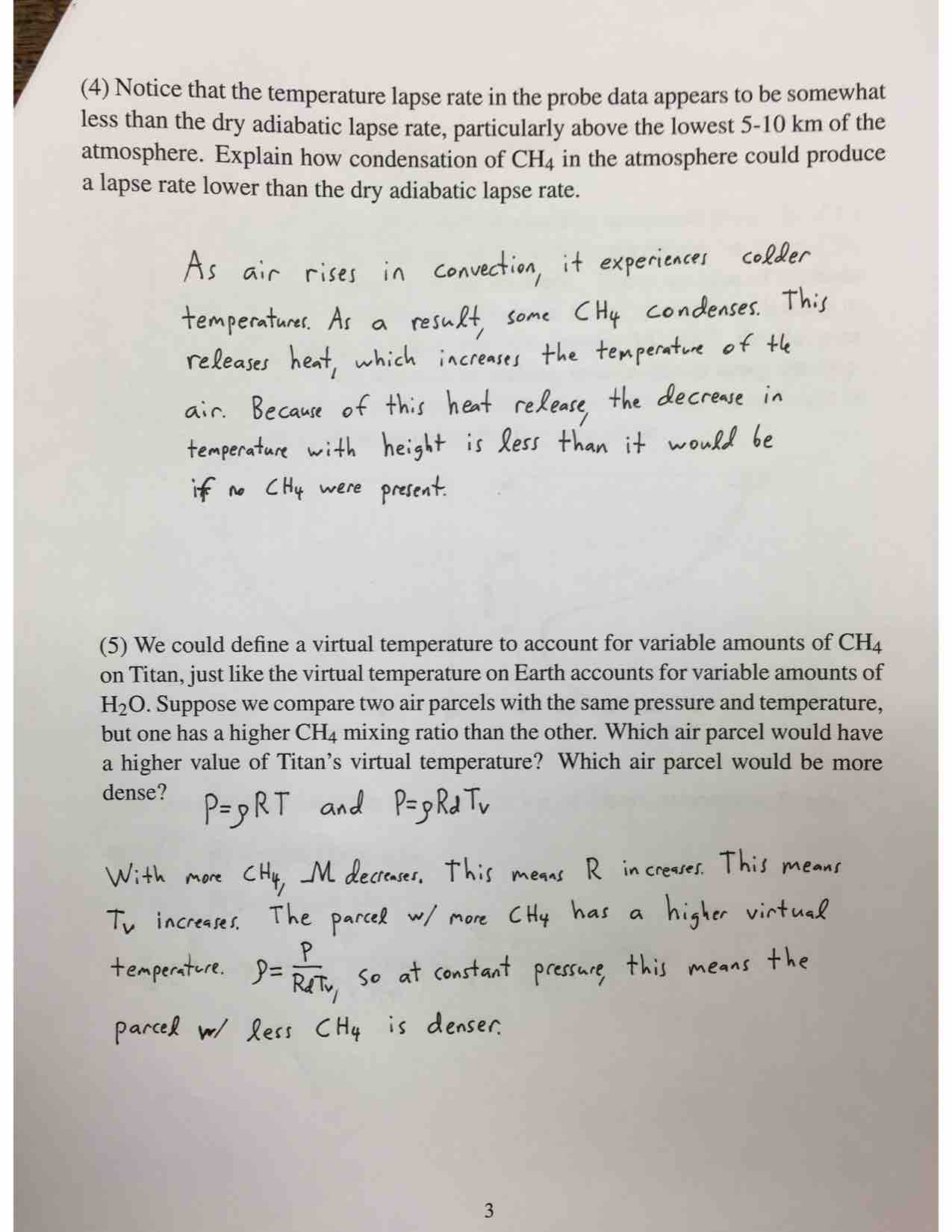}
\end{center}
\end{figure}

\clearpage

\input{Midterms/midterm1_2018.tex}

\clearpage

\begin{figure}[h!]
\begin{center}
  \includegraphics[width=\textwidth]{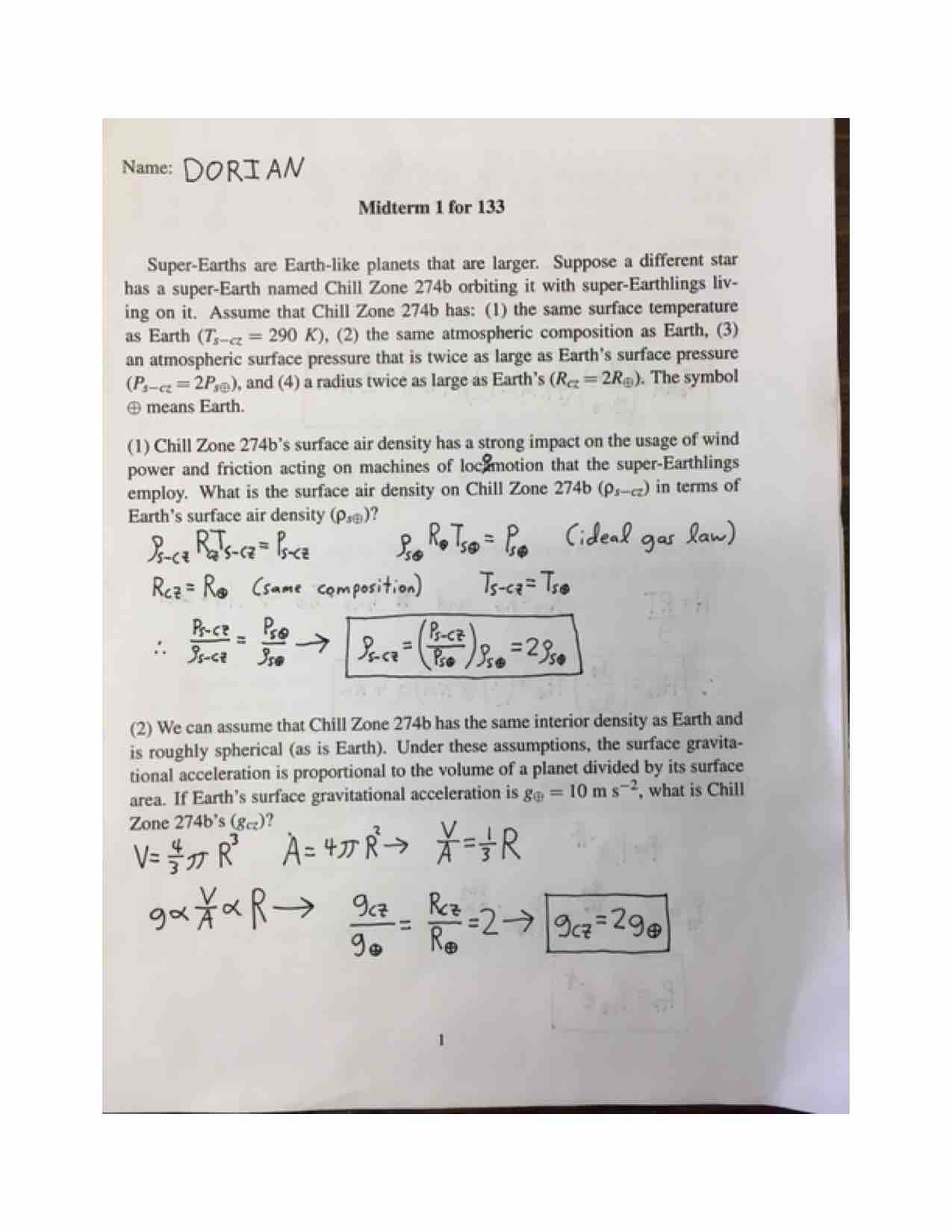}
\end{center}
\end{figure}

\begin{figure}[h!]
\begin{center}
  \includegraphics[width=\textwidth]{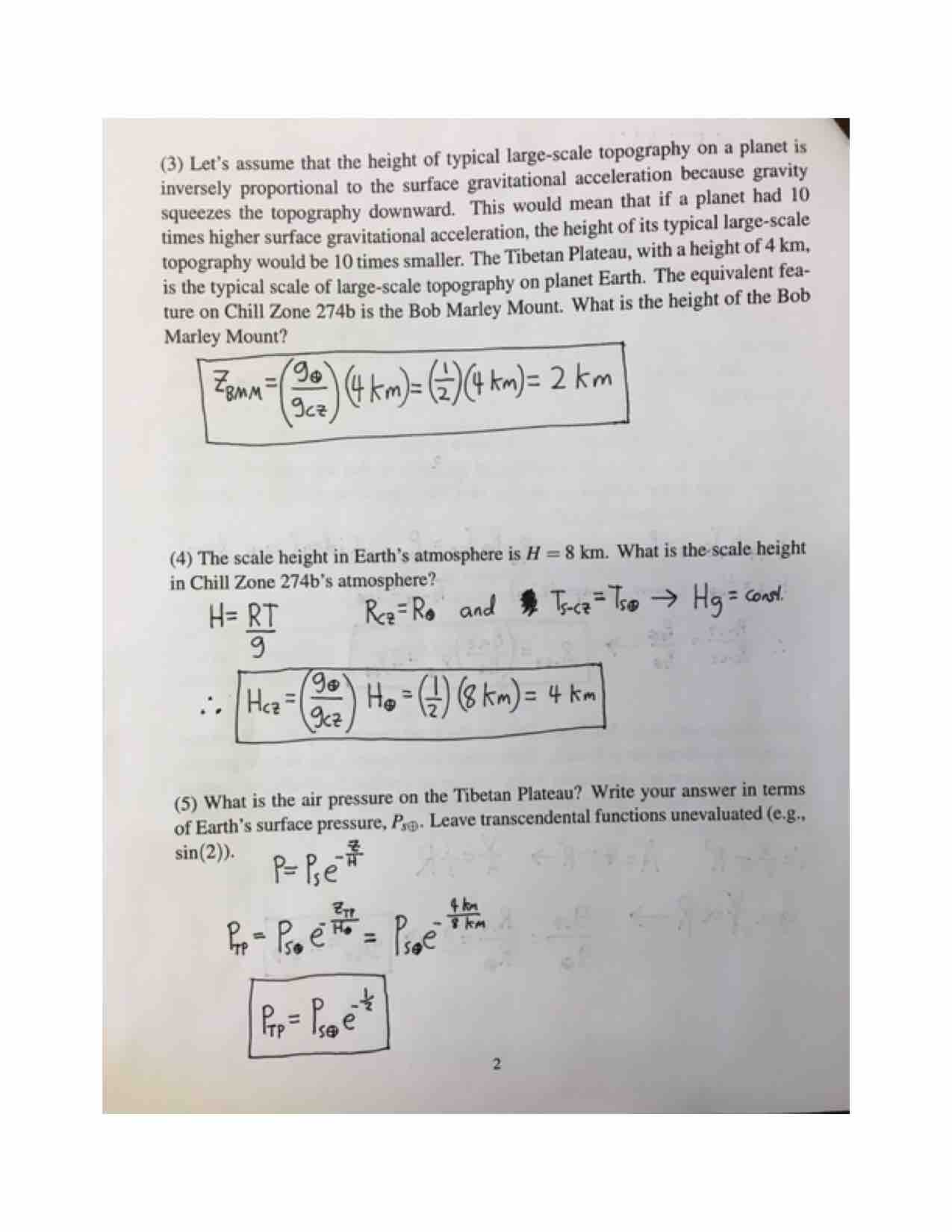}
\end{center}
\end{figure}

\begin{figure}[h!]
\begin{center}
  \includegraphics[width=\textwidth]{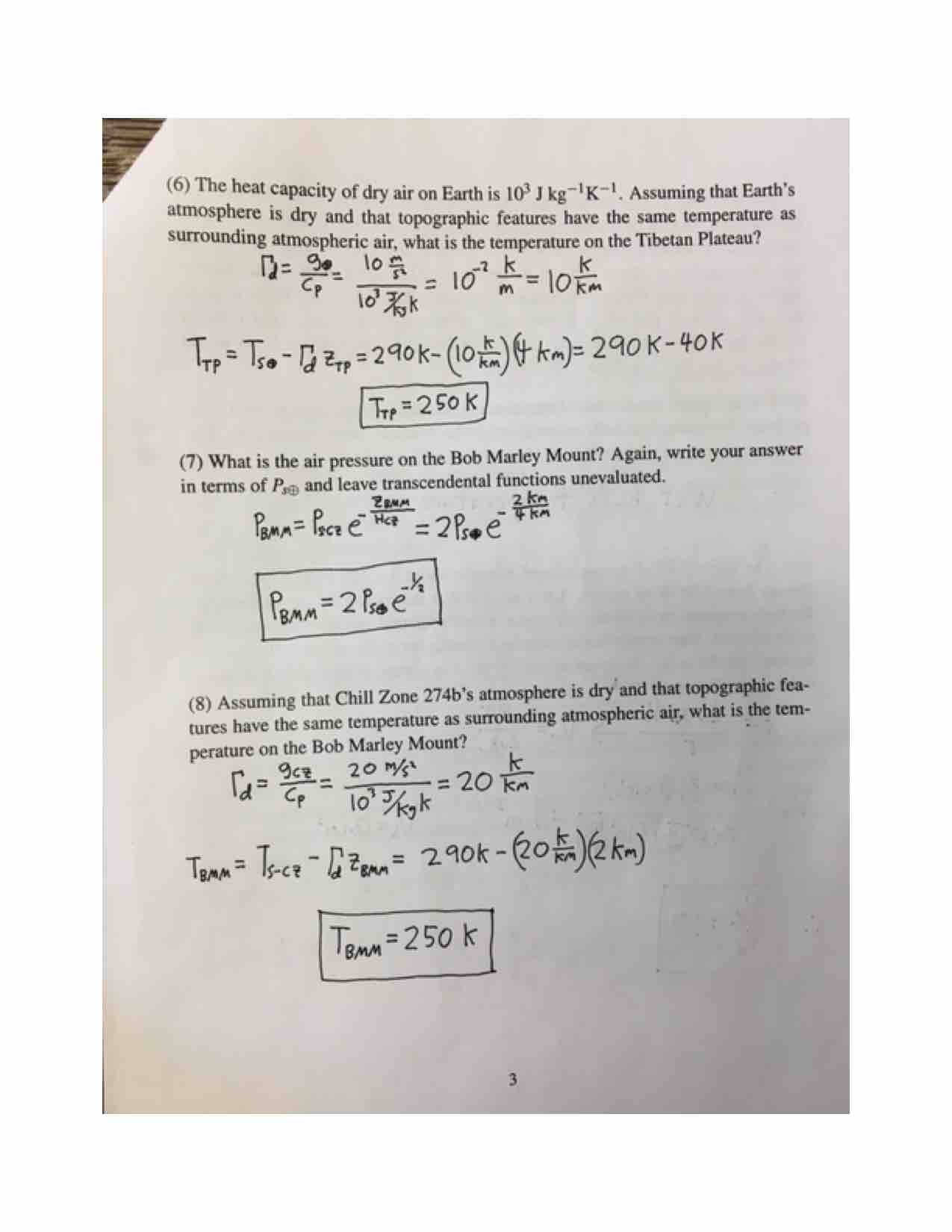}
\end{center}
\end{figure}

\begin{figure}[h!]
\begin{center}
  \includegraphics[width=\textwidth]{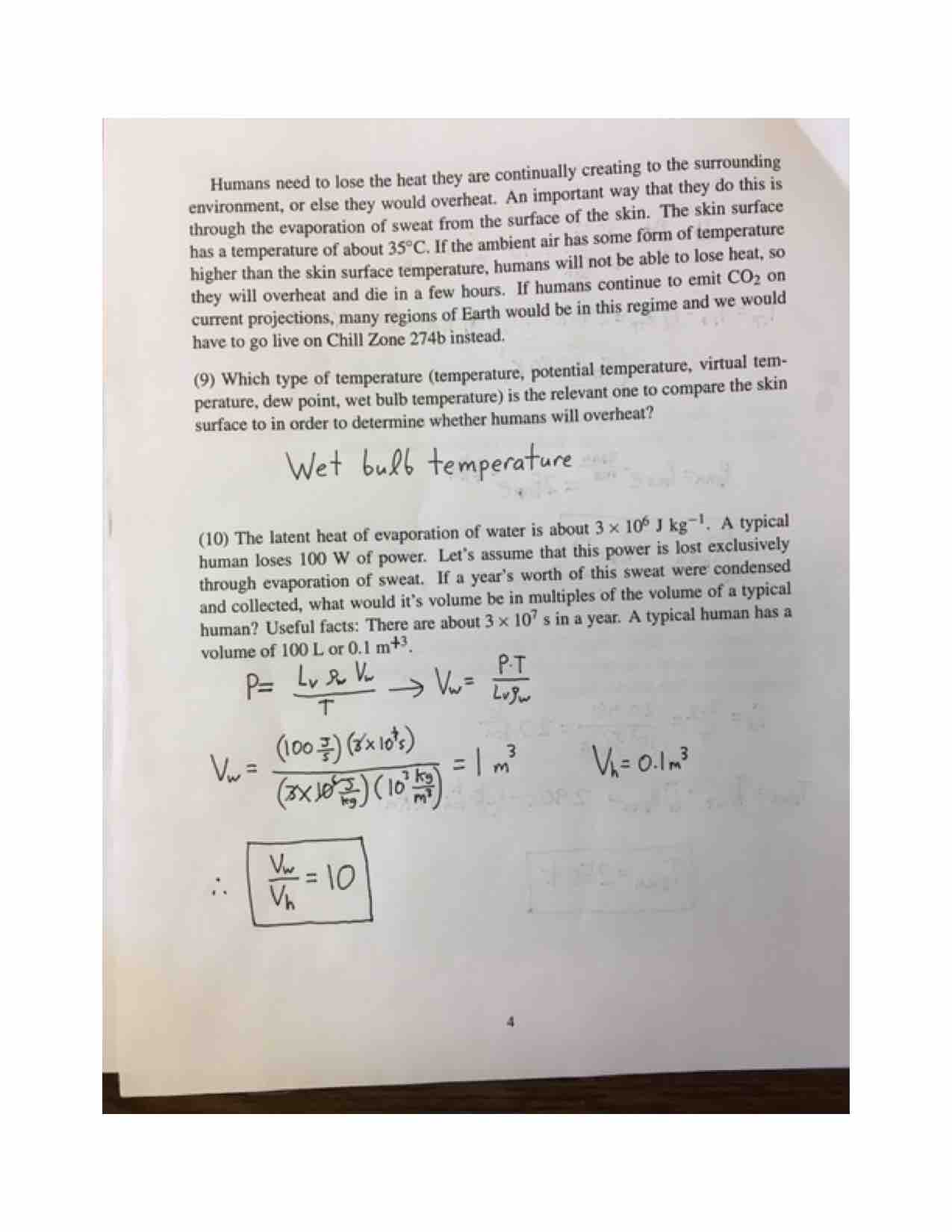}
\end{center}
\end{figure}

\clearpage

\input{Midterms/midterm1_2019.tex}

\clearpage

\begin{figure}[h!]
\begin{center}
  \includegraphics[width=\textwidth]{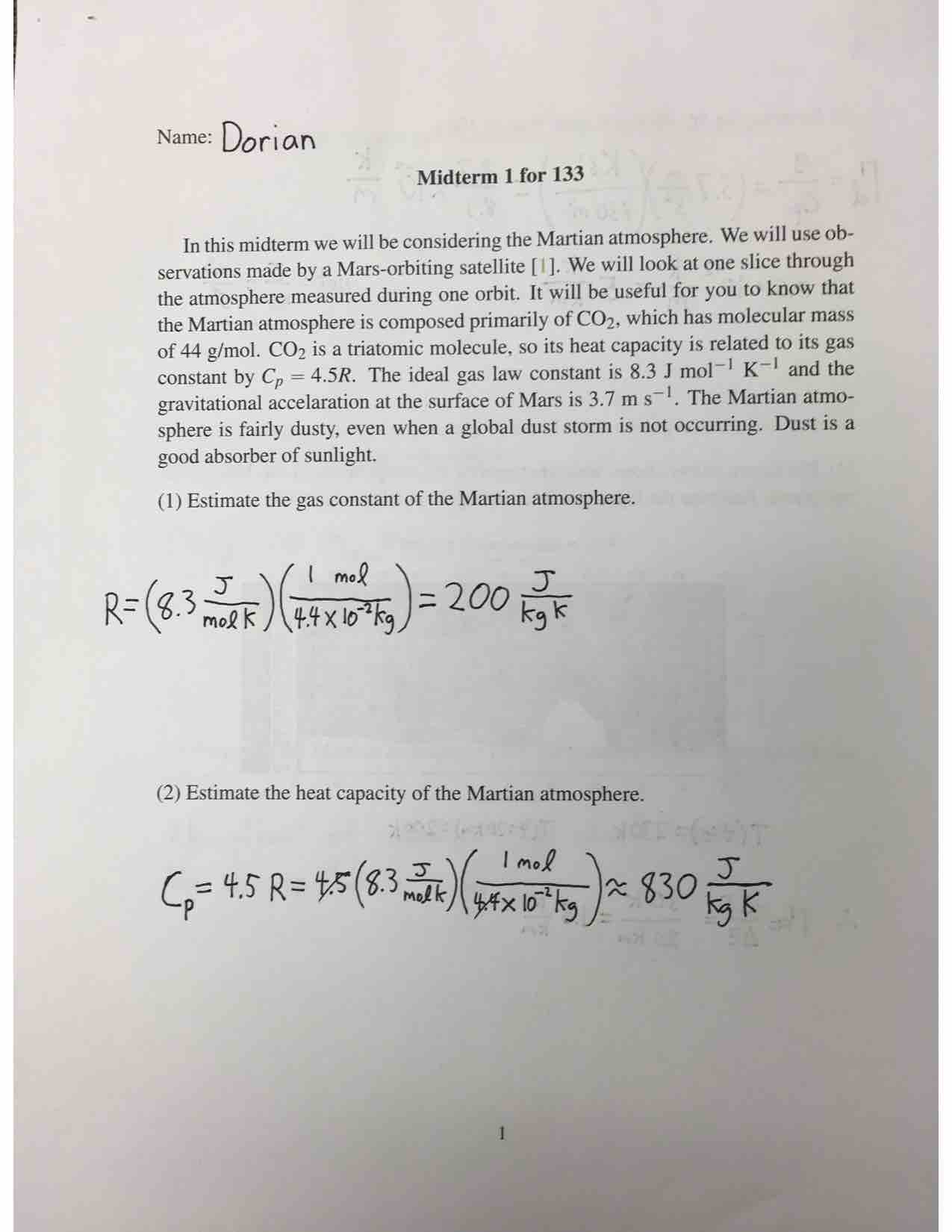}
\end{center}
\end{figure}

\begin{figure}[h!]
\begin{center}
  \includegraphics[width=\textwidth]{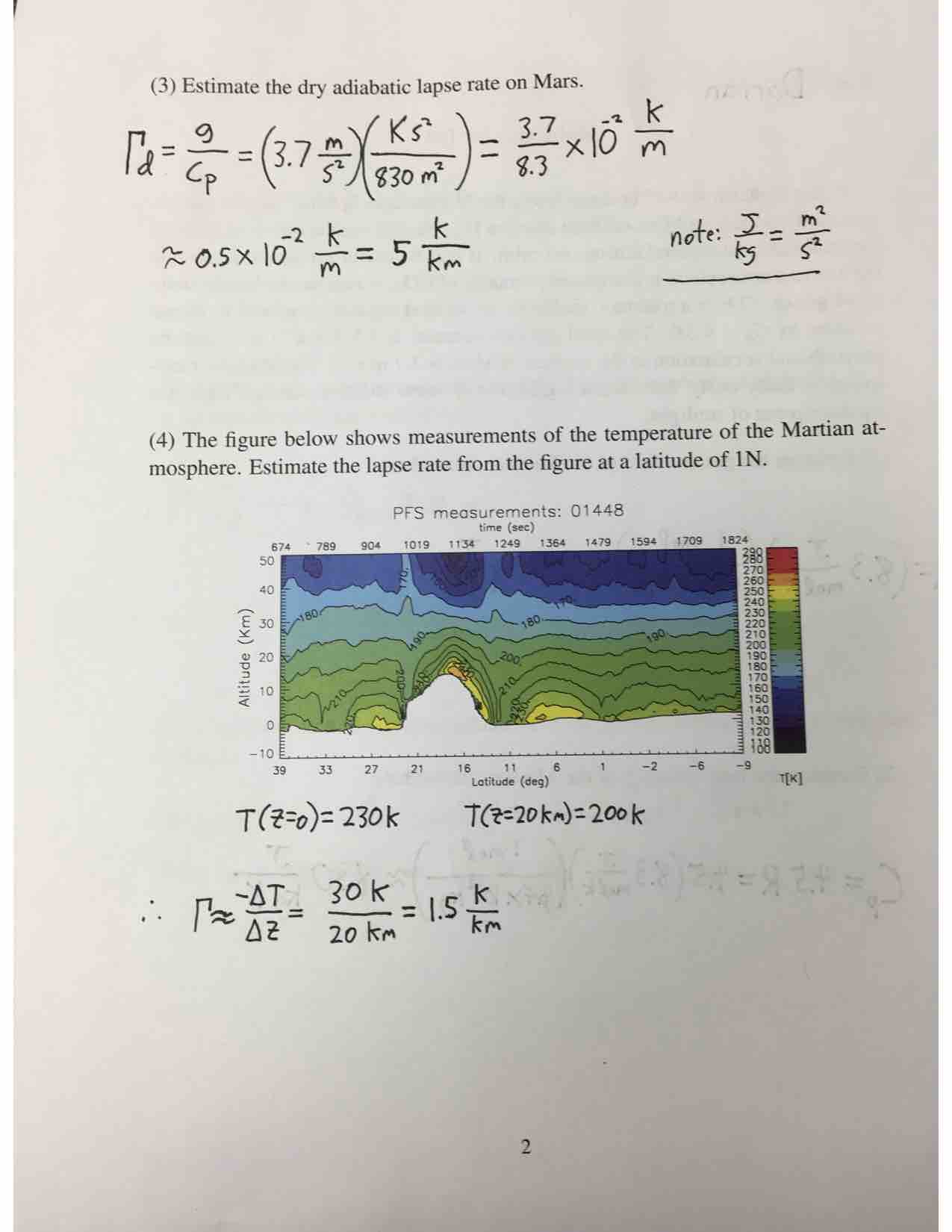}
\end{center}
\end{figure}

\begin{figure}[h!]
\begin{center}
  \includegraphics[width=\textwidth]{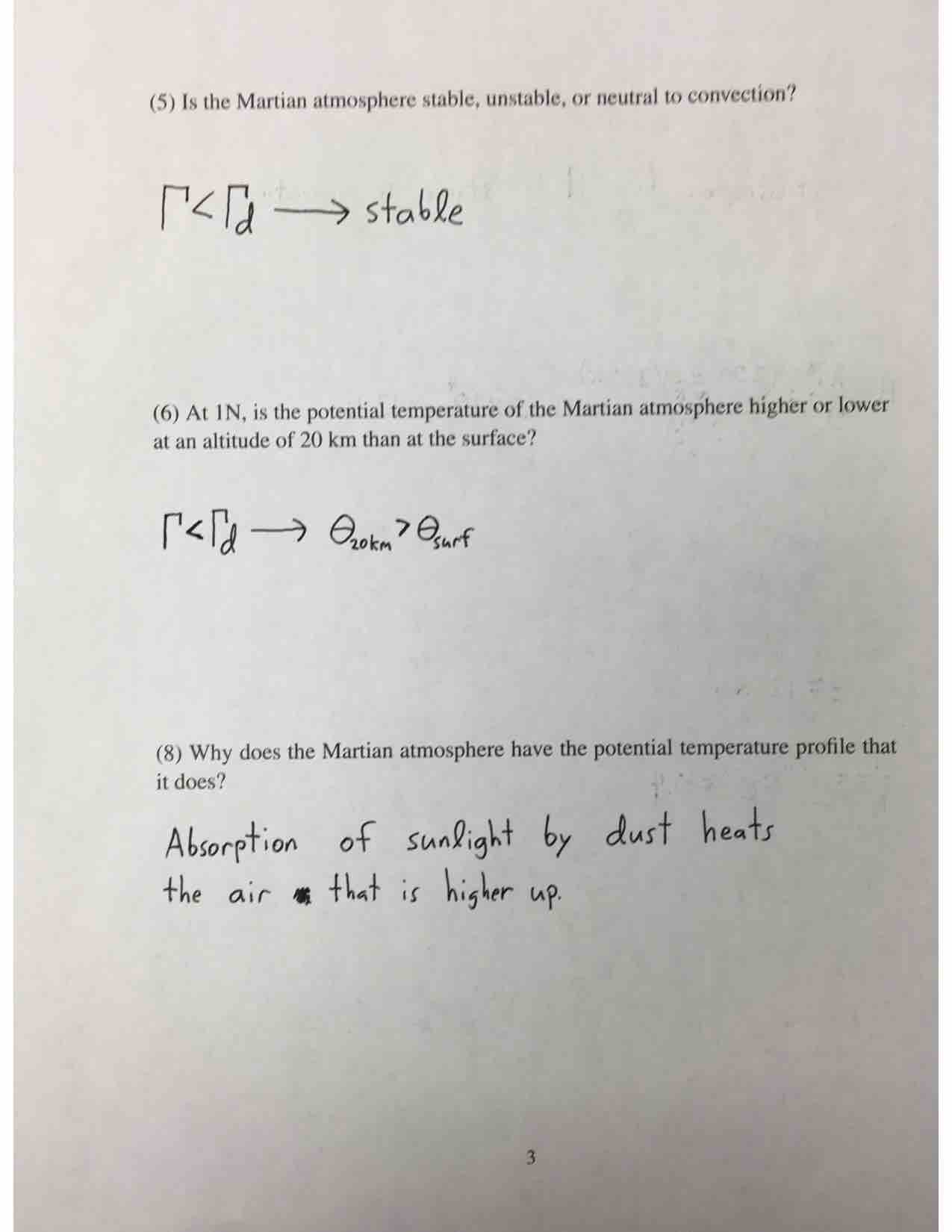}
\end{center}
\end{figure}

\begin{figure}[h!]
\begin{center}
  \includegraphics[width=\textwidth]{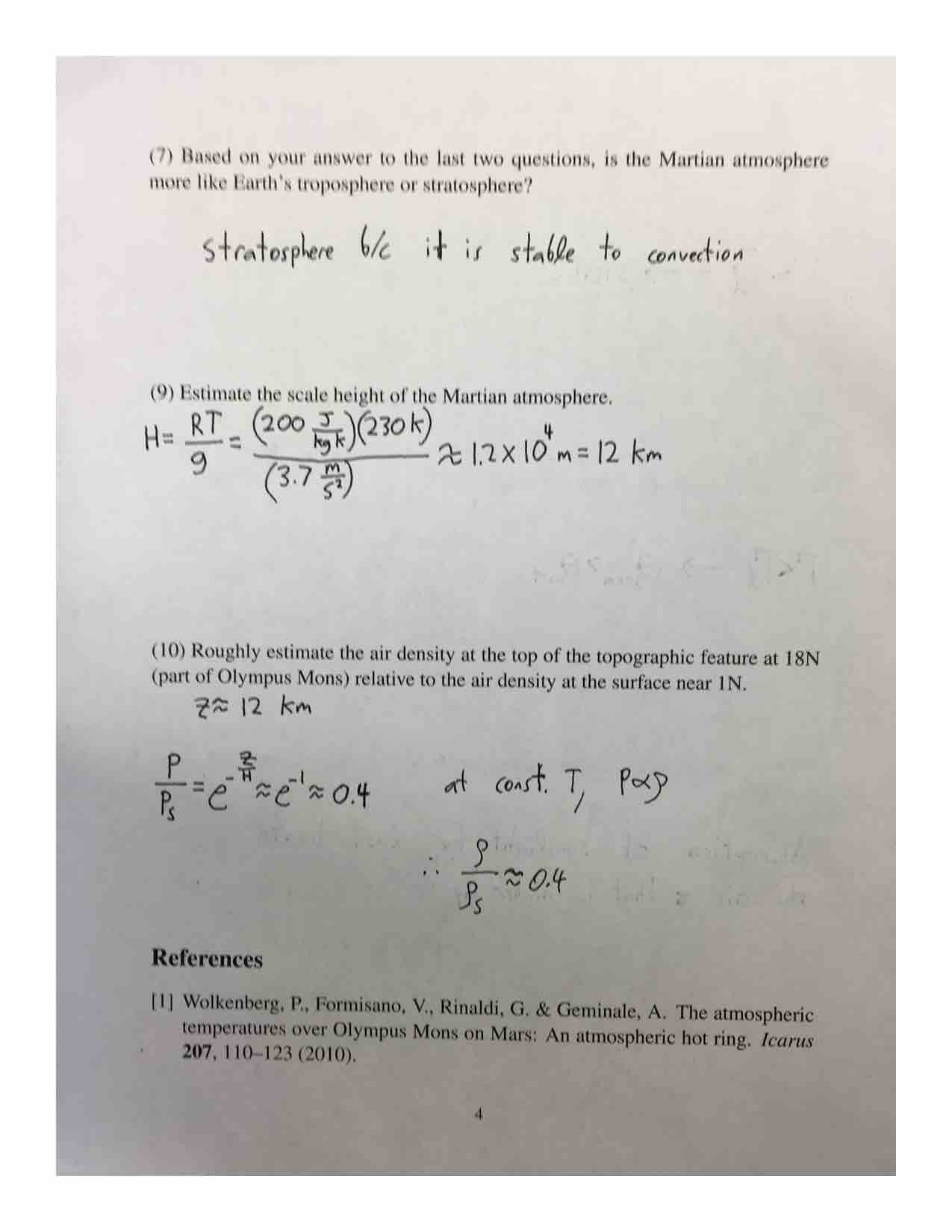}
\end{center}
\end{figure}

\clearpage

\input{Midterms/midterm2_2017.tex}

\clearpage

\begin{figure}[h!]
\begin{center}
  \includegraphics[width=\textwidth]{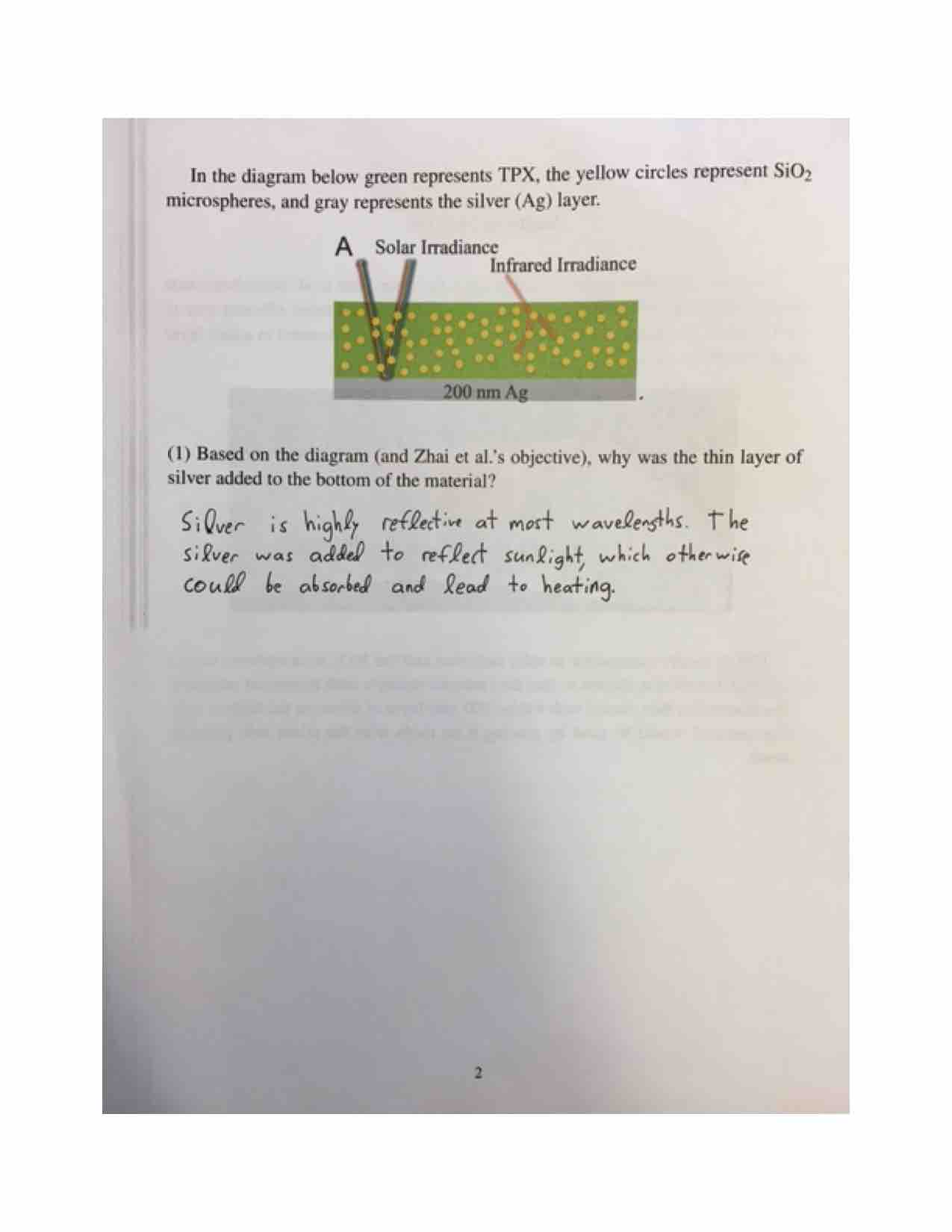}
\end{center}
\end{figure}

\begin{figure}[h!]
\begin{center}
  \includegraphics[width=\textwidth]{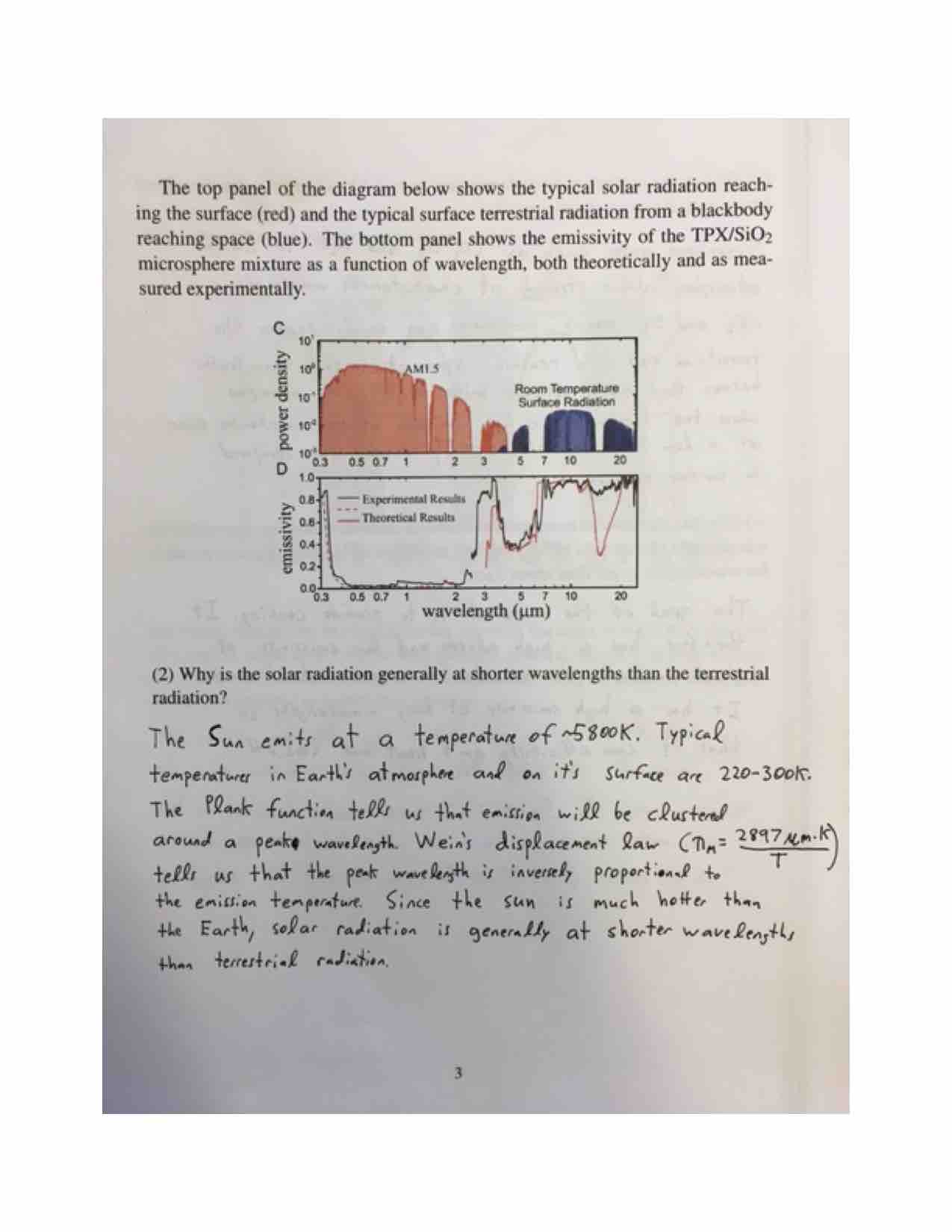}
\end{center}
\end{figure}

\begin{figure}[h!]
\begin{center}
  \includegraphics[width=\textwidth]{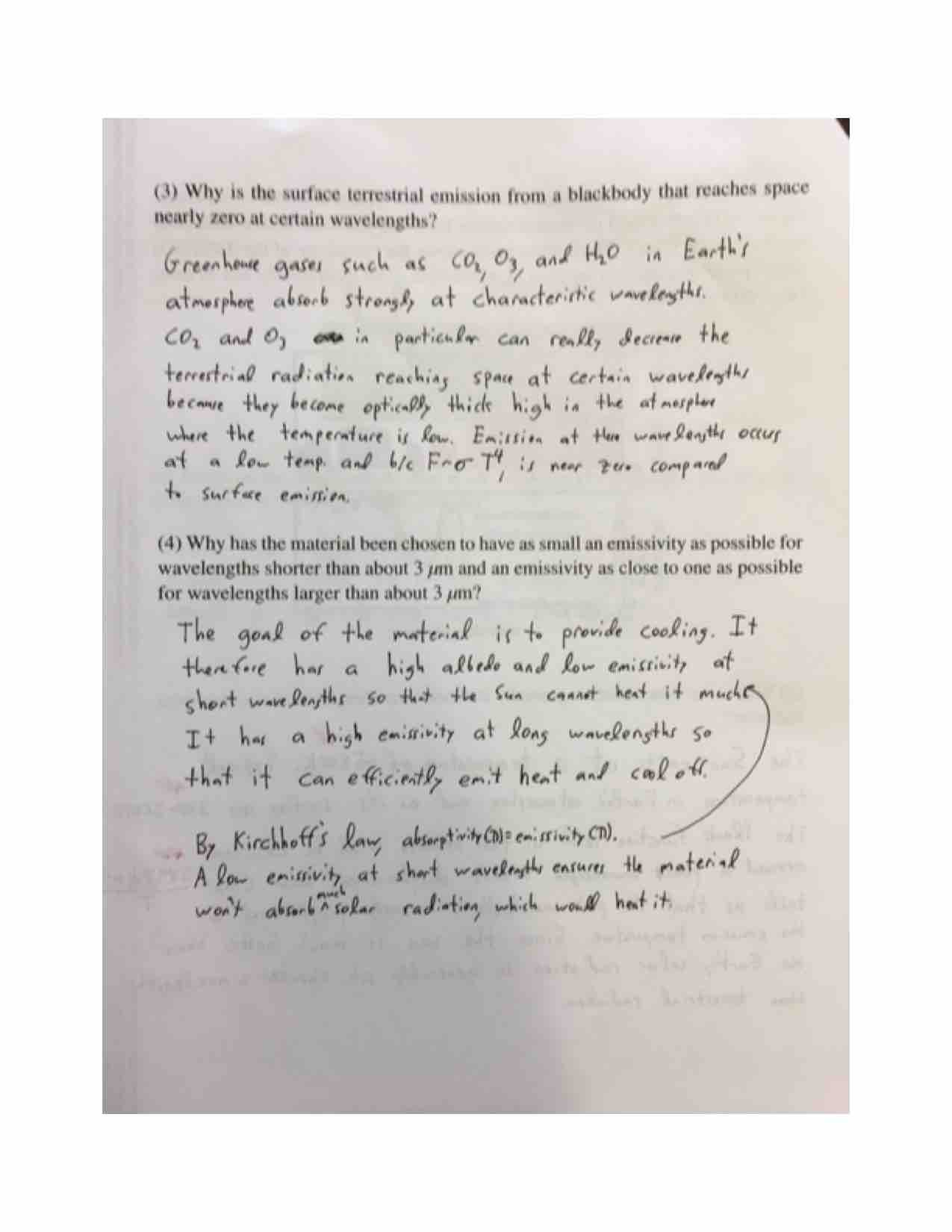}
\end{center}
\end{figure}

\begin{figure}[h!]
\begin{center}
  \includegraphics[width=\textwidth]{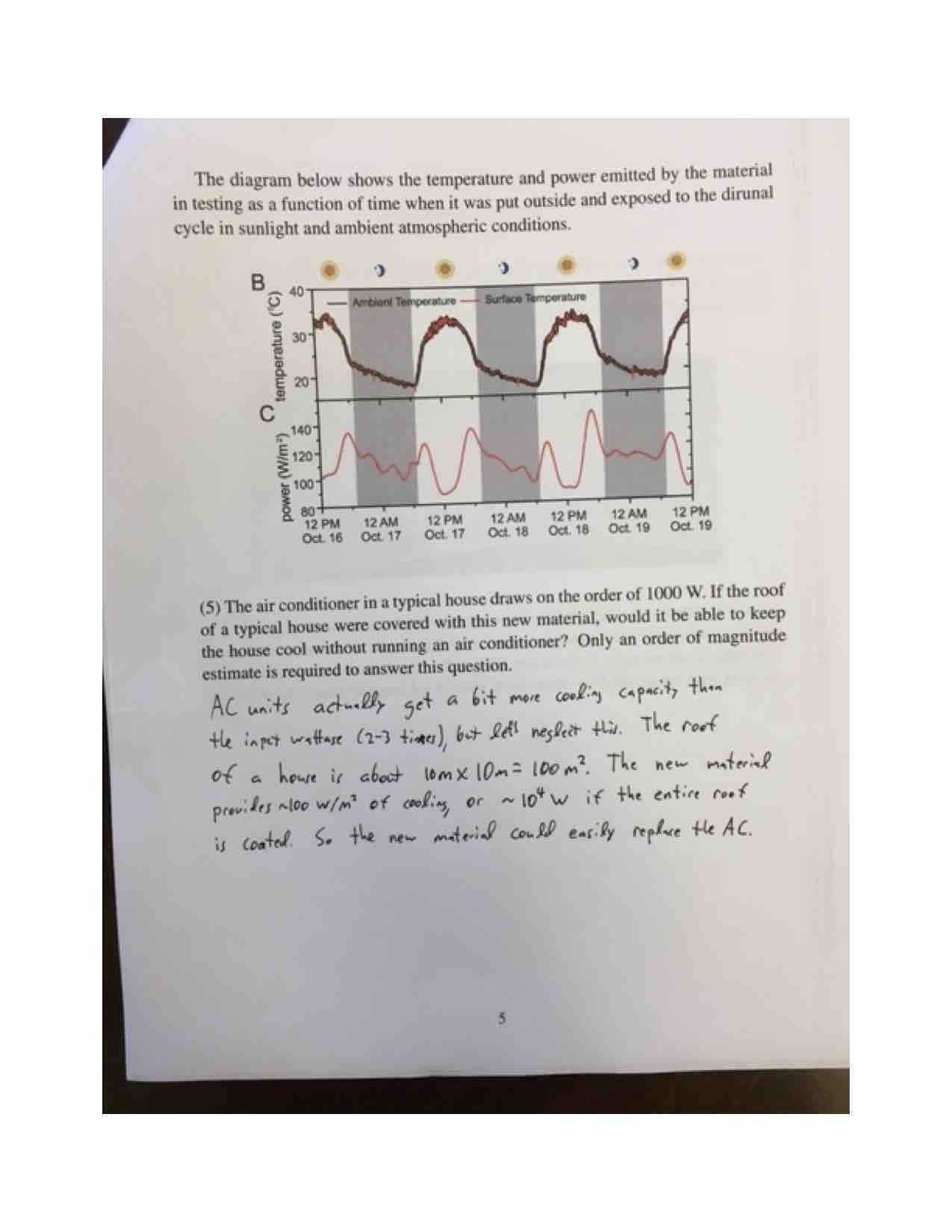}
\end{center}
\end{figure}

\clearpage

\input{Midterms/midterm2_2018.tex}

\clearpage

\begin{figure}[h!]
\begin{center}
  \includegraphics[width=\textwidth]{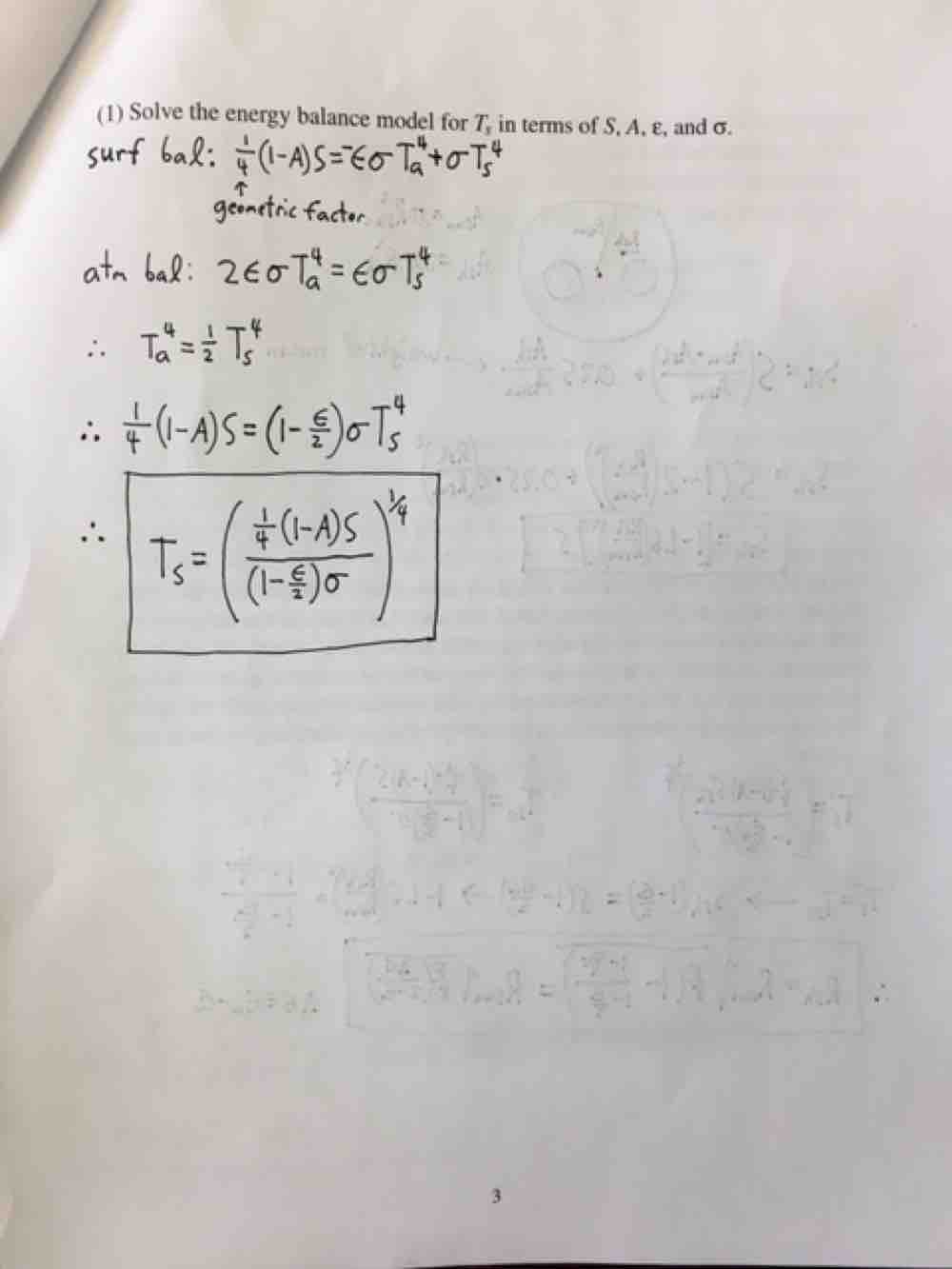}
\end{center}
\end{figure}

\begin{figure}[h!]
\begin{center}
  \includegraphics[width=\textwidth]{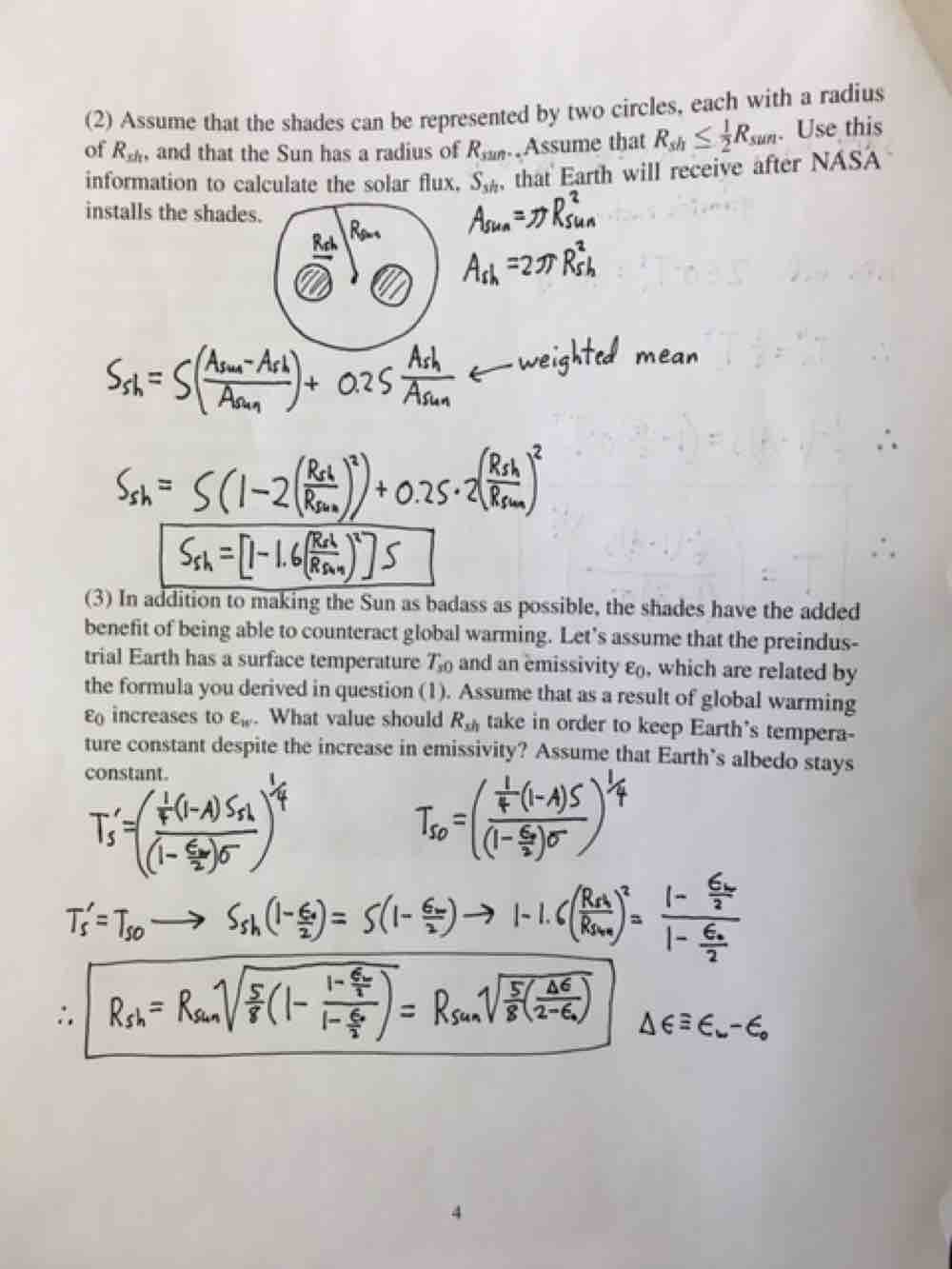}
\end{center}
\end{figure}

\begin{figure}[h!]
\begin{center}
  \includegraphics[width=\textwidth]{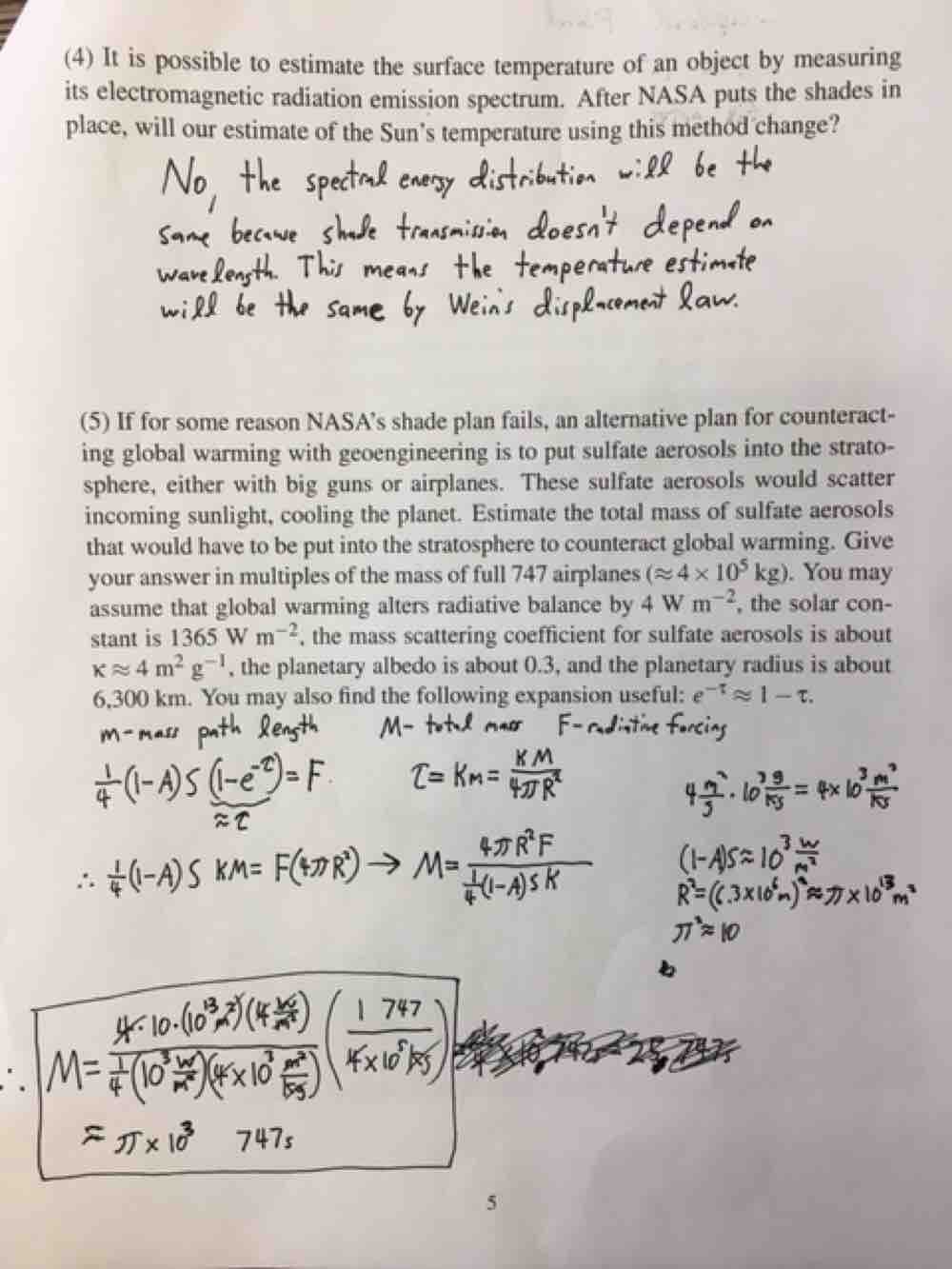}
\end{center}
\end{figure}

\clearpage

\input{Midterms/midterm2_2019.tex}

\clearpage

\begin{figure}[h!]
\begin{center}
  \includegraphics[width=\textwidth]{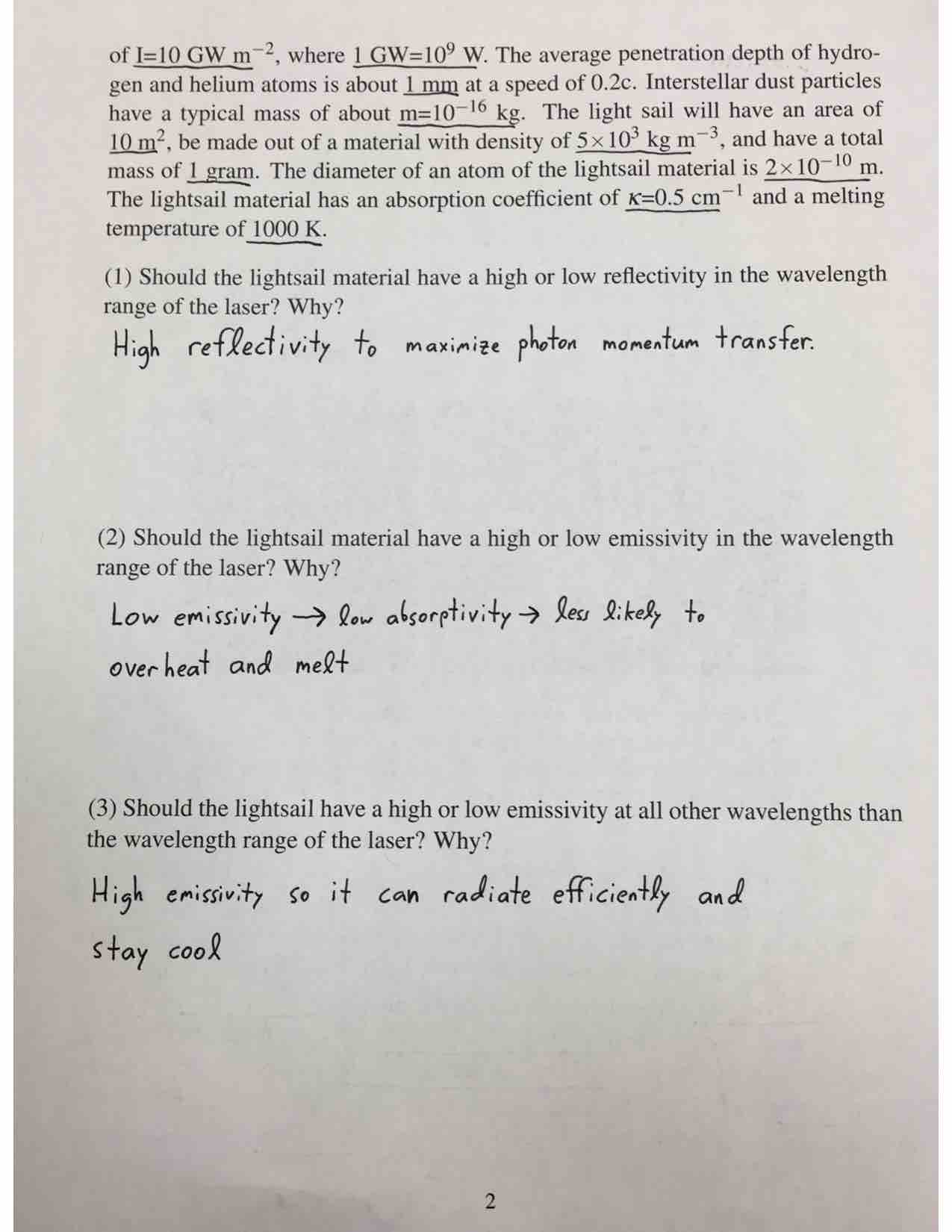}
\end{center}
\end{figure}

\begin{figure}[h!]
\begin{center}
  \includegraphics[width=\textwidth]{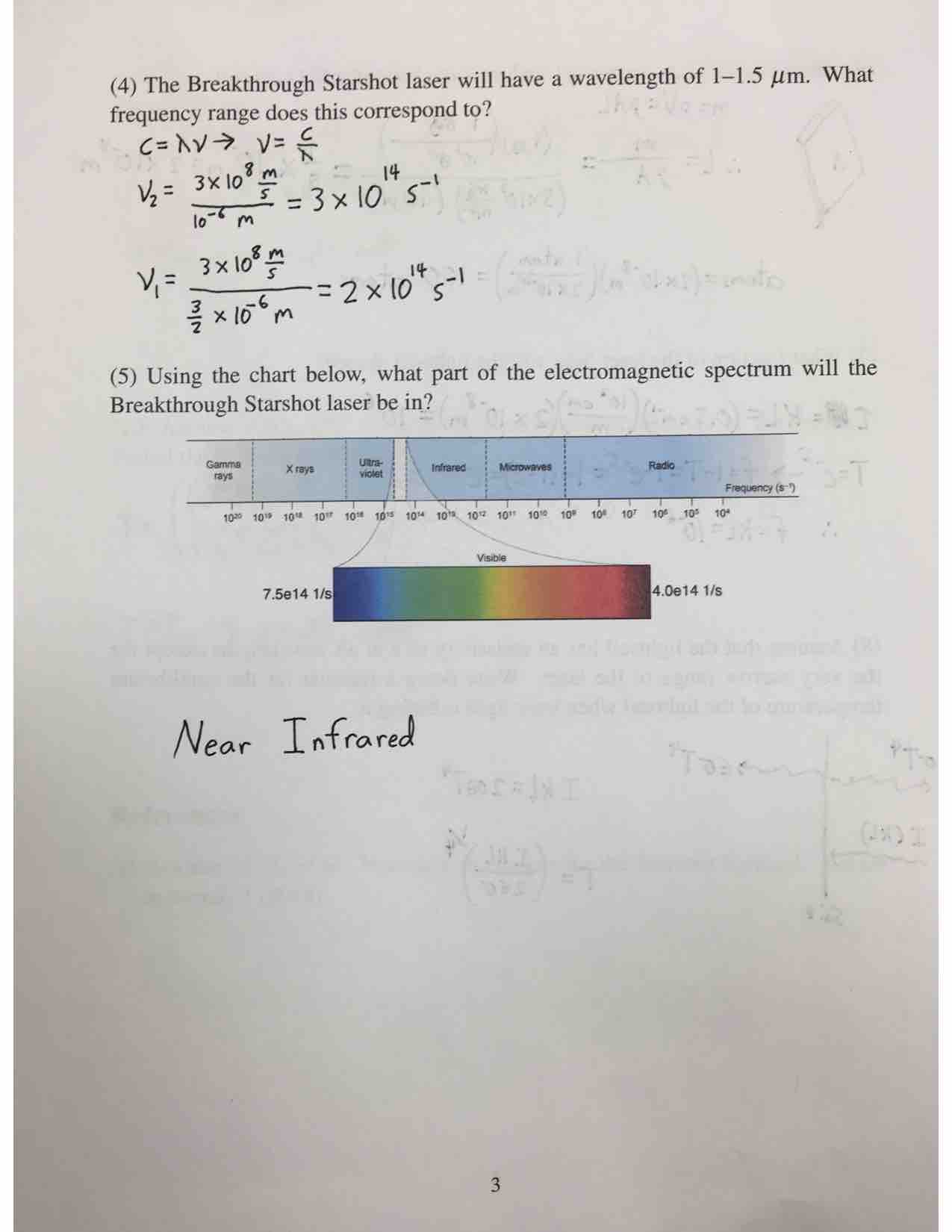}
\end{center}
\end{figure}

\begin{figure}[h!]
\begin{center}
  \includegraphics[width=\textwidth]{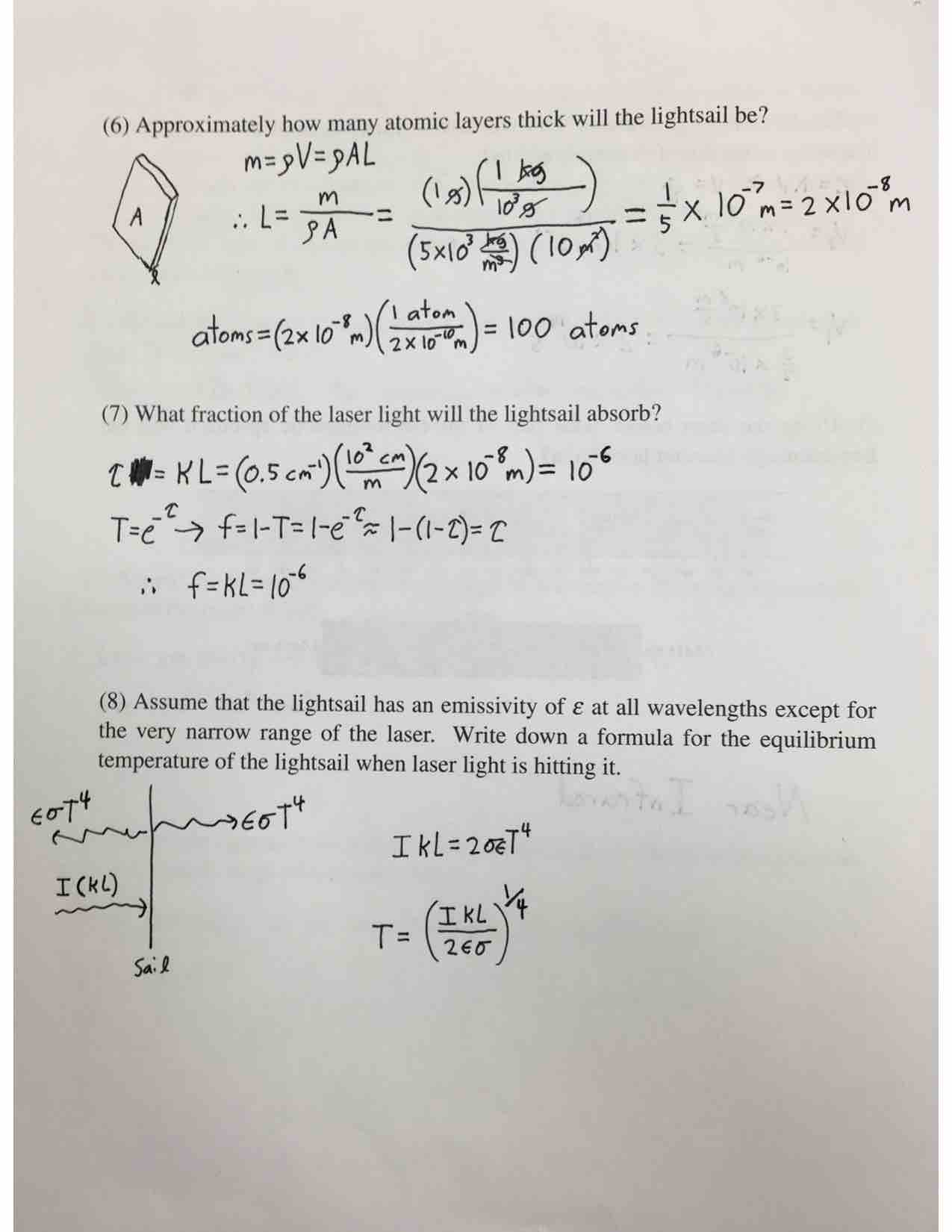}
\end{center}
\end{figure}

\begin{figure}[h!]
\begin{center}
  \includegraphics[width=\textwidth]{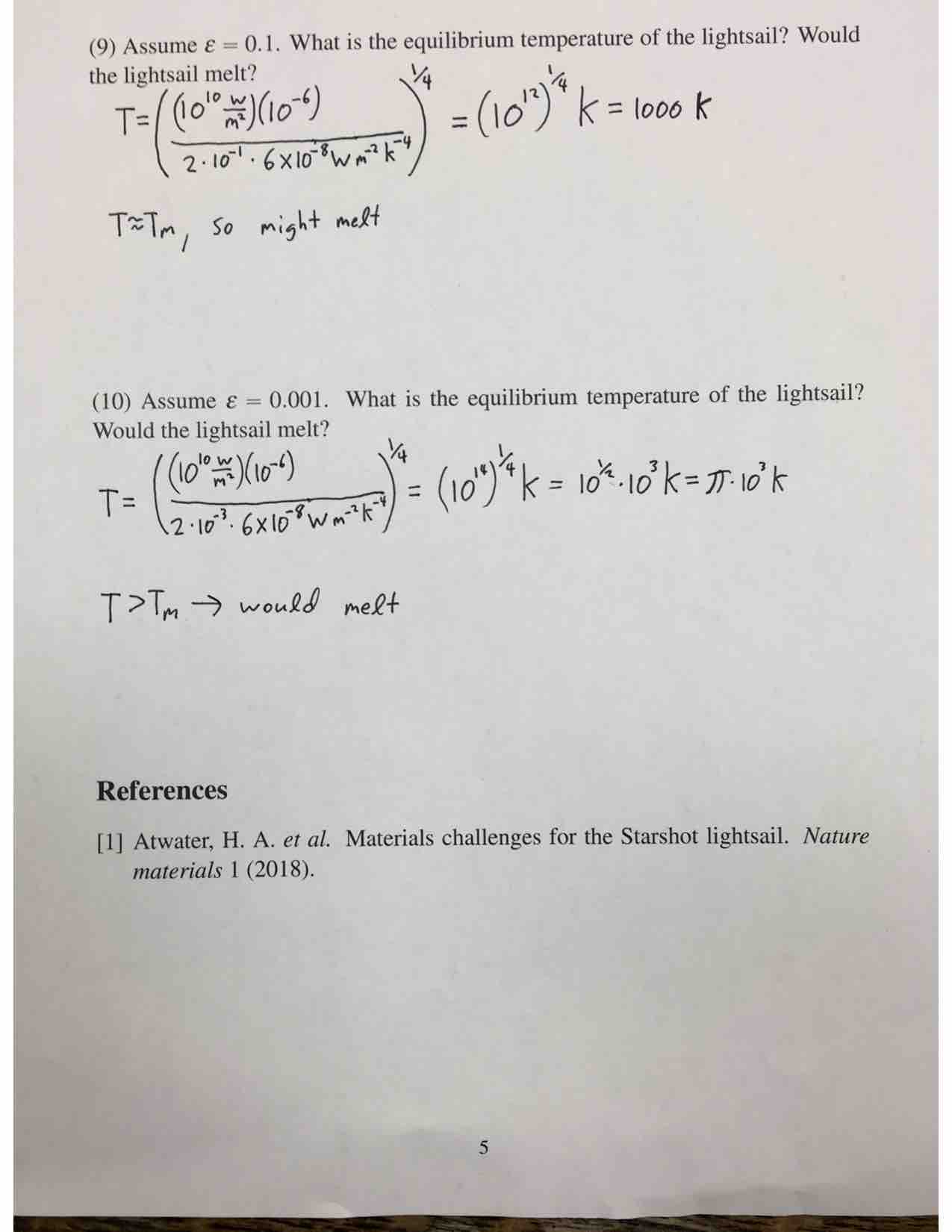}
\end{center}
\end{figure}

\clearpage

\section{Finals}

\clearpage

\input{Finals/final_2017.tex}

\clearpage

\begin{figure}[h!]
\begin{center}
  \includegraphics[width=\textwidth]{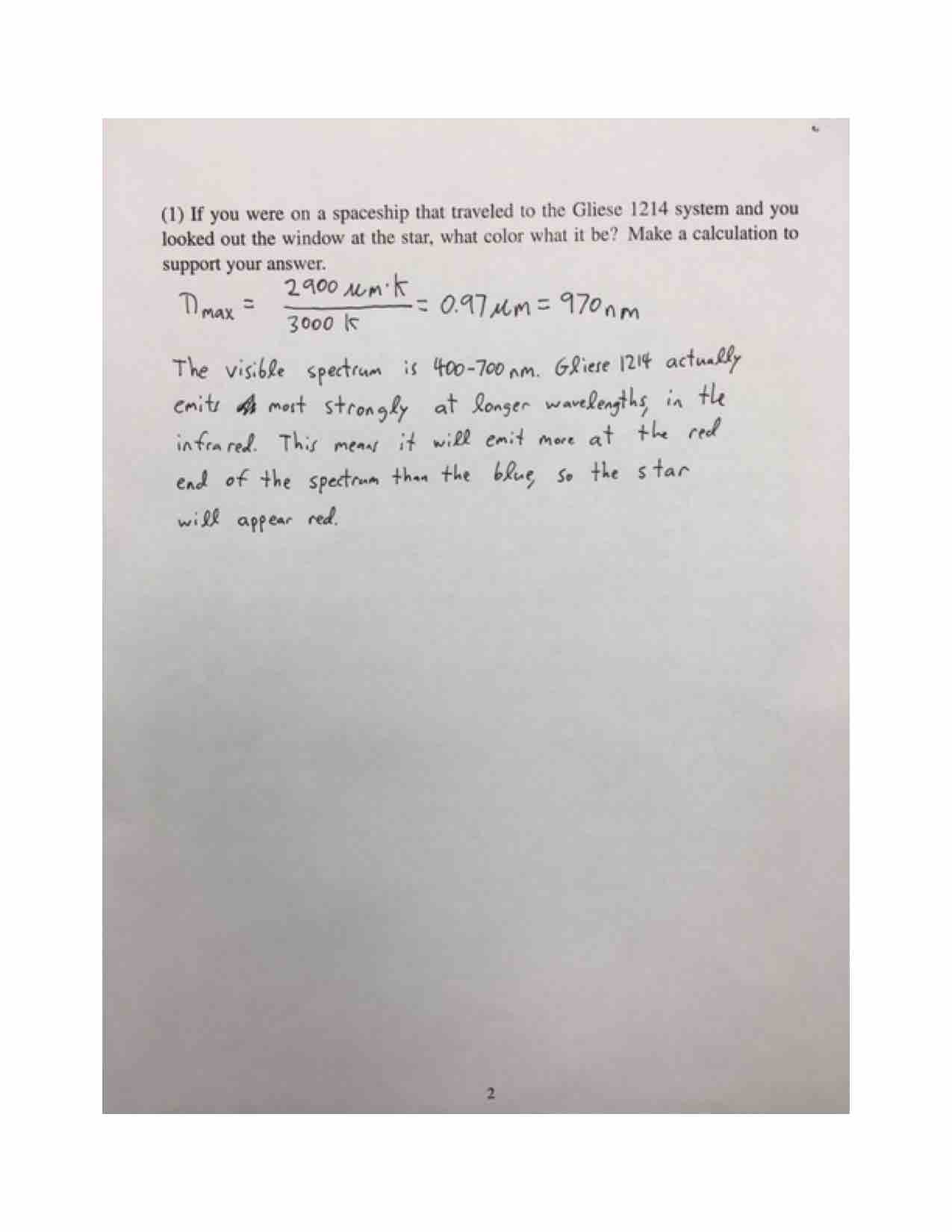}
\end{center}
\end{figure}

\begin{figure}[h!]
\begin{center}
  \includegraphics[width=\textwidth]{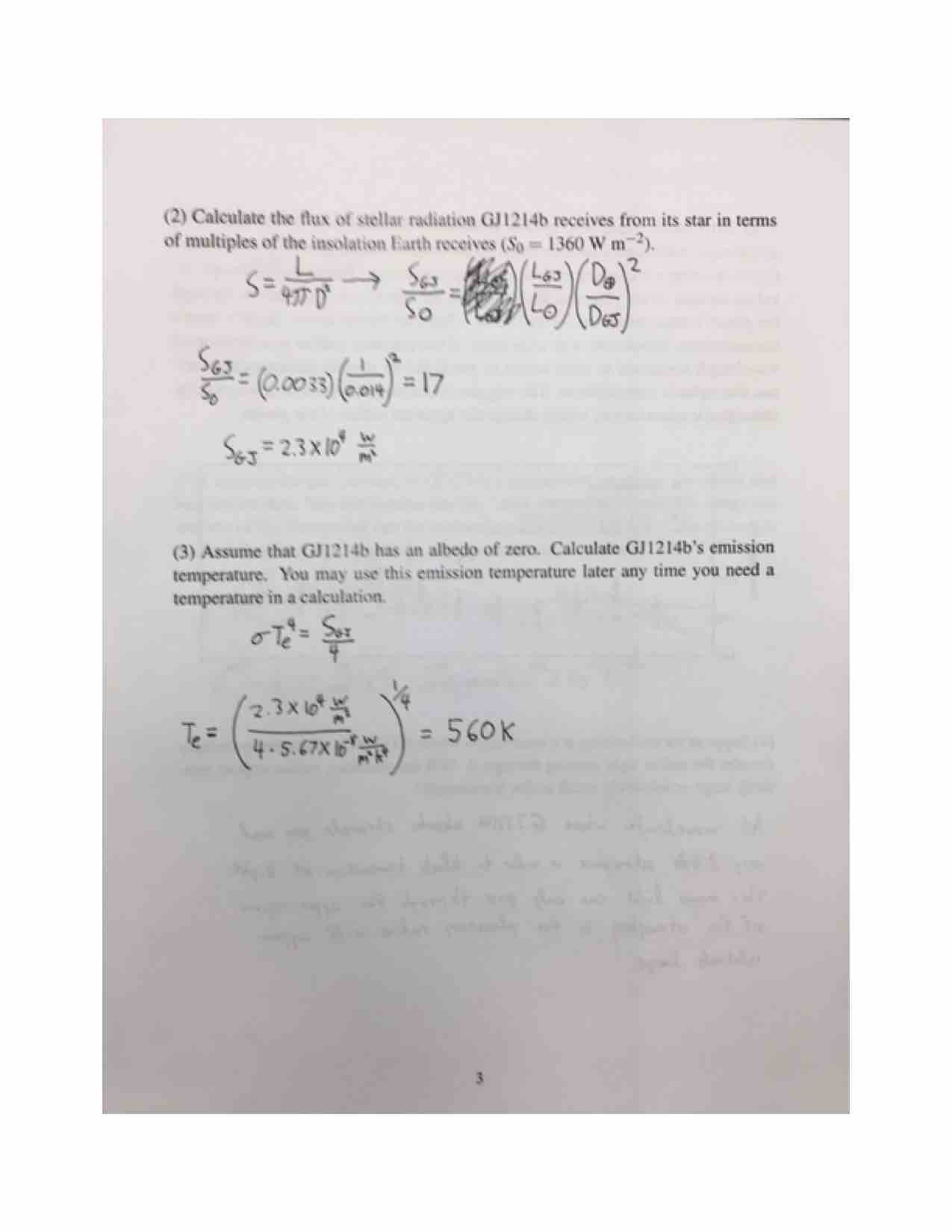}
\end{center}
\end{figure}

\begin{figure}[h!]
\begin{center}
  \includegraphics[width=\textwidth]{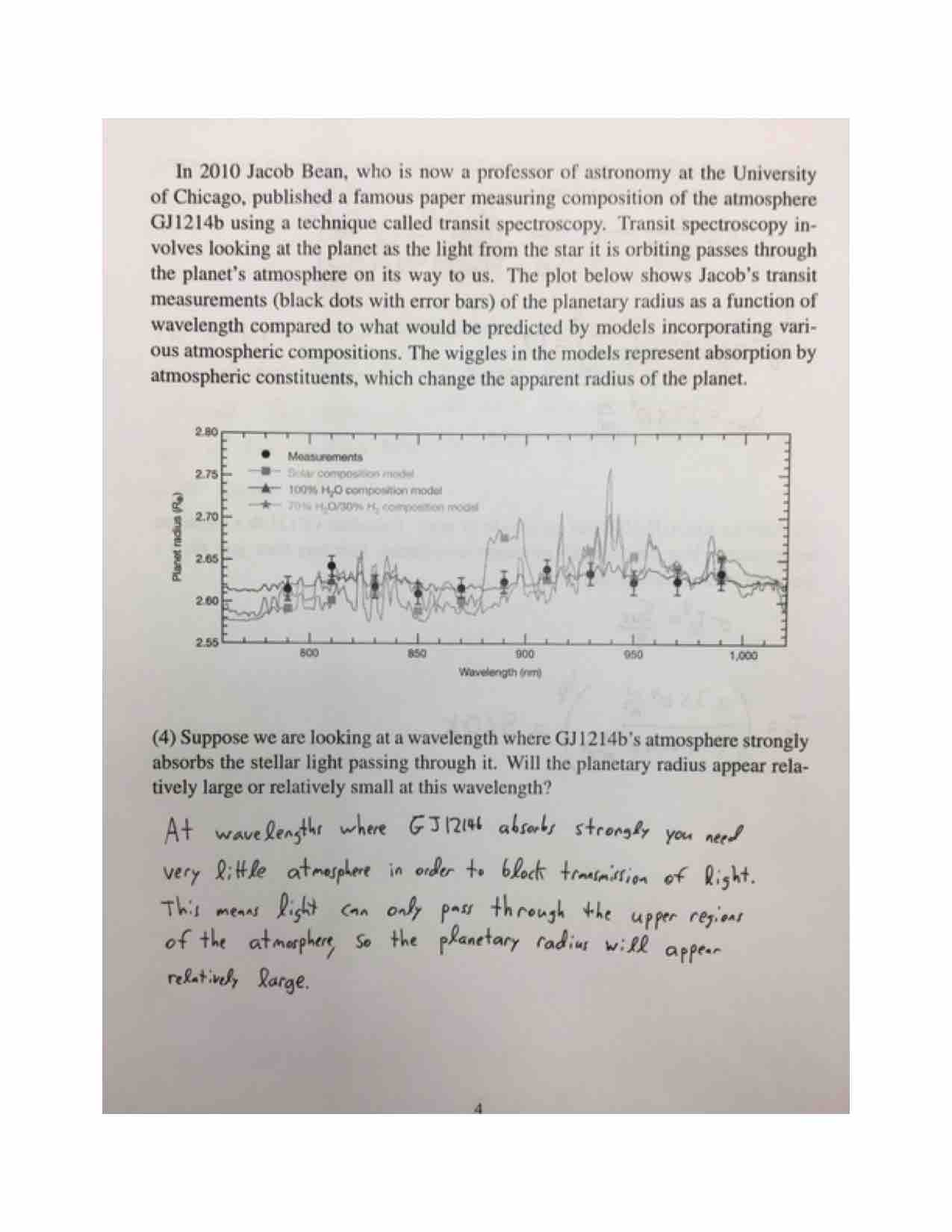}
\end{center}
\end{figure}

\begin{figure}[h!]
\begin{center}
  \includegraphics[width=\textwidth]{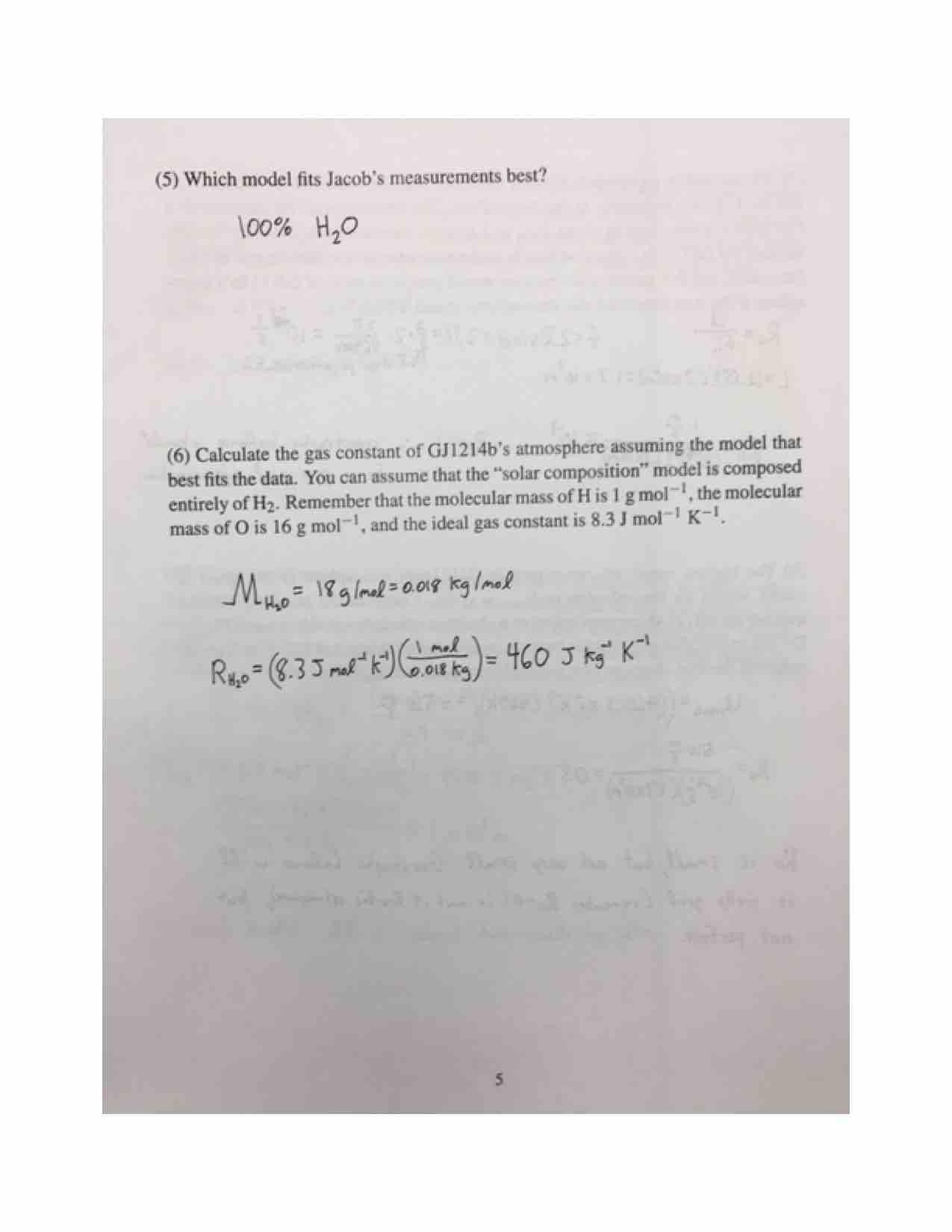}
\end{center}
\end{figure}

\begin{figure}[h!]
\begin{center}
  \includegraphics[width=\textwidth]{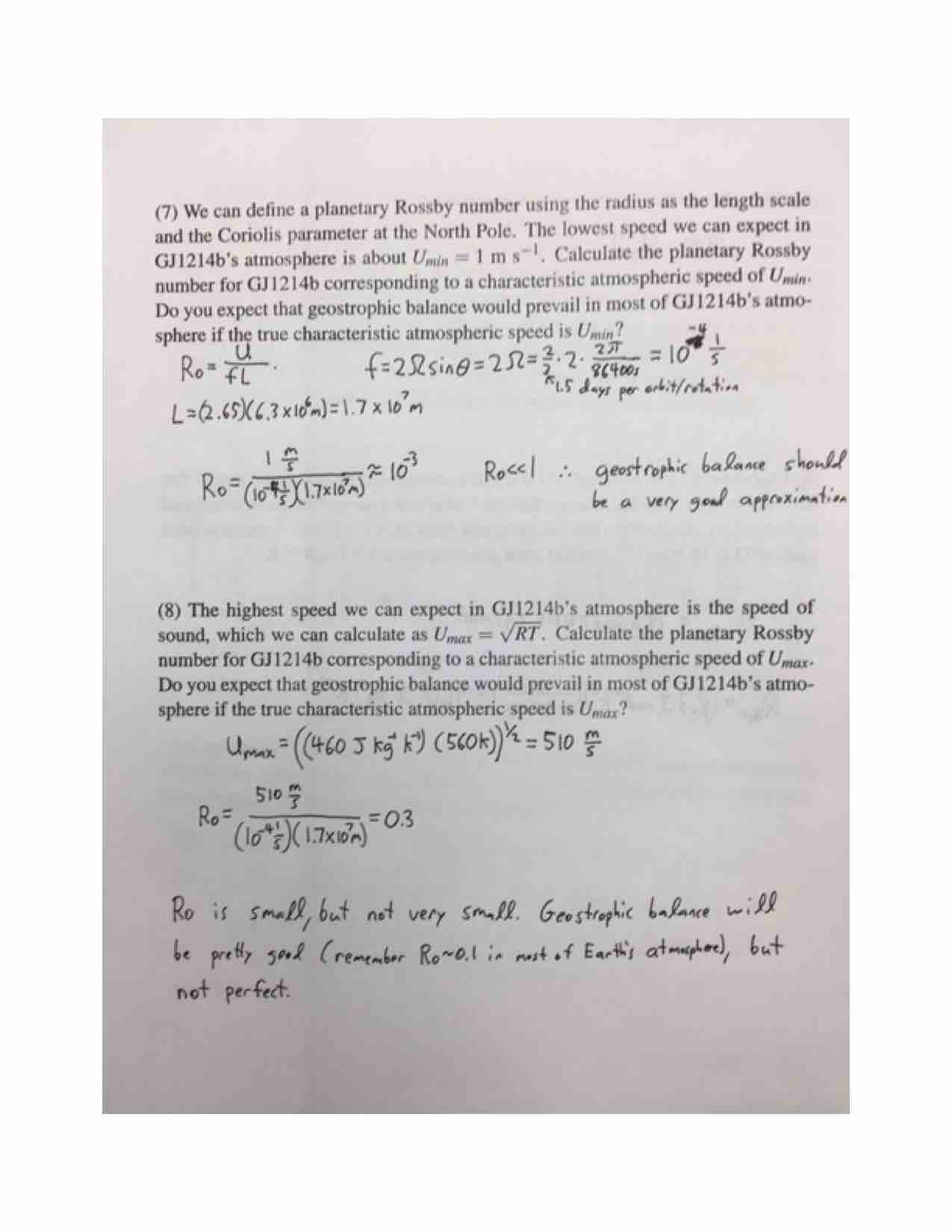}
\end{center}
\end{figure}

\begin{figure}[h!]
\begin{center}
  \includegraphics[width=\textwidth]{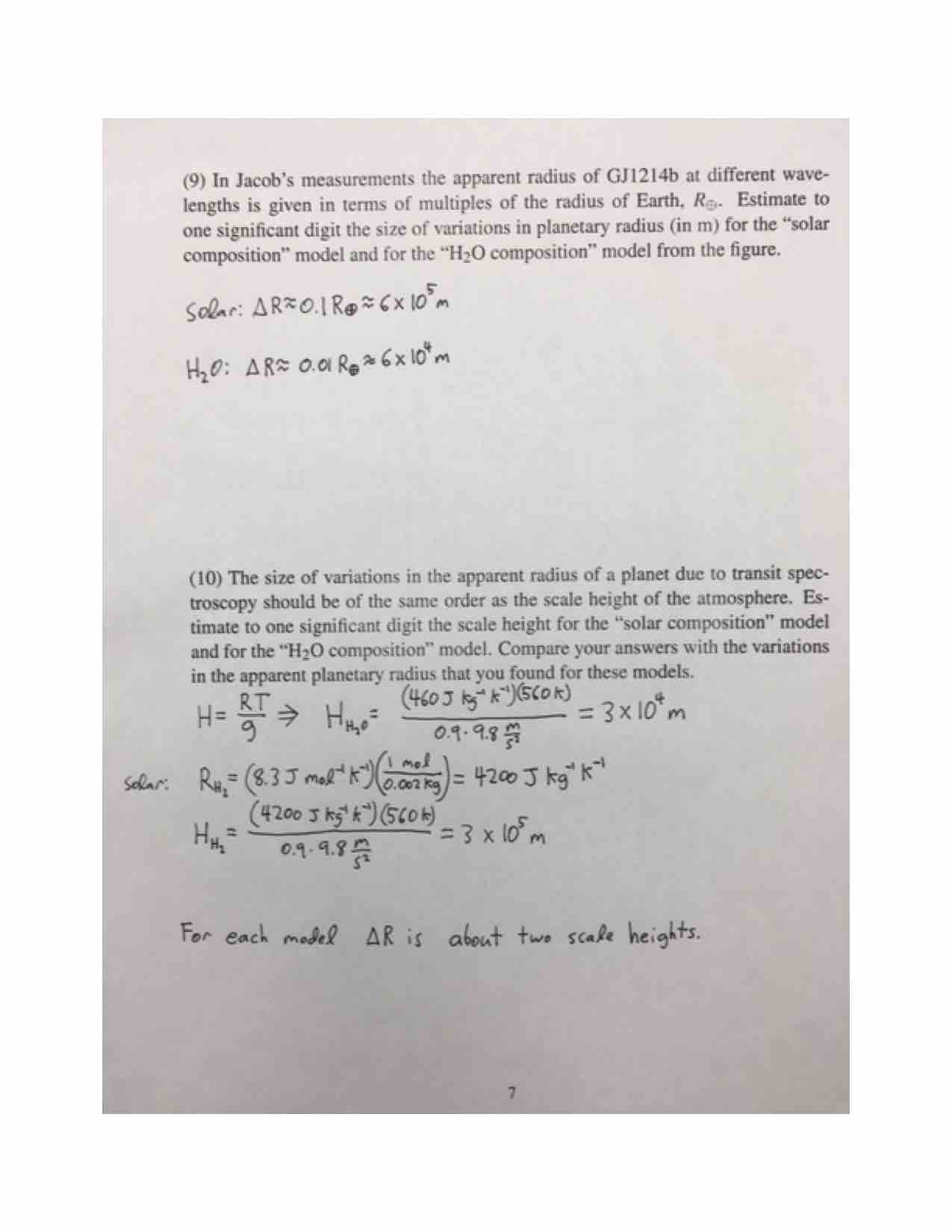}
\end{center}
\end{figure}

\clearpage

\input{Finals/final_2018.tex}

\clearpage

\begin{figure}[h!]
\begin{center}
  \includegraphics[width=\textwidth]{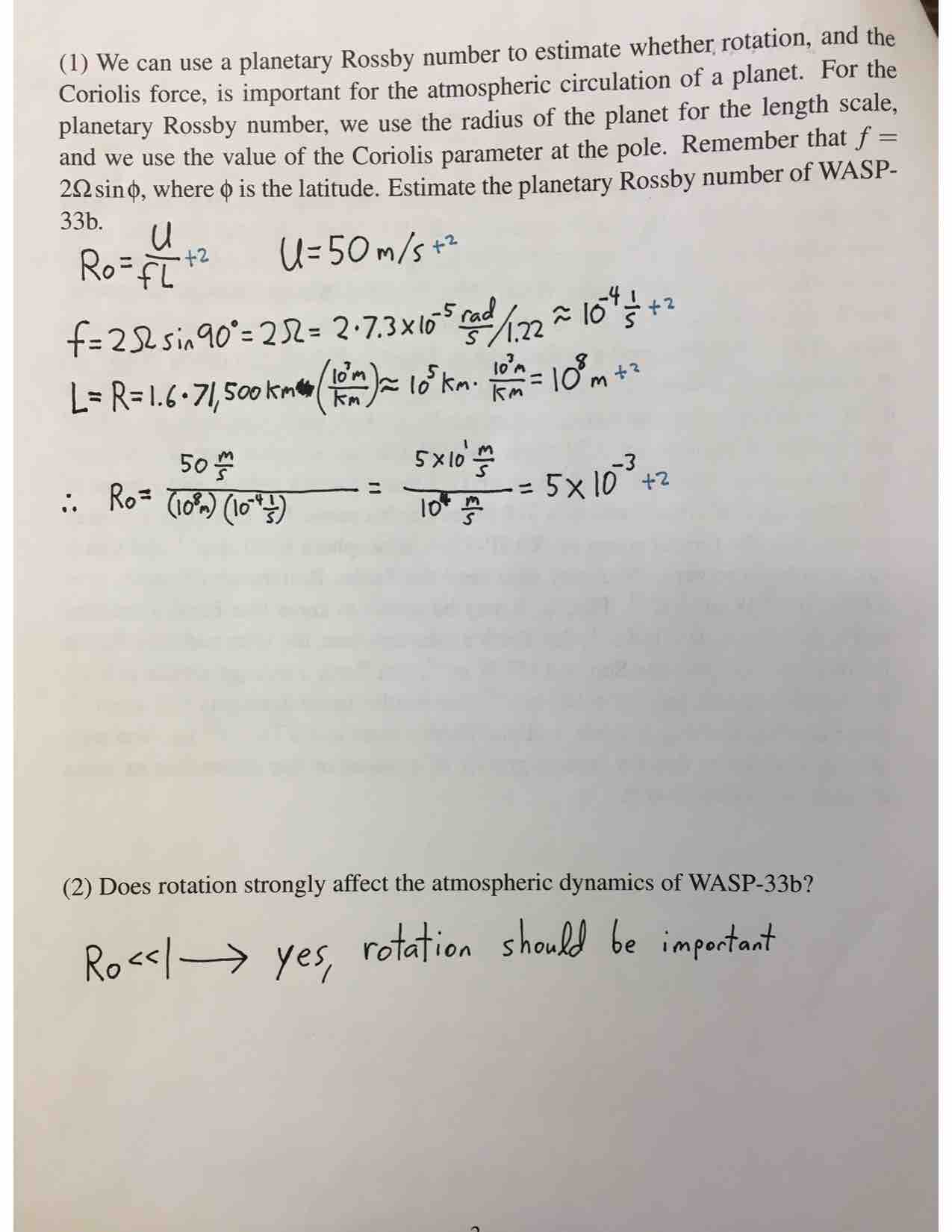}
\end{center}
\end{figure}

\begin{figure}[h!]
\begin{center}
  \includegraphics[width=\textwidth]{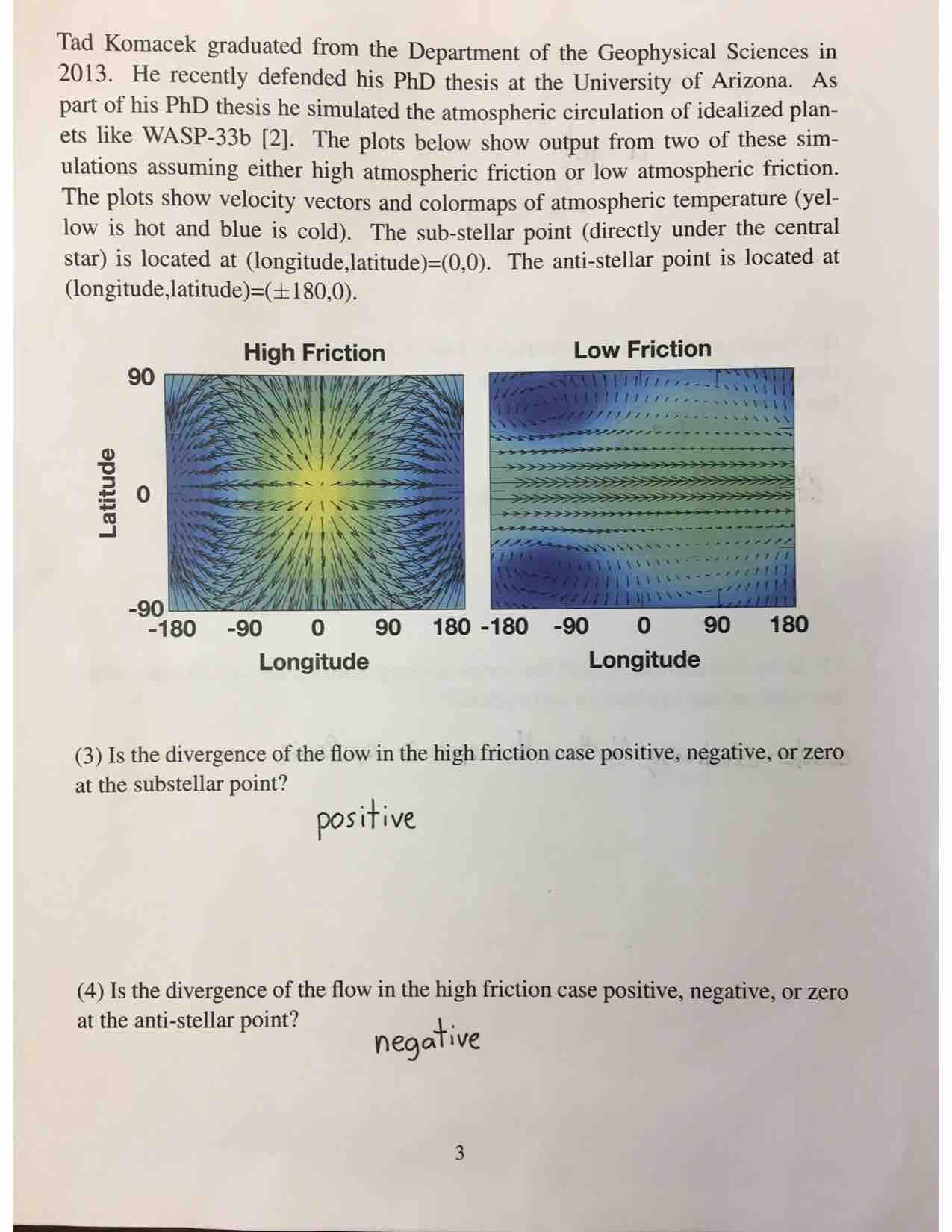}
\end{center}
\end{figure}

\begin{figure}[h!]
\begin{center}
  \includegraphics[width=\textwidth]{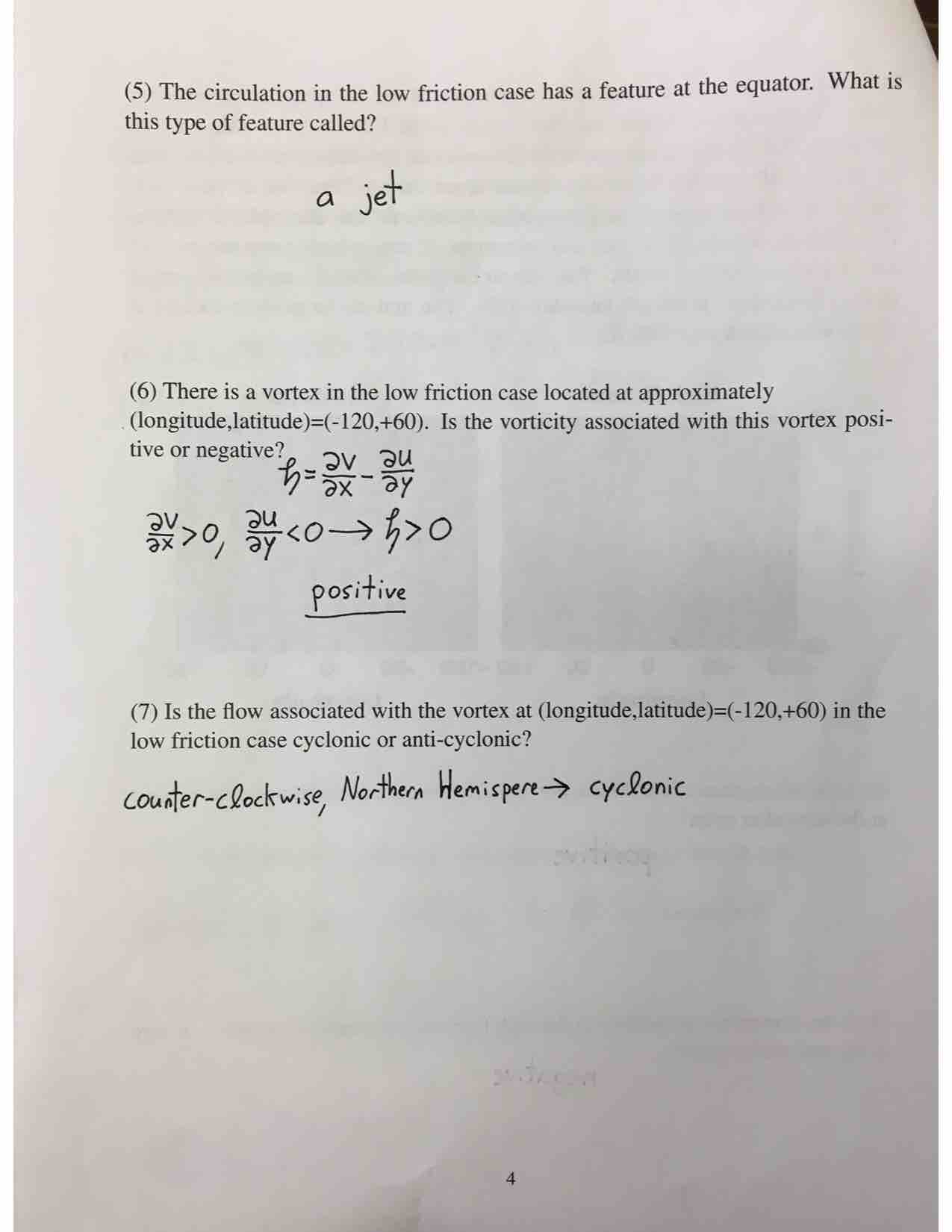}
\end{center}
\end{figure}

\begin{figure}[h!]
\begin{center}
  \includegraphics[width=\textwidth]{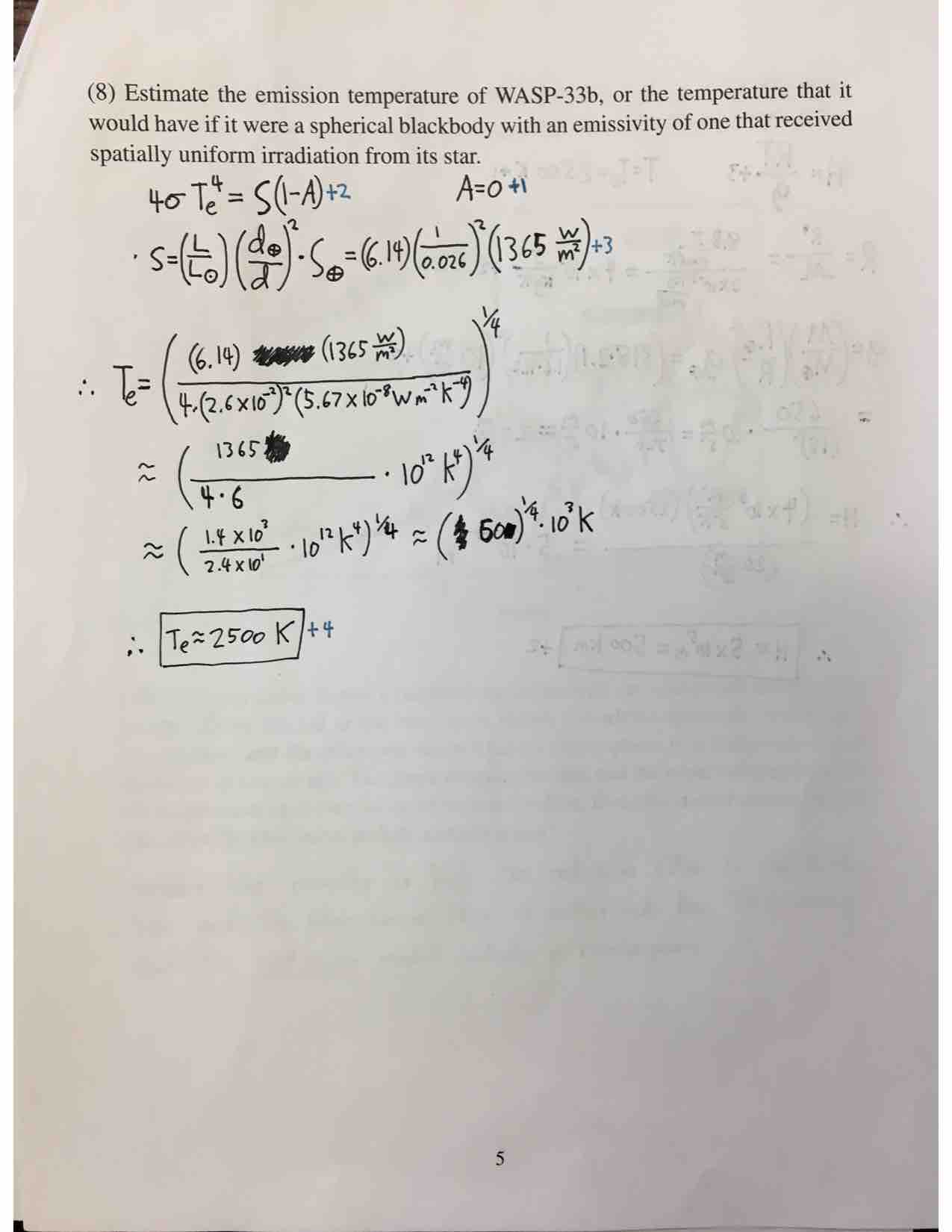}
\end{center}
\end{figure}

\begin{figure}[h!]
\begin{center}
  \includegraphics[width=\textwidth]{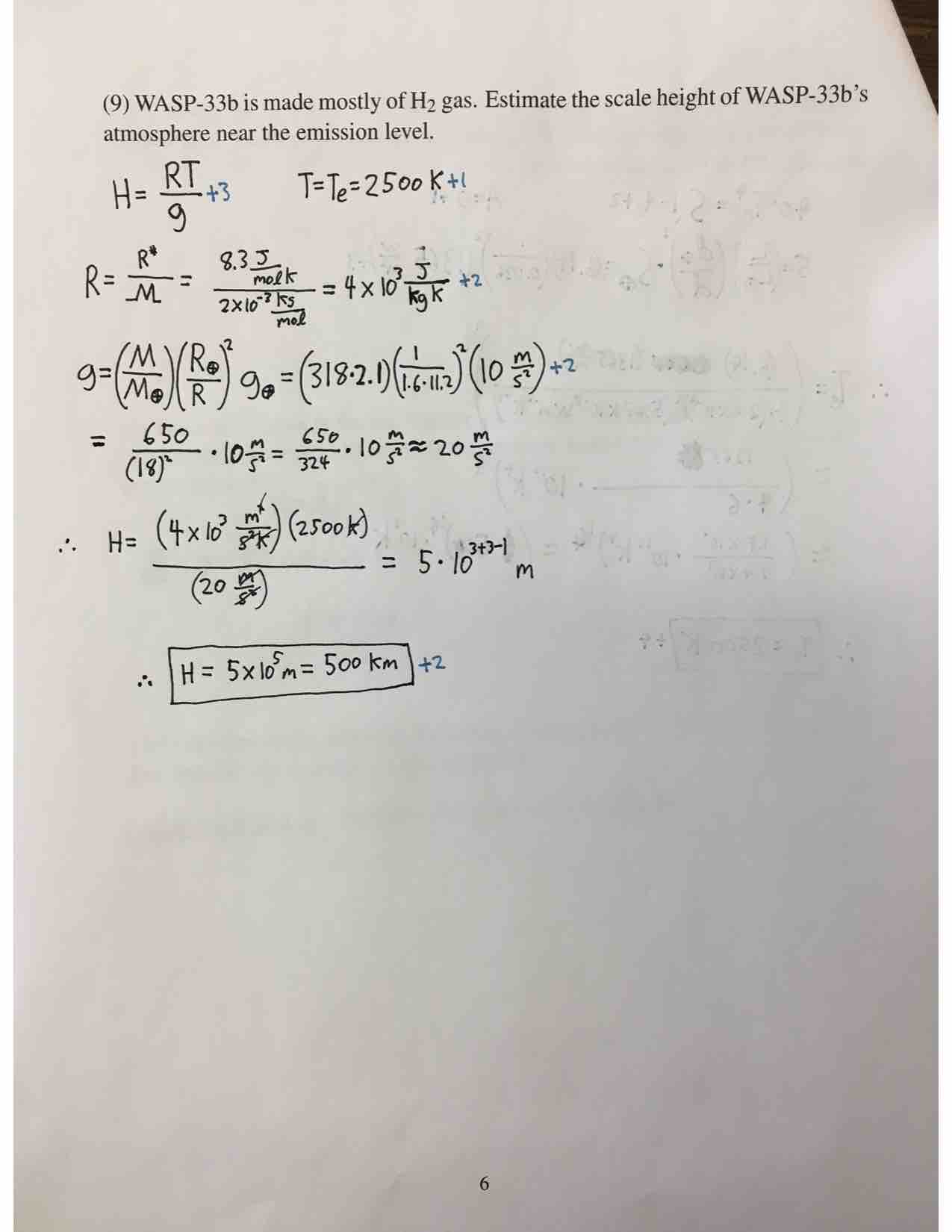}
\end{center}
\end{figure}

\begin{figure}[h!]
\begin{center}
  \includegraphics[width=\textwidth]{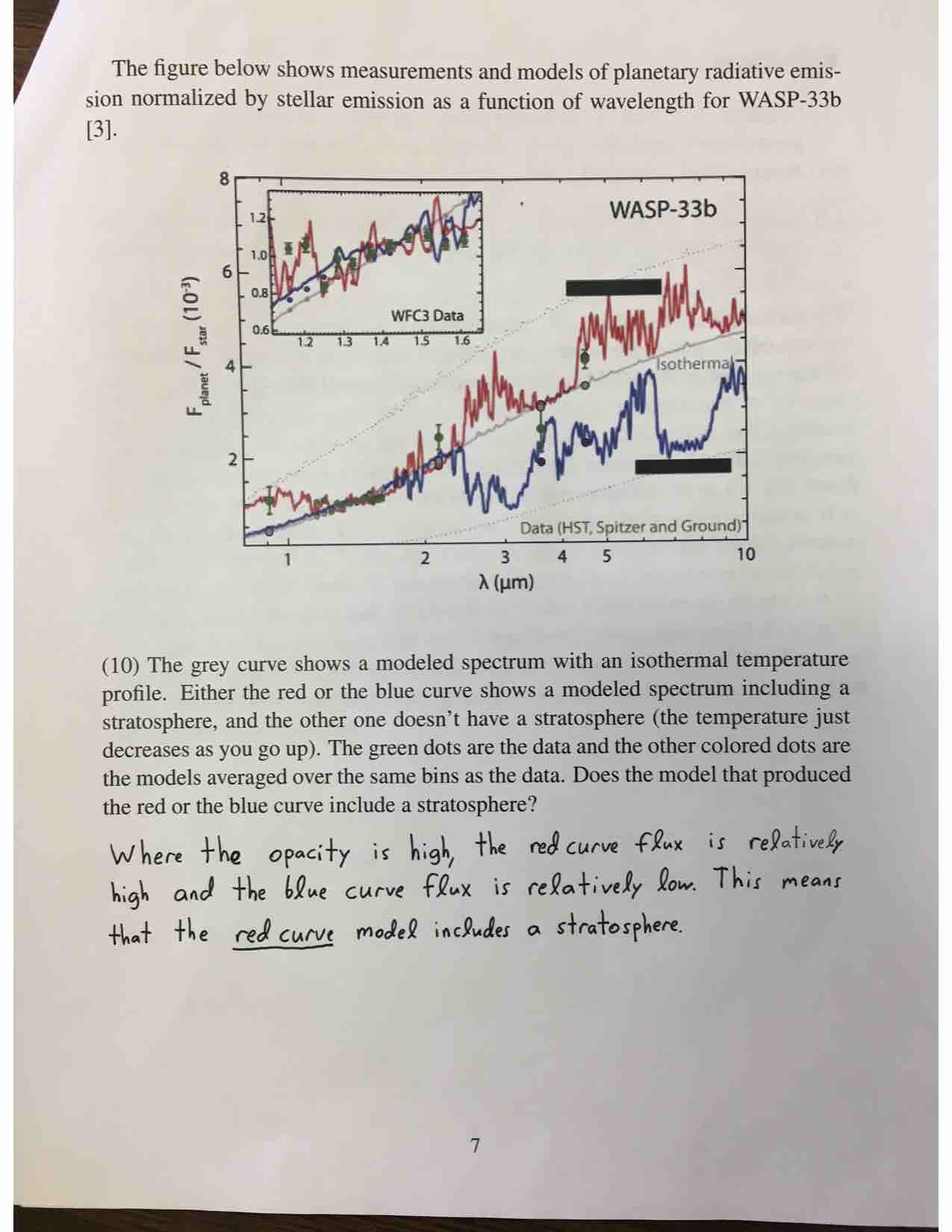}
\end{center}
\end{figure}

\clearpage

\input{Finals/final_2019.tex}

\clearpage

\begin{figure}[h!]
\begin{center}
  \includegraphics[width=\textwidth]{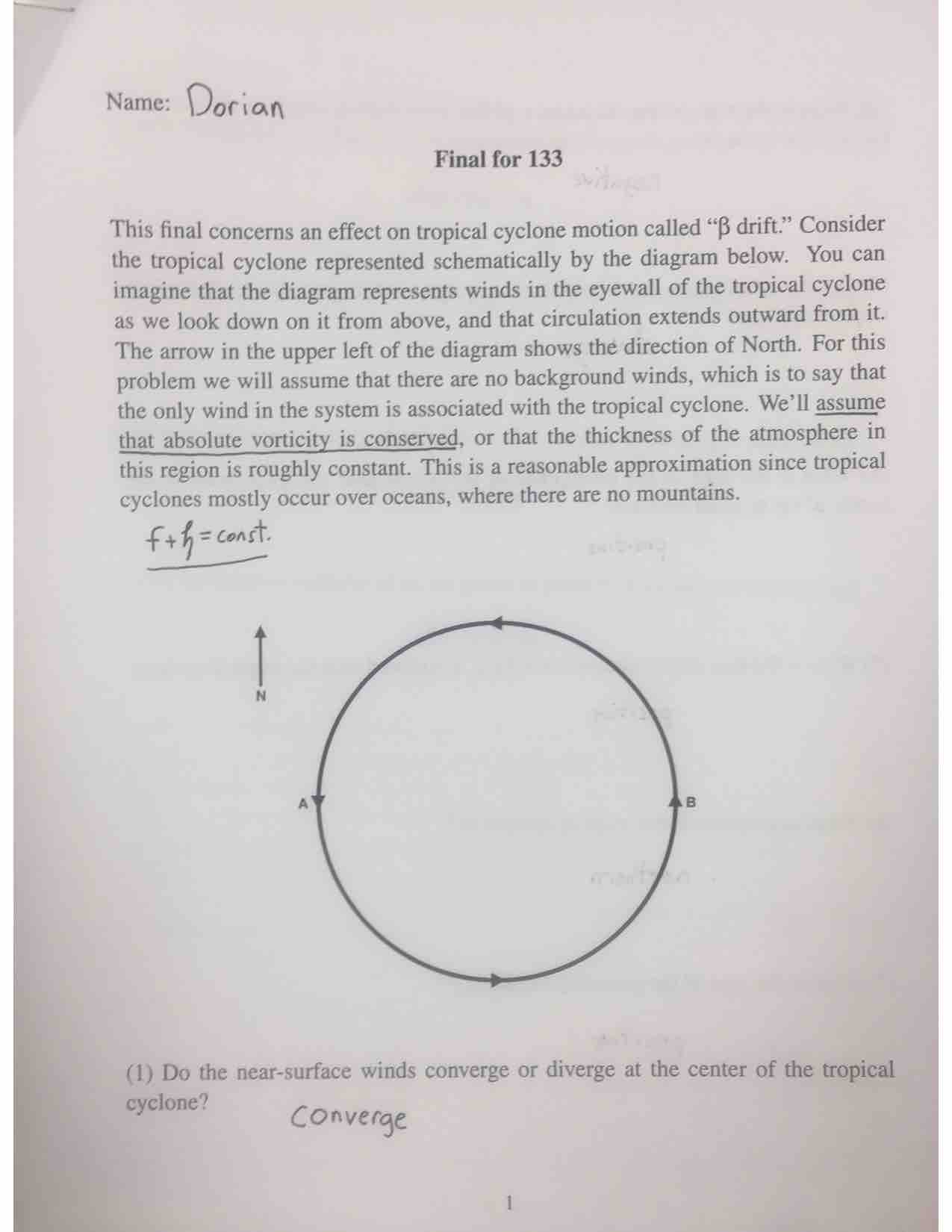}
\end{center}
\end{figure}

\begin{figure}[h!]
\begin{center}
  \includegraphics[width=\textwidth]{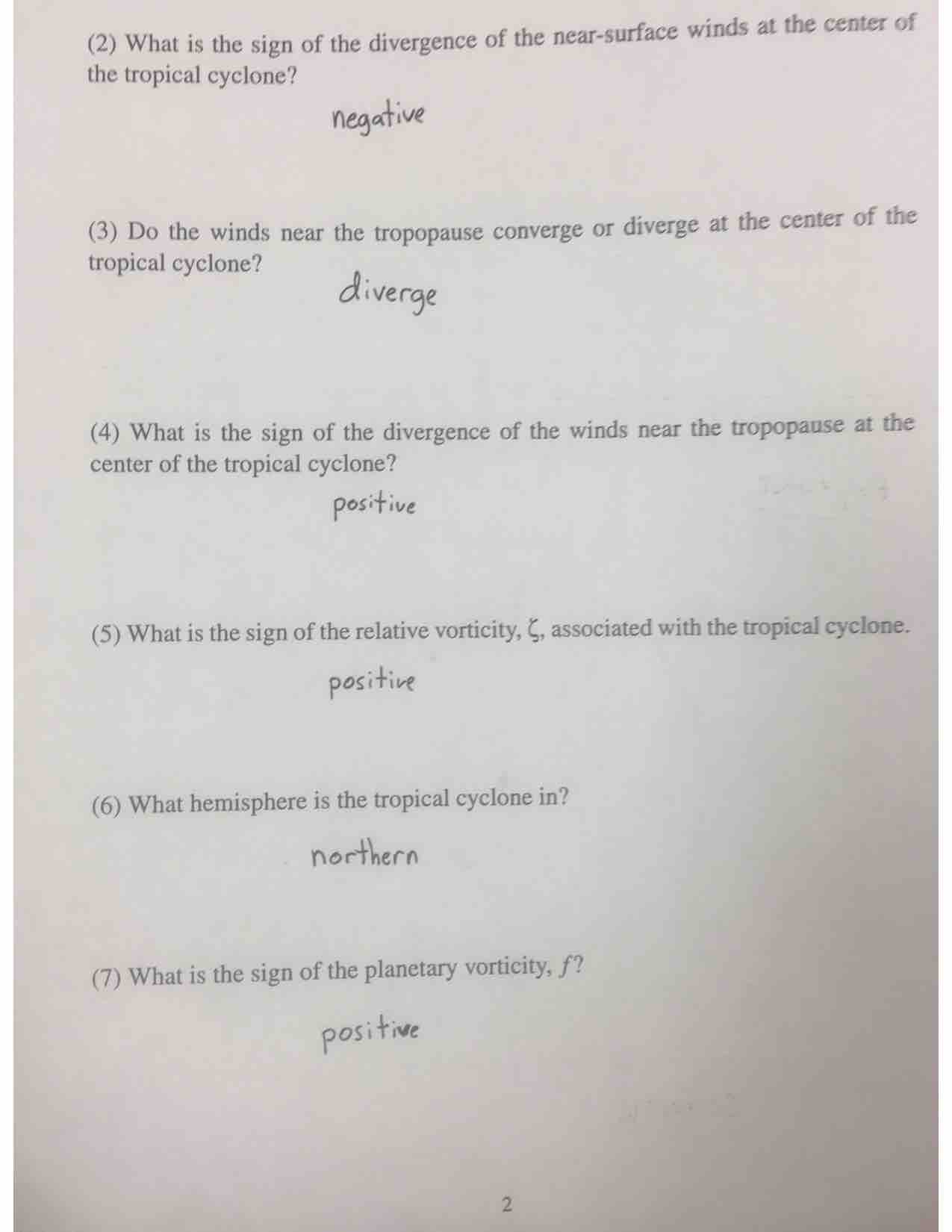}
\end{center}
\end{figure}

\begin{figure}[h!]
\begin{center}
  \includegraphics[width=\textwidth]{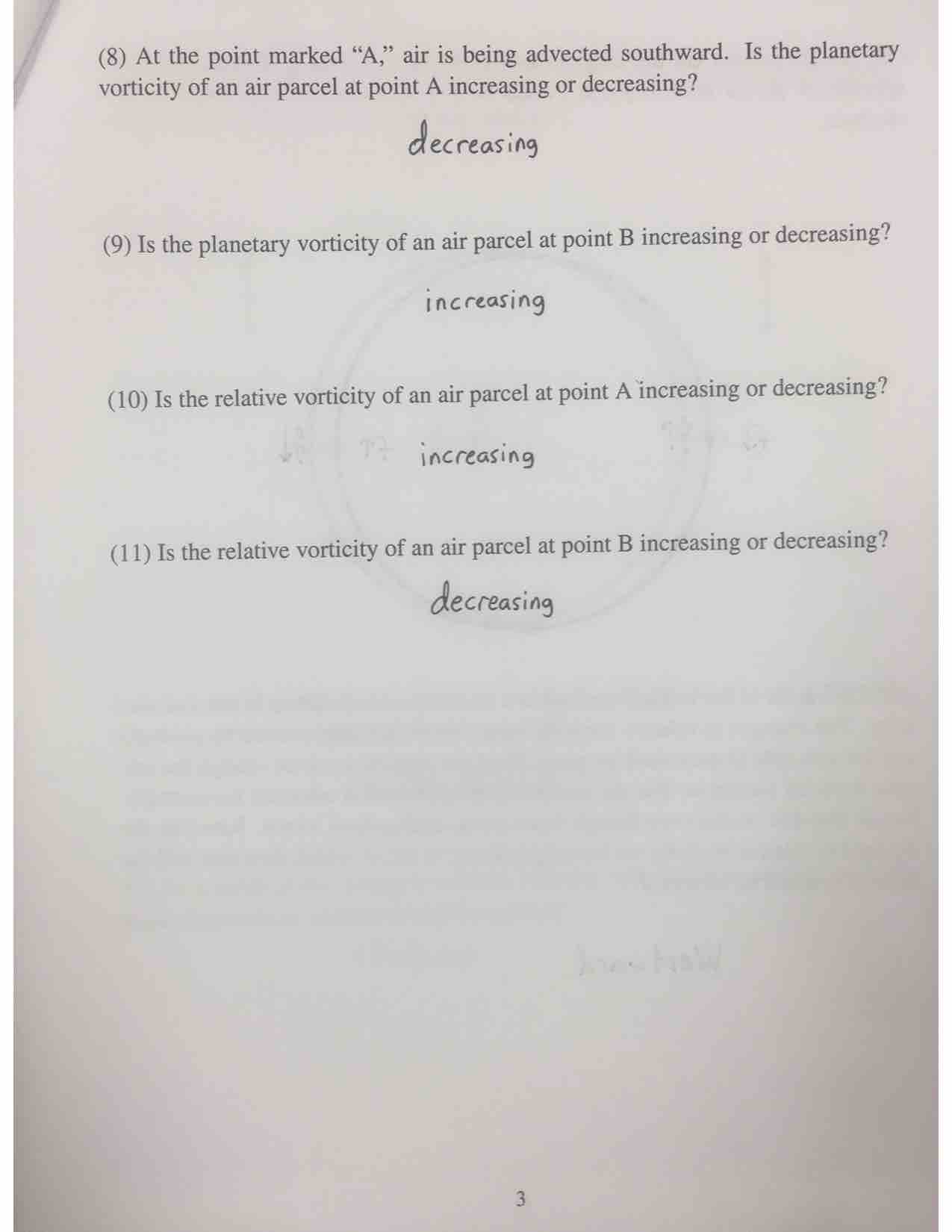}
\end{center}
\end{figure}

\begin{figure}[h!]
\begin{center}
  \includegraphics[width=\textwidth]{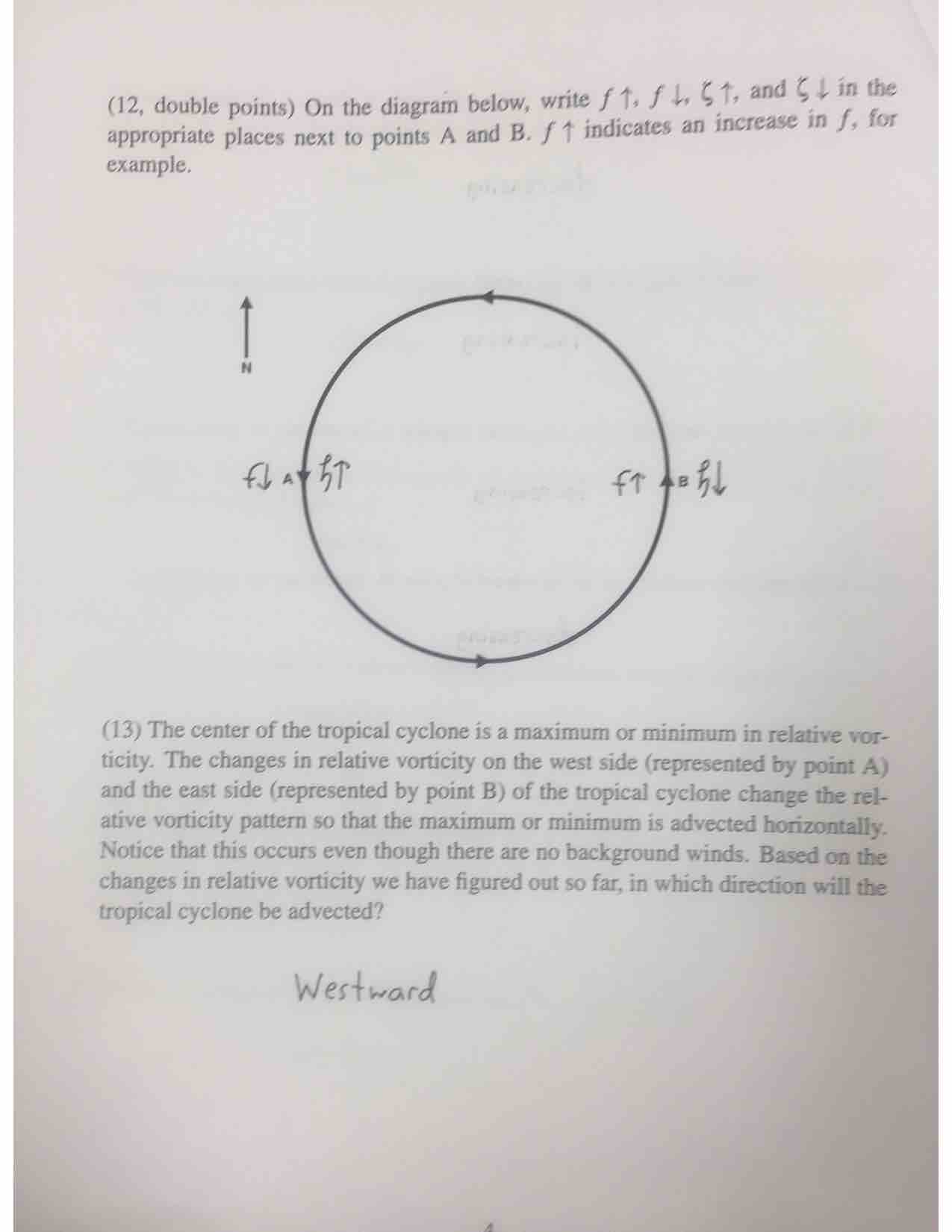}
\end{center}
\end{figure}

\begin{figure}[h!]
\begin{center}
  \includegraphics[width=\textwidth]{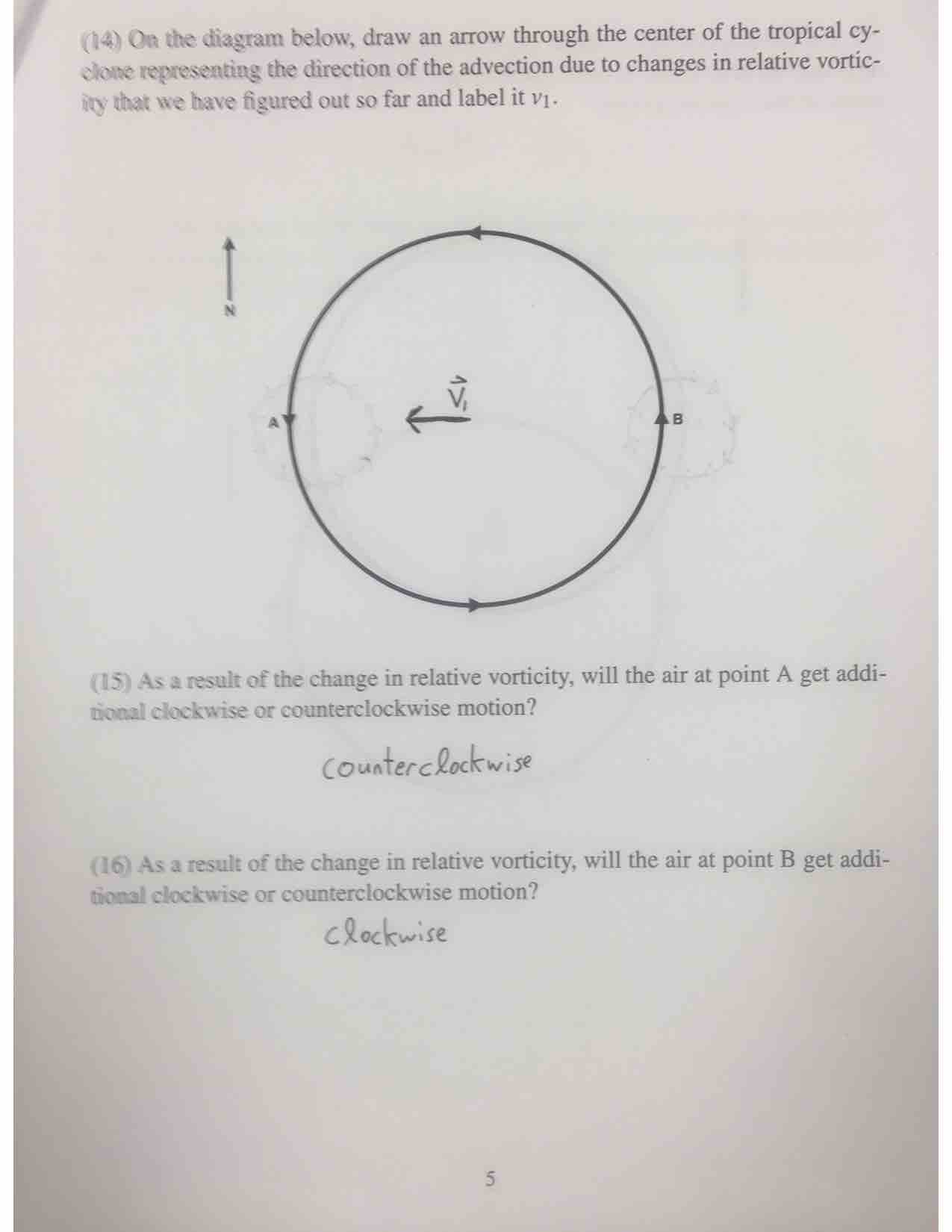}
\end{center}
\end{figure}

\begin{figure}[h!]
\begin{center}
  \includegraphics[width=\textwidth]{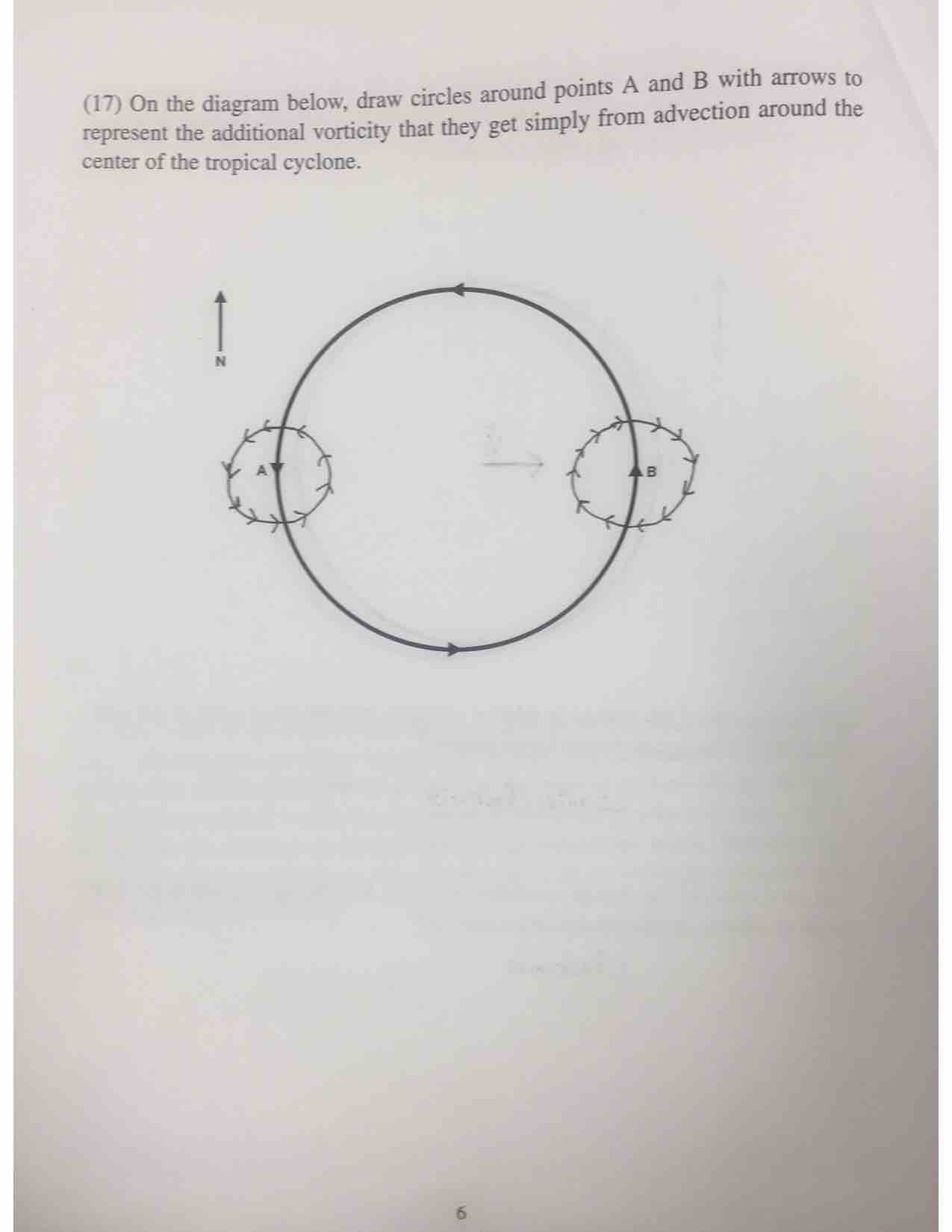}
\end{center}
\end{figure}

\begin{figure}[h!]
\begin{center}
  \includegraphics[width=\textwidth]{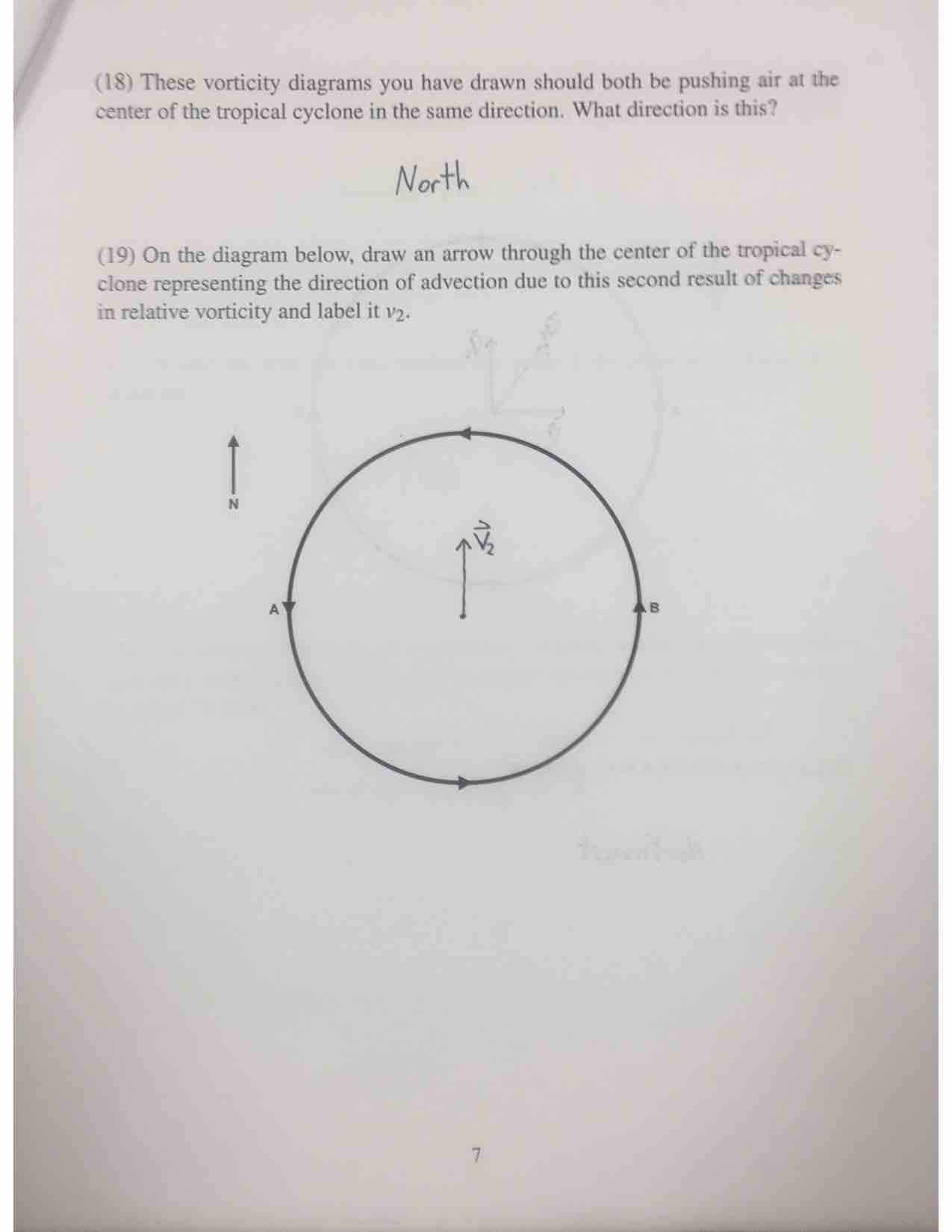}
\end{center}
\end{figure}

\begin{figure}[h!]
\begin{center}
  \includegraphics[width=\textwidth]{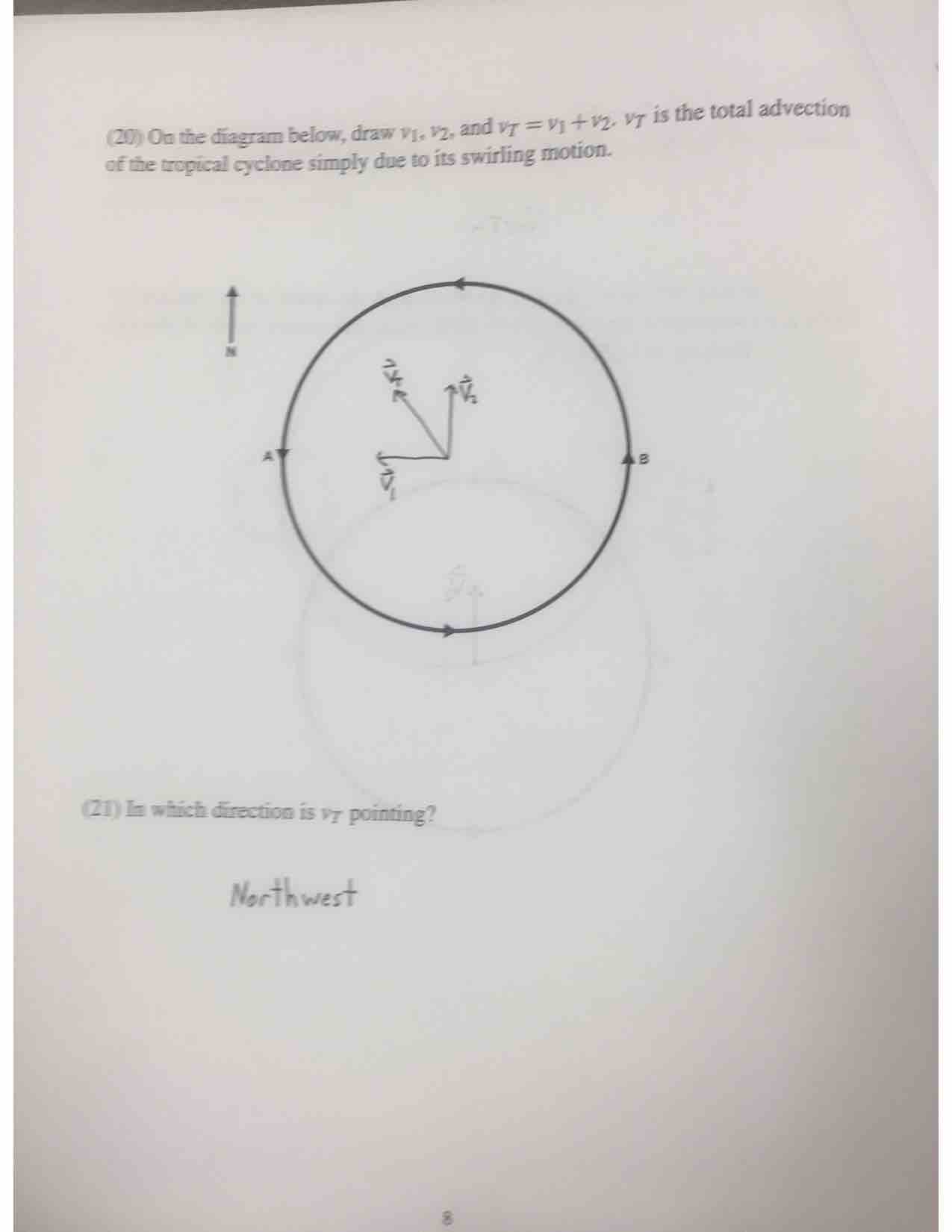}
\end{center}
\end{figure}

\begin{figure}[h!]
\begin{center}
  \includegraphics[width=\textwidth]{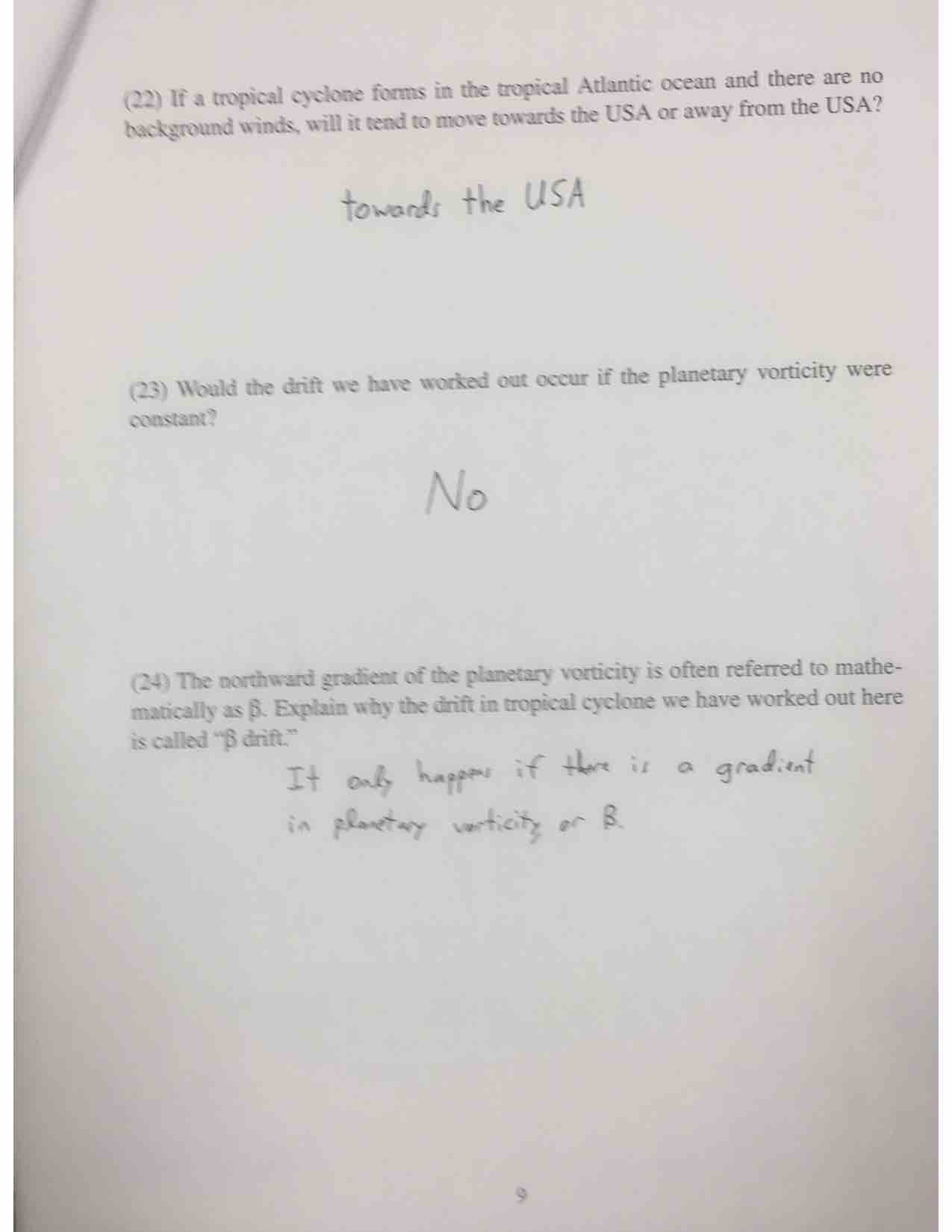}
\end{center}
\end{figure}

\clearpage

\section{References}
\begingroup
\renewcommand{\section}[2]{}


\end{document}

%% file: Introduction/origin.tex
\subsection{Origin of These Notes}

\bigskip

In 2015 our department chair, Michael Foote, asked me to teach ``The
Atmosphere,'' an introductory course in our department required for
all Geophysical Sciences and Environmental Sciences majors. I taught
it using ``Atmospheric Science'' by Wallace and Hobbs
\cite{wallace2006atmospheric}. I had no idea what students learned in
first-year calculus, so I asked them to stop me if I used something
they didn't understand. I used partial derivatives and vector calculus
most days, and six weeks into the course a student asked me after
class what the ``$\partial$'' symbol meant. I realized I was not
teaching effectively! The results on the exams bore this out. After
two years, I decided to try something different.

In 2017 I started teaching the class as a flipped class, where
students work on worksheets in class. I started going around and
interacting with students, and it really helped me understand what
they understood and where they needed extra help. I made notes
motivated by Wallace and Hobbs, but much more informal and expanding
on certain things. My notes also focus more on basic physical concepts
and less on field-specific meteorology and jargon. Wallace and Hobbs is a
great text though, and some students like to read it to go into more
depth. 

After trying it out, I'm a believer in flipped classes. Modern
students are highly motivated and diligent. They don't seem to have
any trouble preparing for class. They also get more out of a class
where they interact with their peers and learn by doing, rather than just
sitting in a lecture. Think of how many times you've fallen asleep in
a department seminar, versus how many times you've fallen asleep
coding or bouncing ideas off your favorite colleagues!

%% file: Introduction/howto.tex
\subsection{How to Use These Notes}

\bigskip

These notes are designed to be used as a flipped course. The students
should read the introduction before every class and ask questions
about anything they didn't understand at the start of class. During
class they should work together in small groups. I encourage students
to work with others, but don't force them if they prefer to work
alone. Some groups start Google Documents where they all add answers
to the exercises for the classes. Some students will finish faster
than others. Remind them that extra class time is a good time to work
on the problem set. Try to keep them from talking loudly about
unrelated stuff, because it distracts other students. I have a
teaching assistant go over the answers to the problems at the end of
each class for 10-15 minutes.

At the University of Chicago we have 10 week-long quarters. My class
meets 3 times per week for 50 minutes.  The lessons in these notes
cover these class meetings, minus days for midterms and holidays. Many
universities have a 13 week long semester. If you want to use these
notes for a semester, you have a few options. First, you could meet
twice per week for 80 minutes and give a 30 minute long lecture on the
reading at the start of each class. Second, you could add more
material. I recommend adding material on clouds, cloud microphysics,
basic atmospheric chemistry, and pollution. I have provided the Tex
files, so this should be possible.

I have also provided 10 weekly problem sets that you can use. These
take the students an average of 1.5-3 hours each, although there is a
fair amount of variation among the students on a given problem set (I
collected data on this one year). Finally, there are old midterms and
finals with solutions that you can use for inspiration and the
students can use to study. I always try to find something fun and
hopefully related to recent research to put on these. The hope is to
show the students that they've learned enough to start understanding
real, current science. I've gone toward not allowing
anything but a writing implement in exams. This forces students to
practice order-of-magnitude thinking.

%% file: Introduction/acknowledgements.tex
\subsection{Acknowledgements}

\bigskip

Daniel Koll and Jonah Bloch-Johnson were my teaching assistants for
the first two years and helped develop many of the questions in the
problem sets. Daniel was particularly good at suggesting
questions. Jonah was my teaching assistant for four years,
and has been extremely helpful in developing the flipped course. He
has excellent physical insight and contributed to the descriptions in
the lesson introductions in a number of places. Jade Checlair, David
Plotkin, and Qi Zhou have also served as teaching assistants and
helped me develop the course a great deal.

I would also like to acknowledge the undergraduate students at the
University of Chicago who have taken the course. The students have
been extremely bright and hard-working. It's been a pleasure to work
with them and seeing them learn has made my work on this course worth
it. Their questions have helped me improve my understanding of the
course material and the descriptions in the lessons tremendously.

I used figures from a variety of sources in these notes: some from
Wallace and Hobbs, some from papers, some from the internet, and some
I made myself. To focus student attention on the most important
references, I didn't reference the source for most figures. But I'd
like to acknowledge the work people put into making them and say
thanks!

Finally, I am grateful to Wallace and Hobbs for developing such an
excellent textbook, which these notes are partially based on.

%% file: Syllabus/syllabus.tex
\subsection{A sample syllabus}

\bigskip

\noindent \textbf{Overview:} This course introduces the physics and
phenomenology of the Earth's atmosphere, with an emphasis on the
fundamental science that underlies atmospheric behavior and
climate. Topics include (1) atmospheric composition, structure, and
thermodynamics; (2) solar and terrestrial radiation in the atmospheric
energy balance; (3) atmospheric dynamics and circulation.

This is a hard course. Understanding the concepts of the
atmosphere requires mathematics. If you struggle with mathematics and
abstraction, expect to devote extra time to this course. That said, no
one who has ever shown up to every class, worked hard, and not
cheated, has ever gotten less than a C. So passing this course is within
all of your capabilities.

\bigskip

\noindent \textbf{Prerequisites:} MATH 13200 (math through basic calculus)

\bigskip

\noindent \textbf{Format:} This will be a ``flipped'' course. Before
each class meeting you should read the introduction of the appropriate
class worksheet. Note that the most important equations will be set
out of the text in numbered equations. If you have any questions on
the material, please ask them at the start of class. In class you will
work in small groups on the problems in the worksheet, and we will be
available for questions. At the end of each class we will discuss the
problems together and give the answers. The worksheets will not be
graded.

In this course we are opening the door to knowledge for you. We cannot
force you through it. If you do not understand something, it is your
responsibility to ask about it. If you didn't understand the reading,
you need to raise your hand at the start of class. If your group is
moving too quickly through a problem that you don't understand, you
need to stop them. If you have questions about course material that
aren't resolved in class, you need to come to office hours and ask
them.

\bigskip

\noindent \textbf{Problem Sets:} There will be one problem set per
week. These are designed to take the average student 1.5-3 hours per
week. There is a fair amount of variation in the time it takes
students, so you will have to calibrate for yourself. Also, some of
the problem sets will take longer than others. Make sure to start all
of them early, and go to office hours if you have questions.

You may discuss problems with each other, but you must each
write up your own solutions separately. It is not appropriate to write
up solutions in the presence of other students, to discuss your
solutions with others as you write them up, or to share your
written-up solutions with other students. Also, if you are asked to
explain something, don't just Google it and copy the first thing that
pops up. Problem set solutions will be due at the start of class on
the specified due date, and will not be accepted late. This is
necessary so that all problem sets can be graded and returned
quickly.

\bigskip

\noindent \textbf{Tests:} There will be two midterms and a final. The
notes provide example tests from previous years. The best way to use
these is to print them out without looking at them, go to the library
with only a pen, and take them for the appropriate amount of time. You
can then grade them yourselves using the solutions, and use the
results to determine how to focus your study.

\bigskip

\noindent \textbf{Grading:} The problem sets will represent 50\% of
the final grade (5\% each), the midterms will be worth 10\% of the
final grade, and the final will be worth 30\% of the final grade. The
grade scheme will be as follows: [90-100]:A, [85-90):A-, [80-85):B+,
    [75-80):B, [70-75):B-, [65-70):C+, [60- 65):C, [50-60):D,
              [0-50):F. In order to be fair to everyone and avoid
                biases I will not round up under any circumstances.

%% file: Flipped/orient_estimation.tex
\subsection{Orientation and Estimation}

We're going to start with some exercises where you will look at
observations to get oriented to the important features of Earth's
atmosphere. You may not know all of the terms involved, but you can
learn by doing the exercises. Here's a diagram of tropical atmospheric
circulation that might be helpful:

\begin{figure}[h!]
\begin{center}
  \includegraphics[width=0.8\textwidth]{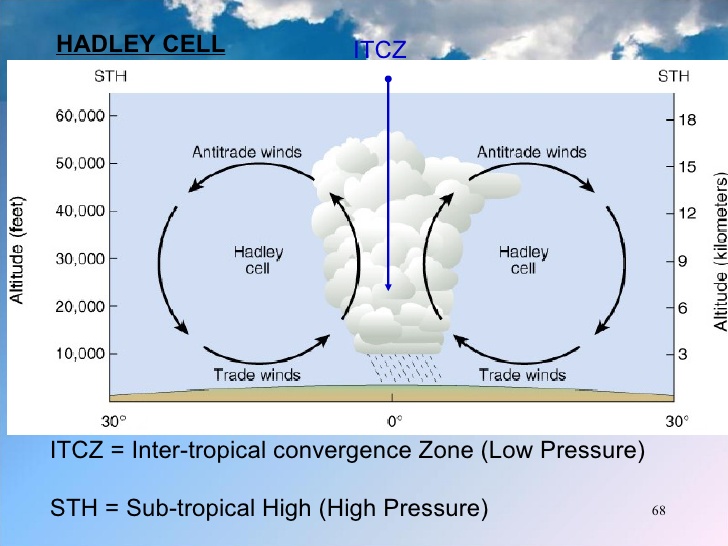}
\end{center}
\end{figure}

An important skill that I hope you will learn a bit about in this
class is the ability to estimate the answer to a problem quickly.
Scientists use this when they encounter a problem to see if a
potential solution is plausible before making a complex calculation.
We also often make complicated calculations on computers nowadays, and
estimation is a great way to check if the computer is telling you
something crazy or maybe true. In this course I may either ask you to
estimate the ``order of magnitude'' or the answer to ``one significant
digit.'' For example, a typical man is about 1.75~m tall. The order of
magnitude of his height is 1~m and his height is 2~m to one
significant digit.

For an example, let's estimate the order of magnitude of the
total mass of beard grown by a man per year. There are many ways to do
this, but let's try the following. The total mass of beard, $m$,
should be the number of beard hairs, $N$, multiplied
by the cross-sectional area of a beard hair, $\sigma$, multiplied by
the length each hair grows in a year, $l$, multiplied by the density
of beard hair $\rho$:
\[m = N \sigma l \rho.\]
To get the answer now we just need to estimate each term in this
equation. The spacing between hairs is about 1~mm, so there are 100
hairs cm$^{-2}$. Consider the head to be a sphere of radius 10~cm. The
area of the beard is a quarter of the total surface area, or
$\frac{1}{4} \times 4 \pi r^2$=$\pi \times 10^2$~cm$^2$. So $N$=($\pi
\times 10^2$~cm$^2 ) \times 10^2$~hairs cm$^{-2}$ = $\pi \times
10^4$~hairs. The smallest thing the human eye can see is about 0.1~mm
(you can check this with a ruler). We can just make out the thickness
of a beard hair, so it's diameter must be about
0.1~mm=$10^{-4}$~m. This means that $\sigma = \pi r^2 = \pi (0.5
\times 10^{-4} m)^2$ hair$^{-1} \approx
10^{-8}$~m$^2$~hair$^{-1}$. Notice that we set $\frac{\pi}{4}\approx
1$. This is appropriate since we are only trying to estimate the order
of magnitude. Hair, including beard hair, grows about 0.5~in per
month, or about 1~cm per month. Many of you probably know this. If you
didn't, you probably know immediately that a beard grows faster than
1~mm per month and slower than 10~cm per month. When you come to
something you really don't know how to estimate, one trick is to find
a lower and upper limit, then take the geometric mean. In this case,
it would give us 1~cm, which turns out to be the right answer. There
are order of 10 months per year, so
$l=10$~cm~year$^{-1}=0.1$~m~year$^{-1}$. Hair sinks in water (if it
breaks the surface tension), but just barely. So the density of hair
must be similar to that of water, or $\rho=10^3$~kg~m$^{-3}$. We can
put this all together to find: $m=$($\pi \times
10^4$~hairs)$\times$($10^{-8}$~m$^2$~hair$^{-1}$)$\times$($10^{-1}$~m~year$^{-1}$)$\times$($10^3$~kg~m$^{-3}$)=$\pi
\times 10^{-2}$~kg~year$^{-1} \approx $ 30~g. For comparison, the mass
of an adult hummingbird is about 3~g.  So a typical man shaves off 10
hummingbirds worth of beard every year! See the picture below for an
example of what happens if you let all that mass accumulate for
multiple years.  Notice that I kept track of the factor of $\pi$ in
this exercise, even though this isn't strictly necessary to estimate
the order of magnitude. Two critical things to take away from this
exercise are: (1) Always check your units and make sure they come out
right, and (2) use scientific notation.

\begin{figure}[h!]
\begin{center}
  \includegraphics[width=0.4\textwidth]{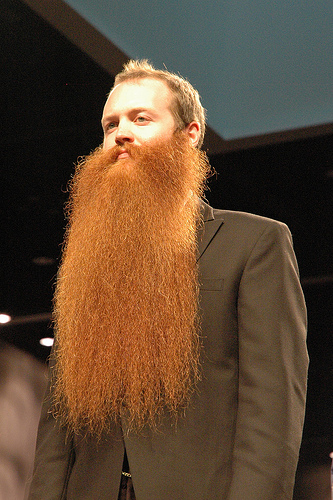}
\end{center}
\end{figure}

\clearpage

\noindent {\large \textbf{Exercises}}
\bigskip

\noindent (1) Watch this video
(\href{https://youtu.be/gpKzp3Br6-k}{https://youtu.be/gpKzp3Br6-k}) a
couple times. It contains pictures of Earth from above taken by an
orbiting satellite stitched together.

\begin{enumerate}
\item Find the extratropical cyclones (eddies), which occur in the
  mid-latitudes. Notice that they are roughly symmetric between the
  northern and southern hemispheres. These are the storms that cause
  most weather in Chicago. Can you tell which direction they spin
  (clockwise or counter-clockwise) in each hemisphere? You can slow
  the video down to make it easier to see what's happening.
\item The Inter-Tropical Convergence Zone (ITCZ) is the region where
  the northern and southern Hadley cells converge, near the equator.
  This causes convection, clouds, and precipitation. Can you find the
  ITCZ? Is it north or south of the equator?
\item Tropical cyclones are called hurricanes, typhoons, or cyclones,
  depending on the ocean basin they occur in. Can you find some
  tropical cyclones? How do they compare with extratropical cyclones?
\item This video goes through about one year. Find at least two ways
  to determine the season in Chicago as time progresses.
\end{enumerate}

\bigskip

\noindent (2) Pressure is force per area. Weight is a force equal to
the gravitational acceleration multiplied by the mass. The surface
pressure of the atmosphere on Earth is about 10$^5$~Pa, where
1~Pa=1~N~m$^{-2}$. Earth's radius is 6,400~km. Estimate the total mass
of Earth's atmosphere without using a calculator. Keep only one
significant digit. The total mass of Earth is 6$\times$10$^{24}$~kg
and the mass of Earth's oceans is 10$^{21}$~kg. How does the mass of
Earth's atmosphere compare to the total mass and the mass of the
oceans?

\bigskip

\noindent (3) As Caesar lay dying in the Theater of Pompey on the Ides
of March 44 BC, he breathed one last breath. As you inhale now, how
many molecules from Caesar's last breath do you breath in? Again, make
an ``order-of-magnitude'' estimate and do not use a calculator.
Remember that Avogadro's number is
$6\times10^{23}$~molecules~mol$^{-1}$. Air has a molecular mass of
about 30~g~mol$^{-1}$. Assume that Caesar's last breath has now been
mixed throughout the entire atmosphere. You will have to estimate the
volume of a breath, and you can assume that the density of air near
the surface is 1~kg~m$^{-3}$.

%% file: Flipped/orientation.tex
\subsection{More Orientation and Estimation}

Today we're going to continue our orientation to Earth's atmosphere
and make some calculations to get a rough understanding of modern
fossil fuel usage. Reread the introduction from last class to make sure
you understand estimation. Remember to use scientific notation and pay
careful attention to units for the estimation problems.

To build up vocabulary, take a look at the figure below. It shows an
idealized typical ``vertical temperature profile'' of Earth's
atmosphere. The lowest region, where we live, is called the
troposphere. The surface is strongly heated by sunlight and transfers
this heat to the lower atmosphere. This causes the air near the
surface to become less dense and rise in a process called
convection. Convection happens throughout the troposphere so that air
in it is well-mixed. Above the troposphere is the stratosphere. The
stratosphere is ``stratified,'' with hotter air on top of colder air
as a result of heating by absorption of sunlight by ozone. This makes
vertical motion and mixing of air much harder in the stratosphere than
in the troposphere. The tropopause separates the troposphere from the
stratosphere. The troposphere and stratosphere contain 99.9\% of the
atmosphere, and will be the focus of this course. The other regions of
the atmosphere only become important in certain specialized areas.

\begin{figure}[h!]
\begin{center}
  \includegraphics[width=0.5\textwidth]{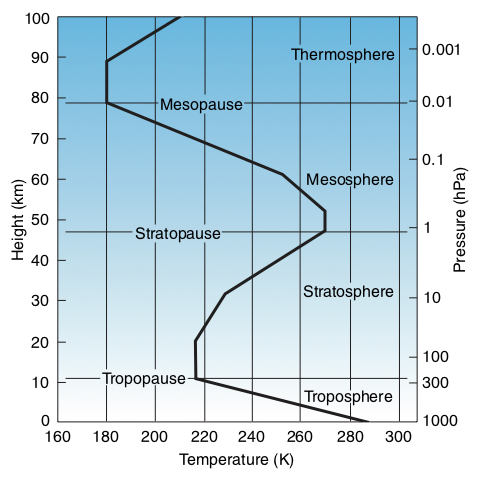}
\end{center}
\end{figure}

\clearpage

\noindent {\large \textbf{Exercises}}
\bigskip

\noindent (1) Watch this video
(\href{https://youtu.be/KtwdaWIGFDQ}{https://youtu.be/KtwdaWIGFDQ}) of
surface temperature from Jan/Feb 2014 and this video
(\href{https://youtu.be/lm5jt3dyYxc}{https://youtu.be/lm5jt3dyYxc}) of
surface temperature from Jul/Aug 2014.

\begin{enumerate}
\item For Jan/Feb, focus on the northern hemisphere, poleward of
  30$^\circ$N. Why are there two large cold regions?
\item Why is there such a persistent blue spot at (30$^\circ$N,90$^\circ$E)?
\item For Jul/Aug, focus on the northern hemisphere, poleward of
  30$^\circ$N. Why are the regions that were previously cold now warm?
\item If you watch the two movies carefully you can detect large, eastwards moving swirls (weather systems) poleward of 30$^\circ$N. When are the swirls most active in each hemisphere?
\end{enumerate}

\bigskip

\noindent (2) What is the total mass of CO$_2$ put into the atmosphere
by a 1-GW coal power plant per year? You may assume that the plant has
an efficiency of 1/3 and that burning coal releases
10$^7$~J~kg$^{-1}$. Remember that coal is basically just carbon. Total
world power consumption is 10$^5$ TW-h per year. How many 1-GW power
stations would be needed to supply this power, assuming that they run
continuously (1~TW=10$^3$~GW)? The total mass of CO$_2$ in the
atmosphere is 3$\times$10$^{15}$~kg. Assuming the entire world's power
supply is provided by coal, how many years would it take to double the
amount of CO$_2$ in the atmsophere? Assume that no other process adds
or removes CO$_2$ from the atmosphere and assume that humanity's power
usage stays constant.

\bigskip

\noindent (3) What is the mass of CO$_2$ emitted by a typical car in
the US per year? Assume it is driven 10$^4$ mi and gets 20 mpg.
You may approximate gasoline as made entirely of carbon. If on average
there is a car for every two people in the US, what is the total mass
of CO$_2$ emitted by cars in the US per year?

%% file: Flipped/pressure_cyclones.tex
\subsection{Pressure and Cyclones}

A cyclone is a relatively small part of the atmosphere (maybe on a
scale of about 1000~km) that rotates around a region of low pressure
(see the diagram below).  This rotation is due to the Coriolis force,
which we will learn about later in the course. The direction of
rotation is counterclockwise in the Northern Hemisphere and clockwise
in the Southern Hemisphere. In either case, this direction is referred
to as ``cyclonic.'' Anticyclones, on the other hand, have high
pressure at their center and rotate in the opposite
direction. Cyclones have air converging near the surface, which
generally leads to air rising, which causes precipitation and the
formation of clouds.  Extratropical cyclones, also called baroclinic
eddies, are the storms we are familiar with in Chicago. They are
particularly common in the winter. The reason is that the North-South
gradient in surface temperature is ultimately what drives them, and
this gradient is stronger in the winter. We usually get a new one
coming through every 3--5 days. Tropical cyclones are driven by a
different mechanism, which we will learn more about later. Tropical
cyclones are called hurricanes in the Atlantic ocean and eastern
Pacific, typhoons in the western Pacific, and cyclones in the Indian
and South Pacific.  Cyclones generally get blown around by the
prevailing winds. There are also larger scale lows and highs in the
atmosphere that persist in roughly the same place for entire
seasons. These can deviate winds and lead to seasonal shifts in
weather.

\begin{figure}[h!]
\begin{center}
  \includegraphics[width=0.5\textwidth]{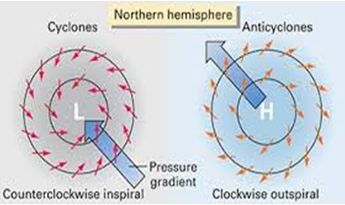}
\end{center}
\end{figure}

\clearpage

\noindent {\large \textbf{Exercises}}
\bigskip

\noindent (1) 

\begin{enumerate}
\item If the surface is roughly at 1000hPa (1 hPa=100 Pa), what
  percentage of the atmosphere's mass is above 500hPa?

\item Watch this video
  (\href{https://youtu.be/skgiZhTFUH8}{https://youtu.be/skgiZhTFUH8})
  of temperature at 500hPa from Jan/Feb 2014. Do you see anomalous
  blobs like we saw in the surface temperature videos from last class?
  Why, or why not?

\item Now watch this video
  (\href{https://youtu.be/VpndGnLym8I}{https://youtu.be/VpndGnLym8I})
  of temperature at 500hPa from Jul/Aug 2014. Compare the two
  movies. What part of the atmosphere is coldest (at 500hPa)?
\end{enumerate}

\bigskip

\noindent (2) 

\begin{enumerate}
\item Watch this video
  (\href{https://youtu.be/rXT8WaYXKtQ}{https://youtu.be/rXT8WaYXKtQ})
  of surface pressure from Jan/Feb 2014. Why is most of the
  continental area distinctly blue?

\item Now compare with this video
  (\href{https://youtu.be/hzQhbfuEyC0}{https://youtu.be/hzQhbfuEyC0})
  of surface pressure from Jul/Aug 2014. Over the ocean, which
  latitude generally has the lowest lows?

\item Focus on the moving lows in the mid-latitudes in Jul/Aug. As
  they move eastwards they tend to rotate. Which direction do they
  rotate in? These lows are called cyclones and the direction they
  rotate in is called cyclonic.

\item Can you identify persistent highs in Jul/Aug? At which latitude
  do you generally find them? Search on google to see what these are
  called.

\item What are the two large lows in the northern hemisphere of the
  Jan/Feb movie called? (google them) How persistent are they in the movie?

\item Focus on the Jul/Aug movie. You will occasionally see small lows
  (blue/white blobs) pop up equatorwards of 30$^\circ$N which then
  move westwards and polewards. These are tropical cyclones. Find and
  track some of them through their life cycle.

\item Watch the blob that hits Japan on July 9, 2014. This blob had a
  name - what is it?
\end{enumerate}

\bigskip

\noindent (3) There are pictures of two tropical cyclones below.
For each storm: Is the pressure at the center of the storm anomalously
high or low? Which hemisphere is it in?

\begin{figure}[h!]
\begin{center}
  \includegraphics[width=0.5\textwidth]{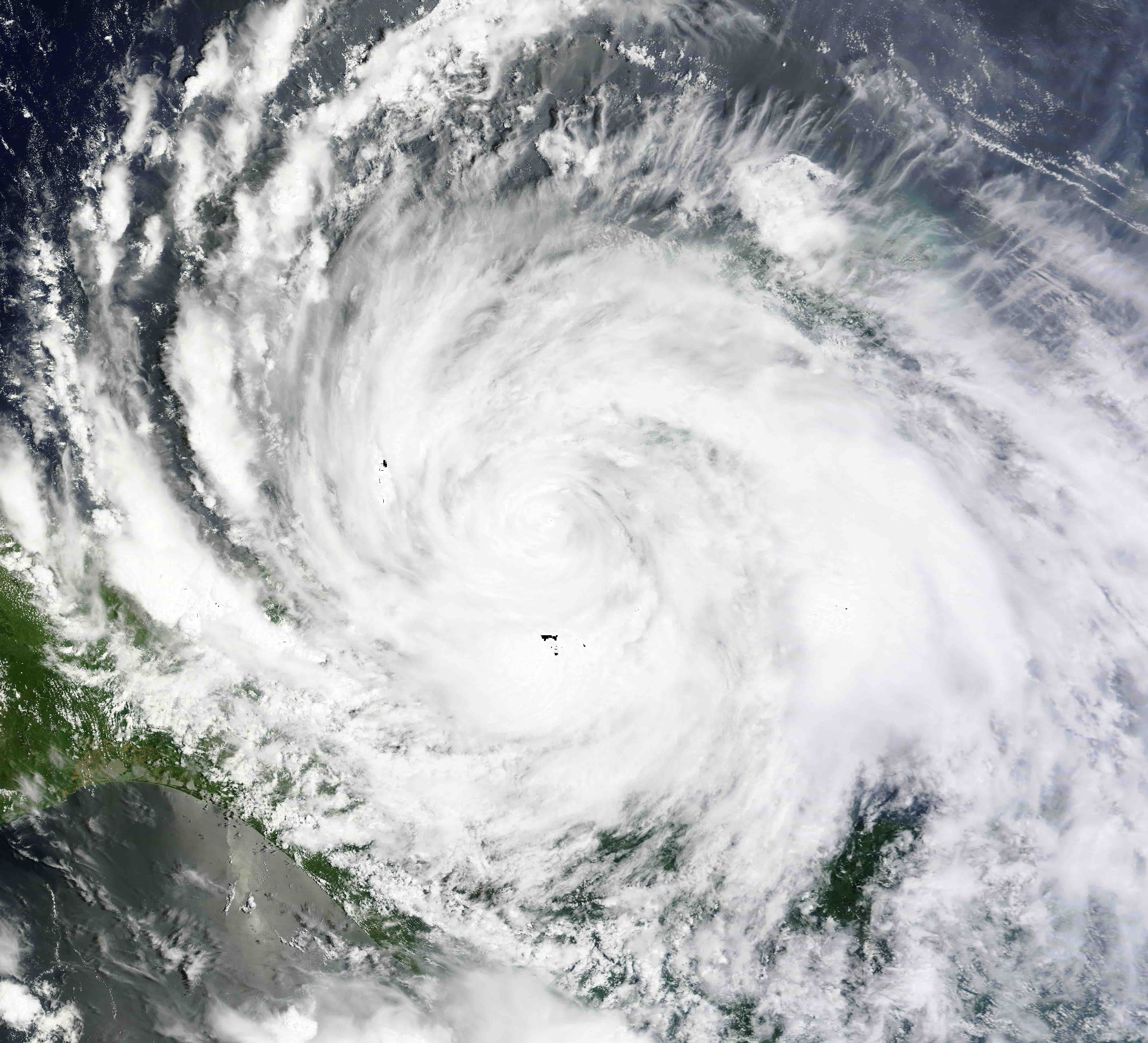}
\end{center}
\end{figure}

\begin{figure}[h!]
\begin{center}
  \includegraphics[width=0.5\textwidth]{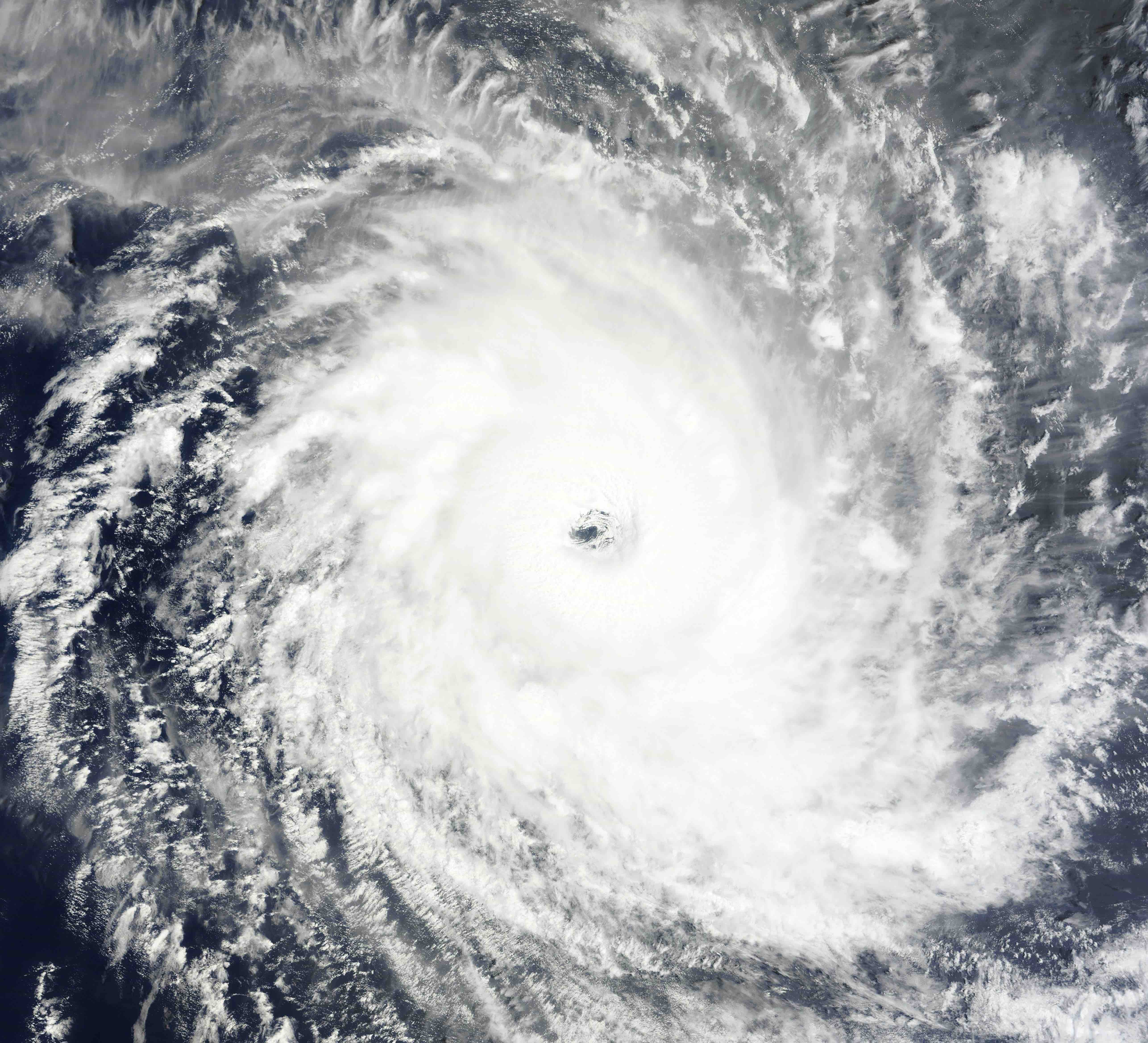}
\end{center}
\end{figure}

%% file: Flipped/composition.tex
\subsection{Compositon and the Ideal Gas Law}

The ideal gas law is 
\begin{equation}
PV=nR^\ast T, 
\end{equation}
where $P$ is the pressure in Pa, $V$ is the volume in m$^3$, $n$ is
the number of moles, $R^\ast$=8.3 J~mol$^{-1}$~K$^{-1}$ is the ideal
gas constant, and $T$ is the temperature in K. In atmospheric science
we usually rewrite this equation as
\begin{equation}
P=\rho R T,
\end{equation}
where $\rho=\frac{n*M}{V}$ is the gas density in kg~m$^{-3}$, $M$ is
the molecular mass of the gas mixture in kg~mol$^{-1}$ (you probably
learned molecular masses in g~mol$^{-1}$, so don't forget to convert
to kg~mol$^{-1}$), and 
\begin{equation}
R=\frac{R^\ast}{M}
\end{equation}
is called the gas constant. The gas constant, $R$, is measured in
J~kg$^{-1}$~K$^{-1}$, and is specific to the particular gas mixture
under consideration (it depends on the molecular mass). If we are
considering a mixture of gases, then the apparent molecular weight is
the molar-fraction-weighted average of the molecular weights
\begin{equation}
M=\sum_i f_iM_i,
\end{equation} where $f_i$ is the molar fraction of the ith
component and $M_i$ is the molecular weight of the ith component.

\clearpage

\noindent {\large \textbf{Exercises}}
\bigskip

\noindent (1) Stars increase their luminosity with time, such that the
Sun was about 25\% less luminous 4 billion years ago than it is today.
Nevertheless, we have strong evidence that Earth's climate was fairly
similar to how it is now, rather than completely frozen over. Carl
Sagan, an alumnus of the University of Chicago, was one of the first
people to point out that this is an interesting mystery, and to try to
solve it \cite{SAGAN:1972p1233}. Most people now think that the
solution is that some greenhouse gases (CO$_2$, CH$_4$, and even
H$_2$) had much higher concentrations on early Earth than they do
today.

\begin{enumerate}
\item The modern atmosphere is composed of about 80\% N$_2$ and 20\%
  O$_2$ by molar fraction or volume when dry. Calculate the mean
  molecular weight of dry air and the dry gas constant. You can use a
  calculator for this (you don't need to estimate).
\item Early Earth had almost no O$_2$. There may have been large
  amounts of CO$_2$ to keep the planet warm. Assume that Early Earth's
  dry atmosphere was composed of 80\% N$_2$ and 20\% CO$_2$. Calculate
  the mean molecular weight of dry air and the dry gas constant.
\item Assume that early Earth had a surface pressure of
  1~bar=10$^5$~Pa. What was the partial pressure of CO$_2$ at the
  surface? Remember that the partial pressure is equal to the total
  pressure times the molar fraction.
\item Calculate the total mass in the entire column of air per unit
  area of this atmosphere.
\item Using the mean molecular mass, calculate the number of moles in
  the entire column of air per unit area of this atmosphere.
\item Calculate the number of moles of CO$_2$ in the entire column of
  air per unit area of this atmosphere.
\item Calculate the mass of CO$_2$ in the entire column of air per
  unit area of this atmosphere. You should have found that the mass of
  CO$_2$ in the entire column of air per unit area is greater than the
  partial pressure of CO$_2$ at the surface divided by the surface
  gravity. The reason is that the partial pressure of CO$_2$ at the
  surface results from the weight of both N$_2$ and CO$_2$ throughout
  the atmosphere, and N$_2$ has a lower molecular mass.
\item  What is the mass fraction of CO$_2$ in the atmosphere?
\item Suppose we want to increase the amount of CO$_2$ so that it
  represents 50\% of the atmosphere by molar fraction. What would the
  new partial pressure of CO$_2$ and total atmospheric pressure have
  to be? 
\item What would the new the mass of CO$_2$ in the entire column of
  air per unit area be? 
\item What would the new mass fraction of CO$_2$ in the atmosphere be?
\end{enumerate}

%% file: Flipped/virtual_temperature.tex
\subsection{Virtual Temperature}

One issue that complicates a lot of things on Earth is that the
atmosphere has a variable amount of water vapor, depending on the
temperature and atmospheric motion. This means that the gas constant
changes with time and as you move around. This makes it complicated to
calculate the density, which we often need to figure out if air is
going to convect, or mix vertically because denser air is above
less dense air (atmospheric convection is slightly more complicated
than this, but we'll get to that later). Here we'll try to deal with
this. First, note that the ideal gas law applies to the dry component
of air: $P_d=\frac{m_d}{V}R_dT$, where $P_d$ is the partial pressure
of dry air, $m_d$ is the mass of dry air, and $R_d$ is the dry air gas
constant. The ideal gas law also applies to the moist component of
air: $e=\frac{m_v}{V}R_vT$, where $e$ is the partial pressure of water
vapor, $m_v$ is the mass of water vapor, and $R_v$ is the gas constant
of water vapor. By Daulton's law, we also know that the total pressure
is the sum of the partial pressures: $P=P_d+e$. Finally, we can write
the total density as: $\rho=\frac{m_d+m_v}{V}$. If we solve these
equations for density we get:
\[\rho=\frac{P_d}{R_dT}+\frac{e}{R_vT}=\frac{P-e}{R_dT}+\frac{e}{R_vT}=\frac{P}{R_dT}\left( 1-\frac{e}{P}+\frac{R_d}{R_v}\frac{e}{P}\right)=\frac{P}{R_dT_v},\]
where we have defined the virtual temperature, $T_v$, as:
\begin{equation}
T_v=\frac{T}{1-\frac{e}{P}(1-\epsilon)},
\end{equation}
where $\epsilon=\frac{R_d}{R_v}=\frac{M_v}{M_d}$. Using the virtual
temperature allows us to use the ideal gas law with a constant gas
constant ($R_d$) rather than a variable gas constant (that depends on
the amount of water vapor) in order to calculate the air density. The
cost is that we have to define a new temperature, the virtual
temperature, which depends on the amount of water vapor. Physically,
the virtual temperature is the temperature dry air would need in order
to have the same density as the moist air under consideration. On
modern Earth $M_v<M_d$, so adding moisture to air makes it less
dense. This means that on modern Earth the virtual temperature is
always greater than or equal to the temperature.

\clearpage

\noindent {\large \textbf{Exercises}}
\bigskip

\noindent (1) Let's think a bit about virtual temperature. 

\begin{enumerate}

\item Explain how the virtual temperature relates to density. If air
  at a constant pressure has a higher virtual temperature, does that
  mean it is more or less dense?

\item On modern Earth $\epsilon$=$\frac{M_w}{M_d}$=0.622. The
  molecular weight of water vapor is less than the molecular weight of
  dry air. Use this fact to explain why at a constant temperature
  adding moisture to the atmosphere should increase the virtual
  temperature.

\item Something we will learn soon is that the amount of water vapor
  that can be present in the atmosphere increases exponentially with
  temperature via the Clausius-Clapeyron equation. Use this fact to
  explain why there are two reasons that the virtual temperature
  should increase with temperature on modern Earth. Use this to
  explain why Kris Bryant is more likely to hit a ball onto Waveland
  Avenue (out of Wrigley to left) in August than in April.

\item Suppose that early Earth was composed of a mixture of 40\% N$_2$
  and 60\% H$_2$. Calculate the new dry gas constant and
  $\epsilon$. You can use a calculator for this.

\item  Assuming that there is surface water on early Earth, explain why now
  as the temperature increases, the virtual temperature can either
  increase or decrease. Explain why this is counterintuitive.

\item Suppose that early Earth was composed of a fraction $\chi$ of
  N$_2$ and $1-\chi$ of H$_2$. For large $\chi$, virtual temperature
  always increases as temperature increases. For small $\chi$, virtual
  temperature can either increase or decrease as temperature
  increases. There is a critical value of $\chi$, $\chi_c$, that
  demarcates these two regions. Calculate $\chi_c$.
\end{enumerate}

%% file: Flipped/hydrostatic_scale_height.tex
\subsection{Hydrostatic Balance and the Scale Height}

First let's derive the hydrostatic equation. This works for an
atmosphere in hydrostatic balance (not accelerating vertically). It
describes a balance between a net effect of pressure pushing up and
the weight (force due to gravity) of air pulling down. Consider the
diagram below:

\begin{figure}[h!]
\begin{center}
  \includegraphics[width=0.4\textwidth]{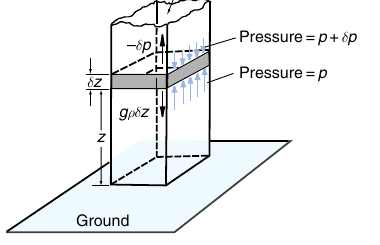}
\end{center}
\end{figure}

Let's do a force balance on the darkened slice of the column of air.
The bottom of this slice has a pressure force pushing up of
$P(z)\times A$, where $A$ is the horizontal area of the column. The
slice has a pressure force pushing down on it of
$-P(z+\delta z)\times A$. The gravitational force of the air pushes
down and is $-\rho \times A \times \delta z \times g$ (mass times
gravitational acceleration). So assuming that there is no net vertical
acceleration, the vertical force balance is:
\[P(z)\times A-P(z+\delta z)\times A-\rho \times A \times \delta z \times g=0.\]
We can cancel $A$ and rearrange this equation as
$\frac{P(z+\delta z)-P(z)}{\delta z}=-\rho g$, or in the infinitesimal
limit 
\begin{equation}
\frac{dP}{dz}=-\rho g.
\end{equation}
 This is the hydrostatic equation and it
describes a force balance between the vertical gradient of pressure
and the weight of the air. Since the pressure goes to zero as the
height goes to infinity, we can integrate vertically to find that
\begin{equation}
P(z)=\int_z^\infty \rho(z) g dz.
\end{equation}

A related concept is the geopotential, $\Phi$, which is defined as the
work that must be done against Earth's gravitational field to raise a
mass of 1~kg from sea level to the height under consideration.
$d \Phi=g dz$, and we can integrate to find that
\begin{equation}
\Phi(z)=\int_0^z g dz\approx gz,
\end{equation}
since $g$ is roughly constant throughout the atmosphere. We will hear
more about the geopotential later in the course.

Finally, let's derive an equation for how the pressure decays with
height. We can combine the hydrostatic equation ($\frac{dP}{dz}=-\rho
g$) and the ideal gas law ($P=\rho R T$) to eliminate $\rho$:
$\frac{dP}{dz}=-\frac{g}{R T}P$. If we assume that $T$ is constant, we
can define 
\begin{equation}
H=\frac{R T}{g}
\end{equation}
and integrate this equation to find
\begin{equation}
P(z)=P(z=0) \exp(-\frac{z}{H}).
\end{equation}
So we can see that pressure decays exponentially in height with a
scale of $H$, as shown in the diagram below. $H$ is therefore called
the atmospheric scale height. We can understand the scale height
physically by noting that the scale height is reached when $gH\approx
R T$. In other words a scale height is where thermal energy roughly
equals graviational potential energy.
 The scale height is telling us
about how well a planet can hold onto it's atmosphere.  If the scale
height is small then the planet is holding it's atmosphere tightly,
but if the scale height is large it is holding it only loosely. Notice
that since $R$ depends on $M$, the molecular mass, the scale height
depends on the molecular mass of the atmosphere.

\begin{figure}[h!]
\begin{center}
  \includegraphics[width=0.8\textwidth]{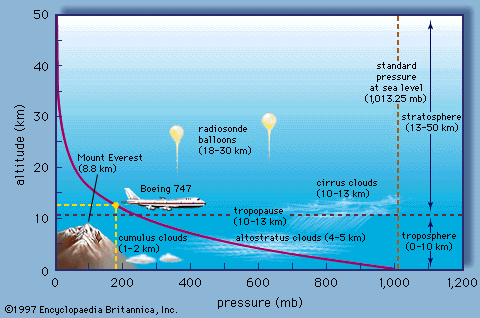}
\end{center}
\end{figure}

\clearpage

\noindent {\large \textbf{Exercises}}
\bigskip

\noindent (1) Show that the units of $H$ are meters.

\bigskip

\noindent (2) Estimate the scale height, $H$, in Earth's lower
atmosphere to one significant digit.

\bigskip

\noindent (3) The Colorado Rockies play at Coors Field in Denver, the
``Mile High City.'' Estimate the ratio of the air density at Coors
Field to that at sea level. Explain why the average baseball game
played at Coors Field has about 11 total runs scored, whereas the
average major league baseball game has about 8 total runs scored.

\bigskip

\noindent (4) A normal human being will become nauseous and may lose
consciousness if oxygen levels drop below 50\% of those at sea level
and is likely to die if oxygen levels drop below 30\%. The summit of
Mount Everest is at 8.85~km. Is it possible for a human to reach the
summit of Mount Everest without an artificial oxygen supply? 

\bigskip

\noindent (5) Olympus Mons on Mars stands 22~km above local relief. If
Earth had a mountain with a summit 22~km above sea level, would it be
possible for a human to climb it without an artificial oxygen supply?

%% file: Flipped/lapse_rate_theta_1.tex
\subsection{The Lapse Rate and Potential Temperature}

Energy comes in different flavors. The first law of thermodynamics
says that energy is conserved. If you lose some in one flavor, you
have to gain it in another flavor. One of the flavors is internal
energy. Even if a parcel of air seems motionless on a large scale, its
molecules still have kinetic and potential energy, which we call
internal energy. Temperature characterizes the kinetic internal energy
of molecules. Other flavors of energy are heat added to the system and
work done by the system. We can write energy conservation in
differential form for a parcel of air as follows
\begin{equation}
dq=du+dw,
\end{equation}
where $dq$ is the heat added to the air parcel, $du$ is the change in
internal energy of the air parcel, and $dw$ is the work done by the
air parcel. This equation says that if we add heat to an air parcel,
it goes into some combination of increasing the temperature of the
parcel or allowing the parcel to do work on its surrounding
environment.

To think about work done by an air parcel, let's consider the air in
the piston shown in the figure below.  The work done by the air in the
chamber is the external force it is pushing against multiplied by the
distance it pushes ($dW=Fdx$). If we assume that the force pushing in
on the chamber is from the pressure of the air outside, then we find
that $F=pA$, where $A$ is the area of the stopper. The change in
volume of the parcel is just its area multiplied by the distance moved
perpendicular to the stopper ($dV=Adx$), so we can write $dW=pdV$. We
won't prove it here, but this also works for an air parcel that isn't
in a chamber. If $w$ is the work per unit mass and
$\alpha=\frac{1}{\rho}$ is the specific volume of the air, or volume
per unit mass, then we can also write this as
$dw=pd\alpha$ and we can rewrite the 1st law of thermodynamics as
\begin{equation}
dq=du+pd\alpha.
\end{equation}

\begin{figure}[h!]
\begin{center}
  \includegraphics[width=0.5\textwidth]{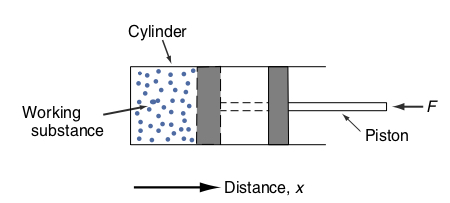}
\end{center}
\end{figure}

The specific heat is the amount of heat we have to add to a substance
to increase its temperature by 1~K. The specific heat at constant
volume is $c_v=(\frac{dq}{dT})_v$, where the subscript $v$ means we
take the derivative at constant volume. At constant volume $dq=du$
(all the heat goes into increasing the temperature, and none goes into
doing work) so that $c_v=(\frac{du}{dT})_v$. For an ideal gas $u$
depends only on $T$ (we didn't show this, but trust me), so we can
drop the constant volume assumption and get $c_v=\frac{du}{dT}$, or
$du=c_v dT$. This means we can rewrite the 1st law as 
\begin{equation}
dq=c_v dT + pd\alpha.
\end{equation}

The specific heat at constant pressure is defined as
$c_p=(\frac{dq}{dT})_p$. It is often more useful than the specific
heat at constant volume in atmospheric sciences because gases in the
atmosphere tend to be at constant pressure rather than volume. Using
the product rule we can rewrite the 1st law as $dq=c_v dT +
d(p\alpha)-\alpha dp$. Using the ideal gas law,
$d(p\alpha)=d(\frac{p}{\rho})=d(RT)=RdT$, so that $dq=(c_v+R)dT-\alpha
dp$. Or at constant pressure ($dp=0$), $dq=(c_v+R)dT$, which means
that $c_p=(\frac{dq}{dT})_p=c_v+R$. This means we can also write the
1st law as 
\begin{equation}
dq=c_pdT-\alpha dp.
\end{equation}

Now let's think about the temperature lapse rate, or decrease in
temperature with height in the atmosphere. Let's consider an air
parcel that changes height with no heat added or subtracted ($dq=0$),
which means the motion is an ``adiabatic process.'' We now have the
tools to infer the lapse rate in this situation, which is called the
``dry adiabatic lapse rate'' and is written
$\Gamma_d=-(\frac{dT}{dz})_{q=0}$. We use the word ``dry'' because the
condensation of moisture releases a lot of heat. The atmosphere tends
to roughly follow this dry adiabatic lapse rate at least sometimes in
regions that are dry like deserts, and cold regions at the poles or
high in the atmosphere. With $dq=0$, we have $0=c_pdT-\alpha
dp$. Using the hydrostatic equation ($dp=-\rho g
dz=-\frac{g}{\alpha}dz$) we find that $0=c_p dT+g dz$ so that we can
write the dry adiabatic lapse rate as
\begin{equation}
\Gamma_d=-(\frac{dT}{dz})_{q=0}=\frac{g}{c_p}.  
\end{equation}
Note that deriving the dry adiabatic lapse rate required the 1st law
of thermodynamics (conservation of energy), the ideal gas law, and the
hydrostatic equation. In Earth's atmosphere, $c_p\approx
10^3$~J~K$^{-1}$~kg$^{-1}$ and $g\approx 10$~m~s$^{-2}$, so that
$\Gamma_d\approx10^{-2}$~K~m$^{-1}=10$~K~km$^{-1}$.

Finally, let's define a new quantity called potential
temperature. Potential temperature is nice because an air parcel has
constant potential temperature if no heat is added or subtracted from
it. Let's return to the 1st law with no heat exchange ($0=c_p
dT-\alpha dp$). The ideal gas law tells us that $\alpha=\frac{RT}{p}$
so that $0=c_p \frac{dT}{T} - R \frac{dp}{p}$.  Integrating we find
that $const. = c_p \log(T)-R \log(p)$ which we can exponentiate to
rewrite as $T=(const.)p^{\frac{R}{c_p}}$. This equation tells us that
for an adiabatic process the temperature and pressure have to be
related in this way for some constant that we can find. At surface
pressure ($p=p_0$), we will define the temperature as $T=\theta$, which
will be the potential temperature. So the constant equals $\theta
p_0^{-\frac{R}{c_p}}$. Notice that since $p_0$ is a constant, $\theta$
needs to be a constant as well. We can rearrange to find 
\begin{equation}
\theta = T (\frac{p_0}{p})^{\frac{R}{c_p}},
\end{equation}
which defines the potential temperature. The bottom line is that any
air parcel that moves adiabatically will have a constant potential
temperature ($\theta$). If it goes to lower pressure and decompresses,
its temperature will decrease, but its potential temperature will
stay constant. If it goes to higher pressure and compresses, its
temperature will increase, but its potential temperature will stay
constant.

\clearpage

\noindent {\large \textbf{Exercises}}
\bigskip

\noindent (1) Explain the following:

\begin{enumerate}
\item Air released from a tire is cooler than its surroundings.
\item The potential temperature in a dry convecting atmosphere is
  constant.
\item In Greek mythology, Icarus falls from the sky because he flew
  too high and close to the Sun so the wings his father, Daedalus,
  made for him get too hot and melt. Assuming that Icarus was flying
  in the troposphere, does this make sense?
\item Students are taught that when air is heated it expands. Then
  they are taught (but not in the same lesson) that when air expands
  it cools. Some students find these two explanations to be
  contradictory. Please resolve the apparent contradiction.
\item An airplane flies in the troposphere of a dry atmosphere. The
  air outside the airplane is very cold. When this air is brought into
  the pressurized cabin of the airplane, it has the same pleasant
  temperature for the passengers that they experienced on the ground
  before taking off. 
\item How does the potential temperature of air at the surface, in the
  atmosphere at the plane's altitude, and inside the pressurized cabin
  compare?
\end{enumerate}

%% file: Flipped/stability.tex
\subsection{Stability}

The lapse rate and potential temperature are important concepts, so
we're going to continue learning about them today. The problems will
continue to focus on these general issues, and these notes will focus
on how you can use the lapse rate and potential temperature to think
about atmospheric stability, or whether vertical convection will
occur.

If an air parcel moves adiabatically vertically, it will follow the
dry adiabatic lapse rate and have a constant potential temperature. If
the actual atmosphere has a lapse rate that is greater than the dry
adiabatic lapse rate, then a parcel lifted from the surface will have
a higher potential temperature than the surrounding atmosphere at
height (and therefore a higher temperature). This means it will be
bouyant (less dense that the surroundings). This is an unstable
situation and convection (strong vertical mixing) will occur until the
lapse rate is equal to the dry adiabatic lapse rate. This means that
an atmosphere cannot sustain a lapse rate higher than the dry
adiabatic lapse rate, or such an atmosphere is unstable (see first figure
below). During the day in the troposphere the Sun is continually
heating the surface, increasing the lapse rate and destabalizing the
atmosphere so that convection occurs. On the other hand, if an
atmosphere has a lapse rate less than the dry adiabatic lapse rate, as
it does in the stratosphere, this is a stable situation and no
convection needs to occur. The tropical troposphere can also have a
lapse rate less than the dry adiabatic lapse rate because of water
vapor condensation, which we will learn about soon. As the second
figure below shows, we can see this as an increase in the potential
temperature with height in the zonal mean (averaged over longitude)
potential temperature profile. This means it is stable to dry
convection, and the potential temperature increases with height. As we
will learn, the tropical troposphere can still be unstable to moist
convection in this configuration. The figure shows that the
extratropics are also generally stable to dry convection, which
results from atmospheric heat transport from lower latitudes and
altitudes to higher latitudes and altitudes.

\begin{figure}[h!]
\begin{center}
  \includegraphics[width=0.5\textwidth]{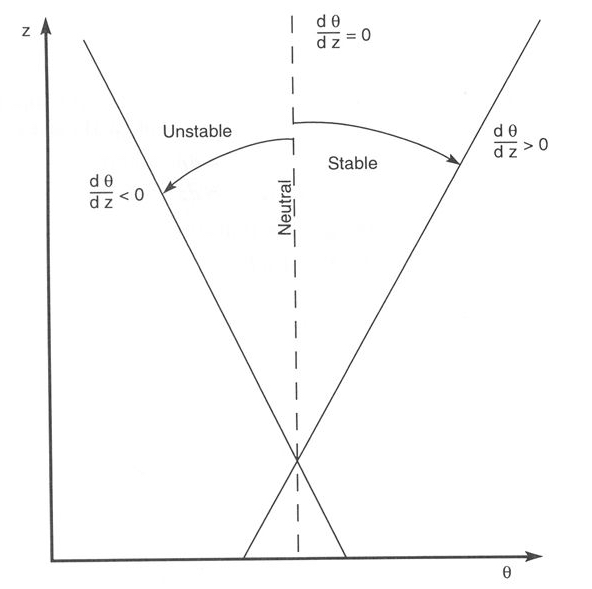}
\end{center}
\end{figure}

\begin{figure}[h!]
\begin{center}
  \includegraphics[width=0.9\textwidth]{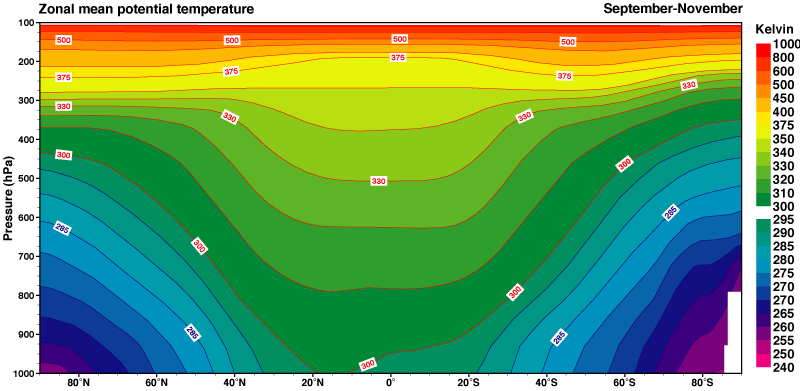}
\end{center}
\end{figure}

\clearpage

\noindent {\large \textbf{Exercises}}
\bigskip

\noindent (1) The Galileo probe descended into Jupiter's atmosphere in
July, 1995 \cite{seiff1996structure}. The figure below shows
temperature data (blue line) retrieved from the probe's descent. The
data start when the probe entered Jupiter's atmosphere. The data end
at the bottom right of the plot because the probe failed. The pressure
was about 23 bar and the temperature was about 425~K when the probe
failed. The notional altitude is taken to be zero at a temperature of
166~K in this plot.  The pressure at this altitude is equal to Earth's
surface pressure.

\bigskip

\begin{figure}[h]
\vspace*{2mm}
\begin{center}
  \includegraphics[width=40pc]{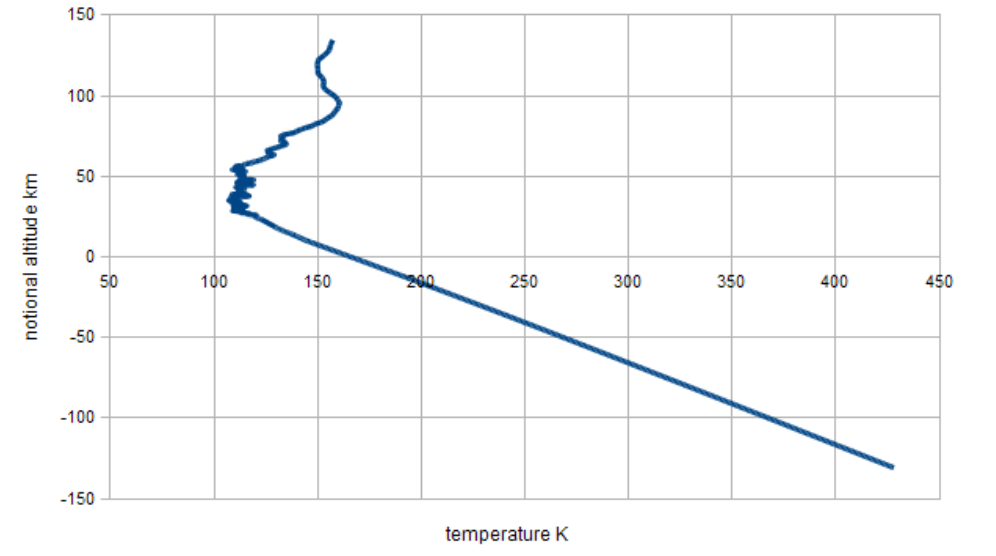} 
\end{center}
\end{figure}

\bigskip

\begin{enumerate}
\item Notice that Jupiter has a stratosphere and a troposphere. At roughly what temperature and notional altitude is the tropopause located?

\item Jupiter's atmosphere is composed mainly of H$_2$. What is the
  dry gas constant for Jupiter's atmosphere?

\item The gravitational acceleration in this region of Jupiter's
  atmosphere is 25~m~s$^{-2}$. Estimate the scale height of Jupiter's
  atmosphere at Earth's surface pressure. Compare this to the scale
  height of Earth's atmosphere near Earth's surface. Also compare to
  the scale height of Jupiter's atmosphere where the probe failed.

\item Notice that we can rewrite the equation for the scale height as
  $M g H = R^\ast T$, where $M$ is the molecular mass and $R^\ast$ is
  the ideal gas law constant in J~mol$^{-1}$~K$^{-1}$. What types of
  energy do the left and right sides of this equation represent? Does
  this help you understand what the scale height represents better?

\item Estimate the pressure at Jupiter's tropopause. How does this
  compare to the pressure of Earth's tropopause?

\item The heat capacity at constant pressure ($C_p$) of an ideal
  diatomic gas in this temperature range is 3.5$R$, where $R$ is the
  (compound specific) gas constant in J~K$^{-1}$~kg$^{-1}$. Make a
  prediction for the temperature lapse rate on Jupiter. Now estimate
  the temperature lapse rate from the plot and compare to your
  prediction.

\end{enumerate}

%% file: Flipped/water_vapor.tex
\subsection{Water Vapor in the Atmosphere}

Today we're going to think more about water vapor in the
atmosphere. Water vapor is important because it affects the air
density and the ability of humans to lose excess heat through
sweating. It is highly variable in the atmosphere and we need to
define variables to keep track of it.

The partial pressure of water vapor in the atmosphere, or vapor
pressure, is often denoted $e$. The mass of water vapor ($m_v$)
divided by the mass of dry air ($m_d$) is called the mixing ratio of
water vapor ($w$),
\begin{equation}
w=\frac{m_v}{m_d}.
\end{equation}
The specific humidity ($q$) is the ratio of water vapor mass to total
air mass, including both dry air and water vapor,
\begin{equation}
q=\frac{m_v}{m_d+m_v}=\frac{w}{1+w}.
\end{equation}
If the air is fairly dry, then $q \approx w$. The vapor pressure is
the molar fraction of water vapor multiplied by the total pressure
($p$), or $e=\frac{n_v}{n_d+n_v}p$. Using the masses (little $m$) and
molecular masses (big $M$), we can rewrite this as
$e=\frac{\frac{m_v}{M_v}}{\frac{m_d}{M_d}+\frac{m_v}{M_v}}p=\frac{\frac{m_v}{m_d}}{\frac{M_v}{M_d}+\frac{m_v}{m_d}}p$, or
\begin{equation}
e=\frac{w}{w+\epsilon}p,
\end{equation}
where $\epsilon=\frac{M_v}{M_d}$.

Liquid water molecules are in constant motion. This kinetic energy is
measured by temperature. Sometimes the molecules burst across the
liquid surface into the surrounding air, which is a process we call
evaporation (left panel of figure below). On the other hand, water
molecules in gaseous form also have kinetic energy, which sometimes
causes them to bash into the liquid water and join it. If we put
liquid water in an enclosed container, these two processes will
eventually balance each other (right panel of figure below). In other
words in any given amount of time, as many liquid water molecules will
be crossing the liquid threshold and becoming gaseous as gaseous water
molecules will be crashing into the liquid and becoming liquid. The
water vapor partial pressure when this equilibrium is established is
called the saturation vapor pressure ($e_s$).

\begin{figure}[h!]
\begin{center}
  \includegraphics[width=0.8\textwidth]{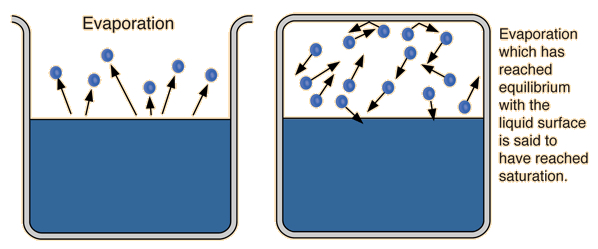}
\end{center}
\end{figure}

The general term for this type of phase transition boundary is a
Clausius-Clapeyron relationship, and this term is used for the water vapor
pressure in Earth's atmosphere. The Clausius Clapeyron relationship
tells us that the saturation vapor pressure is a strong function of
temperature (figure below).  For temperatures below 0$^\circ$C, $e_s$
is very close to zero, and the air is almost completely dry. For a
temperature of 100$^\circ$C, $e_s$ is equal to the standard
atmospheric surface pressure. In fact, this is what defines the
boiling point of water. At normal temperatures on Earth's surface, the
saturation vapor pressure increases about 7\% per degree Celcius of
warming. Another way to measure water is in terms of the relative
humidity, 
\begin{equation}
RH=100\frac{w}{w_s},
\end{equation}
or the percentage of water vapor mixing ratio relative to the
saturation water vapor mixing ratio. Water vapor and dry atmosphere
satisfy the ideal gas law idependently, so
$w=\frac{\rho_{v}}{\rho_d}=\frac{\frac{e}{R_vT}}{\frac{p-e}{R_dT}}=\frac{R_d}{R_v}\frac{e}{p-e}=\epsilon
\frac{e}{p-e}$, where we have used
$\epsilon=\frac{R_d}{R_v}=\frac{M_v}{M_d}$. This means we can rewrite
the relative humidity as
\begin{equation}
RH=100\frac{\frac{e}{p-e}}{\frac{e_s}{p-e_s}}=100\frac{e}{e_s}\frac{p-e_s}{p-e}\approx
100 \frac{e}{e_s}.
\end{equation}

\begin{figure}[h!]
\begin{center}
  \includegraphics[width=0.5\textwidth]{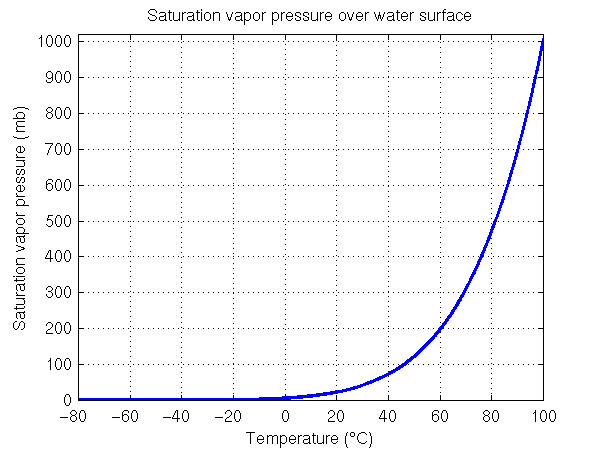}
\end{center}
\end{figure}

The dew point ($T_d$) is the temperature to which air must be cooled
at constant pressure in order for it to be saturated in water
vapor. So for air with a water vapor mixing ratio of $w$,
\begin{equation}
w=w_s(T_d).
\end{equation}
The wet bulb temperature ($T_w$) is the temperature
measured by a thermometer with a wet cloth covering its bulb. If there
is enough circulation for the ambient air to pass through the cloth,
water will evaporate into the ambient air until it is saturated,
lowering the air temperature. So the wet bulb temperature is the
temperature of a parcel of air cooled to saturation by the evaporation
of water into it. In other words, the wet bulb temperature is the
lowest temperature that can be achieved by evaporative cooling of a
surface into ambient air (notice the analogy with human cooling via
sweat). The dark lines in the figure below show contours of the wet
bulb temperature at surface pressure for different ambient
temperatures and relative humidities. As you can see, the drier the
air, the lower the wet bulb temperature. The lower left of this plot
is cut off because the fit being used is unstable in that region, not
for any physical reason.

\begin{figure}[h!]
\begin{center}
  \includegraphics[width=0.6\textwidth]{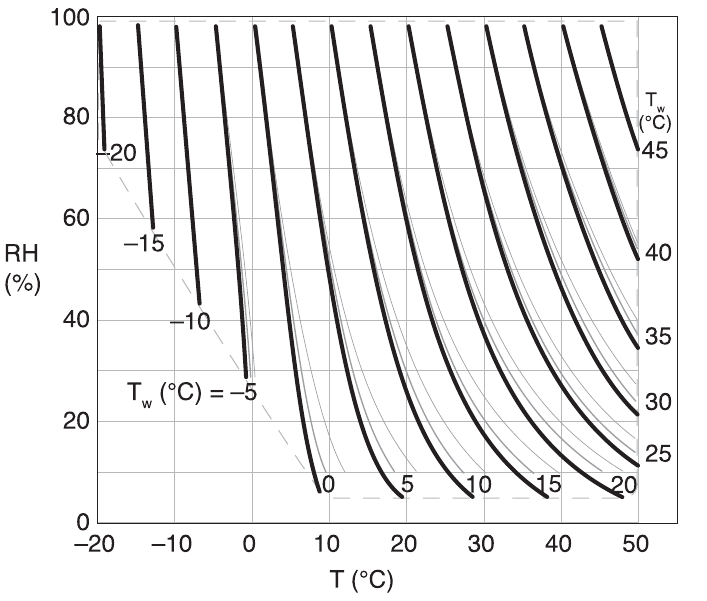}
\end{center}
\end{figure}

To understand the relationship between the dew point and the wet bulb
temperature, let's imagine air with temperature $T$ and a mixing ratio
($w_0$) that is less than saturation approaching a wet cloth. The dew
point is defined by $w_0=w_s(T_d)$. If the air passes slowly through
the cloth, it reaches saturation via evaporation of water into it, and
has a new mixing ratio ($w_f$). Note that $w_0<w_f$. The new
temperature of the air is the wet bulb temperature, which is defined
by $w_f=w_s(T_w)$. Since $w_0<w_f$, $T_d<T_w$. Both the dew point and
the wet bulb temperatures must be less than the initial temperature,
so $T_d<T_w<T$. In the special case where the air was initially
saturated, $T_d=T_w=T$.

\clearpage

\noindent {\large \textbf{Exercises}}
\bigskip

\noindent (1) Explain why hot weather causes more human discomfort
when the air is humid than when it is dry. 

\bigskip

\noindent (2) How much liquid water must evaporate to lower the
temperature of a person by 5$^\circ$C? Give your answer as a
percentage of the mass of the person and assume that the latent heat
of evaporation of water is 2.5$\times$10$^6$~J~kg$^{-1}$ and the
specific heat of the human body is
4.2$\times$10$^3$~J~kg$^{-1}$~K$^{-1}$.

\bigskip

\noindent (3) Solve for vapor pressure in terms of relative humidity,
total pressure, and saturation vapor pressure. Use this to write the
virtual temperature ($T_v=\frac{T}{1-\frac{e}{p}(1-\epsilon)}$) in
terms of relative humidity and saturation vapor pressure, instead of
in terms of vapor pressure. Do this exactly, then make an
approximation assuming air with a low specific humidity.

%% file: Flipped/moist_convection.tex
\subsection{Moist Convection}

Let's consider an air parcel with water vapor in it that rises and
experiences colder temperatures as it does. The vapor pressure is such
a strong function of temperature that this will eventually lead to
condensation if the parcel rises enough. Condensation releases heat,
so the adiabatic assumption we used to derive the dry adiabatic lapse
rate will be violated. In such a situation the actual lapse rate will
be less than the dry adiabatic lapse rate because the heat
released by condensation will raise the temperature of the air aloft
where condensation occurs. We can see this in the plot below, which
shows typical atmospheric temperature profiles in different regions of
Earth and seasons. In all cases the observed lapse rate is less than
the dry adiabatic lapse rate (10~K~km$^{-1}$). We call the lapse rate
in an atmosphere where the only non-adiabatic process we allow is the
condensation of water vapor the moist adiabatic lapse rate. The moist
adiabatic lapse rate is smaller if the water vapor mixing ratio is
higher because more condensation and heating occurs when the air is
lifted.

\begin{figure}[h!]
\begin{center}
  \includegraphics[width=0.8\textwidth]{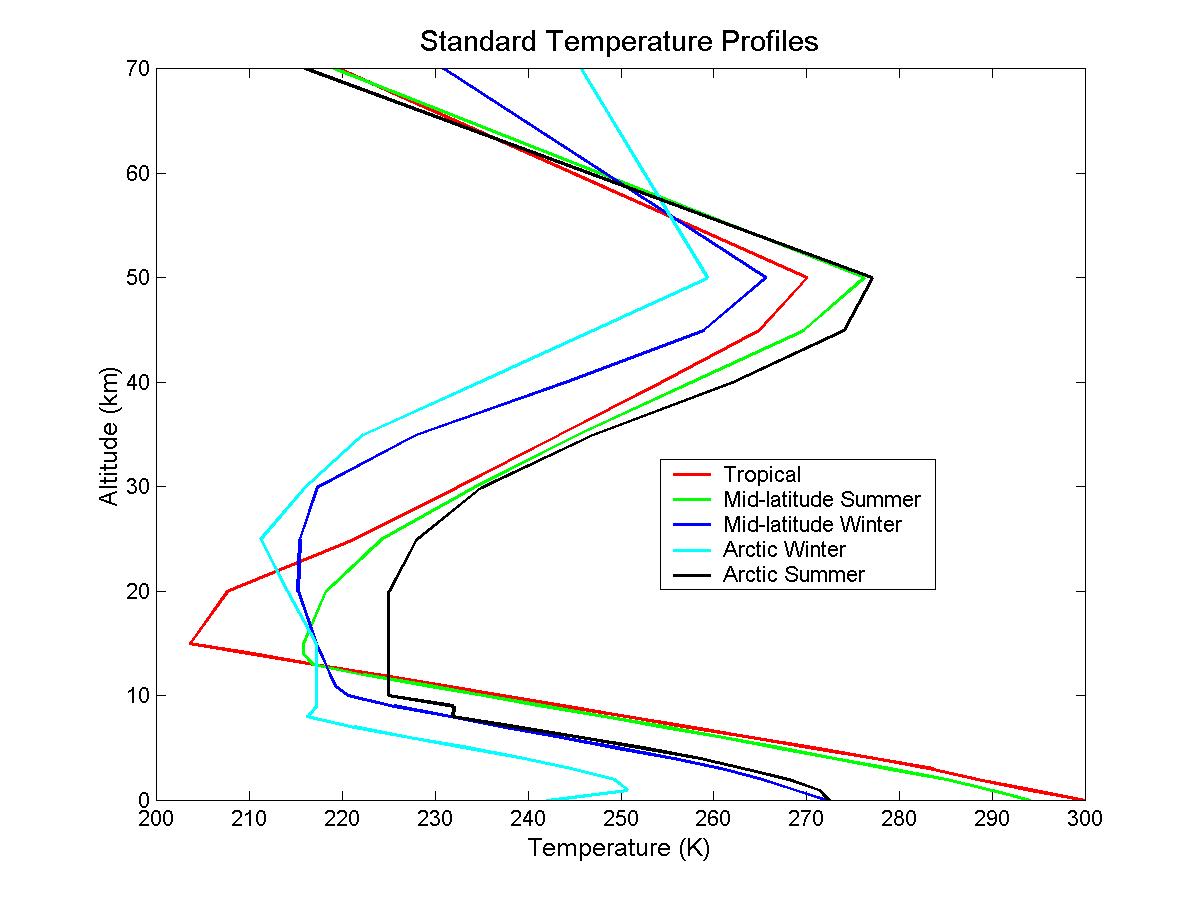}
\end{center}
\end{figure}

Now let's derive a moist analog to the potential temperature. We will
call it the equivalent potential temperature, and it will be roughly
constant in an atmosphere experiencing moist convection. Based on
Lesson 06, we know that the 1st law of thermodynamics can be written
as $\frac{dq}{T}=c_p\frac{dT}{T}-R\frac{dp}{p}$. From $\theta = T
(\frac{p_0}{p})^{\frac{R}{c_p}}$ we can find that
$c_p\frac{d\theta}{\theta}=c_p\frac{dT}{T}-R\frac{dp}{p}$ so that
$\frac{dq}{T}=c_p\frac{d\theta}{\theta}$. If the heat is released by
condensation of water vapor, then $dq = -L_v dw_s$, where $L_v$ is the
latent heat, so that
$\frac{d\theta}{\theta}=-\frac{L_v}{c_pT}dw_s$. Now notice that
$d(\frac{w_s}{T})=\frac{w_s}{T}(\frac{dw_s}{w_s}-\frac{dT}{T})\approx\frac{dw_s}{T}$,
using $\frac{dT}{T}\ll \frac{dw_s}{w_s}$ because a small fractional
change in temperature causes a large fractional change in mixing ratio
along the saturation curve. Therefore, assuming that $L_v$ and $c_p$
are constant, we can write $\frac{d\theta}{\theta}=-d\left(\frac{L_v
  w_s}{c_p T}\right)$, which we can integrate to find $\log \theta +
const. = -\frac{L_v w_s}{c_p T}$. We choose the constant to be $\log
\theta_e$, where $\theta_e$ is the equivalent potential temperature,
using the fact that the $\theta_e=\theta$ if $w_s=0$. Now we can solve
for the equivalent potential temperature as 
\begin{equation}
\theta_e=\theta \exp \left( \frac{L_v w_s}{c_p T} \right).
\end{equation}
The equivalent potential temperature is the potential temperature that
an air parcel would have if all of the water vapor it contains were
condensed and the heat from the condensation used to heat the
parcel. It is also the temperature that an air parcel would have if it
were slowly raised in the atmosphere to a low enough temperature that
essentially all of the water vapor in it had condensed, such that it
retained all the heat that the condensation released, then slowly and
adiabatically lowered to the surface.

\clearpage

\noindent {\large \textbf{Exercises}}
\bigskip

\noindent (1) The plot below shows results from numerical calculations
using a one dimensional (vertical) model of the atmosphere
\cite{manabe1967thermal}. The model can be run with convection
artificially repressed, even if a real atmosphere would be
convectively unstable. This is called radiative equilibrium. The model
can also be run including a simple parameterization of moist
convection. This is called radiative-convective equilibrium.

\begin{enumerate}
\item Is the temperature lapse rate larger in the lowest few km with
  or without convection?
\item Would it be possible for an atmosphere to exist in
the radiative equilibrium state?
\item For the atmosphere with convection, estimate the lapse rate from
  the plot. Is the lapse rate larger or smaller than the dry adiabatic
  lapse rate? Why?
\end{enumerate}

\begin{figure}[h!]
\begin{center}
  \includegraphics[width=0.5\textwidth]{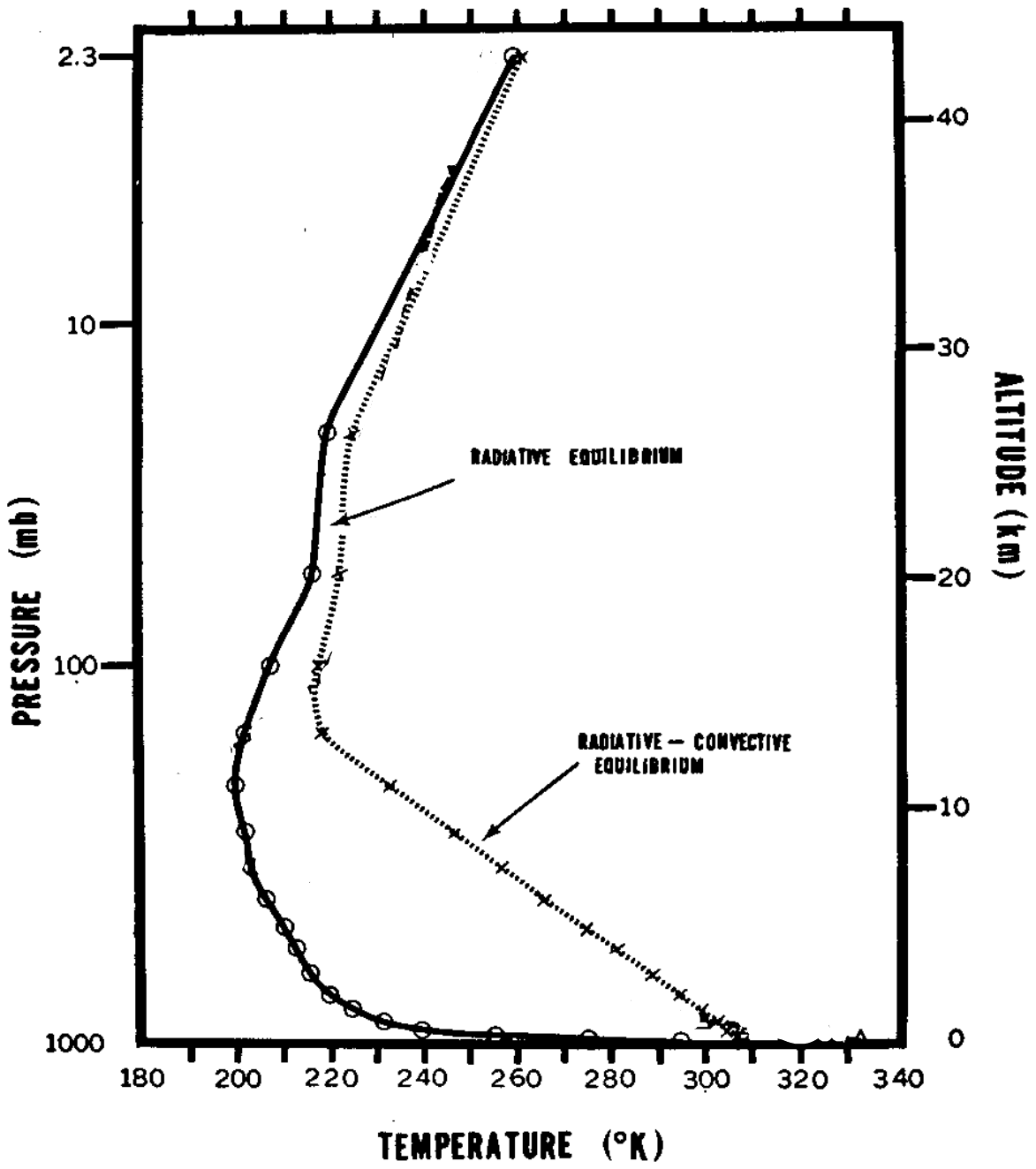} 
\end{center}
\end{figure}

\clearpage

\noindent (2) The figure below shows average profiles of temperature
variables in the tropics.

\begin{enumerate}
\item Explain why the moist potential temperature is roughly constant
  within the troposphere.
\item Explain why the dry potential temperature increases with height.
\item Explain why the moist potential temperature is always at least
  as large as the dry potential temperature.
\item Explain why the moist potential temperature is equal to the dry potential
  temperature above a pressure of about 200~mb.
\item Is the atmosphere stable or unstable to dry convection?
\end{enumerate}

\begin{figure}[h!]
\begin{center}
  \includegraphics[width=0.8\textwidth]{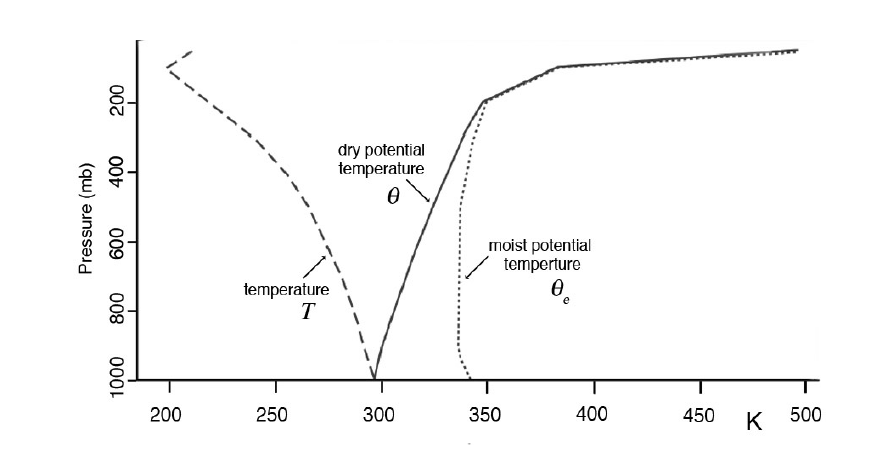} 
\end{center}
\end{figure}

%% file: Flipped/carnot.tex
\subsection{The Carnot Cycle and Hurricanes}

A heat engine is a mechanism that does work by taking heat from a hot
source ($Q_1$) and dumping it in a cold sink ($Q_2$). Heat engines
work through cycles, and all the quantities in this paragraph are
averaged over a cycle. By the conservation of energy, if the internal
energy of the heat engine stays constant, the work it does has to be
equal to the heat it takes from the source minus the heat it dumps
into the sink ($W=Q_1-Q_2$). The efficiency of a heat engine is the
work it can do divided by the heat it takes from the source
($\eta=\frac{W}{Q_1}=\frac{Q_1-Q_2}{Q_1}$).

Carnot imagined an idealized heat engine with the components shown in
the figure below. The upper panel is a cylinder with a conducting base
and insulating sides and top. It has a piston on top that can move
without friction and can either be pushed down or pulled up. There is
a gas inside the cylinder that can expand or contract. The bottom
three panels are a hot reservoir at temperature $T_1$, an insulating
stand that doesn't allow heat transfer, and a cold reservoir at
temperature $T_2$. The reserviors are so large that their temperature
doesn't change if heat is taken out or put into them.

\begin{figure}[h!]
\begin{center}
  \includegraphics[width=0.7\textwidth]{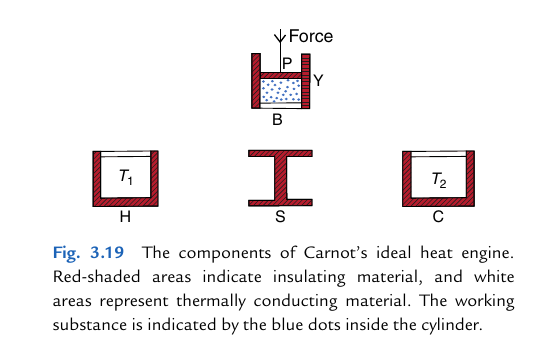} 
\end{center}
\end{figure}

The Carnot cycle is shown in pressure-volume space in the diagram
below. Imagine that we start at point B with the gas compressed (low
volume). We place the cylinder on the hot reservoir, and it expands
isothermally (at a constant temperature $T_1$) to point C, drawing
heat $Q_1$. We then place it on the insulating platform and allow it
to expand adiabatically (no heat exchange) to point D, when it has
temperature $T_2$. Then we move it to the cold reservoir and it
contracts isothermally at temperature $T_2$, dumping heat $Q_2$ to the
cold reservoir. Finally, we move it back to the insulating platform
and it contracts adiabatically until it has temperature $T_1$. We can
then repeat the cycle if we want to. We already saw that $dW=pdV$, so
the work done by or to the cylinder on each component is $W=\int
pdV$. This means that the total work done by the cylinder over the
cycle is represented by the blue area in the diagram. It turns out
that the Carnot cycle represents the most efficient heat engine
possible. That means that the efficiency of the Carnot cycle is the
theoretical maximum efficiency of a heat engine.

\begin{figure}[h!]
\begin{center}
  \includegraphics[width=0.7\textwidth]{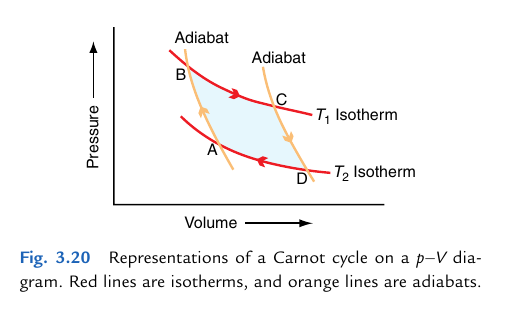} 
\end{center}
\end{figure}

Now let's calculate the efficiency of the Carnot cycle. The ideal gas
law can be written $pV=m R T$, where $m$ is the mass of gas in the
cylinder and we have used $\rho=\frac{m}{V}$. Two components of the
cycle occur at constant temperature (isothermal), and the mass of gas
in the cylinder is constant. This means that the quantity $pV$ is
constant for the isothermal components of the cycle, so that $p_B V_B
= p_C V_C$ and $p_A V_A = p_D V_D$. On the adiabatic components of the
cycle the potential temperature, $\theta$, is a constant. So $\theta =
T\left( \frac{p_0}{p} \right)^\frac{R}{c_p}=\left(\frac{p}{\rho R}
\right) \left( \frac{p_0}{p} \right)^\frac{R}{c_p}$ is constant, where
we have used the ideal gas law. Since the density is inversely
proportional to the volume ($V=\frac{const.}{\rho}$) for a constant
amount of gas, $Vp^{1-\frac{R}{c_p}}=Vp^\frac{c_v}{c_p}$ is constant,
where we have used $c_p=c_v+R$. We can rewrite this to find that
$pV^\gamma$ is constant, with $\gamma \equiv \frac{c_p}{c_v}$. This
means $p_A V_A^\gamma=p_B V_B^\gamma$ and $p_C V_C^\gamma=p_D
V_D^\gamma$. We now have four equations that we can use to eliminate
the pressures and find that $\frac{V_C}{V_B}=\frac{V_D}{V_A}$.

Returning to the 1st law of thermodynamics, $dq=c_vdT+dW$, we note
that for the isothermal components of the cycle $dq=dW$. This means
$Q_1=\int_{V_B}^{V_C} p dV=\frac{R T_1}{m}\int_{V_B}^{V_C}
\frac{1}{V}dV$. We can integrate to find $Q_1=\frac{R T_1}{m}\log
\left( \frac{V_C}{V_B} \right)$ and similarly $Q_2=\frac{R T_2}{m}\log
\left( \frac{V_D}{V_A} \right)$. Since
$\frac{V_C}{V_B}=\frac{V_D}{V_A}$,
$\frac{Q_2}{Q_1}=\frac{T_2}{T_1}$. So the efficiency of the Carnot
cycle is 
\begin{equation}
\eta = 1-\frac{Q_2}{Q_1} = 1-\frac{T_2}{T_1}. 
\end{equation}
This is the maximum possible efficiency of a heat engine, and it is
not 100\% efficient. The smaller the temperature of the cold reservior
($T_2$) and the larger the temperature of the hot reservior ($T_1$),
the more efficient the Carnot cycle is.

\clearpage

\noindent {\large \textbf{Exercises}}
\bigskip

The Carnot cycle can be used as a model to understand hurricane
strength \cite{emanuel1986air}. The diagram below shows the radial
motion of air in a hurricane. Imagine that the hurricane is axially
symmetric like a doughnut. The diagram is a slice through the doughnut
from the center to the outside. The horizontal axis is radius and the
vertical axis is height. The clouds at small radius in the diagram
represent the hurricane eye wall.

\bigskip

\begin{figure}[h!]
\begin{center}
  \includegraphics[width=0.8\textwidth]{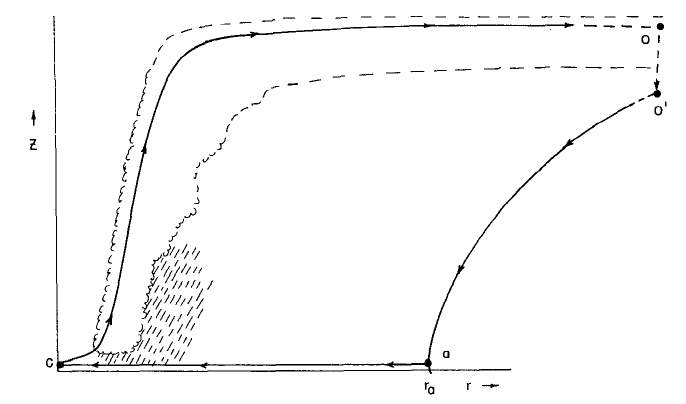} 
\end{center}
\end{figure}

\bigskip

\begin{enumerate}
\item Assume this hurricane is in the northern hemisphere. Based on
  your study of cyclones in the first week, will the surface winds of
  the storm be into or out of the page?
\item The steps of the Carnot cycle are represented by the lines with
  arrows on them connecting the letters c, o, o', and a. Identify which
  steps are approximately adiabatic and which are approximately
  isothermal.
\item During which stage of the cycle is energy added to the
  hurricane? Can you guess the way energy is transferred to the
  hurricane? This is hard, but try guessing. 
\item During which stage of the cycle is energy taken away from the
  hurricane? Can you guess the way energy is taken from the
  hurricane?
\item Carnot engines produce work. What happens to the work produced
  by the hurricane?
\item Assume that the colder isothermal stage of the cycle stays at
  the same temperature during during global warming. Based on the
  Carnot efficiency, what does this imply about how much work can be
  done by the hurricane? What does this say about how we should expect
  hurricanes to change as a result of global warming?
\end{enumerate}

%% file: Flipped/electromagnetic.tex
\subsection{Electromagnetic Radiation}

Energy can be transferred between objects by a variety of
mechanisms. The diagram below shows four types of energy transfer that
occur when a pot is heated on a stove. We've learned about convection
and latent heat in the context of the atmosphere.  If you don't know
what conduction is, look it up online now. We're going to start
talking about electromagnetic radiation now.

\begin{figure}[h!]
\begin{center}
  \includegraphics[width=0.55\textwidth]{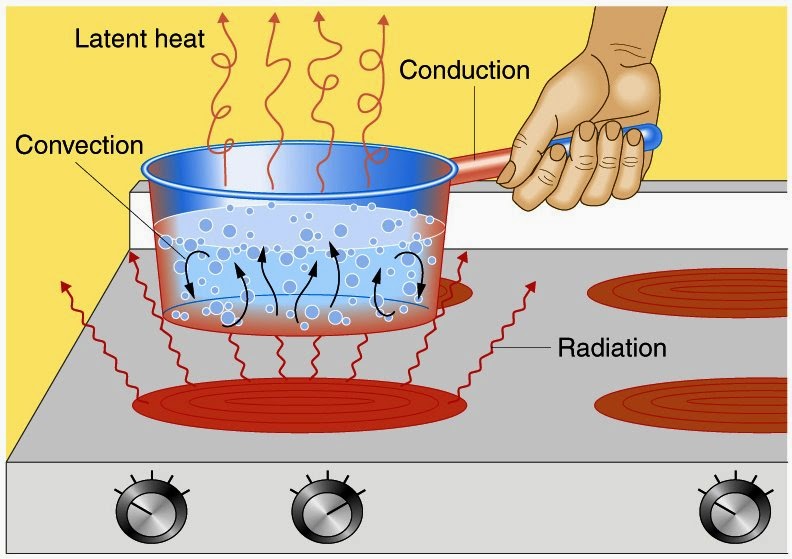}
\end{center}
\end{figure}

Electromagnetic radiation is propagating oscillations in electric and
magnetic fields. As we will learn over the next few weeks, everything
emits electromagnetic radiation all the time. The amount and type of
electromagnetic radiation that an object emits depend on its
temperature and radiative properties. Electromagnetic radiation has a
frequency ($\nu$) and a wavelength ($\lambda$). In a vacuum, it
travels at a constant speed of 
\begin{equation}
c=\lambda \nu = 3.0 \times 10^8\ m \ s^{-1}.
\end{equation}
There are many types of electromagnetic radiation (see figure
below). We can use the wavelength of electromagnetic radiation, or
light, to classify it. Visible light has a wavelength of
400--750~nm. Ultraviolet radiation has a shorter wavelength and
infrared radiation has a longer wavelength. The Sun's radiation is
mainly in the visible and near infrared (almost visible) part of the
spectrum, and is often referred to as shortwave radiation in Earth
science. The radiation Earth emits is mainly in the infrared part of
the spectrum, and is often referred to as longwave radiation in Earth
science.

\begin{figure}[h!]
\begin{center}
  \includegraphics[width=0.9\textwidth]{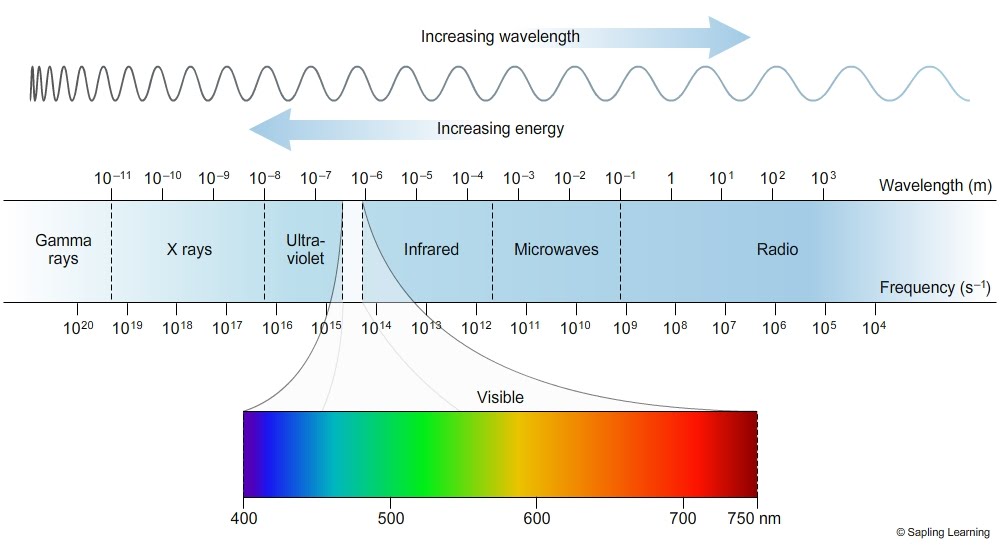}
\end{center}
\end{figure}

Quantum mechanics tells us that we can also think of electromagnetic
radiation as a particle, a photon, in addition to as a wave. The
energy of a photon is proportional to its frequency. Thinking in terms
of photons with discrete energy will be necessary for us to understand
the greenhouse effect.

\clearpage

\noindent {\large \textbf{Exercises}}
\bigskip

\noindent (1) This question concerns the pot in the first figure of
this handout.

\begin{enumerate}
\item What is the difference between conduction and convection?
\item The pot is being heated from below and convection in the water
  is moving this heat upward. If the pot were heated from above, would
  it move energy downward through conduction or convection? Why?
\item Is the troposphere or the stratosphere more like a pot being
  heated from above?
\item The right hand of the person holding the pot experiences heat by
  conduction through the metal of the handle. Suppose the person puts
   their left hand over the pot. Describe mechanistically how
  heat can be moved from the hot pot to the left hand.
\item In the picture heat is leaving the burners via radiation. If the
  person stuck their left hand over a burner, would they sense heat?
  Explain qualitatively how radiation can move energy between two
  objects that don't touch each other.
\item Given that we can see a red color coming from the burners, in
  what part of the spectrum are they emitting radiation?
\item If you touch the ground on a typical day, does it feel hotter or
  colder than the burner?
\item The part of the spectrum that something emits in is related to
  the temperature it emits at. We can see the ground on a typical day
  in various colors. Combine this fact with your answer from the
  previous part to argue whether we can see direct emission of
  radiation from the ground. If we can't see direct emission from the
  ground, how do we see it?
\item Is the radiation emitted by the ground considered longwave or
  shortwave? Is the radiation you see from the ground considered
  longwave or shortwave?
\end{enumerate}

%% file: Flipped/blackbody.tex
\subsection{Blackbody Emission}

A blackbody is a perfect emitter and absorber of electromagnetic
radiation. A blackbody emits electromagnetic radiation at different
wavelengths according to the Planck function, as shown in the diagram
below. The most important things to remember about the Planck function
is that hotter bodies emit more radiation and have a peak emission at
a shorter wavelength.

\begin{figure}[h!]
\begin{center}
  \includegraphics[width=0.55\textwidth]{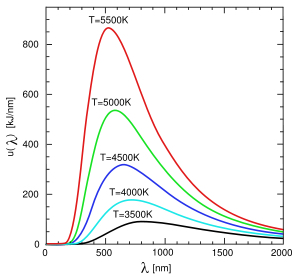}
\end{center}
\end{figure}

The decrease in the wavelength of peak emission ($\lambda_m$) as a
function of temperature is captured in Wein's displacement law, which
states that 
\begin{equation}
\lambda_m T = 2897\ \mu m \cdot K.
\end{equation} 
You can use the figure above to check that this works (just remember
to convert between nm and $\mu$m). The radiative flux emitted by an
object is the integral over wavelength of the Planck curves shown in
the figure above. The increase in radiative flux ($F$, in W~m$^{-2}$)
emitted as a function of temperature is captured in the
Stefan-Boltzmann law, 
\begin{equation}
F=\sigma T^4, 
\end{equation}
where $\sigma = 5.67 \times 10^{-8}$~W~m$^{-2}$~K$^{-4}$. Notice that
the radiative flux increases with the fourth power of temperature, so
hotter objects emit much more radiation. To use both of these laws,
always remember to convert the temperature into degrees Kelvin.

Most objects are not blackbodies. Instead they emit and absorb
radiation less efficiently than a blackbody. At a given wavelength, a
real object emits a fraction $\epsilon_\lambda$ of the radiation a
blackbody would emit. $\epsilon_\lambda$ is called the
emissivity. Similarly, at a given wavelength a real object absorbs a
fraction $a_\lambda$ of the radiation a blackbody would
absorb. $a_\lambda$ is called the absorptivity. Something called
Kirchhoff's law tells us that the absorptivity equals the emissivity at
every wavelength 
\begin{equation}
\epsilon_\lambda=a_\lambda.
\end{equation}

\clearpage

\noindent {\large \textbf{Exercises}}
\bigskip

\noindent (1) 

\begin{enumerate}
\item Your body's skin temperature is about 95$^\circ$F. Using Wien's
  displacement law, what wavelength of electromagnetic radiation do
  you tend to give off the most (in meters)?
\item Using the Stefan-Boltzmann law (assuming you, like most
  liquids/solids, have an emissivity of approximately 1), estimate the
  radiative energy flux emitted by your body in W~m$^{-2}$.
\item Estimate the surface area of your body. 
\item Based on your previous two answers, what is the power you are
  losing, in W, all the time?
\item How does the power you are losing compare to the power
  consumption of a typical incandescent light bulb? What about a
  typical fluorescent light bulb? What about your laptop?
\item A human typically consumes about 2000 Cal per day. 1 Cal=1000
  cal=4184 J. Estimate the power produced by a human due to food
  consumption, which ultimately is radiated away as heat. How does
  this compare to the total power lost by a human. If you get a value
  that is larger or smaller, how is energy balance (and a roughly
  constant temperature) maintained?
\item Earth has an emission temperature of 255~K and a radius of
  6,400~km. What is the radiative energy flux of Earth and the total
  power emitted by Earth? 
\item How does the power emitted by Earth compare to the power emitted
  by all of the human beings on planet Earth?
\end{enumerate}

%% file: Flipped/greenhouse.tex
\subsection{The Greenhouse Effect}

This is a link to a cool video
(\href{https://www.youtube.com/watch?v=c-AHtUsO6Wc}{https://www.youtube.com/watch?v=c-AHtUsO6Wc})
that demonstrates a lot about radiation.  It is filmed with infrared
cameras that are super-cooled so they don't measure themselves. The
cameras measure infrared as black and white and the colors are added
falsely. The dancers are in a cold barn during winter to maximize the
temperature contrast between their bodies and the surroundings.

Watch the video and try to notice as many interesting things as you
can. Here are some observations I made: You can see hot and cold areas
of their bodies. I think the hot corresponds to areas where blood is
near the skin surface. Muscle and fat deposits show up as cold. You
can also see the arms get progressively colder toward the hand,
reflecting cooling as the blood flows out from the heart. There are a
couple times when the dancers lay on the floor and when they move the
floor stays hot for a while. The walls of the barn are lined with
aluminum foil, which reflects infrared. At some point they dance near
the walls and you can see their heat reflected off the crinkly
surface. I think the ``trippy tracers'' are due to an extended
exposure rather than a physical effect. I think the vertical
reflections are a post-processing effect, rather than due to an
infrared mirror. Let me know if you see anything else cool worth
mentioning.

\clearpage

\noindent {\large \textbf{Exercises}}
\bigskip

Today we're going to build a simple model for Earth's greenhouse
effect. This is a long, but important worksheet. The figure below is a
simple schematic of a one-layer model of the greenhouse effect. S is
the solar flux, A is the surface albedo, G is the ground thermal flux,
and H is the thermal flux emitted by the one atmospheric layer. The
ground directly absorbs a solar flux of (1-A)S. All fluxes are in
units of W~m$^{-2}$.

\begin{figure}[h!]
\begin{center}
  \includegraphics[width=0.5\textwidth]{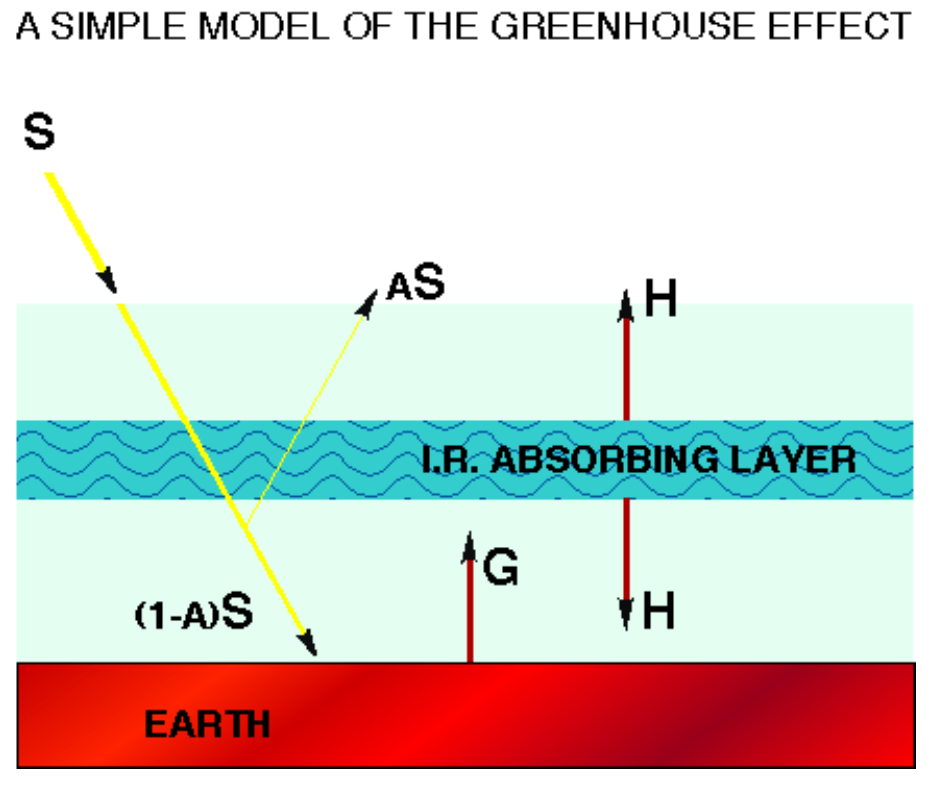}
\end{center}
\end{figure}

\noindent (1)

\begin{enumerate}
\item Find a formula for G in terms of $T_s$, the surface temperature.
  Assume that the ground has an emissivity of 1.
\item Assume the atmospheric layer has an emissivity of $\epsilon$ and
  a temperature of $T_a$. Find a formula for H.
\item S is the solar flux in W~m$^{-2}$ at the distance of Earth's
  orbit. So S Watts pass through 1 m$^2$ of area oriented
  perpendicular to the Sun's rays at Earth's distance from the Sun.
  Write the total amount of solar energy absorbed by Earth's surface
  ($S_{abs}$, in W) in terms of S, the surface albedo (A), and Earth's
  radius (R). Hint: Think about the area of sunlight that the Earth
  blocks (Earth's shadow), not the surface area of Earth.
\item Write the total amount of energy emitted by Earth's surface
  ($E_{em}$) in terms of $T_s$, $\sigma$, and R. 
\item Earth's surface can also absorb energy from the atmospheric
  layer (flux H). This is the Greenhouse effect. Write the total
  amount of atmospheric radiation energy absorbed by Earth's surface ($E_{abs}$)
  in terms of $\epsilon$, $T_a$, $\sigma$, and R.
\item For Earth to be in equilibrium we must have
  $S_{abs}+E_{abs}=E_{em}$. Plug in your formulas for each of these
  terms to determine a formula for the surface energy balance. R
  should cancel, but a geometric factor of $\frac{1}{4}$ should remain
  on one of the terms.
\item Calculate the total energy emitted by the atmospheric layer in
  terms of $\epsilon$, $T_a$, $\sigma$, and R.
\item Calculate the total energy absorbed by the atmospheric layer in
  terms of $\epsilon$, $T_s$, $\sigma$, and R. You must use
  Kirchhoff's law, which states that the emissivity of a material
  equals its absorbtivity. Also, notice that we are assuming that
  solar energy passes directly through the atmospheric layer without
  being absorbed.
\item For Earth to be in equilibrium the energy emitted and absorbed
  by the atmospheric layer must be equal. Use this to find a formula
  for the atmospheric energy balance. R should cancel.
\item Combine your formulas for surface and atmsopheric energy balance
  to eliminate $T_a$ and solve for $T_s$.
\item If the CO$_2$ is increased, which term in your solution for
  $T_s$ does this effect, and how? Use your model to predict what will
  happen to Earth's surface temperature if the CO$_2$ is increased.
\end{enumerate}

%% file: Flipped/abs_scattering.tex
\subsection{Absorption and Scattering}

As light passes through the atmosphere, it can be absorbed, scattered
in different directions, or transmitted unaffected. Often some
combination of the three occurs.  Atmospheric gases, clouds, and
aerosols (solids and liquids floating in the atmosphere) can all lead
to absorption and scattering of light. Absorption by a particular
species (a gas, cloud, or aerosol) at a particular wavelength is
characterized by the mass absorption coefficient ($\kappa_\lambda^a$),
which is measured in m$^2$~kg$^{-1}$. You can think of
$\kappa_\lambda^a$ as the area that 1~kg of the particles of the
material would take up if they were laid out in a single layer on a
flat surface. This isn't exactly right because individual particles
can weirdly cast a shadow that is not always equal to their geometric
cross section, but it's approximately correct.  What's important to
remember is that the higher $\kappa_\lambda^a$ is, the more that
species absorbs at that wavelength. Similarly, scattering by a
particular species at a particular wavelength is characterized by the
mass scattering coefficient ($\kappa_\lambda^s$). For either
absorption or scattering, we can define something called optical
thickness for each species (labled $i$)
\begin{equation}
\tau_{\lambda i}=\kappa_{\lambda i} M_i, 
\end{equation}
where $M_i$ is the mass path length, or total amount of mass of the
species in the way of the light. We have also indexed $\kappa_{\lambda
  i}$ to make it clear that absorption and scattering coefficients are
different for different species. $M_i$ is measured in kg~m$^{-2}$ and
we can calculate it between two heights as
\begin{equation}
M_i=\int_{z_1}^{z_2}\rho_i dz. 
\end{equation}
$\rho_i$ here is the density of the species under consideration, not
the total air density. If we wanted to consider the mass path length of
the entire atmosphere, then we would write $M=\int_{0}^{\infty}\rho_i
dz$. Now suppose we are considering the special case of a well-mixed
gas (such as CO$_2$) with a mass fraction of $r$, so that at any level
$\rho_i=r \rho$, where $\rho$ is the total density of air. The total
path of this gas over the entire atmosphere is 
\begin{equation} 
M=\int_{0}^{\infty}r \rho dz= -r
\int_{P_s}^{0}\frac{1}{g}dp=r\frac{P_s}{g},
\end{equation}
where we have used the hydrostatic relation ($dp=-\rho g dz$) and
$P_s$ is the surface pressure. If there are multiple species that can
absorb and/or scatter, we just have to add up the individual optical
thicknesss to get the total optical thickness 
\begin{equation}
\tau_\lambda=\sum_i \tau_{\lambda i}.
\end{equation}
We use the optical thickness in something called Beer's law,
\begin{equation}
T_\lambda=e^{-\tau_\lambda},
\end{equation}
where $T_\lambda$ is the fraction of light transmitted. In words, the
fraction of light transmitted through the atmosphere at a particular
wavelength is an exponentially decaying function of optical thickness
at that wavelength.

Absorption by gases depends strongly on wavelength. Quantum mechanics
tells us that gases can only absorb and emit electromagnetic radiation
at certain discrete energies. We can think of electromagnetic
radiation as particles (photons) in addition to waves. The energy of
photons is propotional to their frequency. This means that only
particular frequencies, and wavelengths, of light can interact with a
particular gas. We can see this in the diagram below. If we take light
from a blackbody and shine it through a prism, the colors are
separated out. If we heat up a gas to the same temperature as the
blackbody and shine the light it emits through a prism, only certain
colors will be visible. The wavelengths of the colors we see in this
case correspond to the energies that the gas can emit at. This is how
neon lights work (different gas mixtures are used to make different
colors). If instead we shine the original blackbody through the same
gas at a colder temperature and then through the prism, we see
absorption features at the same wavelengths where we previously saw
emission in the resulting spectrum. Absorption by clouds and aerosols
tends to not have as fine structure in wavelength as absorption by
gases, although there are often broad differences between absorption
of longwave and shortwave radiation.

\begin{figure}[h!]
\begin{center}
  \includegraphics[width=0.7\textwidth]{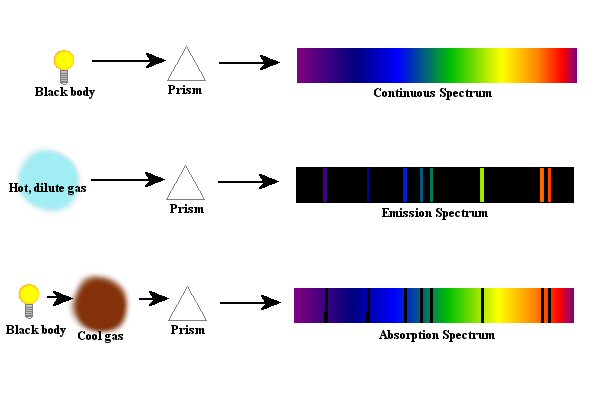}
\end{center}
\end{figure}

Scattering by particles much smaller than the wavelength of
electromagnetic radiation is called Rayleigh scattering. Atmospheric
gases Rayleigh scatter sunlight. Rayleigh scattering is proportional
to the inverse of the wavelength of light to the fourth power
\begin{equation}
\kappa_\lambda^{Rayleigh} \propto \lambda^{-4}.
\end{equation} 
This means that light at shorter wavelengths is scattered much more
than light at longer wavelengths. Given that blue light has a
wavelength of about 400~nm and red light has a wavelength of about
700~nm, this means that the atmosphere should scatter blue light about
$\left( \frac{7}{4} \right)^4=9.4$ times more than red light. As the
diagram below shows, this is why the sky appears blue. Sunlight that
is not pointed directly at us can be scattered toward us by the
atmosphere. This scattering is much more efficient for blue light than
red light, so we see the atmosphere as blue. Larger particles, such as
those that make up clouds and aerosols, scatter light through a
process called Mie scattering. The dependence of Mie scattering on
wavelength and direction is much more complicated than Rayleigh
scattering, and is not something we have time to get into. Just
remember that it exists and that scattering by clouds and aerosols can
have a large effect on the climate of a planet.

\begin{figure}[h!]
\begin{center}
  \includegraphics[width=0.5\textwidth]{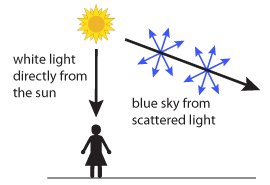}
\end{center}
\end{figure}

\clearpage

\noindent {\large \textbf{Exercises}}
\bigskip

\noindent (1) The figure on the next page shows absorption
coefficients for CO$_2$ and H$_2$O as a function of wavenumber (one
divided by wavelength). It also shows a blackbody curve at 260~K,
which gives you a sense of which wavenumbers are important for
terrestrial radiation.

\begin{enumerate}
\item An atmosphere is generally considered optically thick at a given
  wavenumber if $\tau \gtrsim 1$ (optically thick means the atmosphere
  interacts strongly with radiation). Approximately what fraction of
  radiation makes it through an atmosphere if $\tau=1$?
\item On a typical day water vapor represents on the order of 1\% of
  the mass of Earth's atmosphere. Use this to estimate the order of
  magnitude of $M$, in kg~m$^{-2}$ for water vapor. What value of
  $\kappa$ is necessary for water vapor to make Earth's atmosphere
  optically thick at a given wavenumber?
\item The mass fraction of CO$_2$ in the atmosphere is about
  $6\times10^{-4}$. What value of $\kappa$ is necessary for CO$_2$ to
  make Earth's atmosphere optically thick at a given wavenumber?
\item Based on these calculations and the figure, do you think CO$_2$
  or H$_2$O provides more greenhouse warming for Earth?
\item Use the figure to explain why CO$_2$ is an important greenhouse
  gas on Earth.
\end{enumerate}

\bigskip

\noindent (2) Explain why the Sun appears red during sunrise and sunset.

\bigskip

\noindent (3) Large, explosive volcanic eruptions loft sulfate
aerosols into the stratosphere. Particularly vibrant sunsets are often
observed around the world after such events. Explain this.

\clearpage

\begin{figure}[h!]
\begin{center}
  \includegraphics[width=0.9\textwidth]{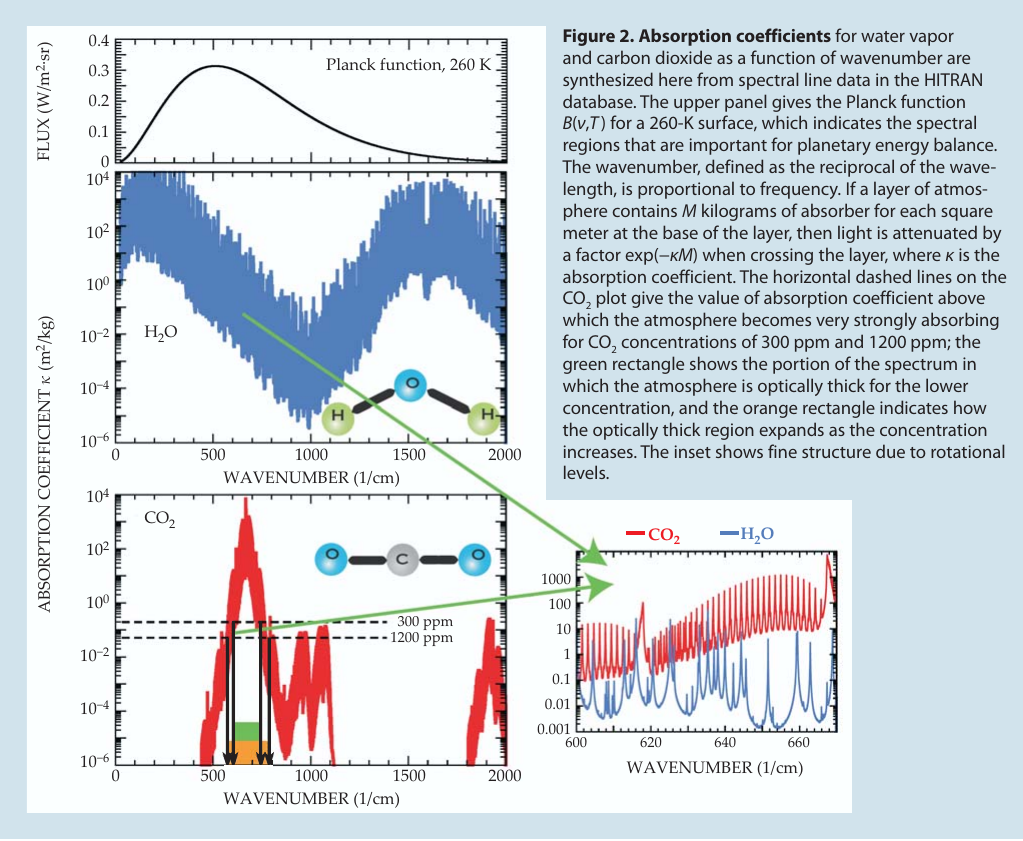}
\end{center}
\end{figure}

%% file: Flipped/rad_vertical.tex
\subsection{Radiation and Vertical Structure}

We can measure electromagnetic radiation emitted by a planet as a
function of wavelength to produce a spectrum. We might do this for
Earth using a satellite instrument, or another planet in the solar
system using an instrument on a fly-by mission or that we have put
into orbit around that planet. We can even make crude measurements of
the electromagnetic spectra of extrasolar planets orbiting stars light
years away using fancy space telescopes. Last time we saw how the
electromagnetic spectrum of a planet can reveal crucial information
about the gases present in the planet's atmosphere. Today we are going
to learn how we can use the electromagnetic spectrum of a planet to
learn about its vertical temperature structure.

The mass absorption/emission coefficients ($\kappa_\lambda^a$) of the
gases in a planet's atmosphere depend strongly on wavelength. This
means that the mass necessary for an atmosphere to be optically thick
($\tau \approx 1$) depends strongly on wavelength. Emission to space
is dominated by the region of the atmosphere where $\tau \approx
1$. At higher altitudes there isn't enough of the emitting gas for
much emission to space. At lower altitudes almost all of the emitted
radiation is absorbed by the atmosphere before it reaches space. The
emitted radiation we measure from space at a given wavelength ends up
looking like blackbody emission with a temperature corresponding to
the atmospheric temperature where $\tau \approx 1$. We call this
equivalent blackbody temperature the brightness temperature, and it is
very important to remember that it depends on wavelength. At
wavelengths where the atmosphere absorbes well (large
$\kappa_\lambda^a$), $\tau \approx 1$ will occur high up in the
atmosphere and the brightness temperature will correspond to the
atmospheric temperature at these high altitudes. At wavelengths where
the atmosphere absorbs poorly (small $\kappa_\lambda^a$), $\tau
\approx 1$ will occur low in the atmosphere and the brightness
temperature will correspond to the atmospheric temperature at these low
altitudes. There may even be regions of the spectrum where the
atmosphere absorbs so poorly ($\kappa_\lambda^a \approx 0$) that we
can see emission from the planet's surface. In this way we can use a
measurement of the electromagnetic spectrum of a planet to infer
things such as its vertical temperature lapse rate and whether it has
a stratosphere.

\clearpage

\noindent {\large \textbf{Exercises}}
\bigskip

\noindent (1) The figure below shows measured radiative emission as a
function of wavenumber for Earth, Mars, and Venus as well as reference
blackbody curves.

\begin{enumerate}
\item Why does Earth's emission in the main CO$_2$ band overlap with the
  220~K backbody?
\item Between about 800--1000 1/cm, why does Earth's emission overlap
  with the 285~K blackbody?
\item Why does Earth's emission in regions where H$_2$O absorbs
  correspond to a higher brightness temperature than in regions where
  CO$_2$ absorbs?
\item If you were going to design an efficient greenhouse gas to add
  to Earth's atmosphere, at what wavenumbers would you want it to
  absorb?
\item At the very center of the CO$_2$ and O$_3$ bands in Earth's
  emission spectrum the brightness temperature abruptly
  increases. This is marked by arrows in the top panel. Can you
  explain this? Note that the absorption/emission coefficient is much
  stronger right at the center of a band than a little bit off the
  center. Also think carefully about the vertical temperature
  structure in Earth's atmosphere.
\item Why do Mars's and Venus's infrared emission spectrums look so
  similar even though Mars has a mean surface temperature of about
  220~K and Venus has a mean surface temperature of about 735~K?
\item In 1960, as part of his PhD thesis at the University of Chicago,
  Carl Sagan estimated the surface temperature of Venus
  \cite{Sagan-1960}. He did this using observations of the planet in the
  microwave part of the electromagnetic spectrum. Explain why
  microwave emission might be different for Venus and Mars, despite
  the fact that infrared emission is so similar.
\end{enumerate}

\begin{figure}[h!]
\begin{center}
  \includegraphics[width=0.79\textwidth]{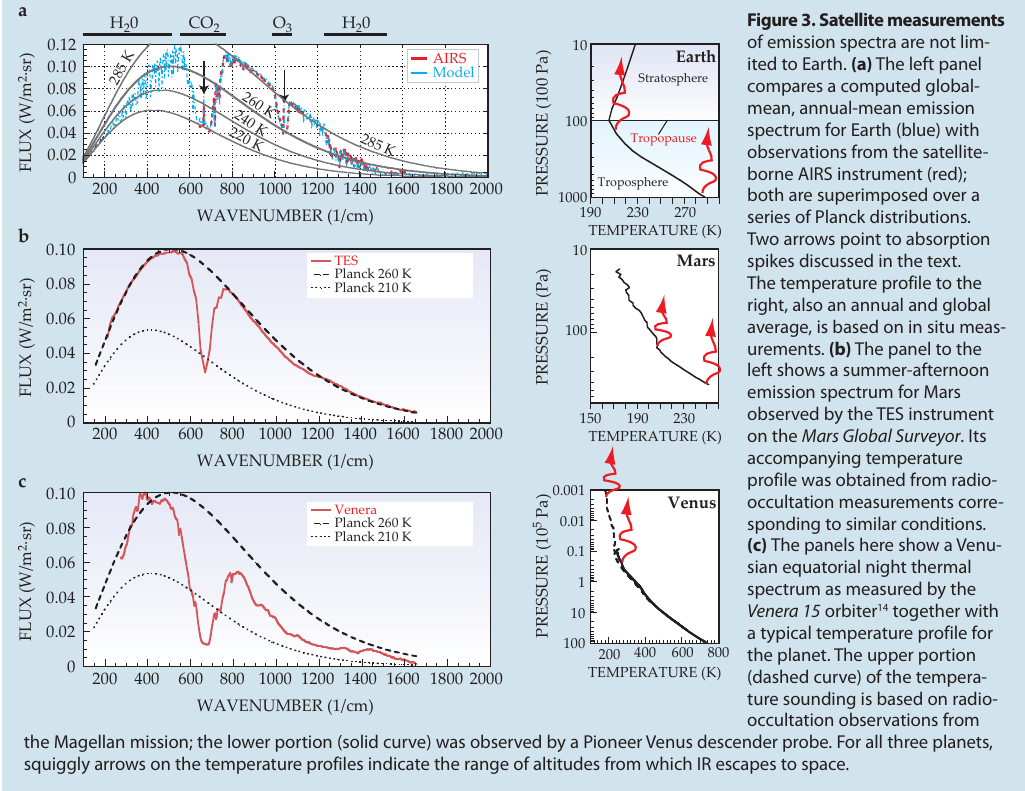}
\end{center}
\end{figure}

%% file: Flipped/energy_balance.tex
\subsection{Global Energy Balance}

The figure below is a diagram of Earth's global-mean energy budget
both at the surface and at the top-of-atmosphere (TOA). This budget
was put together through a painstaking compilation of available
surface, aircraft, balloon, and satellite measurements. Study the
diagram a bit to get a sense of what the big players in Earth's energy
budget are.

\begin{figure}[h!]
\begin{center}
  \includegraphics[width=0.8\textwidth]{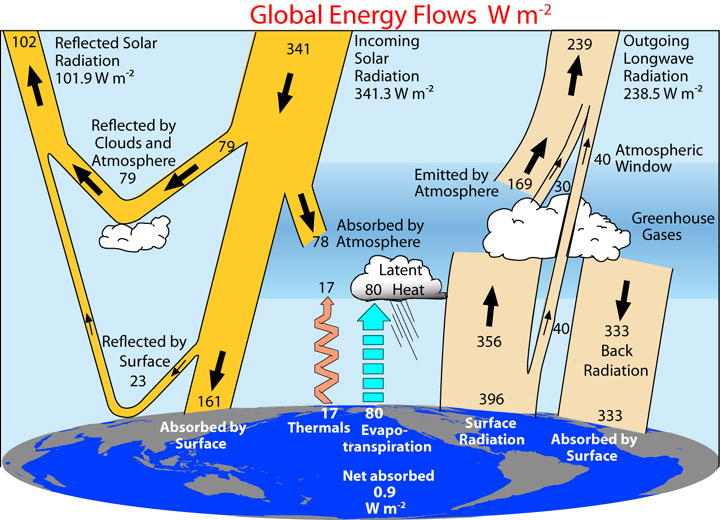}
\end{center}
\end{figure}

Today we're going to think about annual-mean maps of TOA radiation
measurements from the CERES satellite instrument. We can look at these
maps to understand a lot about Earth. The maps show longwave emission,
net shortwave (downward minus upward), and net radiation (net
shortwave minus longwave). Longwave radiation is directly emitted by
terrestrial sources. Shortwave radiation is emitted by the Sun and
impinges on Earth. Some shortwave radiation is reflected back to
space, and the plot shows the net shortwave. The net radiation tells
us about the overall impact of both shortwave and longwave
radiation. Where the net radiation is positive, radiation has a
warming effect on the planet in the annual-mean. Where the net
radiation is negative, radiation has a cooling effect on the planet in
the annual mean.

\begin{figure}[h!]
\begin{center}
  \includegraphics[width=0.8\textwidth]{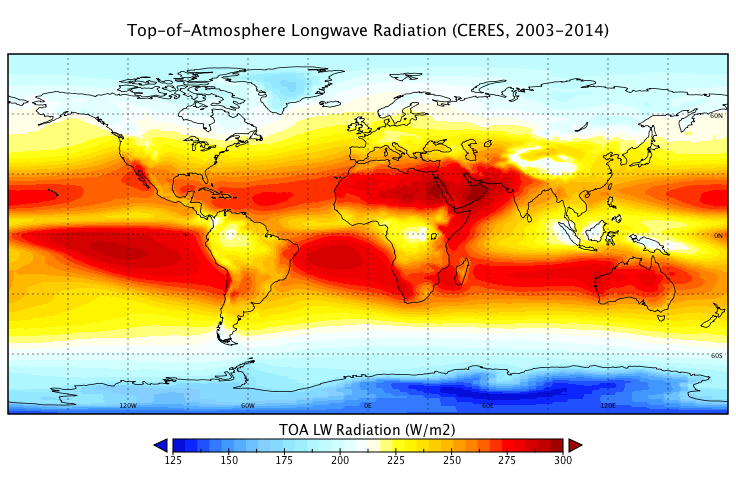}
\end{center}
\end{figure}

\begin{figure}[h!]
\begin{center}
  \includegraphics[width=0.8\textwidth]{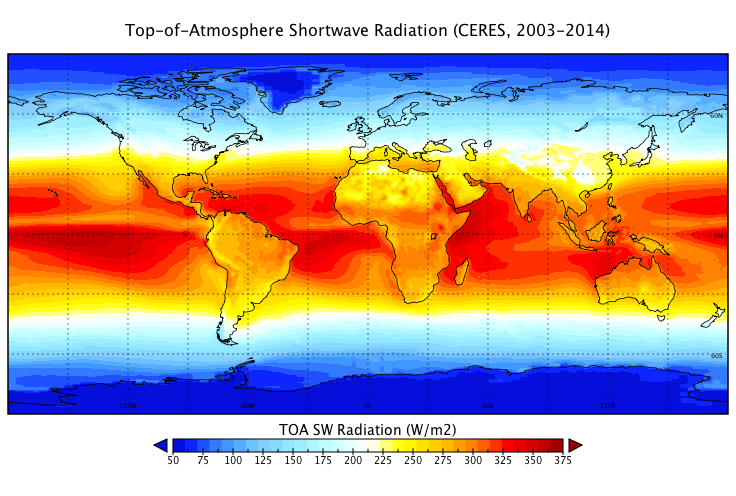}
\end{center}
\end{figure}

\begin{figure}[h!]
\begin{center}
  \includegraphics[width=0.8\textwidth]{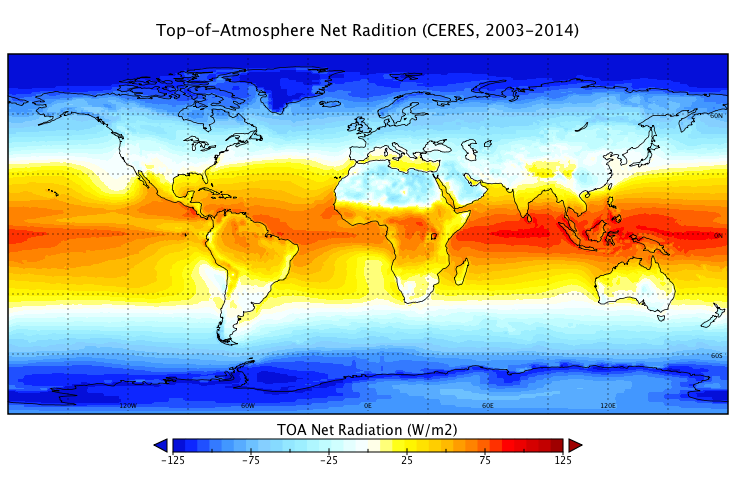}
\end{center}
\end{figure}

\clearpage 

\noindent {\large \textbf{Exercises}}
\bigskip

\noindent (1) Consider the maps of annual-mean top-of-atmosphere (TOA)
radiation measured by the CERES satellite instrument.

\begin{enumerate}

\item If we take a broad average, does longwave emission tend to be higher
  in the tropics or near the poles? Why?
\item Why is the longwave emission over Indonesia small even though the
  surface temperature is high? 
\item The annual mean incoming shortwave does not depend on longitude.
  The annual mean TOA shortwave plotted here does depend on longitude.
  Explain.
\item The TOA Net map shows that if we take a broad average, energy
  enters the planet in the tropics and leaves it near the poles.
  Explain why this means there must be heat continually transported
  from the tropics to the poles.
\item What are two reasons the Sahara and the
  Arabian Peninsula might tend to give off more energy than they
  receive? Using the longwave and shortwave maps, which effect do you think is
  dominant? (Hint: what is the most powerful greenhouse gas in Earth's
  atmosphere, and how does it show up in these maps?)
\item Why are there blobs of net radiation loss to the west off Baja
  California and Peru? Why don't these blobs show up on the longwave map?
\end{enumerate}

%% file: Flipped/partial_derivative_streamfunction.tex
\subsection{Partial Derivatives and the Streamfunction}

Today we will start thinking about atmospheric dynamics. In order to
understand atmospheric dynamics we will need to review some concepts from
calculus. The partial derivative is the derivative of a function of
multiple variables with respect to just one of them. For example, if
we have a function of two variables $f(x,y)$, we can write the partial
derivative with respect to $x$ as $\frac{\partial f(x,y)}{\partial x}$
and the partial derivative with respect to $y$ as $\frac{\partial
  f(x,y)}{\partial y}$. When we take $\frac{\partial f(x,y)}{\partial
  x}$, we treat $y$ as if it were just another constant. Similarly,
when we take $\frac{\partial f(x,y)}{\partial y}$, we treat $x$ as if
it were just another constant.

One useful place partial derivatives come up is with something called
the streamfunction. Under certain conditions a streamfunction can be
found for a two-dimensional flow. The streamfunction ($\psi$) is
defined such that 
\begin{eqnarray}
u &=& -\frac{\partial \psi}{\partial y}, \\
v &=& \frac{\partial \psi}{\partial x}, 
\end{eqnarray}
where $u$ is the velocity in the $x$ direction and $v$ is the velocity
in the $y$ direction. The velocity vector points along contours of a
streamfunction (lines in $(x,y)$ space where the streamfunction is
constant), so the streamfunction is a scalar field that can be used to
represent a vector field.

An analogy to help you think about the streamfunction is as
follows. Imagine a wanderer wandering in hilly terrain. If we choose
the way she walks carefully, then the topography would be the
streamfunction of her flow. In particular, imagine that the wanderer
always wanders at constant altitude, or at a constant value of the
streamfunction in this analogy. She always wanders with her right hand
facing the slope, and she moves with a speed proportional to the
steepness of the slope. If she moves like this, then the topography
is her streamfunction. You have probably seen a topographic map
before. A topographic map is an example of a contour plot, and we will
be working with contour plots of the streamfunction below.

\clearpage

\noindent {\large \textbf{Exercises}}
\bigskip

\noindent (1) Let's practice taking partial derivatives.
\begin{enumerate}
\item $f(x,y)=x^2y^3$, calculate $\frac{\partial f}{\partial x}$ and
  $\frac{\partial f}{\partial y}$.
\item $f(x,y)=\frac{\cos(x)}{y^2+\sin(x)}$, calculate
  $\frac{\partial f}{\partial x}$ and $\frac{\partial f}{\partial y}$.
\item $u(x,y)=x \sin(x y)$ and $v(x,y)=\sin(x) \cos(y)$, calculate
  $\frac{\partial u}{\partial x}+\frac{\partial v}{\partial y}$.
\item $u(x,y)=x \sin(x y)$ and $v(x,y)=\sin(x) \cos(y)$, calculate
  $\frac{\partial v}{\partial x}-\frac{\partial u}{\partial y}$.
\end{enumerate}

\bigskip

\noindent (2) The figure on the next page is a contour plot of the
streamfunction ($\psi$) of a flow. You may have seen a topographic map
before, and this contour plot is kind of like that. Positive contours
are represented by solid lines (and you can imagine them coming out of
the page) and negative contours are represented by dotted lines (and
you can imagine them going into the page). Remember, the velocity
points along the direction of constant $\psi$, or along the
streamfunction contours.  There are four maximums or minimums of the
streamfunction, which are labeled by letters. Each maximum and minimum
in the streamfunction represents a vortex with velocity vectors
pointing around it in either the clockwise or counterclockwise
direction.

\begin{enumerate}
\item Using the definition of the streamfunction, you can infer the
  direction of flow around the vortexes. Which vortexes represent
  clockwise flow?
\item Which vortexes represent counterclockwise flow?
\item The speed is the magnitude of the velocity vector. Is the speed
  larger in vortex \textbf{A} or vortex \textbf{C}?
\item Is the speed
  larger in vortex \textbf{A} or vortex \textbf{B}?
\end{enumerate}

\clearpage

\begin{figure}[h!]
\begin{center}
  \includegraphics[width=0.9\textwidth]{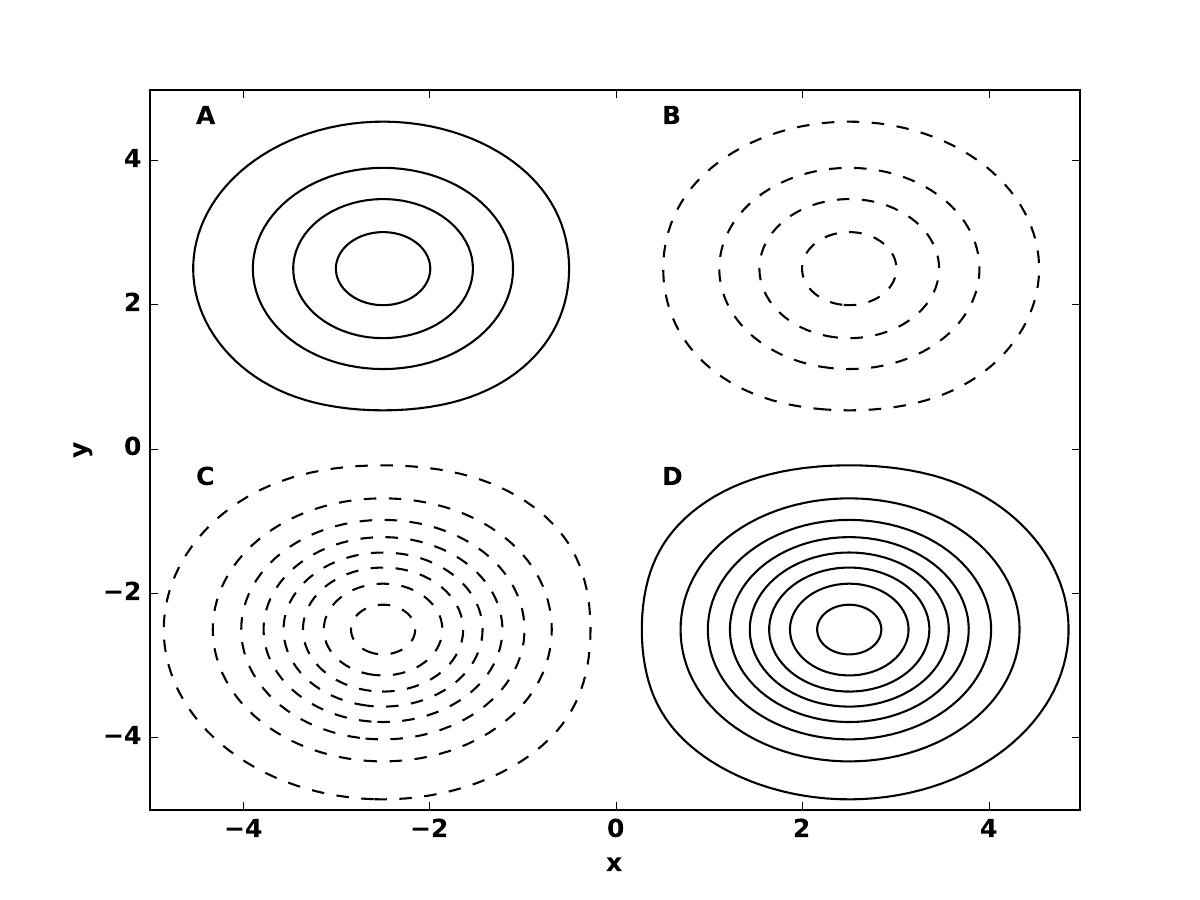}
\end{center}
\end{figure}

%% file: Flipped/divergence.tex
\subsection{The Divergence}

The divergence is an operator on a vector field that
tells us about the source of the field. We will consider
two-dimensional flow, so we can think of a velocity vector $\vec{v}$
with two components $(u,v)$. The divergence can be written as
\begin{equation}
\vec{\nabla} \cdot \vec{v} = \frac{\partial u}{\partial x} +
\frac{\partial v}{\partial y}.
\end{equation}
If $\vec{\nabla} \cdot \vec{v} > 0$ at a point then the field is
diverging and we can think of the flow as spreading out from that
point. If $\vec{\nabla} \cdot \vec{v} < 0$ at a point then the field
is converging and we can think of the flow as flowing toward that
point.

To consider a concrete example, let's think about the bathtub pictured
below. Consider the horizontal, two-dimensional flow of water in the
bathtub. If the drain is plugged and we turn the spigot on, the
horizontal flow will have to diverge from where the water from the
spigot enters the tub because this is a source of water. The
divergence of the horizontal flow will be positive where the water
enters the tub. Now imagine we fill the tub and turn the spigot
off. If we unplug the drain, the drain will be a sink of water, and
the horizontal flow will have to converge toward it. The divergence of
the horizontal flow will be negative where the water enters the drain.

\bigskip 

\begin{figure}[h!]
\begin{center}
  \includegraphics[width=\textwidth]{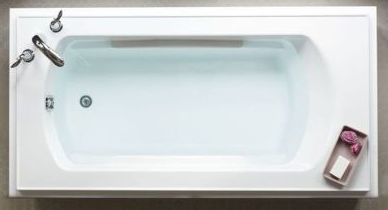}
\end{center}
\end{figure}

\clearpage

\noindent {\large \textbf{Exercises}}
\bigskip

\noindent (1) The plot below shows the 2D velocity vector field
$\vec{v}=(u,v)=(4,2)$. 

\begin{enumerate}
\item From looking at the plot, is $\vec{\nabla} \cdot \vec{v}$
  positive, negative, or zero at $(x,y)=(0,0)$?
\item Calculate $\vec{\nabla} \cdot \vec{v}$ at $(x,y)=(0,0)$.
\end{enumerate}

\begin{figure}[h!]
\begin{center}
  \includegraphics[width=\textwidth]{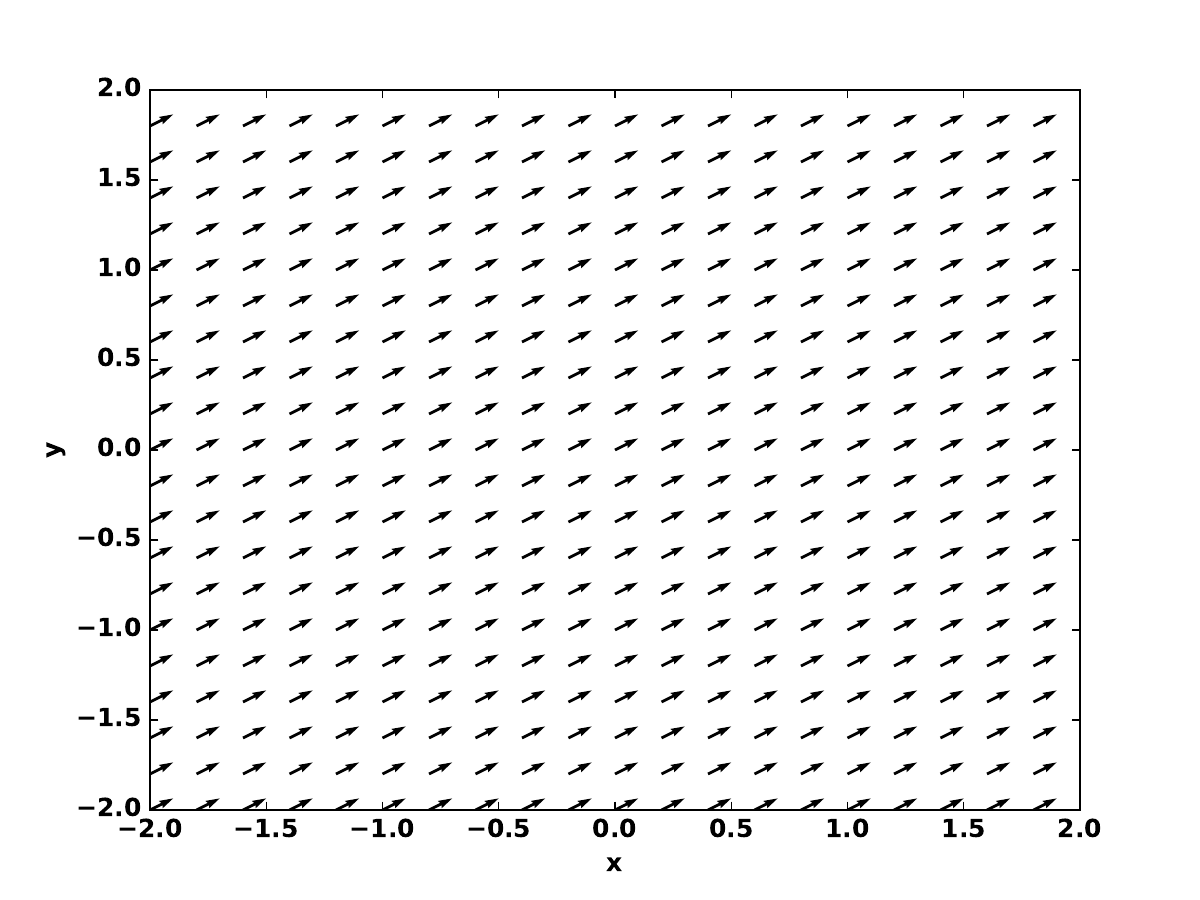}
\end{center}
\end{figure}

\clearpage

\noindent (2) The plot below shows the 2D velocity vector field
$\vec{v}=(u,v)=(\sin(x),0)$. 

\begin{enumerate}
\item From looking at the plot, is $\vec{\nabla} \cdot \vec{v}$
  positive, negative, or zero at $(x,y)=(0,0)$?
\item Calculate $\vec{\nabla} \cdot \vec{v}$ at $(x,y)=(0,0)$.
\end{enumerate}

\begin{figure}[h!]
\begin{center}
  \includegraphics[width=\textwidth]{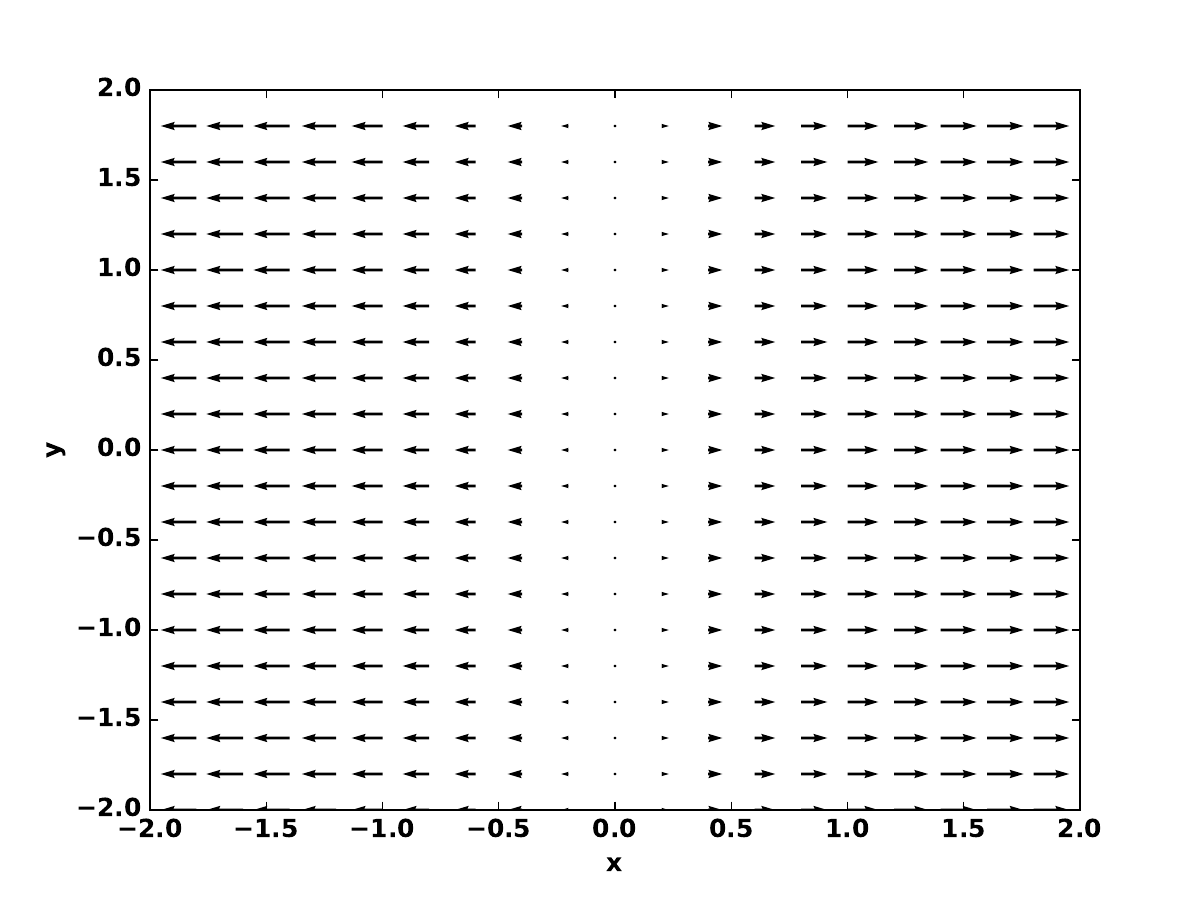}
\end{center}
\end{figure}

\clearpage

\noindent (3) The plot below shows the 2D velocity vector field
$\vec{v}=(u,v)=(y,-x)$. 

\begin{enumerate}
\item From looking at the plot, is $\vec{\nabla} \cdot \vec{v}$
  positive, negative, or zero at $(x,y)=(0,0)$?
\item Calculate $\vec{\nabla} \cdot \vec{v}$ at $(x,y)=(0,0)$.
\end{enumerate}

\begin{figure}[h!]
\begin{center}
  \includegraphics[width=\textwidth]{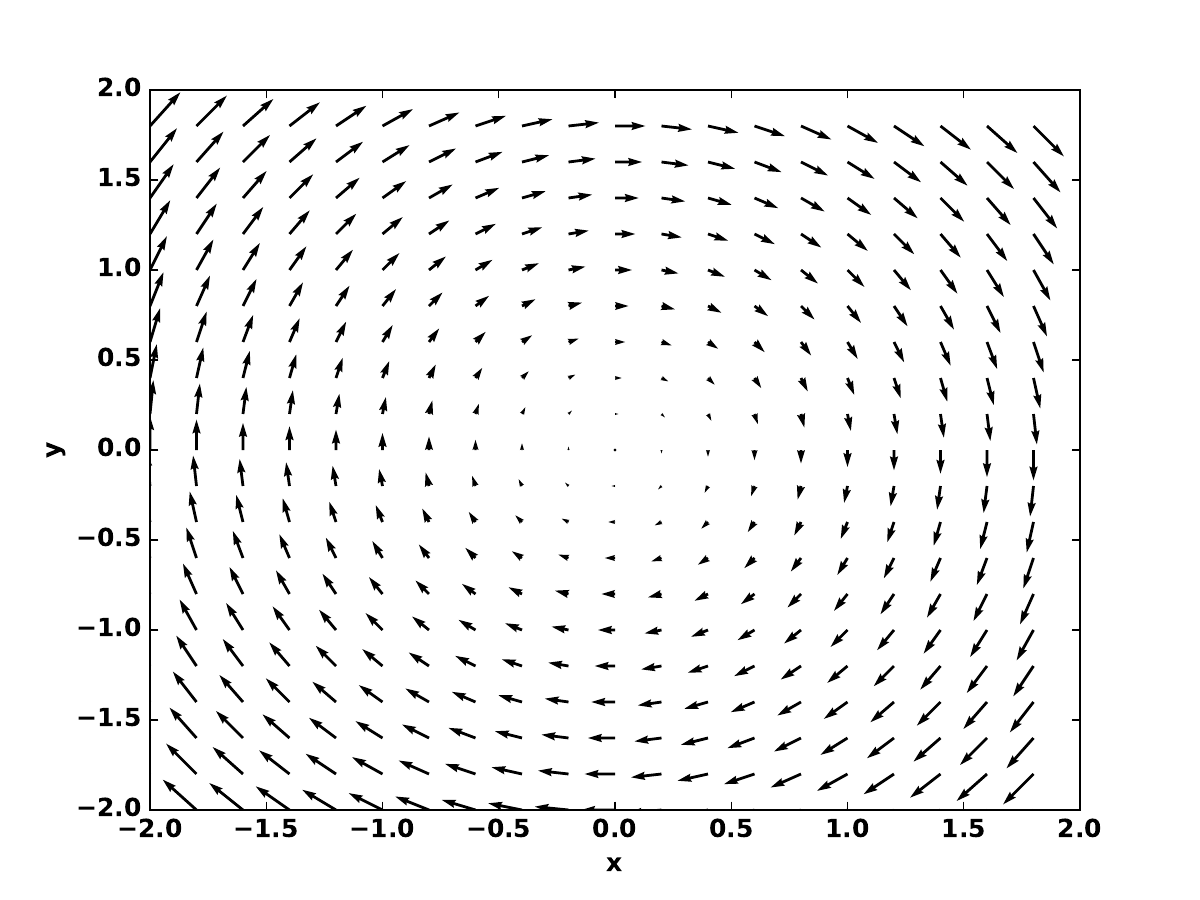}
\end{center}
\end{figure}

\clearpage

\noindent (4) The plot below shows the 2D velocity vector field
$\vec{v}=(u,v)=(y-x,-x-y)$. 

\begin{enumerate}
\item From looking at the plot, is $\vec{\nabla} \cdot \vec{v}$
  positive, negative, or zero at $(x,y)=(0,0)$?
\item Calculate $\vec{\nabla} \cdot \vec{v}$ at $(x,y)=(0,0)$.
\end{enumerate}

\begin{figure}[h!]
\begin{center}
  \includegraphics[width=\textwidth]{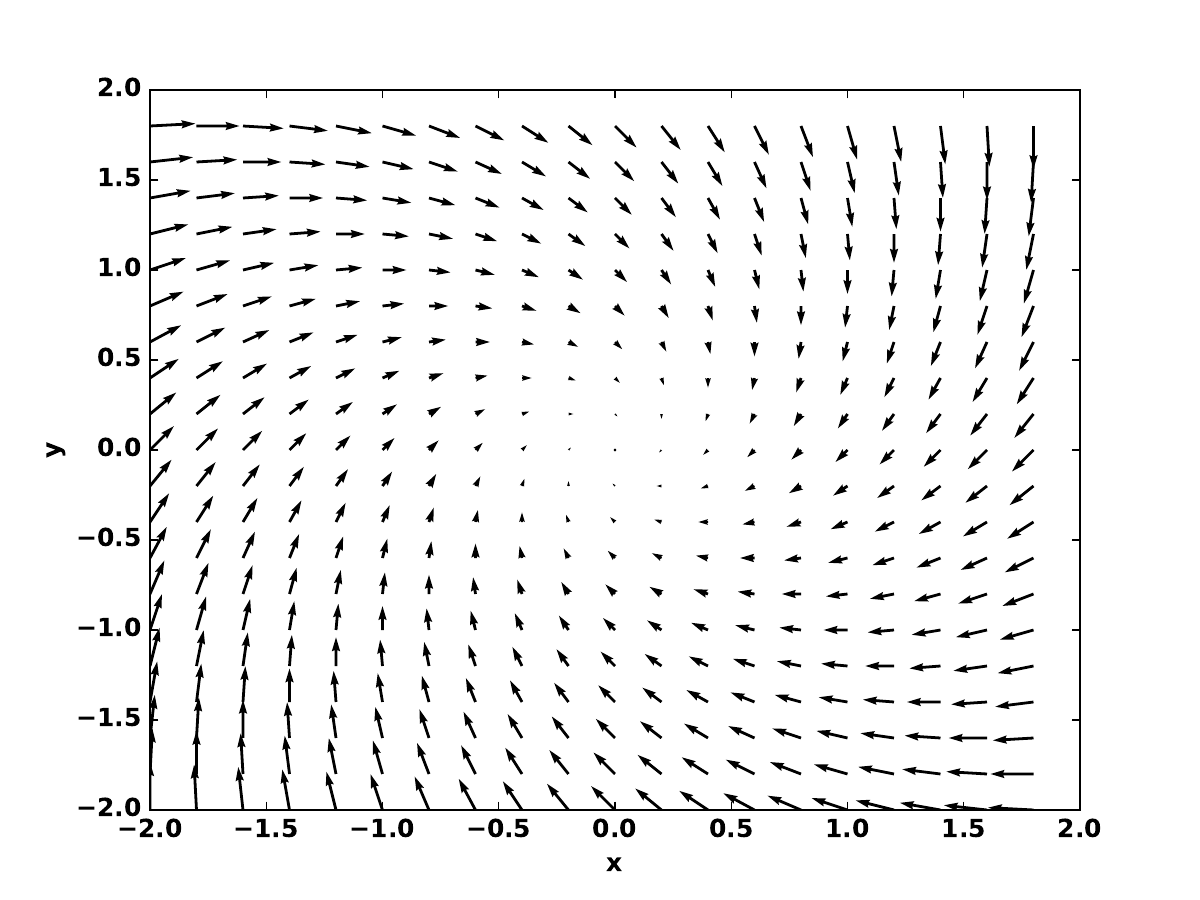}
\end{center}
\end{figure}

\clearpage

\noindent (5) The plot below shows the 2D velocity vector field
$\vec{v}=(u,v)=(y,0)$. 

\begin{enumerate}
\item From looking at the plot, is $\vec{\nabla} \cdot \vec{v}$
  positive, negative, or zero at $(x,y)=(0,0)$?
\item Calculate $\vec{\nabla} \cdot \vec{v}$ at $(x,y)=(0,0)$.
\end{enumerate}

\begin{figure}[h!]
\begin{center}
  \includegraphics[width=\textwidth]{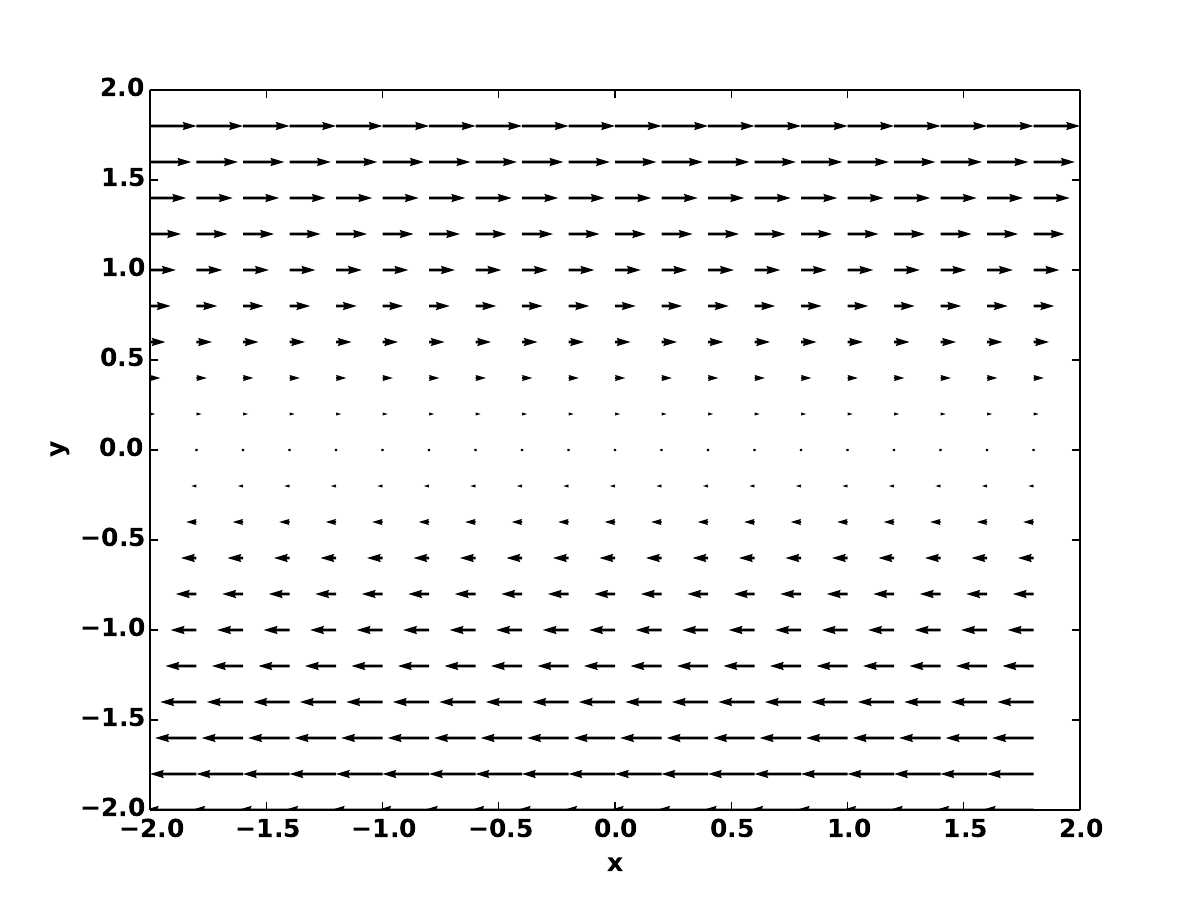}
\end{center}
\end{figure}

\clearpage

\noindent (6) In the last class we learned that for certain flows we
can define a streamfunction ($\psi$) such that
$u=-\frac{\partial \psi}{\partial y}$ and
$v=\frac{\partial \psi}{\partial x}$. Calculate the divergence
$\vec{\nabla} \cdot \vec{v}$ in terms of the streamfunction, $\psi$.
Your answer should indicate that flows for which we can define a
streamfunction have a special property. What is it?

%% file: Flipped/curl.tex
\subsection{The Curl and Vorticity}

The curl is an operator on a vector field that tells us about the
infinitesimal rotation at different points in the field.
Qualitatively, the curl of a velocity field at a particular point
tells us how much ``spinniness'' or ``whirling'' there is at that
point. If we imagine putting a tiny water wheel or turnstyle into the
velocity field at a given point, the curl of the velocity field would
tell us which direction and how fast the water wheel would turn. The
curl of a velocity field is also called the vorticity. The vorticity
is incredibly important for understanding atmospheric circulation
because cyclones are clumps of spinniness in the flow. We will
consider two-dimensional flow, so we can think of a velocity vector
$\vec{v}$ with two components $(u,v)$. For two-dimensional flow, the
curl of the velocity vector can be written 
\begin{equation}
\vec{\nabla} \times \vec{v} = \frac{\partial v}{\partial x} -
\frac{\partial u}{\partial y}.
\end{equation}

A trick we can use to assess the sign of the vorticity is the
``right-hand rule.'' To use the right-hand rule, start with a vector
plot of velocity. Put your right hand down where you want to assess
the vorticity. Fold your fingers toward the palm of your hand in the
direction that the velocity is swirling. If your thumb is pointed away
from the page, then the vorticity is positive. If your thumb is
pointed toward the page, then the vorticity is negative. If the
velocity isn't swirling, then the vorticity is zero.

\clearpage

\noindent {\large \textbf{Exercises}}
\bigskip

\noindent (1) The plot below shows the 2D velocity vector field
$\vec{v}=(u,v)=(4,2)$. 

\begin{enumerate}
\item From looking at the plot, is $\vec{\nabla} \times \vec{v}$
  positive, negative, or zero at $(x,y)=(0,0)$?
\item Calculate $\vec{\nabla} \times \vec{v}$ at $(x,y)=(0,0)$.
\end{enumerate}

\begin{figure}[h!]
\begin{center}
  \includegraphics[width=\textwidth]{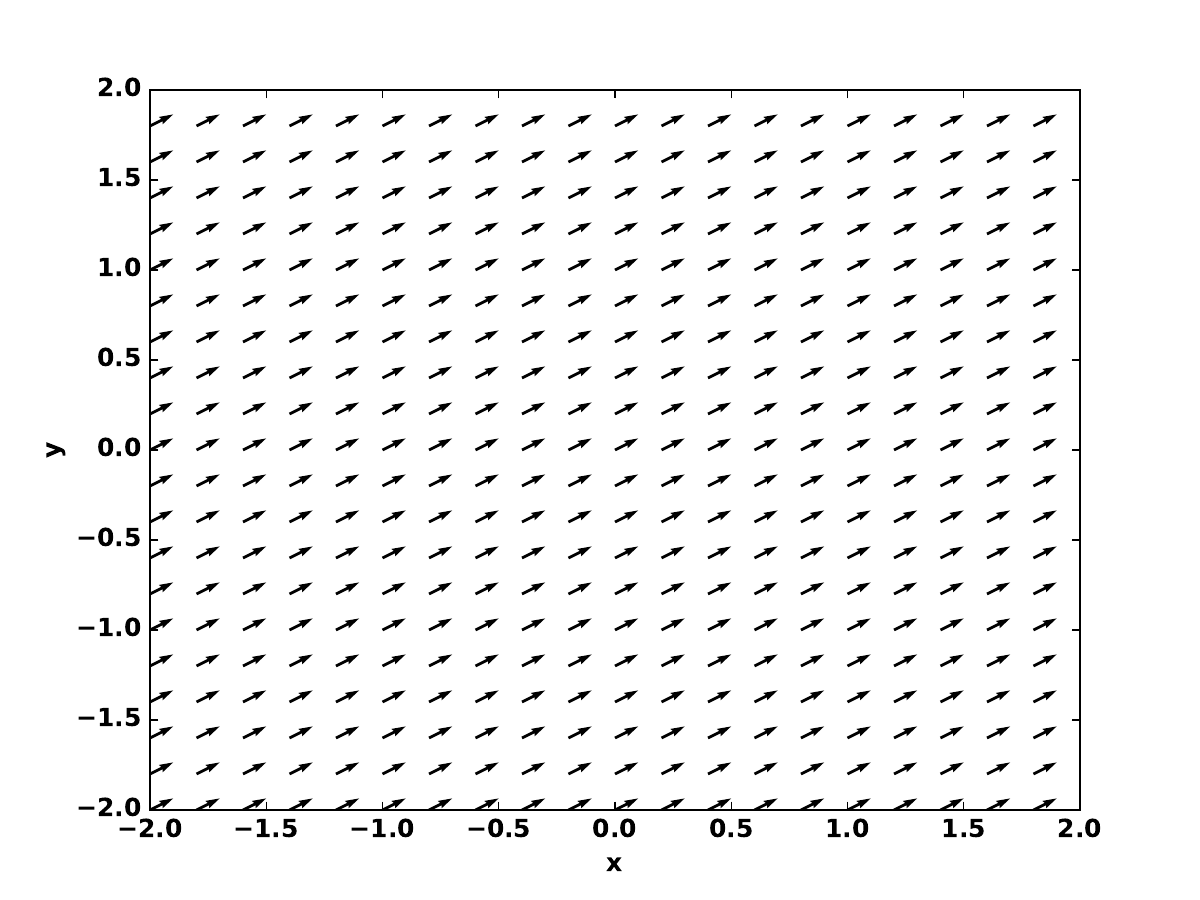}
\end{center}
\end{figure}

\clearpage

\noindent (2) The plot below shows the 2D velocity vector field
$\vec{v}=(u,v)=(\sin(x),0)$. 

\begin{enumerate}
\item From looking at the plot, is $\vec{\nabla} \times \vec{v}$
  positive, negative, or zero at $(x,y)=(0,0)$?
\item Calculate $\vec{\nabla} \times \vec{v}$ at $(x,y)=(0,0)$.
\end{enumerate}

\begin{figure}[h!]
\begin{center}
  \includegraphics[width=\textwidth]{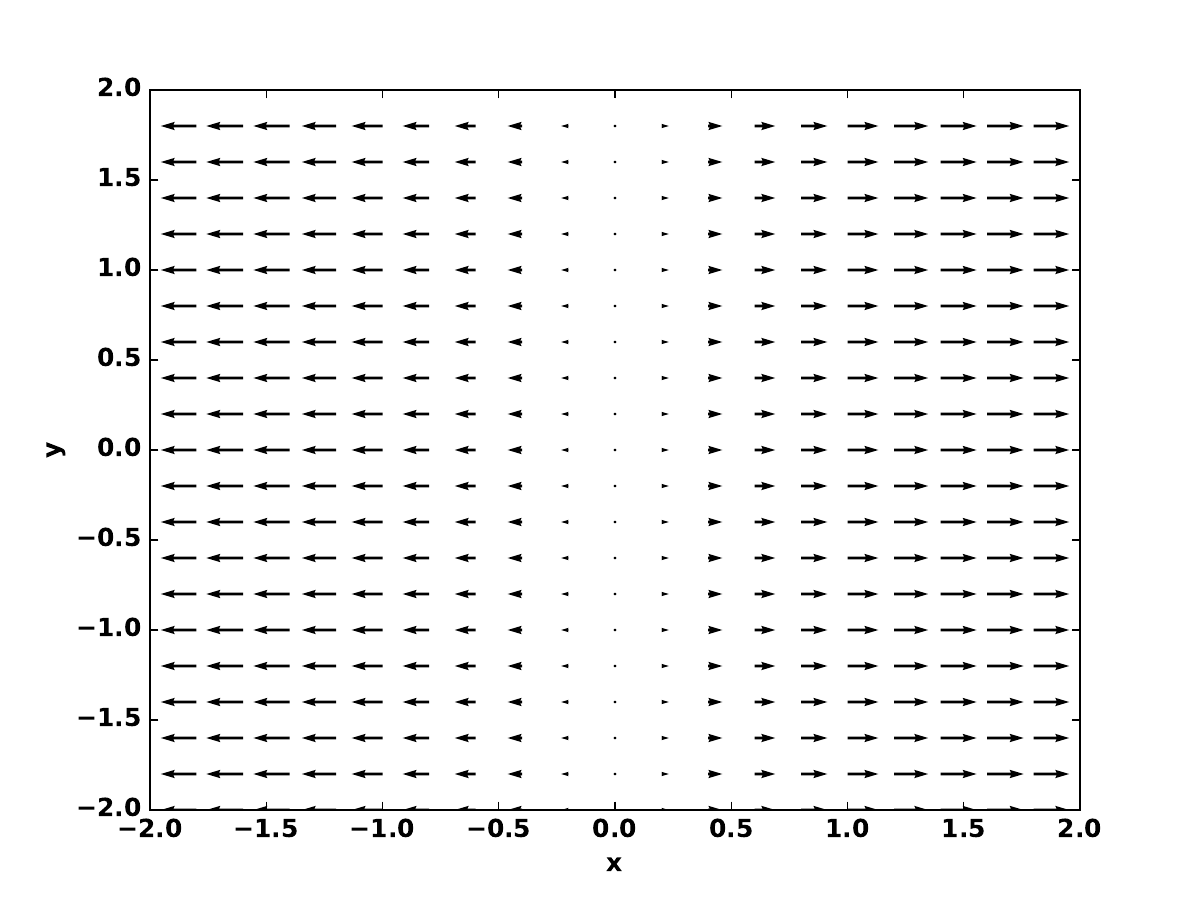}
\end{center}
\end{figure}

\clearpage

\noindent (3) The plot below shows the 2D velocity vector field
$\vec{v}=(u,v)=(y,-x)$. 

\begin{enumerate}
\item From looking at the plot, is $\vec{\nabla} \times \vec{v}$
  positive, negative, or zero at $(x,y)=(0,0)$?
\item Calculate $\vec{\nabla} \times \vec{v}$ at $(x,y)=(0,0)$.
\item Calculate $\vec{\nabla} \times \vec{v}$ at $(x,y)=(1,1)$. Does
  it make sense from the plot why it is the same or different as $\vec{\nabla} \times \vec{v}$ at $(x,y)=(0,0)$?
\end{enumerate}

\begin{figure}[h!]
\begin{center}
  \includegraphics[width=\textwidth]{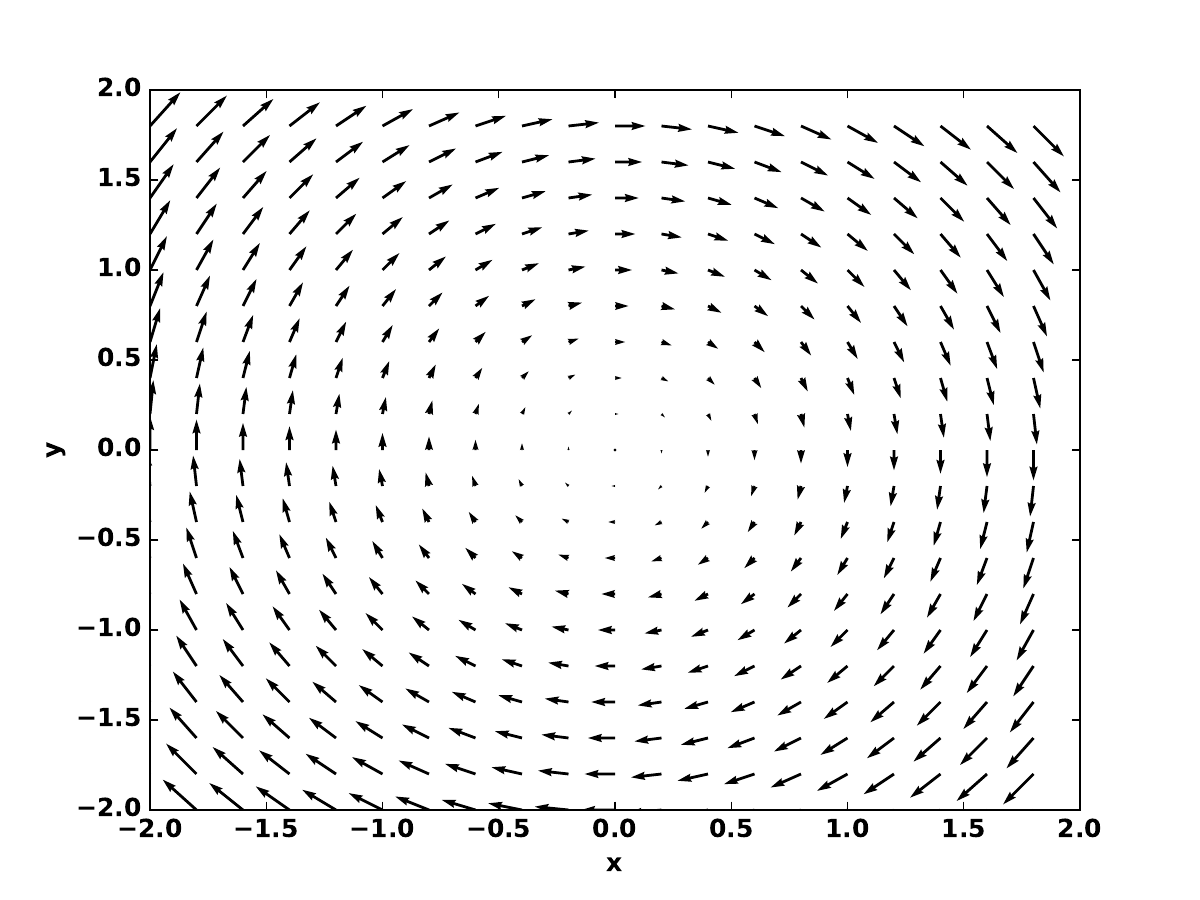}
\end{center}
\end{figure}

\clearpage

\noindent (4) The plot below shows the 2D velocity vector field
$\vec{v}=(u,v)=(y-x,-x-y)$. 

\begin{enumerate}
\item From looking at the plot, is $\vec{\nabla} \times \vec{v}$
  positive, negative, or zero at $(x,y)=(0,0)$?
\item Calculate $\vec{\nabla} \times \vec{v}$ at $(x,y)=(0,0)$.
\item Calculate $\vec{\nabla} \times \vec{v}$ at $(x,y)=(1,1)$. Does
  it make sense from the plot why it is the same or different as
  $\vec{\nabla} \times \vec{v}$ at $(x,y)=(0,0)$?
\end{enumerate}

\begin{figure}[h!]
\begin{center}
  \includegraphics[width=\textwidth]{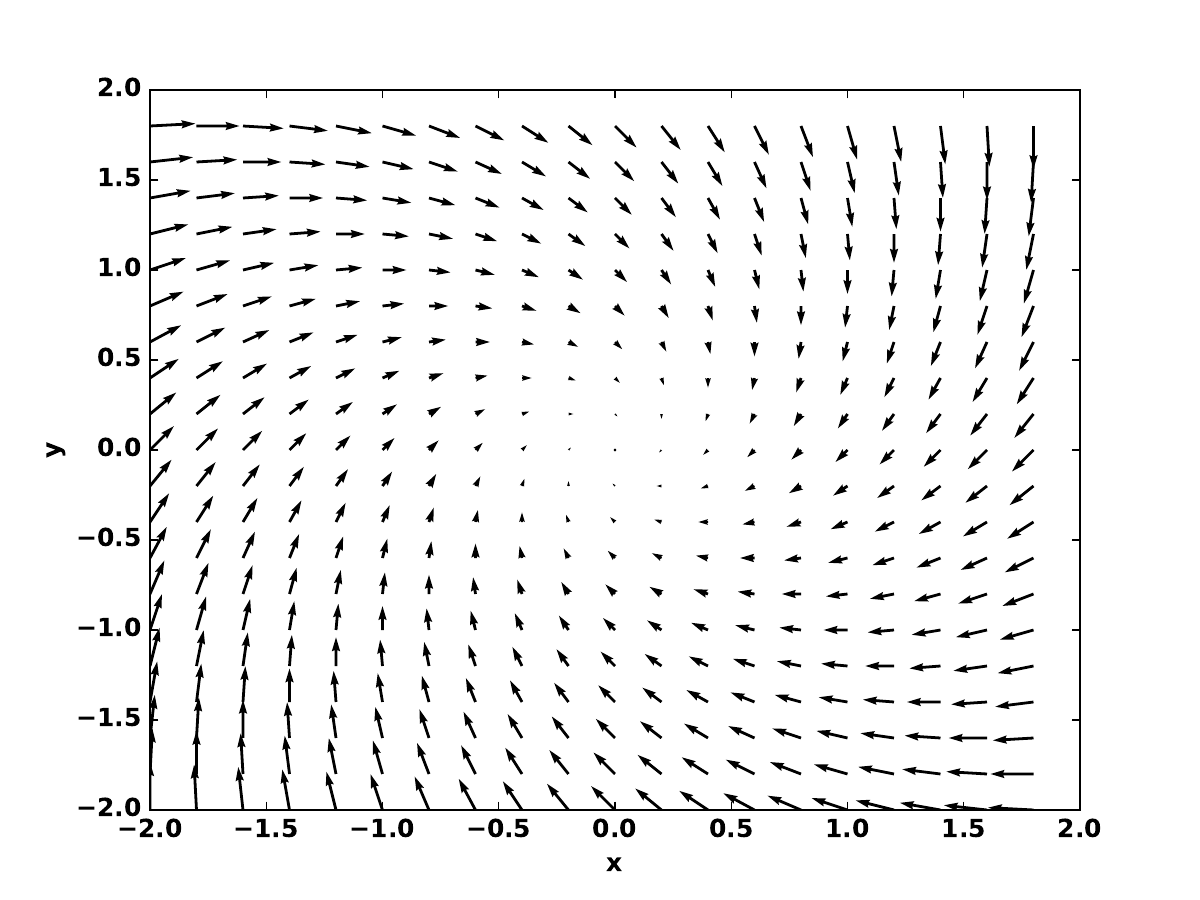}
\end{center}
\end{figure}

\clearpage

\noindent (5) The plot below shows the 2D velocity vector field
$\vec{v}=(u,v)=(y,0)$. 

\begin{enumerate}
\item From looking at the plot, is $\vec{\nabla} \times \vec{v}$
  positive, negative, or zero at $(x,y)=(0,0)$?
\item Calculate $\vec{\nabla} \times \vec{v}$ at $(x,y)=(0,0)$.
\end{enumerate}

\begin{figure}[h!]
\begin{center}
  \includegraphics[width=\textwidth]{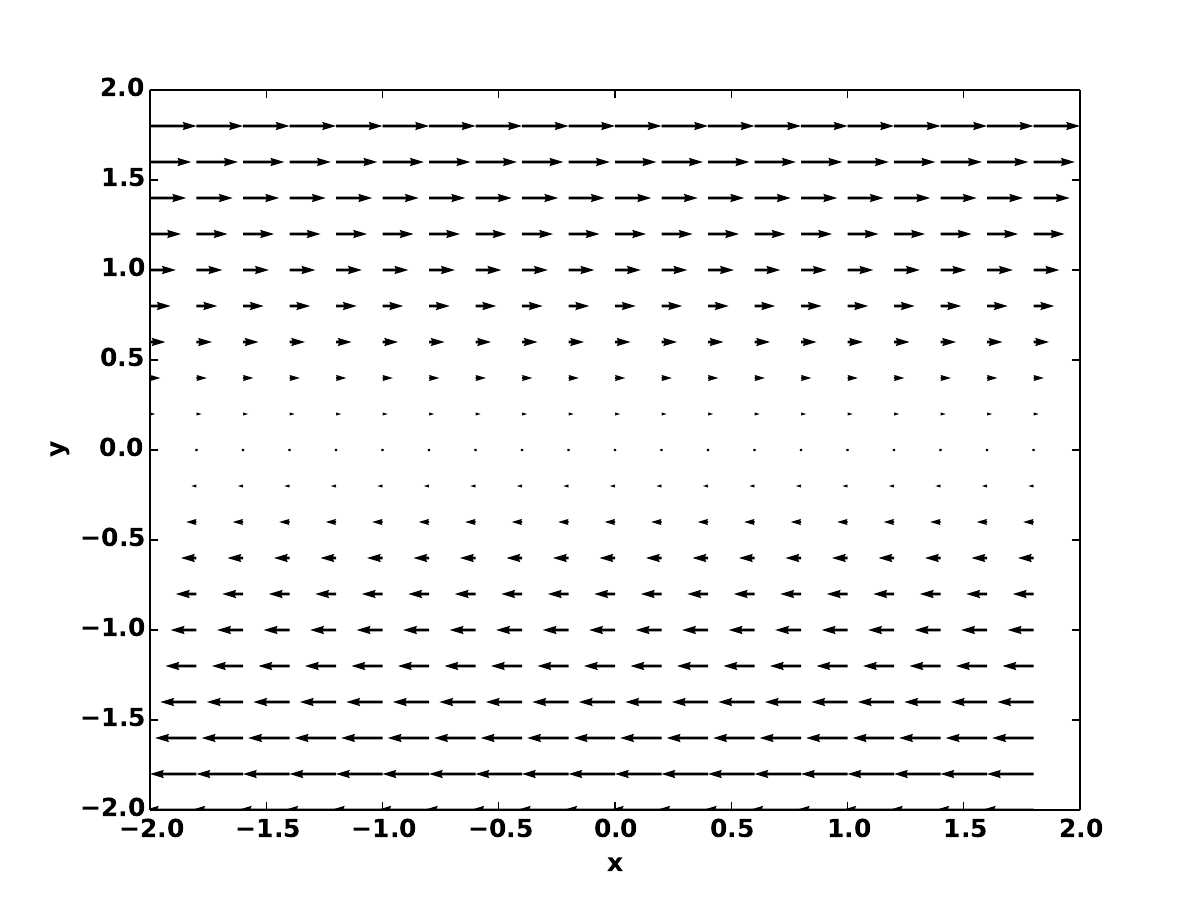}
\end{center}
\end{figure}

\clearpage

\noindent (6) Our old friend the streamfunction ($\psi$) is defined
such that $u=-\frac{\partial \psi}{\partial y}$ and
$v=\frac{\partial \psi}{\partial x}$. Calculate the vorticity in terms
of the streamfunction, $\psi$.

\bigskip

\noindent (7) Consider a cyclonic storm in the midlatitudes of the
northern hemisphere. Is the vorticity positive or negative?

%% file: Flipped/coriolis.tex
\subsection{The Coriolis Force}

The Coriolis force is an apparent force felt in a rotating reference
frame. Watch this video
(\href{https://youtu.be/dt_XJp77-mk}{https://youtu.be/dt\_XJp77-mk}) to
understand the Coriolis force better. The Coriolis force is equal to
\begin{equation}
\vec{F}_c=-f \hat{k} \times \vec{v}
\end{equation} 
on a planet, where $\hat{k}$ represents the vertical direction, 
\begin{equation}
f=2 \Omega \sin \phi
\end{equation}
is the Coriolis parameter, $\phi$ is the latitude, and $\Omega = 2
\pi$~rad~day$^{-1} = 7.3 \times 10^{-5}$~s$^{-1}$ on Earth. We can
write the Coriolis force in the $x$ direction ($F_c^x$) and the $y$
direction ($F_c^y$) as
\begin{eqnarray}
F_c^x &=& fv, \\
F_c^y &=& -fu.
\end{eqnarray}
Notice that a positive velocity in the $x$ direction, $u$, causes a
Coriolis force in the negative $y$ direction (as long as we are in the
Northern Hemisphere, so $f$ is positive). Also, a positive velocity in
the $y$ direction, $v$, causes a Coriolis force in the positive $x$
direction. Notice that the Coriolis force is always perpendicular and
to the right of the flow in the Northern Hemisphere (see the diagram
below). In the Southern Hemisphere the Coriolis force is always
perpendicular and the the left of the flow, since $f$ is negative.

\begin{figure}[h!]
\begin{center}
  \includegraphics[width=0.35\textwidth]{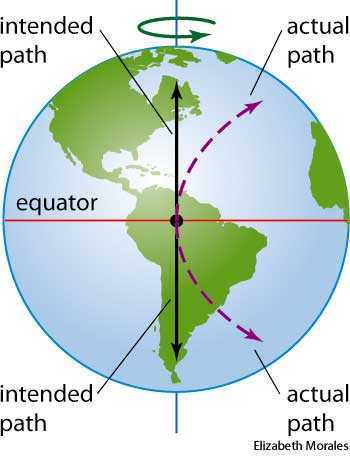}
\end{center}
\end{figure}

We can also use the right-hand rule and the formula for the Coriolis force
($\vec{F}_c=-f \hat{k} \times \vec{v}$) to determine the direction it
acts in. To use the right-hand rule for a cross product ($\vec{A}
\times \vec{B} = \vec{C}$), point the fingers of your right hand
toward vector $\vec{A}$ and curl them toward vector $\vec{B}$. Your
thumb will now point in the direction of vector $\vec{C}$. In the
Northern Hemisphere, $f$ is positive, so the right-hand rule tells us
that the Coriolis force pushes an object to the right of its direction
of travel (see the diagram above, and remember the negative sign in the
formula for the Coriolis force). In the Southern Hemisphere, $f$ is
negative and the Coriolis force bends an object to the left of its
direction of travel.

The Rossby number ($Ro$) is used to characterize the
importance of the Coriolis force in atmospheric
flows. 
\begin{equation}
Ro=\frac{U}{fL}, 
\end{equation}
where $U$ is the typical speed of air or the object under
consideration and $L$ is the typical length scale of the flow. The
Coriolis force is dominant if $Ro\ll 1$ and the Coriolis force can be
neglected if $Ro\gg 1$. If $Ro \approx 1$, then the Coriolis force is
important, but not dominant.

The Rossby number is extremely important. It is named after
Carl-Gustaf Rossby, one of the most important atmospheric scientists
of all time. He was so amazing that he appeared on the cover of Time
magazine! Rossby was the chair of the Department of Meteorology at the
University of Chicago in the 1940's. The Department of Meteorology
merged with the Department of Geology in the 1960's to form the
Department of the Geophysical Sciences, and Hinds building was built
to house it.

\begin{figure}[h!]
\begin{center}
  \includegraphics[width=0.6\textwidth]{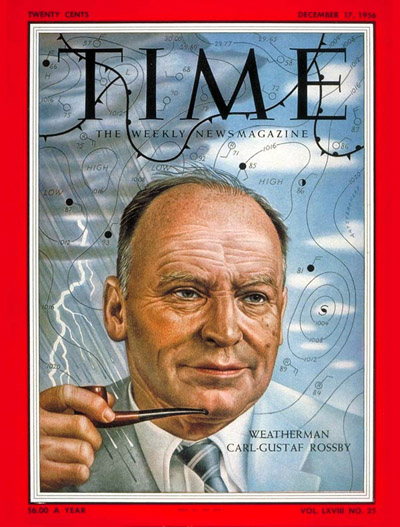}
\end{center}
\end{figure}

\clearpage

\noindent {\large \textbf{Exercises}}
\bigskip

\noindent (1) Jon Lester pitches a fastball to Willson Contreras in
Wrigley field. Estimate the Rossby number. Will the Coriolis force
significantly affect how Jon aims his pitch?

\bigskip 

\noindent (2) Suppose you are crossing the Equator headed directly
south at 10~m~s$^{-1}$. How much of a Coriolis force would you feel,
and in what direction? What about if you are headed south from the
North Pole?

\bigskip

\noindent (3) Suppose you try to travel in a straight line a distance
of $L$ on Earth's surface at a velocity of $U$. As a result of the
Coriolis force, you would deviate to the right (in the Northern
Hemisphere) by a distance $d$ by the time you have traveled forward by
a distance of $L$, assuming that friction with the ground doesn't keep
you from deviating side-to-side. Let's assume that $d \ll L$, so that
we can approximate the path traveled as always in the same direction,
even though the true path deviates to the right. Calculate the time
($T$) it takes to travel the distance $L$.

\bigskip 

\noindent (4) Calculate the Coriolis force acting on you as you
move. Note that what we have been calling the Coriolis ``force'' is
really an acceleration.

\bigskip

\noindent (5) Calculate $d$, the distance you would deviate to the right
based on the acceleration you feel to the right and the amount of time
you feel it.  How does your answer relate to the Rossby number? Why
does this make sense?

\bigskip

\noindent (6) How far off course would you be if you jog at
3~m~s$^{-1}$ in a straight line from the University of Chicago to the
Bean statue downtown (about 10~km)? What if you drive at
30~m~s$^{-1}$?

\bigskip

\noindent (7) How far off course would you be if you drive at
30~m~s$^{-1}$ in a straight line from Chicago to Little Rock (about
1000~km)? What about if you fly at 300~m~s$^{-1}$?

%% file: Flipped/pressure_grad.tex
\subsection{The Pressure Gradient Force}

Today we're going to talk about the pressure gradient force. The
pressure gradient force is the force that pushes from high pressures
toward low pressures. First, we'll derive a formula for the pressure
gradient force in one dimension, then we'll apply our knowledge of the
pressure gradient force to weather maps. To do this, we'll need to use
the gradient operator. The gradient is the multivariable version of
the derivative. It points in the direction of the steepest slope and
it's magnitude is the magnitude of the steepest slope (see the diagram
below). The gradient of the function $f(x,y)$ is defined as
\begin{equation}
\vec{\nabla} f(x,y)=\left( \frac{\partial f}{\partial x},
\frac{\partial f}{\partial y} \right). 
\end{equation}
In words, the gradient of $f(x,y)$ is a vector that has a magnitude of
$\frac{\partial f}{\partial x}$ in the x-direction and a magnitude of
$\frac{\partial f}{\partial y}$ in the y-direction.

\begin{figure}[h!]
\begin{center}
  \includegraphics[width=0.5\textwidth]{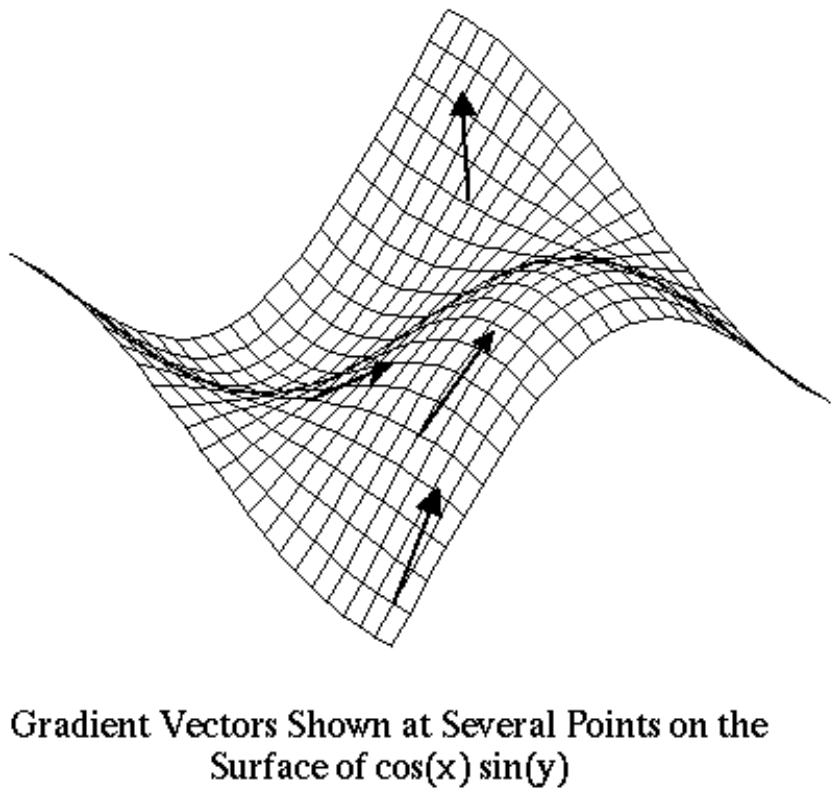}
\end{center}
\end{figure}

\clearpage

\noindent {\large \textbf{Exercises}}
\bigskip

\noindent (1) Let's derive the pressure gradient force in the $x$
direction. The diagram below shows an infinitesimal cube of air with
pressure pushing on it from the right and left. The left face of the
cube is at $x$, and the right face is at $x+\delta x$.

\begin{figure}[h!]
\begin{center}
  \includegraphics[width=0.5\textwidth]{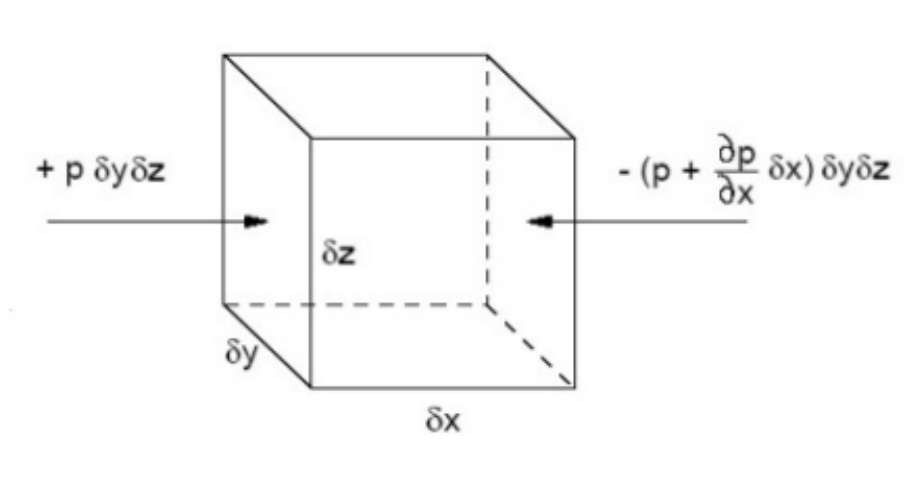}
\end{center}
\end{figure}

\begin{enumerate}
\item Show that the force pushing on the cube from the left in the $x$
  direction is $P(x) \delta y  \delta z$.
\item Show that the force pushing on the cube from the right in the
  $x$ direction is
  $-P(x+\delta x)  \delta y  \delta z$.
\item The total force on the cube in the x direction is
  $F_x=P(x) \delta y \delta z-P(x+\delta x) \delta y \delta z$.
  Simplify $F_x$ using
  $P(x+\delta x)\approx P(x)+\frac{\partial P}{\partial x} \delta x$.
\item Show that the total mass of the cube is
  $m=\rho \delta x \delta y \delta z$, where $\rho$ is the density of
  air in the cube.
\item Using Newton's second law ($F=ma$), show that
  $\frac{du}{dt}=-\frac{1}{\rho} \frac{\partial P}{\partial x}$, where
  $u$ is the velocity in the $x$ direction.
\item We call $-\frac{1}{\rho} \frac{\partial P}{\partial x}$ the
  pressure gradient force, even though it is really the force per unit
  mass. Explain why there is a negative sign in the pressure
  gradient force.
\end{enumerate}

\noindent (2) Now let's think about the pressure gradient force in two
dimensions. The pressure gradient force can operate both in the $x$
direction and the $y$ direction. We can write $\vec{F}_p=\left(
-\frac{1}{\rho} \frac{\partial P}{\partial x},-\frac{1}{\rho}
\frac{\partial P}{\partial y} \right) = -\frac{1}{\rho} \vec{\nabla}
P$.  The figure below shows a surface pressure contour map of
hurricane Sandy as it approaches New York City, as well as colored
contours of surface wind speed.

\begin{enumerate}
\item The low located at the ``L'' at roughly 70W, 39N is the center
  of hurricane Sandy. Does the pressure gradient point toward or away
  from the center of hurricane Sandy?
\item Does the pressure gradient force point toward or away from the
  center of hurricane Sandy?
\item Is the pressure gradient force larger near Chicago or near New
  York City?
\end{enumerate}

\begin{figure}[h!]
\begin{center}
  \includegraphics[width=0.8\textwidth]{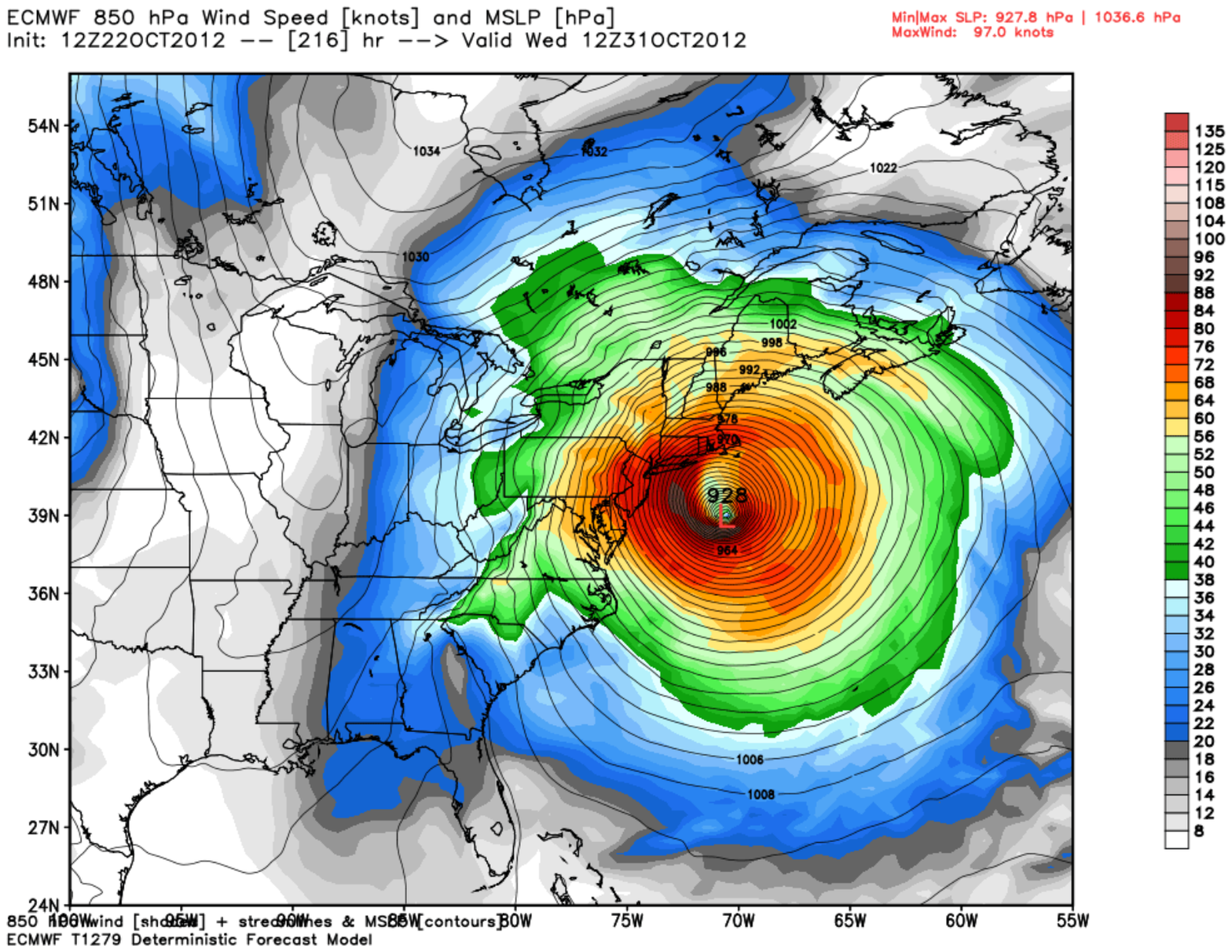}
\end{center}
\end{figure}

\clearpage

\noindent (3) Through a coordinate transformation, we can rewrite the
pressure gradient force as
$\vec{F}_p=-g\vec{\nabla}z=-\vec{\nabla}\Phi$, where $\Phi$ is the
geopotential. $z$ is the height of a given pressure level, sometimes
called the geopotential height. The plot below shows the
climatological mean geopotential height at a pressure of 500~mb over
North America in the month of October.

\begin{enumerate}
\item Over the US and Canada, does the pressure gradient force tend to
  point more toward the north or more toward the south?
\item What is the direction of the pressure gradient force at the
  border of North Dakota and Canada?
\item Is the magnitude of the pressure gradient force larger over
  Miami or Toronto?
\end{enumerate}

\begin{figure}[h!]
\begin{center}
  \includegraphics[width=\textwidth]{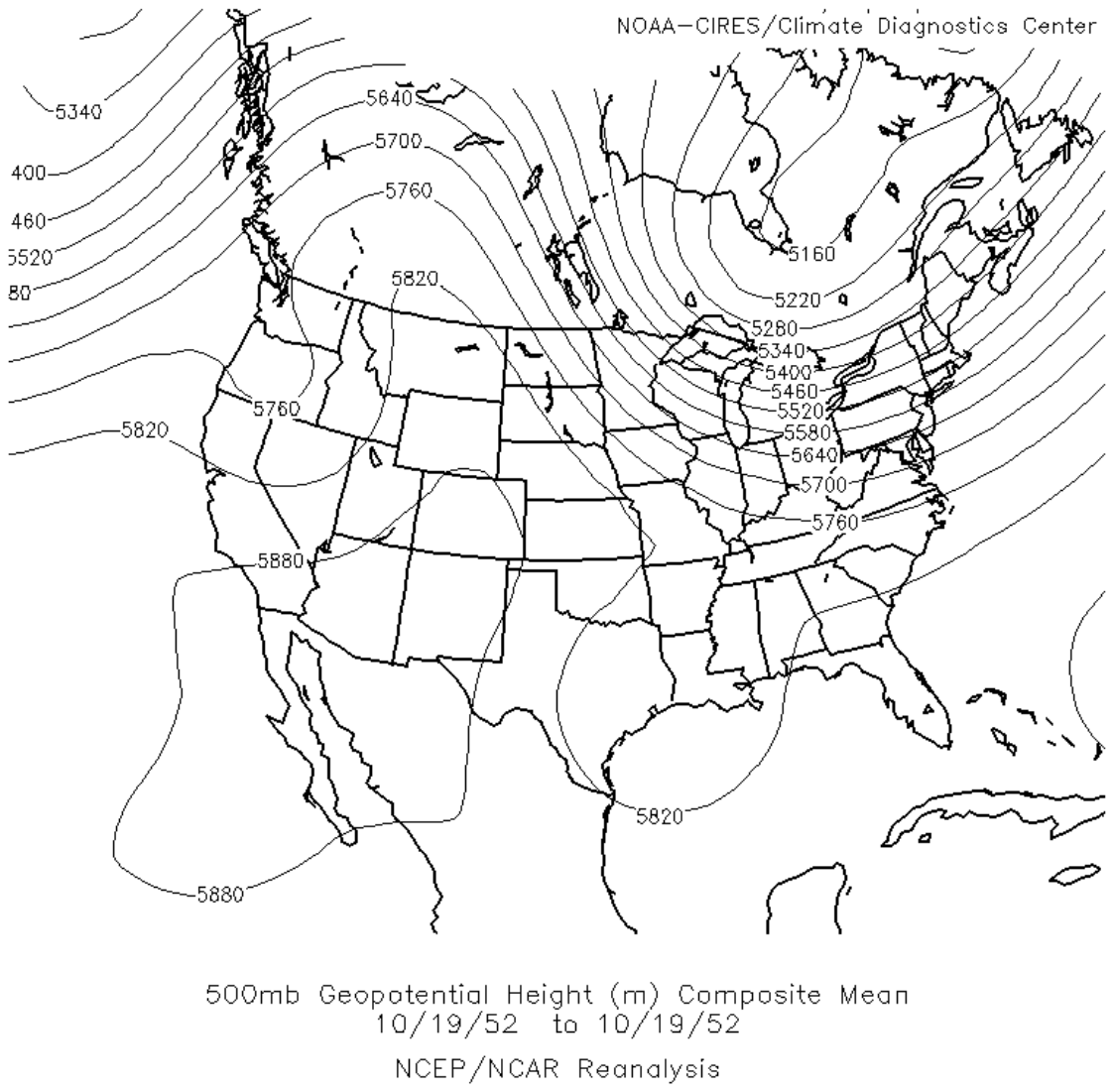}
\end{center}
\end{figure}

%% file: Flipped/geostrophic.tex
\subsection{Geostrophic Balance 1}

Above the atmospheric boundary layer friction and acceleration are
generally negligible. The dominant horizontal force balance is between
the pressure gradient force and the Coriolis force. In this limit the
Rossby number is small. This is called ``geostrophic balance'' and it
is an extremely important concept in atmospheric science. Remember,
the pressure gradient force flows down gradient from high to low
pressure, and the Coriolis force points 90$^\circ$ to the right of the
velocity vector in the Northern Hemisphere. This means that
geostrophic balance is established with the velocity flowing at right
angles to the pressure gradient (see the figure below)!

\begin{figure}[h!]
\begin{center}
  \includegraphics[width=0.4\textwidth]{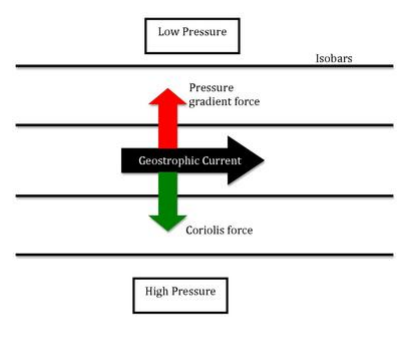}
\end{center}
\end{figure}

Mathematically, we can write the pressure gradient force in the x
direction as $-\frac{1}{\rho} \frac{\partial P}{\partial x}$ or as
$-\frac{\partial \Phi}{\partial x}$, where $\Phi = gz$ is the
geopotential ($g$=9.8~m~s$^{-2}$). We can write the Coriolis force in
the x direction as $fv$, where $f=2 \Omega \sin \phi$ is the Coriolis
parameter, $\phi$ is the latitude, and $\Omega = 2 \pi$~rad~day$^{-1}
= 7.3 \times 10^{-5}$~s$^{-1}$ on Earth. To check that we have the
sign right, consider a northward flow (positive $v$) in the Northern
Hemisphere. The Coriolis force should point to the right of this flow,
or in the positive x direction, so we have the sign right. In order
for the Coriolis force to balance the pressure gradient force, we need
their sum to be zero: $0=fv-\frac{\partial \Phi}{\partial x}$, where
we are using the geopotential version of the pressure gradient
force. This can also be written as $fv=\frac{\partial \Phi}{\partial
  x}$. Similarly, we can perform a force balance in the y direction to
find that $fu=-\frac{\partial \Phi}{\partial y}$.  $f$ depends on
latitude, and therefore on $y$, but let's only consider small enough
deviations in latitude that $f$ is roughly constant. We can therefore
define a streamfunction, 
\begin{equation}
\psi=\frac{\Phi}{f}=\frac{gz}{f},
\end{equation} 
such that
\begin{eqnarray}
u &=& -\frac{\partial \psi}{\partial y}, \\
v &=& \frac{\partial \psi}{\partial x}, 
\end{eqnarray}
This is why we can infer velocities from maps of
geopotential height! As a reminder, notice that if we want $u$, the
velocity in the x direction, we need to look at the derivatives of the
streamfunction in the y direction. Similarly, if we want $v$, the
velocity in the y direction, we need to look at derivatives in the x
direction. That's how a streamfunction works, and it ultimately works
for our particular streamfunction, which is proportional to pressure,
because the Coriolis force points to the right of the velocity.

\clearpage

\noindent {\large \textbf{Exercises}}
\bigskip

\noindent (1) The figure below is the weather map from Problem Set
07. The solid contours show the geopotential height ($z$) of the
500~hPa pressure surface on a particular day at a particular time. The
numbers are contour levels in decameters (570 decameters = 5700
meters) and the contour spacing is 60~m. Bands of latitude and
longitude are shown as dotted lines.

\begin{figure}[h!]
\begin{center}
  \includegraphics[width=0.48\textwidth]{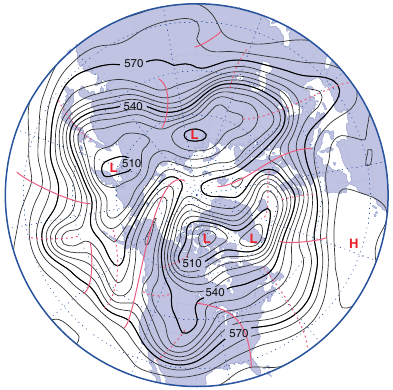}
\end{center}
\end{figure}

\begin{enumerate}
\item Estimate the speed of the jet stream over the North Atlantic.
  You may use the fact that the latitude bands are in 15$^\circ$
  increments and that one degree of latitude corresponds to 111~km.
\item Convert your estimate from m~s$^{-1}$ to mph. 
\item The distance from Chicago to Berlin is about 4400 mi. A Boeing
  747 has a cruising speed of 550 mph through the air. Assume a flight
  from Chicago to Berlin has a constant tail wind equal to your
  estimate of the jet stream speed. How many hours does the flight
  take?
\item Assume a flight from Berlin to Chicago has a constant head wind
  equal to your estimate of the jet stream speed. How many hours does
  the flight take?
\end{enumerate}

%% file: Flipped/geostrophic2.tex
\subsection{Geostrophic Balance 2}

Geostrophic balance is really important for understanding atmospheric
flow in the midlatitudes. It's so important that we're going to spend
another class getting familiar with it! Last time we found that we can
define a geostrophic streamfunction as
$\psi=\frac{\Phi}{f}=\frac{gz}{f}$ such that $u=-\frac{\partial
  \psi}{\partial y}=$ and $v=\frac{\partial \psi}{\partial x}$, where
$z$ is the geopotential height. Alternatively, we can write the
pressure gradient force as $\vec{F}_p=\left( -\frac{1}{\rho}
\frac{\partial P}{\partial x},-\frac{1}{\rho} \frac{\partial
  P}{\partial y} \right) = -\frac{1}{\rho} \vec{\nabla} P$. Performing
a force balance between the pressure gradient force and the Coriolis
force leads to the definition of an alternative streamfunction,
\begin{equation}
\psi=\frac{P}{\rho f}. 
\end{equation}
This streamfunction assumes that the density is approximately constant
horizontally at a given level, and again neglects variations in $f$,
the Coriolis parameter, in the y direction.

\clearpage

\noindent {\large \textbf{Exercises}}
\bigskip

\noindent (1) Watch this movie
(\href{https://youtu.be/n4rJX4IIRaU}{https://youtu.be/n4rJX4IIRaU}).
Focus on the extratropical cyclone above Chicago. Estimate the
pressure differential between the storm center and far away from the
center.  How big is this differential compared to standard sea level
pressure?  Hints: You may assume an air density of 1~kg~m$^{-3}$. You
will need to estimate the distances involved and read the video
caption to estimate the amount of time that elapses over the course of
the video.

\bigskip

\noindent (2) Consider an intense tropical cyclone located at
15$^\circ$ latitude. It's maximum azimuthal wind speed is
60~m~s$^{-1}$, which occurs just outside the eyewall (10~km from the
storm center).

\begin{enumerate}
\item Calculate the Coriolis force (really an acceleration)
  experienced by the air as it rotates around the cyclone's center. In
  which direction (toward or away from the center) does the Coriolis
  force point?
\item Calculate the centrifugal acceleration ($\frac{v^2}{r}$)
  experienced by the air as it rotates around the cyclone's center. In
  which direction (toward or away from the center) does the
  centrifugal acceleration point?
\item Is one of these accelerations larger than the other? What is it
  balanced by?
\item How is this different from the situation for the extratropical
  cyclone in the last question?
\end{enumerate}

%% file: Flipped/potential_vorticity.tex
\subsection{Potential Vorticity}

Today we're going to learn about the ``potential vorticity.'' It is a
conserved quantity for fluid in a rotating reference frame. Potential
vorticity conservation is the fluid equivalent of angular momentum
conservation for solids. To remember how the conservation of angular
momentum works, watch this video
(\href{https://youtu.be/FmnkQ2ytlO8}{https://youtu.be/FmnkQ2ytlO8}).
Conservation of angular momentum allows the skater in the video to
increase her angular velocity (spininess) by pulling her legs and arms
into her body.

For fluids we can define a quantity called the potential vorticity,
\begin{equation}
Q=\frac{f+\zeta}{H}, 
\end{equation}
where $f$ is the Coriolis parameter and
represents planetary vorticity, 
\begin{equation}
\zeta=\hat{k} \cdot \vec{\nabla} \times \vec{v} = \frac{\partial v}{\partial x}-\frac{\partial u}{\partial y},
\end{equation} 
is the relative vorticity (the vorticity of the fluid itself, also
$\zeta$ is pronounced ``zeta''), and $H$ is the thickness of the fluid
(the vertical distance, or height). Technically this definition
applies for a constant density fluid, but it gives us a good
qualitative picture of what happens in the atmosphere, and can be
generalized to the atmospheric case with a bit more math. The point is
that the potential vorticity, $Q$, is conserved (constant) during
flow. This is a very powerful statement because it allows us to
predict aspects of the flow easily.  Atmosphere geeks get so excited
about potential vorticity, or PV, that they even write songs about it
(\href{https://youtu.be/g2-RpvGGQU4}{https://youtu.be/g2-RpvGGQU4}).
To use potential vorticity in the atmosphere, we just have to think
about the thickness, $H$, as the thickness between two layers of
constant potential temperature.

The figure below shows how potential vorticity conservation works.
Assume we have a parcel of air with some vorticity and thickness
associated with it. For now let's assume that all of it's vorticity is
relative and none of it is planetary (say the parcel is at the
equator), so you can see all of the spinning directly. This air parcel
is represented on the left of the figure below. Now let's assume that
the air is stretched vertically. To conserve mass it will have to
squeeze in toward it's center. Just like how the skater spun up as she
pulled her arms and legs toward her center, the air spins up as it
stretches vertically and pulls in toward its center. This is
represented by the right diagram in the figure below. The higher
vorticity on the right is represented by more lines curling around the
air parcel. The fact that the two parcels have the same potential
vorticity, even though they have different vorticities, is the
statement of potential vorticity conservation that we can use to
understand atmospheric flow. We just have to remember that
$Q=\frac{f+\zeta}{H}$ is conserved.

\begin{figure}[h!]
\begin{center}
  \includegraphics[width=0.6\textwidth]{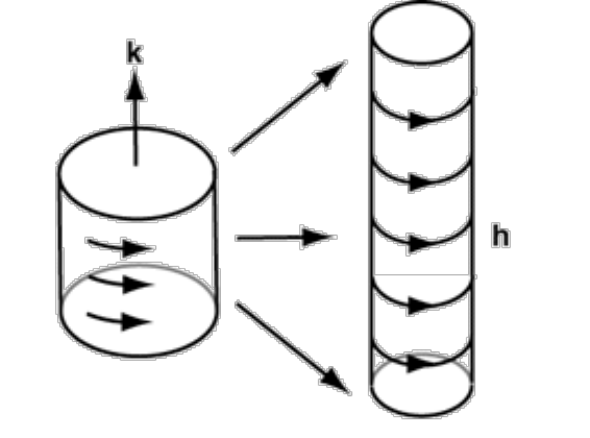}
\end{center}
\end{figure}

\clearpage

\noindent {\large \textbf{Exercises}}
\bigskip

\noindent (1) A cylindrical column of air with a radius of 100~km
starts at the equator at rest. It moves northward to Chicago
maintaining a constant height.

\begin{enumerate}
\item Is the relative vorticity,
  $\zeta=\frac{\partial v}{\partial x}-\frac{\partial u}{\partial y}$,
  of the air positive or negative when it reaches Chicago? 
\item Does the column of air spin clockwise or counter-clockwise when
  it reaches Chicago?
\item Explain physically how the conservation of potential vorticity
  leads to the column spinning in the direction it does. Hint:
  Consider a column of fluid in a frictionless jar at rest on top of a
  turntable record player. How does it's rotation change if you turn
  the record player on so that it starts spinning?
\item Let's assume that the column has a tangential velocity of $U$ in
  the counter-clockwise direction at it's radius, $R$. The tangential
  velocity is zero at the center of the column. Show that we can
  approximate $\frac{\partial v}{\partial x} \approx \frac{U}{R}$ and
  $\frac{\partial u}{\partial y} \approx -\frac{U}{R}$.
\item Use this to estimate the mean tangential velocity at the
  perimeter of the air column when it reaches Chicago.
\end{enumerate}

\bigskip

\noindent (2) An air column with zero relative vorticity ($\zeta=0$)
stretches from the surface to the tropopause, which we assume is a
rigid lid, at $10$~km. The column follows the jet stream and moves
zonally onto the Rocky Mountains, which we idealize as a big plateau.

\begin{enumerate}
  \item Which direction does the column start spinning in?
  \item Assume that the air column starts at $60^{\circ}$~N and moves
    directly eastward onto the plateau, which is $2.5$~km high. What
    is its relative vorticity once it moves onto the plateau?
  \item Suppose it then moves southwards to $30^{\circ}$~N while still
    on the plateau. What is its relative vorticity now?
\end{enumerate}

%% file: Flipped/tornadoes.tex
\subsection{Tornadoes}

Today we're going to learn how tornadoes form using our knowledge of
potential vorticity. The relative vorticity is so high in tornadoes
that we can neglect the planetary vorticity in this discussion. There
are three steps to the formation of tornadoes. First, a large amount
of vorticity needs to be generated. This is done by a wind shear
(winds that are higher up in the atmosphere than at the
surface). It is natural for a wind shear to develop near the surface
because surface friction brings the wind to near zero right at the
surface. The special condition we need is a large wind shear, so
really fast horizontal winds relatively close to the surface. The diagram below
shows how a wind shear leads to vorticity. Notice that it is
horizontal vorticity.

\begin{figure}[h!]
\begin{center}
  \includegraphics[width=0.5\textwidth]{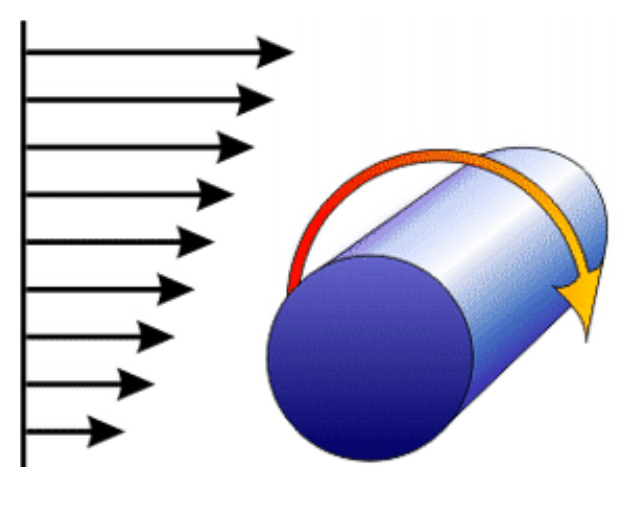}
\end{center}
\end{figure}

Next this horizontal vorticity is tilted vertically. This is
accomplished by convection. See the diagram below. To understand how
this works, watch this video
(\href{https://youtu.be/sb82tVOq2dY}{https://youtu.be/sb82tVOq2dY}) of
high divers. At the start of most of these jumps the high divers cause
rotation of their body around the axis from their feet to their head.
They then allow this axis to tilt in a range of directions. The body
keeps spinning around the axis of the trunk, but the trunk can be
pointed in different directions. A similar thing is happening when
horizontal vorticity is tilted upward by convection to become vertical
vorticity.

\begin{figure}[h!]
\begin{center}
  \includegraphics[width=0.8\textwidth]{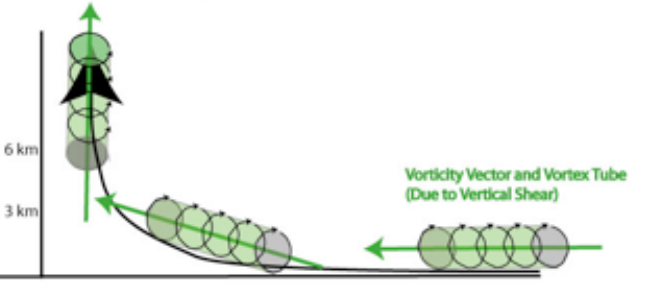}
\end{center}
\end{figure}

Finally, the air column is stretched vertically, which increases its
vorticity. We can understand this through the conservation of
potential vorticity (see the diagram below, which is from last class).
The convection that caused the tilting also causes the stretching.

\begin{figure}[h!]
\begin{center}
  \includegraphics[width=0.6\textwidth]{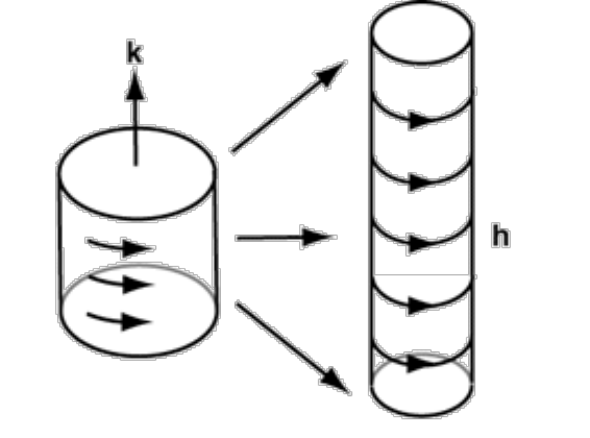}
\end{center}
\end{figure}

The diagram below puts the three steps together. In panel A wind shear
generates horizontal vorticity. In panel B vertical convection
transforms the horizontal vorticity into vertical vorticity. In panel
C the convection stretches the vertical vorticity, which causes the
vertical vorticity to increase drastically, which we observe as a
tornado.

\clearpage 

\begin{figure}[h!]
\begin{center}
  \includegraphics[width=0.9\textwidth]{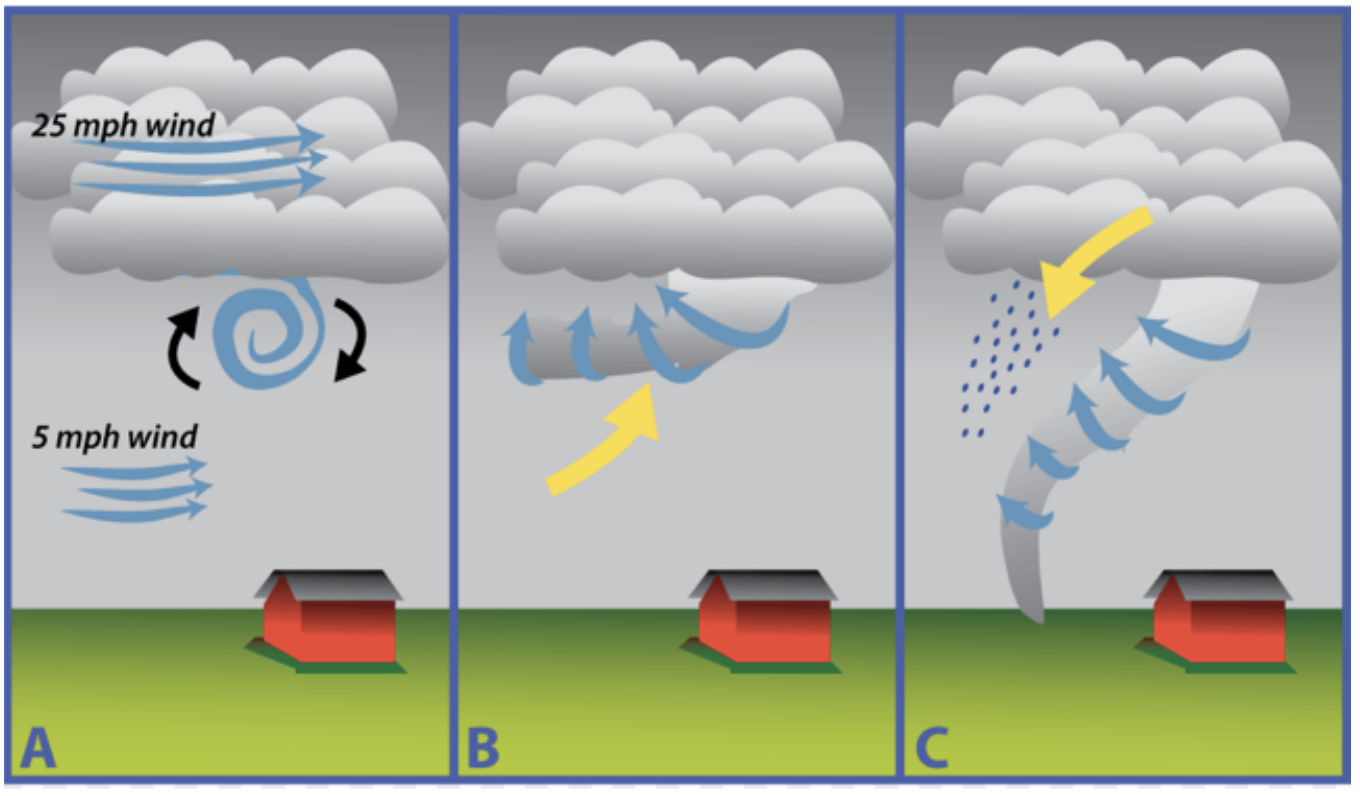}
\end{center}
\end{figure}

Finally, you may have heard of the ``Fujita scale,'' which is used to
measure tornado intensity. It is named after Ted Fujita, who was a
member of the Department of Meteorology and later the Department of
the Geophysical Sciences at the University of Chicago from the 1950's
until his death just before the turn of the century. Fujita was a
young professor in Japan during the second World War in the mid-20th
century. After the United States dropped an atomic bomb on the city of
Hiroshima, Japan, as part of that war, the Japanese government sent
Fujita in to figure out what had happened. By carefully analyzing burn
marks on objects such as gravestones around the city, he was able to
reconstruct the three-dimensional location and power of the explosion
fairly accurately. He later used similar methods to study tornadoes in
North America by analyzing the wreckage they create. This led to the
development of his famous scale, as well as his nickname,
``Mr. Tornado.'' You can see Mr. Tornado in action (and some vintage
footage of Hinds building as well as 20th century technology) in this
video
(\href{https://youtu.be/YGrhwlYSWUU}{https://youtu.be/YGrhwlYSWUU}).

\clearpage

\noindent {\large \textbf{Exercises}}
\bigskip

\noindent (1) Let's try to put some numbers on the qualitative
explanation above.

\begin{enumerate}
\item Suppose we are out in the Great Plains and there is a strong
  wind shear. The wind ($u$) is 0~m~s$^{-1}$ at the surface and linearly
  increases to 50~m~s$^{-1}$ at a height of 1~km, above which it has
  a constant value of 50~m~s$^{-1}$. Calculate the horizontal vorticity,
  which in this case will be $\frac{\partial u}{\partial z}$.
\item Now suppose that a 400~m long stretch of this rotating air is
  tilted vertically by convection to form a column of rotating air.
  What is the radius of the column of rotating air?
\item What is the vertical (relative) vorticity ($\zeta$) of the
  column of rotating air?
\item The potential vorticity is $Q=\frac{f+\zeta}{H}$. Compare $f$ to
  $\zeta$ to show that we can drop $f$.
\item Suppose that now convection stretches the air vertically by a factor
  10, so that the height of our rotating column increases from 400~m
  to 4~km. What is the vertical vorticity now?
\item If we approximate the air as constant density, then the volume
  of an air column must be conserved as it is stretched. What is the
  new radius of the column of rotating air after it has been stretched
  vertically by a factor of 10? 
\item What is the tangential velocity at the perimeter of the rotating
  column of air, which we now recognize as a tornado?
\item Calculate the Coriolis acceleration ($fv$) at the perimeter of the
  tornado.
\item Calculate the centrifugal acceleration ($\frac{v^2}{r}$) at the
  perimeter of the tornado.
\item Using the symbolic formulas for centrifugal and Coriolis
  accelerations, what is the ratio of the centrifugal to the Coriolis
  acceleration? 
\item This formula has a special name. What is it?
\item Use this to again argue that planetary rotation is not important
  for the tornado.
\end{enumerate}

%% file: Flipped/clouds.tex
\subsection{Clouds}

Today is going to be a fun day where we learn about the types of
clouds. \textbf{Cirrus} clouds are high wispy clouds that can precede
storms but do not lead to precipitation themselves. Cirrus means lock
of hair. \textbf{Stratus} clods are low clouds that form at a constant
layer and stretch for large horizontal distances and can be associated
with light rain. Stratus means layer. \textbf{Cumulus} clouds are
usually at low levels, look like cotton balls, are not associated with
rain, and often (but not always) come in 1D or 2D patterns. Cumulus
means pile. Some clouds appear to be combinations of two types, so we
just merge the names. We can also add \textbf{alto}, means high, or
\textbf{nimbus}, which means rainstorm, to the name of a cloud as
appropriate. See the diagram below for schematic examples.

\begin{figure}[h!]
\begin{center}
  \includegraphics[width=0.85\textwidth]{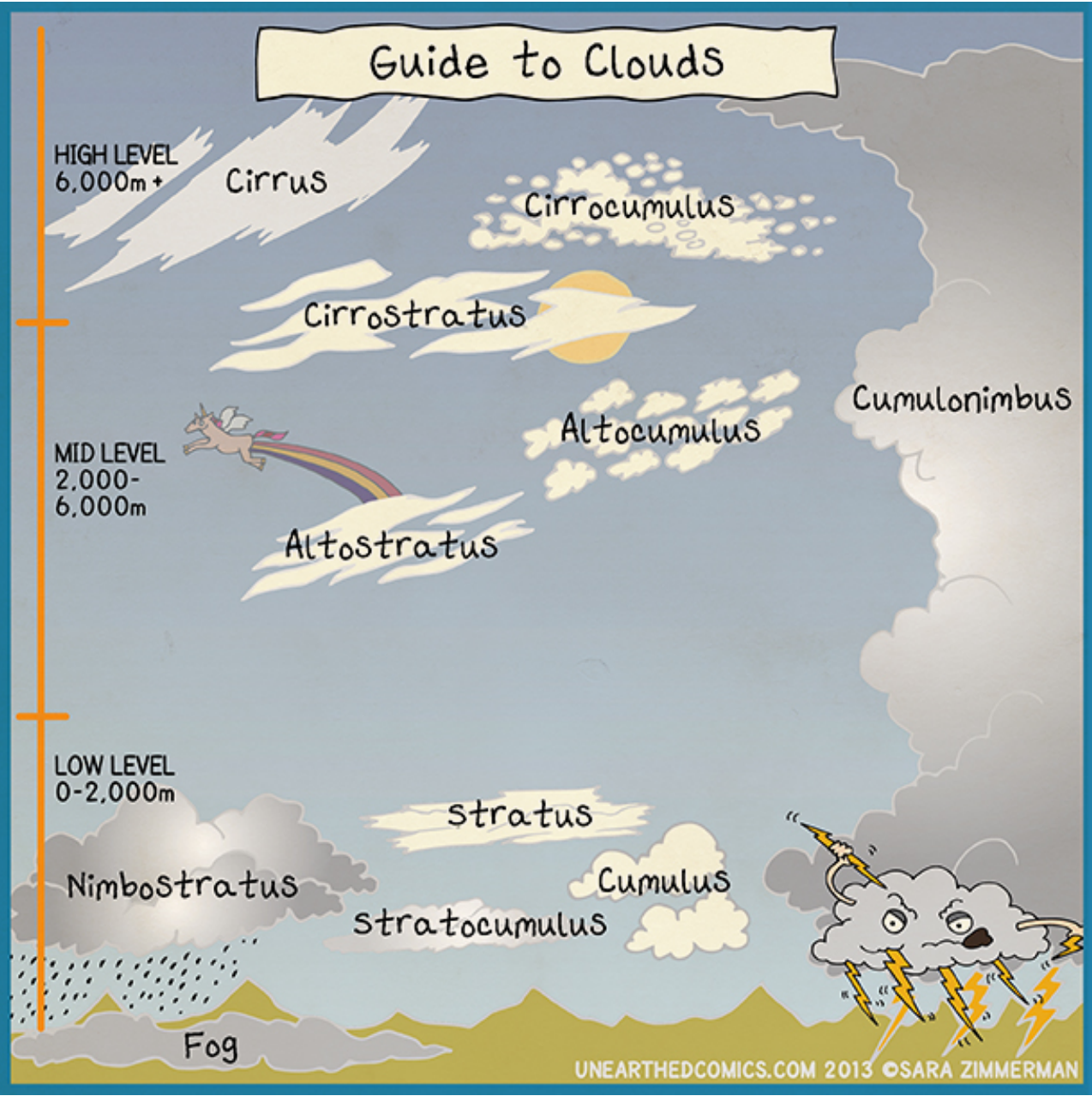}
\end{center}
\end{figure}

\clearpage

\noindent {\large \textbf{Exercises}}
\bigskip

\noindent (1) Play the game at this link
(\href{https://www.purposegames.com/game/cloud-types-quiz}{https://www.purposegames.com/game/cloud-types-quiz}.
Keep playing until you can consistently get 100\%.

\bigskip

\noindent (2) Now try this game
(\href{http://www.mnn.com/earth-matters/climate-weather/quiz/can-you-name-these-clouds}{http://www.mnn.com/earth-matters/climate-weather/quiz/can-you-name-these-clouds}),
which shows some real cloud pictures. There are some weird ones in
there that are cool and rare.

\bigskip 

\noindent (3) Let's go outside and try to identify some clouds!

%% file: PSets/pset01.tex
\subsection{Problem Set 1}

\bigskip

\noindent (1) Watch this movie
(\href{https://youtu.be/KSzElTYS9qM}{https://youtu.be/KSzElTYS9qM}),
which shows two satellite movies side-by-side.

\begin{enumerate}
\item One of the movies shows pictures of Earth during Jan/Feb, the
  other shows Earth during Jul/Aug. Which of the movies corresponds to
  which season, and why? Name several reasons.
\item Over the continents, which regions/countries are associated with
  the downwelling branch of the Hadley circulation? Based on the
  movies, which vegetation patterns do you find in these regions and
  why?
\end{enumerate}

\bigskip

\noindent (2) Watch these movies of surface winds from Jan/Feb
(\href{https://youtu.be/YveJAU5sQiI}{https://youtu.be/YveJAU5sQiI})
and Jul/Aug
(\href{https://youtu.be/u2g0X1mDjW0}{https://youtu.be/u2g0X1mDjW0}).
 
\begin{enumerate}
\item Where are surface winds generally stronger - over the ocean or over land? Why?
\item Look at India. For each season, does the surface wind blow from ocean to land, or the opposite? Why?
\item The Pacific Ocean got its name from Ferdinand Magellan, who
  reached it after passing through the Strait of Magellan at the end
  of 1520. In your opinion, why did Magellan call the ocean
  ``Pacific''?

\item Now compare the surface wind movies with the surface
  pressure movies from class. Are the mid-latitude blobs of high
  surface winds associated with high or low surface pressure?

\end{enumerate}

%% file: PSets/pset02.tex
\subsection{Problem Set 2}

\bigskip

\noindent (1) Estimate the mass of CO$_2$ that a new forest can remove
from the atmosphere per year per square kilometer. 

\bigskip

\noindent (2) Estimate the mass of CO$_2$ humans have added to the
atmosphere through deforestation throughout history. Compare this to
the mass of CO$_2$ added to the atmosphere by cars in the US each
year.

%% file: PSets/pset03.tex
\subsection{Problem Set 3}

\bigskip

\noindent (1) This problem is based on a method that has been used to
reconstruct the atmospheric pressure on ancient Earth
\cite{som2016earth}. Vesicles in crystallized lava flows form because
air bubbles were trapped in the lava as it crystallized. The figure
shows a schematic diagram of this.
\begin{figure}[h!]
\begin{center}
  \includegraphics[width=0.2\textwidth]{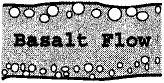} 
\end{center}
\end{figure}

\begin{enumerate}
\item The diagram shows that vesicles near the top of a
lava flow will tend to have a larger volume than vesicles near the
bottom. Explain physically why this is. 

\item Suppose that the atmospheric pressure is $P_a$ and the lava
  density is $\rho_l$. Derive a formula for the pressure felt by the
  air bubbles when the vesicles are formed as a function of depth
  ($z$) in the lava.

\item You can apply your formula from part B to determine
the air pressure inside a bubble near the top of a lava flow and near the
bottom (at a depth of $d$). Use this to find an equation that yields
atmospheric pressure as a function of the ratio of vesicle volume near
the top of the lava flow ($V_t$) to that near the bottom of the lava
flow ($V_b$). You may assume that the number of moles of air per bubble is
 constant and that the temperature at which crystallization occurs is
the same near the top and bottom of the lava flow. 

\item Let the atmospheric scale height be $H$ and the
atmospheric surface pressure be $P_s$. Derive a formula for altitude
as a function of the atmospheric pressure. 

\item Suppose that we wish to calibrate this methodology by
studying recent lava flows on an active volcano, such as Mauna Loa in
Hawaii. Combine your results from (C) and (D) to derive a formula for
the altitude of a lava flow as a function of the ratio
$\frac{V_t}{V_b}$, which we can measure. For lava flows of equal
depth, how will $\frac{V_t}{V_b}$ change as you consider flows of
increasingly large altitudes? 

\end{enumerate}

%% file: PSets/pset04.tex
\subsection{Problem Set 4}

\bigskip

\noindent (1) Explain why a liquid boils when its saturation vapor
pressure equals ambient atmospheric pressure. Imagine we bring water
to a boil in a pot in Chicago and in Denver. We then put and egg into
the boiling water. In which city will the egg be fully cooked (hard
boiled) first? Why?

\bigskip

\noindent (2) The figure below shows the saturation vapor pressure
of water as a function of temperature. When you exhale, the air coming
from your lungs is at body temperature and close to being saturated
(point ``A''). Now you go outside on a warm day (when ambient air is
at point ``B''), and on a cold day (when ambient air is at point
``C''). Do you expect to see your breath on either day? Why? Hint:
consider what the mixing of two parcels of air would look like on this
plot.
\begin{figure}[h!]
\begin{center}
  \includegraphics[width=0.5\textwidth]{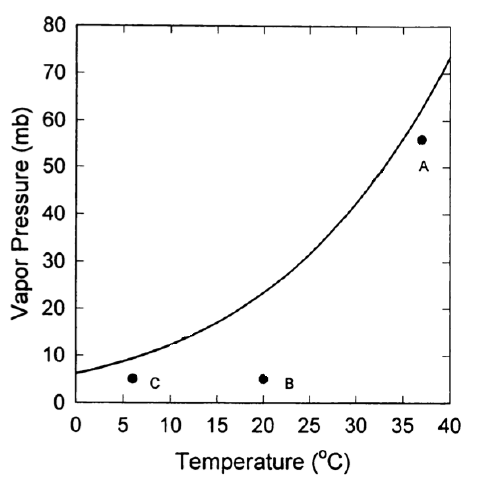} 
\end{center}
\end{figure}

\bigskip

\noindent (3) A modern, pressurized jet airplane is flying in the upper
troposphere in the tropics. Air is brought into the cabin from the
atmosphere outside the airplane. For the passengers to experience a
similar cabin air temperature to the surface air temperature, does the
air brought into the cabin from outside need to be cooled, heated, or
neither? If you think the air needs to be cooled or heated, estimate
by how many degrees to one significant digit.

%% file: PSets/pset05.tex
\subsection{Problem Set 5}

\bigskip

\noindent (1) The figures on the next page show the spectrums of a
Sun-like star (a G-star) and a smaller M-star. M-stars represent about
75\% of stars in the galaxy, and possible terrestrial planets in the
habitable zone of nearby M-stars (Proxima Centauri and TRAPPIST-1)
have been found recently. The spectrums are only approximate black
bodies because of non-uniform absorption and emission by gases in the
upper atmospheres of the stars. For this problem, however, you can
assume that the spectrums are blackbodies.

\begin{enumerate}
\item Estimate the photospheric (where the light is emitted to space)
  temperatures of the Sun-like star and the M-star.
\item Estimate the energy flux emitted by each star.
\item Estimate the power emitted by each star (the luminosity). You
  may assume that the Sun-like star has a radius of 700,000~km and the
  M-star has a radius of 430,000~km.

\item The Sun has a habitable planet (Earth) orbiting it at a distance
  of 1 astronomical unit. For a planet orbiting the M-star to receive
  the same amount of stellar radiation (in W~m$^{-2}$) as Earth, at
  what distance (in astronomical units) would it need to be orbiting
  the M-star? Compare this to the distances of planets in the solar
  system to the Sun. \textit{Hint:} A star emits a roughly constant
  amount of energy per unit time (it's luminisity). As you go to
  larger distances from the star, this energy is spread out over the
  surface of an imaginary sphere of larger radius.
\end{enumerate}

\begin{figure}[h!]
\begin{center}
  \includegraphics[width=0.75\textwidth]{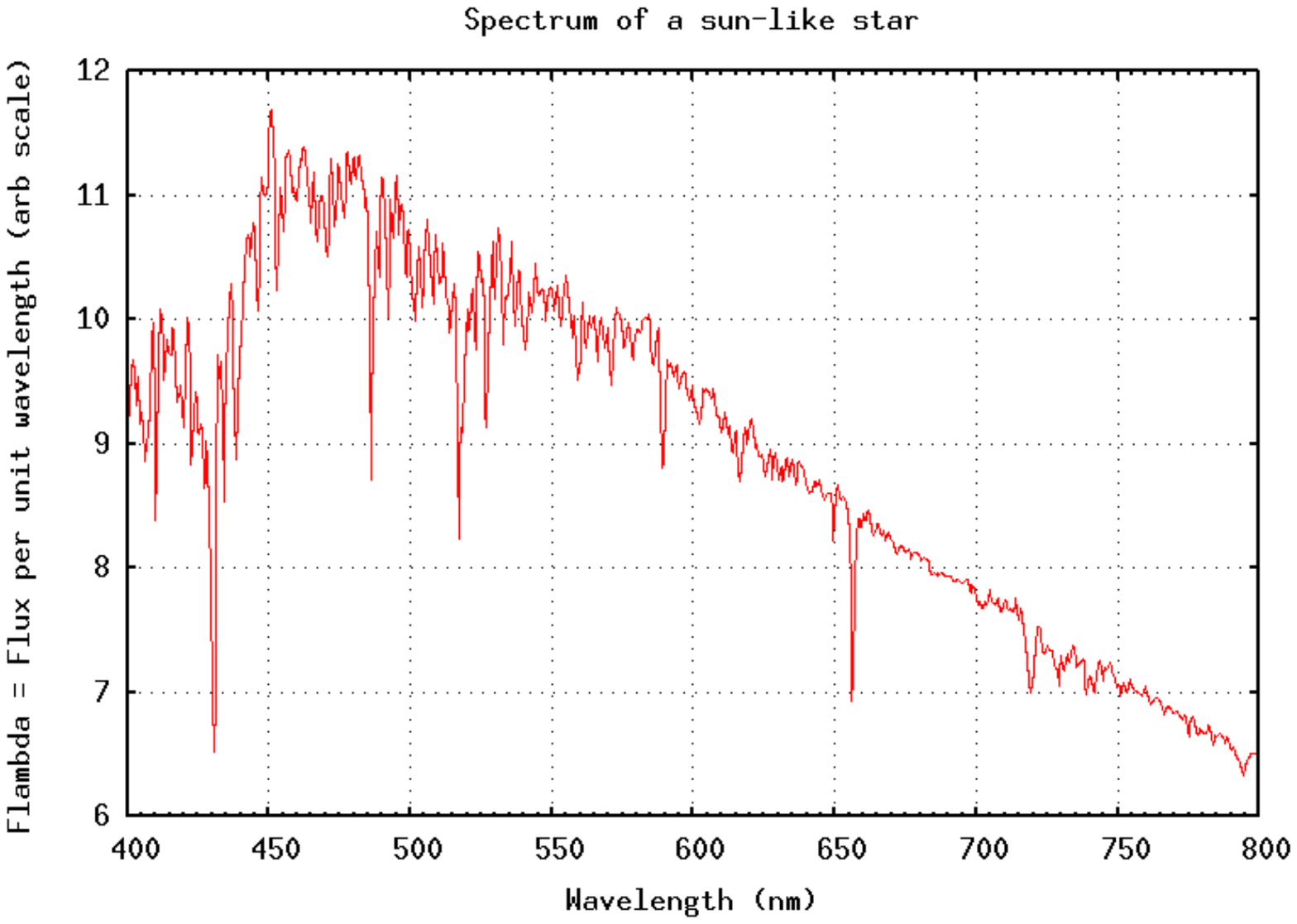}
  \includegraphics[width=0.75\textwidth]{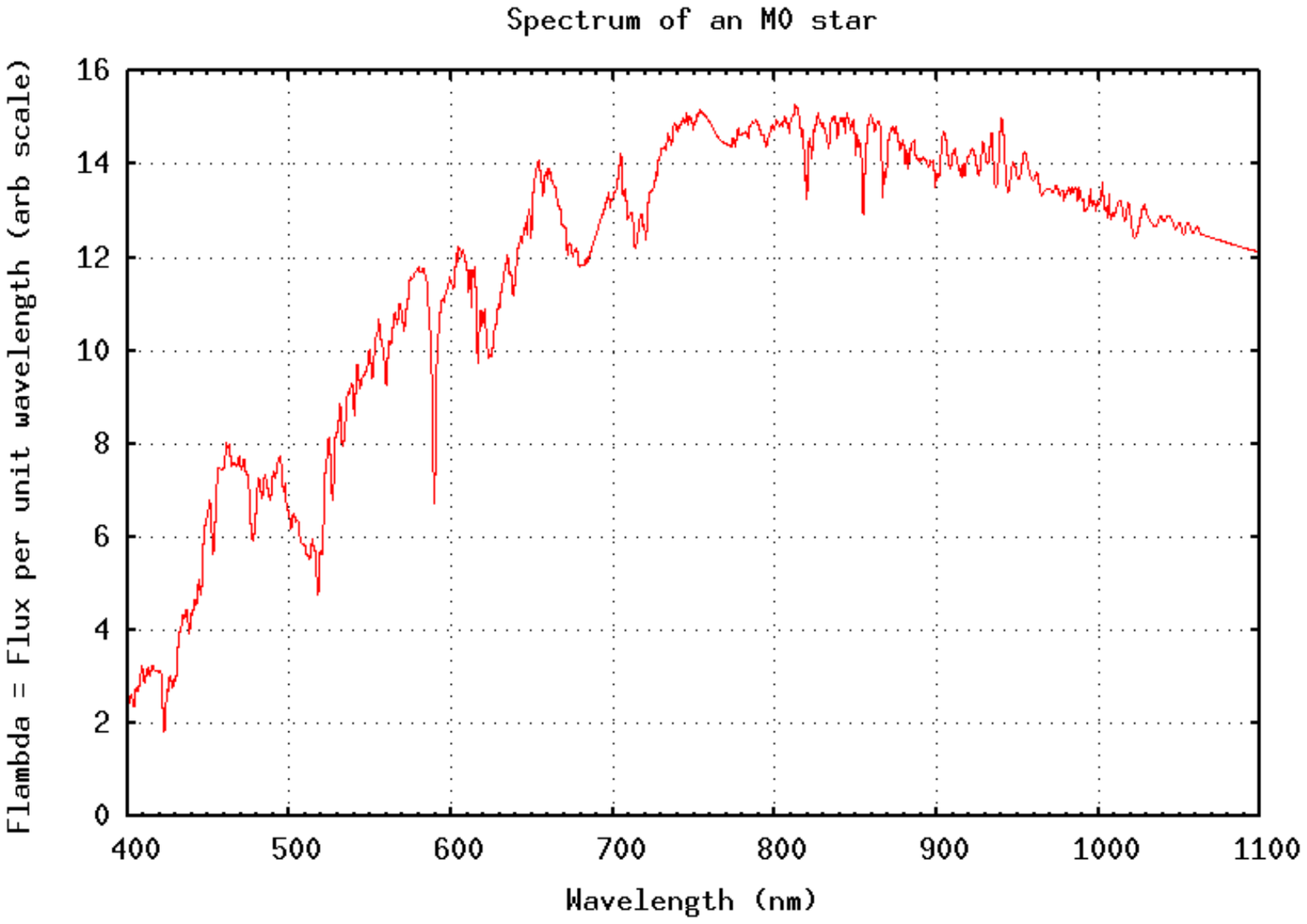}
\end{center}
\end{figure}

%% file: PSets/pset06.tex
\subsection{Problem Set 6}

\bigskip

\noindent (1) The flux of sunlight that strikes both the Earth and
the Moon when the Sun is directly overhead is about 1360 W~m$^{-2}$.
The moon has a very low albedo (about .1), very little thermal
inertia, and very little ability to move energy horizontally, so that
every point on the surface immediately comes into equilibrium with the
sunlight that is striking it. The Earth's surface has a large thermal
inertia, and the oceans and atmosphere have the ability to transport
heat horizontally. As a result, it can smooth out differences in
incoming sunlight between different locations. One example of this is
the fact that the temperature on Earth only drops a little over night
even though no sunlight is being absorbed, so that the equilibrium
nighttime temperature is near absolute zero.

\begin{enumerate}
\item What temperature do you expect the hottest point of the Moon to have (in K and in $^\circ$F)?
\item Why do Astronauts wear space suits that are white?
\item Suppose it's currently noon, and that we're considering the
  equinox. Roughly what flux of sunlight (in W~m$^{-2}$) strikes the
  Earth over Chicago ($\approx$42$^\circ$N)? How about for noon on the winter
  and summer solstices (given that the Earth's obliquity is
  $\approx$23$^\circ$)?
\item What's the flux of sunlight (in W~m$^{-2}$) hitting the
  Earth when averaged over Earth's entire surface?
\item If the Earth had no greenhouse gases and a uniform surface
  temperature, what surface temperature would the planet approach to
  balance this incoming sunlight, assuming the planet absorbed all
  incoming sunlight? What would the surface temperature be if the
  planet had an average albedo of 0.3?
\item Consider a model in which the Earth is surrounded by a single-layer
  atmosphere with emissivity $\epsilon$ in the longwave, but
  completely transparent in the shortwave. Assume an albedo
  of 0.3. What value of $\epsilon$ would the atmosphere need to have
  to give Earth its current surface temperature of 287~K?
\item Using Kirchhoff's law, what is the longwave absorptivity of this
  single-layer atmosphere? Assuming no longwave scattering, what is
  the longwave transmissivity? Using Beer's law what is the optical
  thickness of this atmosphere?
\end{enumerate}

%% file: PSets/pset07.tex
\subsection{Problem Set 7}

\bigskip

\noindent (1) The figure below is the type of map used in weather
forecasting. It is a contour plot of the geopotential height of the
500~hPa pressure surface on a particular day at a particular time. If
we imagine a surface through the atmosphere that has a constant
pressure of 500~hPa, sometimes this surface will go up and sometimes
it will go down as we move horizontally. The plot shows the height of
this surface above sea level. To orient you, the plot is a polar
projection, with the north pole at the center. Continents are shown in
blue and oceans in white. \textit{The reason this plot is useful is that the
500~hPa geopotential height is approximately a streamfunction of the
atmospheric flow.} So we can directly infer atmospheric velocities by
looking at it. The numbers are contour levels in decameters (570
decameters = 5700 meters) and the contour spacing is 60~m. ``H''
represents a local maximum of the height field and ``L'' represents a
local minimum.

\begin{figure}[h!]
\begin{center}
  \includegraphics[width=0.48\textwidth]{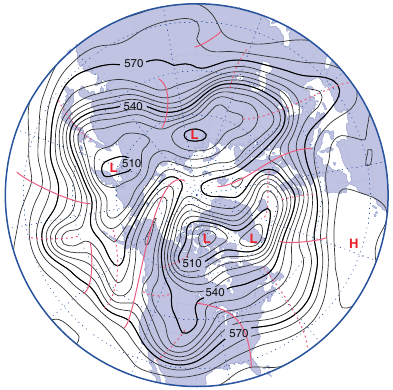}
\end{center}
\end{figure}

\begin{enumerate}
\item The jet stream is the strong upper level (e.g., 500~hPa) flow in
  the midlatitudes. What feature of this plot tells us where the jet
  stream is? Hint: what feature of a streamfunction tells you that the
  speed is high?
\item The jet stream meanders north and south, but on average is the
  flow ``westerly'' (from the west) or ``easterly'' (from the east)?
\item Over the western United States, does the jet stream appear to be
  meandering northward or southward? How should this affect the
  temperature in that region?
\item Over the eastern United States, does the jet stream appear to be
  meandering northward or southward? How should this affect the
  temperature in that region?
\item Is the circulation around the lows clockwise or counterclockwise?
\item Is the circulation around the highs clockwise or counterclockwise?
\end{enumerate}

%% file: PSets/pset08.tex
\subsection{Problem Set 8}

\bigskip

\noindent (1) The figure below is a diagram of a northern hemisphere
tropical cyclone, and the winds within it, viewed in a radial cross
section.

\begin{figure}[h!]
\begin{center}
  \includegraphics[width=\textwidth]{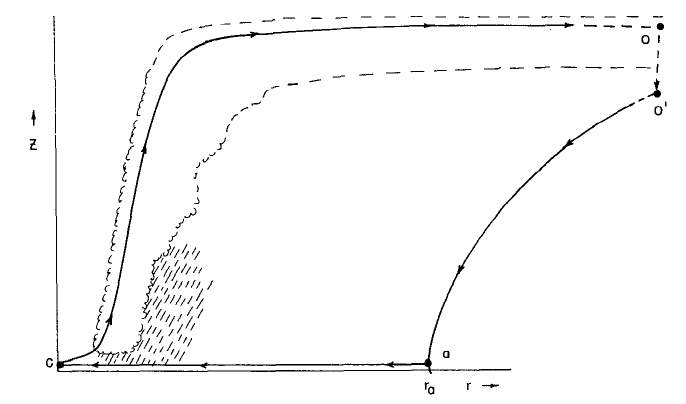}
\end{center}
\end{figure}

\begin{enumerate}
\item Is the divergence of the horizontal velocity field near the
  surface at the center of the tropical cyclone positive, negative, or
  zero?
\item Is the divergence of the horizontal velocity field at the center
  and top of the tropical cyclone positive, negative, or zero?
\item In the eyewall air is mostly convecting vertically. Use the
  conservation of air mass to argue that the sign of the divergence at
  the center and top of the cyclone must be opposite of that at the
  center near the surface.
\item What are the units of the divergence of the horizontal velocity
  field?
\item Suppose that the diameter of the eye of this tropical cyclone is
  50~km and the radial wind speed in the eyewall is 1~m~s$^{-1}$.
  Calculate the divergence of the velocity field near the surface at
  the center of the tropical cyclone.
\item Is the vorticity at the center of
  the tropical cyclone positive, negative, or zero?
\item What are the units of the vorticity?
\item Suppose that the diameter of the eye of this tropical cyclone is
  50~km and the cyclonic wind speed in the eyewall is 60~m~s$^{-1}$.
  Calculate vorticity at the center of the tropical cyclone.
\item The vorticity due to the rotation of the planet can be
  calculated as $2 \Omega \sin(\theta_l)$, where $\theta_l$ is the
  latitude. Suppose that the hurricane is at a latitude of
  20$^\circ$N. Calculate the planetary vorticity and compare it to the
  local, relative vorticity of the tropical cyclone, which you
  calculated in the last part.
\item A streamfunction can be found for the non-divergent component of
  the velocity field in a tropical cyclone. To a good approximation,
  this streamfunction is
  $\psi = 2 u_0 r_0^{\frac{1}{2}}\left( x^2 + y^2
  \right)^{\frac{1}{4}}$,
  where $u_0$ is the cyclonic wind speed in the eyewall and $r_0$ is
  the radius of the eye. Calculate the horizontal components of the
  non-divergent wind velocity, $u$  and $v$.
\item If we define $r$ as the radius from the center of tropical
  cyclone and $\theta$ as the counter-clockwise angle from the x-axis,
  then $x=r \cos( \theta )$ and $y = r \sin( \theta )$. Rewrite $u$
  and $v$ in terms of $r$ and $\theta$.
\item How do the velocities decay as a function of the distance from
  the center of the tropical cyclone ($r$)?
\end{enumerate}

%% file: PSets/pset09.tex
\subsection{Problem Set 9}

\bigskip

\noindent (1) A major attraction in Nanyuki, Kenya, is watching water
drain out of a bucket on both sides of the equator and observing which
way the water spins. Watch this video
(\href{https://youtu.be/X0GLILQck-g}{https://youtu.be/X0GLILQck-g}).

\begin{enumerate}
\item Estimate the Rossby number of the flow in the bucket of water.
  Is the Coriolis force strong enough to influence the direction the
  bucked drains?
\item Estimate how slowly the water would have to spin to ensure that
  the flow is definitely influenced by the Coriolis force. 
\item Would it matter if you did this experiment at the North Pole
  instead of near the equator?
\end{enumerate}

\bigskip

\noindent (2) Cyclone Yasi hit Australia on February 3, 2011 and
caused \$3.6 billion in damage. Consider the map of Australia and the
surface pressure map of Yasi just before it hit Australia on the
following pages.  In this problem you may assume that air near the
surface has a density of about 1~kg~m$^{-3}$ and use the fact that a
degree of latitude measures 111~km.

\begin{enumerate}
\item Did the air rotate clockwise or counterclockwise in Cyclone Yasi? 
\item Estimate the wind speed assuming geostrophic balance. Compare
  your answer to the speed of sound at sea level (340~m~s$^{-1}$). Is
  your answer reasonable?
\item Now estimate the wind speed by balancing both the Coriolis and
  centrifugal accelerations against the pressure gradient force.
\end{enumerate}

\begin{figure}[h!]
\begin{center}
  \includegraphics[width=\textwidth]{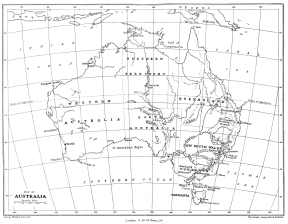}
\end{center}
\end{figure}

\begin{figure}[h!]
\begin{center}
  \includegraphics[width=\textwidth]{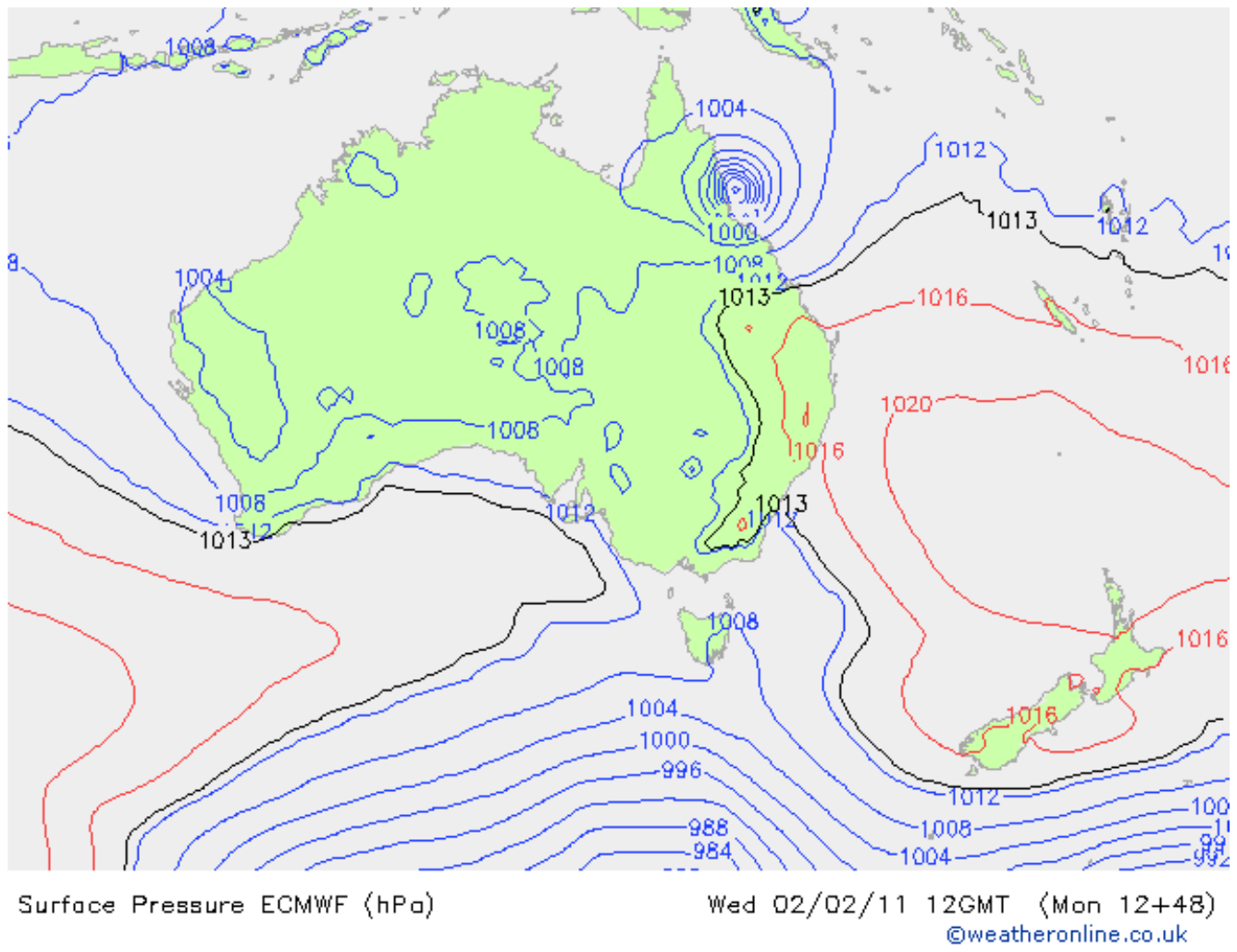}
\end{center}
\end{figure}

%% file: PSets/pset10.tex
\subsection{Problem Set 10}

\bigskip

\noindent (1) A cylindrical column of air at $30^{\circ}$ latitude
with radius 100~km expands horizontally to twice its original
radius. The air is initially at rest. You can assume that the volume
of the air column is conserved.

\begin{enumerate}
\item After the expansion, what is the mean tangential velocity at the
  perimeter?
\item The air column that was initially at rest now has relative
  vorticity. Where did this vorticity come from?
\item Calculate the Rossby number for the flow.
\item Is planetary rotation important for this flow?
\end{enumerate}

\bigskip 

%
%

\noindent (2) Assume there is tornado in Kansas. 
\begin{enumerate}
\item Calculate the Rossby number if the tornado has a tangential
  velocity of 50~m~s$^{-1}$ at a radius of 100~m. Will planetary
  rotation be important for the flow?
\item Now suppose the tornado has a tangential velocity of
  70~m~s$^{-1}$ at a radius of 1~km. What is the Rossby number?
\item For this stronger storm, how small would the velocity have to be
  for planetary rotation to significantly affect the flow? Would this
  be a tornado?
\end{enumerate}

%% file: Midterms/midterm1_2017.tex
\subsection{Midterm 1, 2017}

\bigskip
\noindent Titan is a moon of Saturn that has a thick atmosphere
composed primarily of N$_2$ with a small amount of CH$_4$. Because of
the cold temperatures on Titan, CH$_4$ has a condensation cycle
similar to water on Earth. There are seas of methane on the surface
which can evaporate and condense in the atmosphere, leading to methane
clouds and methane precipitation. On January 14, 2005, the Huygens
probe parachuted into Titan's atmosphere making measurements along the
way \cite{fulchignoni2005situ}. The plots below show data for pressure
and temperature as a function of height in Titan's lower atmosphere as
measured by the Huygens probe.

\begin{figure}[h!]
\begin{center}
  \includegraphics[width=0.45\textwidth]{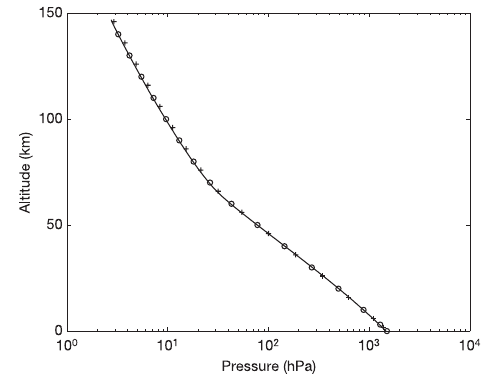} 
  \includegraphics[width=0.45\textwidth]{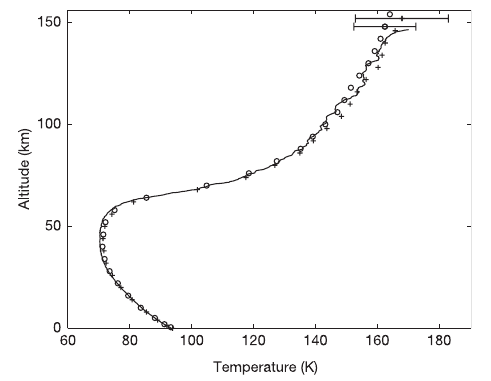} 
\end{center}
\end{figure}

\noindent (1) Estimate the scale height in the lowest 50~km of Titan's
atmosphere from the probe data.

\clearpage

\noindent (2) Use your scale height and the fact that
$g=1.3$~m~s$^{-2}$ on Titan to estimate the average molecular mass of
Titan's atmosphere. Is your estimate roughly consistent with N$_2$
being the dominant atmospheric constituent?

\vspace{3.5in}

\noindent (3) Calculate the dry adiabatic temperature lapse rate
using $C_p=3.5R$. Estimate the lapse rate
in the lowest 25~km of Titan's atmosphere from the probe data. Is your
calculation roughly correct?

\clearpage

\noindent (4) Notice that the temperature lapse rate in the probe data
appears to be somewhat less than the dry adiabatic lapse rate,
particularly above the lowest 5-10~km of the atmosphere. Explain how
condensation of CH$_4$ in the atmosphere could produce a lapse rate
lower than the dry adiabatic lapse rate.

\vspace{3.5in}

\noindent (5) We could define a virtual temperature to account for
variable amounts of CH$_4$ on Titan, just like the virtual temperature
on Earth accounts for variable amounts of H$_2$O. Suppose we compare
two air parcels with the same pressure and temperature, but one has a
higher CH$_4$ mixing ratio than the other. Which air parcel would have
a higher value of Titan's virtual temperature? Which air parcel would
be more dense?

%% file: Midterms/midterm1_2018.tex
\subsection{Midterm 1, 2018}
\bigskip

%
%
%

%
%
%
%
%
%
%
%

Super-Earths are Earth-like planets that are larger. Suppose a
different star has a super-Earth named Chill Zone 274b orbiting it
with super-Earthlings living on it. Assume that Chill Zone 274b has:
(1) the same surface temperature as Earth ($T_{s-cz}=290~K$), (2) the
same atmospheric composition as Earth, (3) an atmospheric surface
pressure that is twice as large as Earth's surface pressure
($P_{s-cz}=2P_{s\oplus}$), and (4) a radius twice as large as Earth's
($R_{cz}=2R_\oplus$). The symbol $\oplus$ means Earth.

\bigskip 

\noindent (1) Chill Zone 274b's surface air density has a strong
impact on the usage of wind power and friction acting on machines of
locomotion that the super-Earthlings employ.  What is the surface air
density on Chill Zone 274b ($\rho_{s-cz}$) in terms of Earth's surface
air density ($\rho_{s\oplus}$)?

\vspace{2in}

\noindent (2) We can assume that Chill Zone 274b has the same interior
density as Earth and is roughly spherical (as is Earth). Under these
assumptions, the surface gravitational acceleration is proportional to
the volume of a planet divided by its surface area. If Earth's surface
gravitational acceleration is $g_\oplus=10$~m~s$^{-2}$, what is Chill
Zone 274b's ($g_{cz}$)?

\clearpage

\noindent (3) Let's assume that the height of typical large-scale
topography on a planet is inversely proportional to the surface
gravitational acceleration because gravity squeezes the topography
downward. This would mean that if a planet had 10 times higher surface
gravitational acceleration, the height of its typical large-scale
topography would be 10 times smaller. The Tibetan Plateau, with a
height of 4~km, is the typical scale of large-scale topography on
planet Earth. The equivalent feature on Chill Zone 274b is the Bob
Marley Mount. What is the height of the Bob Marley Mount?

\vspace{2in}

\noindent (4) The scale height in Earth's atmosphere is $H=8$~km. What
is the scale height in Chill Zone 274b's atmosphere?

\vspace{2in}
\noindent (5) What is the air pressure on the Tibetan Plateau? Write
your answer in terms of Earth's surface pressure, $P_{s\oplus}$. Leave
transcendental functions unevaluated (e.g., $\sin(2)$).

\clearpage

\noindent (6) The heat capacity of dry air on Earth is
$10^3$~J~kg$^{-1}$K$^{-1}$. Assuming that Earth's atmosphere is dry
and that topographic features have the same temperature as surrounding
atmospheric air, what is the temperature on the Tibetan Plateau?

\vspace{2in}

\noindent (7) What is the air pressure on the Bob Marley Mount? Again,
write your answer in terms of $P_{s\oplus}$ and leave transcendental
functions unevaluated.

\vspace{2in}

\noindent (8) Assuming that Chill Zone 274b's atmosphere is dry and
that topographic features have the same temperature as surrounding
atmospheric air, what is the temperature on the Bob Marley Mount?

\clearpage

Humans need to lose the heat they are continually
creating to the surrounding environment, or else they would
overheat. An important way that they do this is through the
evaporation of sweat from the surface of the skin. The skin surface
has a temperature of about 35$^\circ$C. If the ambient air has some
form of temperature higher than the skin surface temperature, humans
will not be able to lose heat, so they will overheat and die in a few
hours. If humans continue to emit CO$_2$ on current projections, many
regions of Earth would be in this regime and we would have to go live
on Chill Zone 274b instead. 

\bigskip

\noindent (9) Which type of temperature (temperature, potential
temperature, virtual temperature, dew point, wet bulb temperature) is
the relevant one to compare the skin surface to in order to determine
whether humans will overheat?

\vspace{1in}

\noindent (10) The latent heat of evaporation of water is about $3
\times 10^6$~J~kg$^{-1}$. A typical human loses 100~W of power. Let's
assume that this power is lost exclusively through evaporation of
sweat. If a year's worth of this sweat were condensed and collected,
what would it's volume be in multiples of the volume of a typical
human? Useful facts: There are about $3 \times 10^7$~s in a year. A
typical human has a volume of 100~L or 0.1~m$^{3}$.

%% file: Midterms/midterm1_2019.tex
\subsection{Midterm 1, 2019}

\bigskip

In this midterm we will be considering the Martian atmosphere. We will
use observations made by a Mars-orbiting satellite
\cite{wolkenberg2010atmospheric}. We will look at one slice through
the atmosphere measured during one orbit. It will be useful for you to
know that the Martian atmosphere is composed primarily of CO$_2$,
which has molecular mass of 44 g/mol. CO$_2$ is a triatomic molecule,
so its heat capacity is related to its gas constant by $C_p=4.5R$. The
ideal gas law constant is 8.3~J~mol$^{-1}$~K$^{-1}$ and the
gravitational accelaration at the surface of Mars is
3.7~m~s$^{-1}$. The Martian atmosphere is fairly dusty, even when a
global dust storm is not occurring. Dust is a good absorber of
sunlight.

\bigskip

\noindent (1) Estimate the gas constant of the Martian atmosphere.

\vspace{2.5in}

\noindent (2) Estimate the heat capacity of the Martian atmosphere.

\clearpage

\noindent (3) Estimate the dry adiabatic lapse rate on Mars.

\vspace{3in}

\noindent (4) The figure below shows measurements of the temperature
of the Martian atmosphere. Estimate the lapse rate from the figure at
a latitude of 1N.

\begin{figure}[h!]
\begin{center}
  \includegraphics[width=0.8\textwidth]{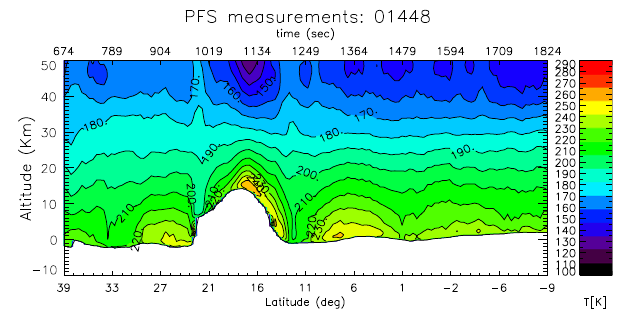}
\end{center}
\end{figure}

\clearpage

\noindent (5) Is the Martian atmosphere stable, unstable, or neutral
to convection?

\vspace{2.5in}

\noindent (6) At 1N, is the potential temperature of the Martian
atmosphere higher or lower at an altitude of 20~km than at the
surface?

\vspace{2.5in}

\noindent (8) Why does the Martian atmosphere have the potential
temperature profile that it does?

\clearpage

\noindent (7) Based on your answer to the last two questions, is the
Martian atmosphere more like Earth's troposphere or stratosphere?

\vspace{1.75in}

\noindent (9) Estimate the scale height of the Martian atmosphere.

\vspace{2.25in}

\noindent (10) Roughly estimate the air density at the top of the
topographic feature at 18N (part of Olympus Mons) relative to the air
density at the surface near 1N.

%% file: Midterms/midterm2_2017.tex
\subsection{Midterm 2, 2017}

\bigskip In a paper published in the journal \textit{Science} this
year, Zhai et al. described a new material they had invented that
might lead to a cheaper and more efficient way to cool homes
\cite{zhai2017scalable}. The material consists of SiO$_2$ microspheres
embedded in a thin layer of polymethylpentene (TPX) (see figure
below).

\begin{figure}[h!]
\begin{center}
  \includegraphics[width=0.8\textwidth]{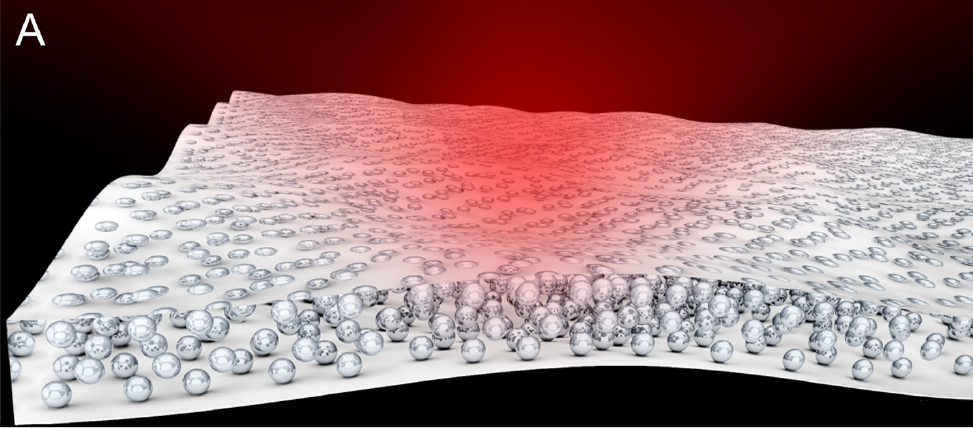} 
\end{center}
\end{figure}

TPX is nearly transparent to solar radiation and the SiO$_2$
microspheres have a size and distribution chosen so that they interact
strongly with terrestrial radiation. The material is then coated with a
thin (200~nm) layer of silver on the bottom side. The material would
be used by placing it on roofs with the silver side pointing down.

\clearpage

In the diagram below green represents TPX, the yellow
circles represent SiO$_2$ microspheres, and gray represents the
silver (Ag) layer. 

\begin{figure}[h!]
\begin{center}
  \includegraphics[width=0.5\textwidth]{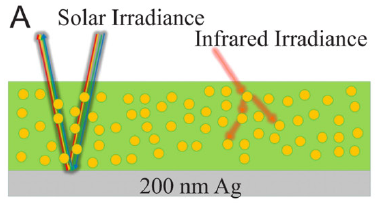} 
\end{center}
\end{figure}

\noindent (1) Based on the diagram (and Zhai et al.'s objective), why
was the thin layer of silver added to the bottom of the material?

\clearpage

The top panel of the diagram below shows the typical solar radiation
reaching the surface (red) and the typical surface terrestrial
radiation from a blackbody reaching space (blue). The bottom panel
shows the emissivity of the TPX/SiO$_2$ microsphere mixture as a
function of wavelength, both theoretically and as measured
experimentally.

\begin{figure}[h!]
\begin{center}
  \includegraphics[width=0.6\textwidth]{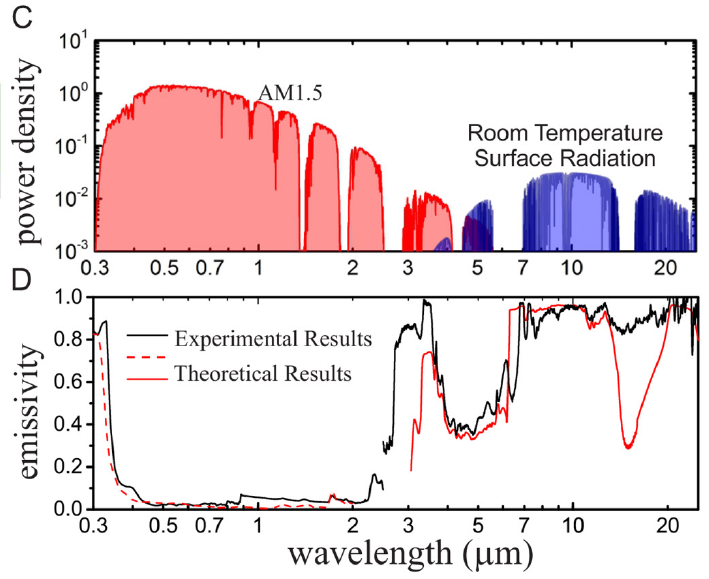} 
\end{center}
\end{figure}

\noindent (2) Why is the solar radiation generally at shorter wavelengths
than the terrestrial radiation?

\clearpage

\noindent (3) Why is the surface terrestrial emission from a blackbody
that reaches space nearly zero at certain wavelengths?

\vspace{3in}

\noindent (4) Why has the material been chosen to have as small an
emissivity as possible for wavelengths shorter than about 3~$\mu$m and
an emissivity as close to one as possible for wavelengths larger than
about 3~$\mu$m?

\clearpage

The diagram below shows the temperature and power emitted by the
material in testing as a function of time when it was put outside and
exposed to the dirunal cycle in sunlight and ambient atmospheric
conditions.

\begin{figure}[h!]
\begin{center}
  \includegraphics[width=0.8\textwidth]{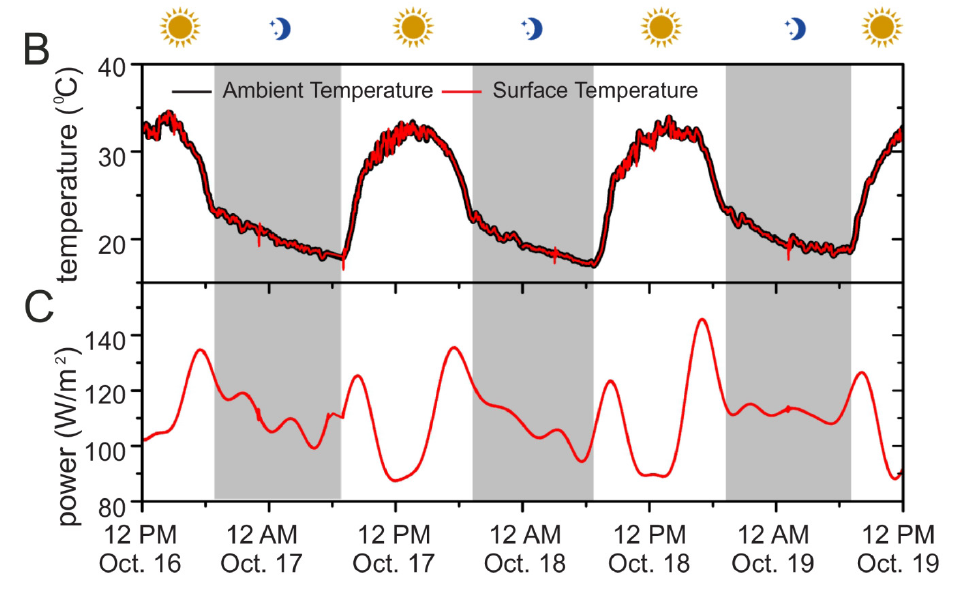} 
\end{center}
\end{figure}

\noindent (5) The air conditioner in a typical house draws on the
order of 1000~W. If the roof of a typical house were covered with this
new material, would it be able to keep the house cool without running
an air conditioner? Only an order of magnitude estimate is required to
answer this question.

%% file: Midterms/midterm2_2018.tex
\subsection{Midterm 2, 2018}
\bigskip

Consider the following news article, which appeared in the science
section of the esteemed national newspaper, \textit{The Onion} on
4/20/18:

\bigskip

``WASHINGTON-In an effort to make the solar system's central star look
as badass as possible, NASA officials announced Friday the agency's
plans to place a 864,600-mile-wide pair of shades on the sun. ``With
this mission, we'll be taking a great leap forward in our
understanding of how cool and chilled-out our sun really is,'' said
NASA acting administrator Robert M. Lightfoot Jr., noting that the
eyewear's design was based in part on a 1588 drawing by Galileo, who
was branded a heretic at the time for suggesting the sun wasn't as
laid-back as previously thought. ``For generations to come, every time
we look up at the sky, we'll be reminded that the sun is this super
rad celestial body that's rocking a sweet pair of American-built
shades.'' Lightfoot added that if the mission was a success, NASA
would expedite its plans to launch a giant, ice-cold glass of lemonade
for the sun to sip.''

\begin{figure}[h!]
\begin{center}
  \includegraphics[width=\textwidth]{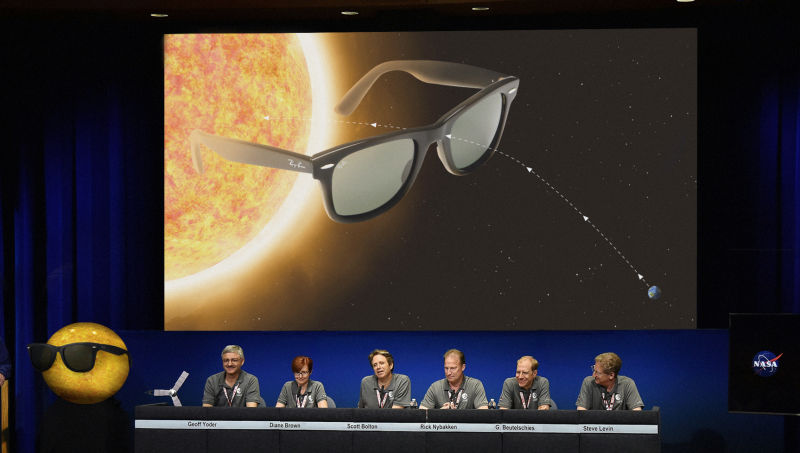}
\end{center}
\end{figure}

\clearpage

The article doesn't specify this, but a friend at NASA told me that
the shades will be set up to rotate so that they will always be
pointed directly at Earth as it orbits the Sun. She also told
me that the shades will transmit 20\% of light that hits them, equally
at all wavelengths. 

Interestingly, we may have already observed an example of an alien
civilization trying to increase the dankitude of its star with some
shades. The star KIC 8462852, also known as Boyajian's Star, is famous
for having strange dips in its lightcurve. This means that it gets
much darker sometimes for unknown reasons, possibly because of an
alien superstructure such as some sweet shades.

We'll be using the simple energy balance model of climate from class
to understand the effects of NASA's mission on planet Earth. To remind you,
the model has a surface and one atmospheric layer. The surface has a
longwave emissivity of 1 and a shortwave albedo of $A$. The atmosphere
has a longwave emissivity of $\epsilon$ and is transparent to
shortwave radiaton. Here is the diagram of the model:

\begin{figure}[h!]
\begin{center}
  \includegraphics[width=0.5\textwidth]{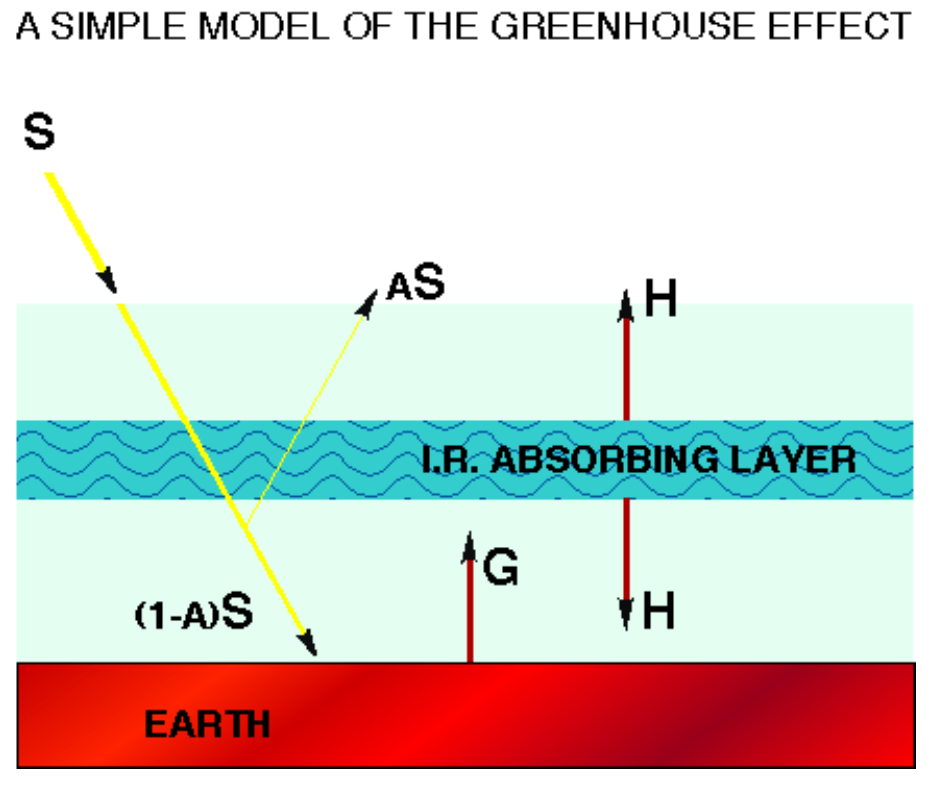}
\end{center}
\end{figure}

In class we rewrote this model in terms of the solar flux ($S$, in
W~m$^{-2}$), the surface temperature ($T_s$, in $K$), the atmospheric
temperature ($T_a$, in $K$), and the Stefan-Boltzmann constant
($\sigma=5.67 \times 10^{-8}$~W~m$^{-2}$).

\clearpage

\noindent (1) Solve the energy balance model for $T_s$ in terms of
$S$, $A$, $\epsilon$, and $\sigma$.

\clearpage

\noindent (2) Assume that the shades can be represented by two
circles, each with a radius of $R_{sh}$, and that the Sun has a radius
of $R_{sun}$. Assume that $R_{sh} \leq \frac{1}{2}R_{sun}$. Use this
information to calculate the solar flux, $S_{sh}$, that Earth will
receive after NASA installs the shades.

\vspace{3in}

\noindent (3) In addition to making the Sun as badass as possible, the
shades have the added benefit of being able to counteract global
warming. Let's assume that the preindustrial Earth has a surface
temperature $T_{s0}$ and an emissivity $\epsilon_0$, which are related
by the formula you derived in question (1). Assume that as a result of
global warming $\epsilon_0$ increases to $\epsilon_w$. What value
should $R_{sh}$ take in order to keep Earth's temperature constant
despite the increase in emissivity?  Assume that Earth's albedo stays
constant.

\clearpage

\noindent (4) It is possible to estimate the surface temperature of an
object by measuring its electromagnetic radiation emission
spectrum. After NASA puts the shades in place, will our estimate of
the Sun's temperature using this method change?

\vspace{2in}

\noindent (5) If for some reason NASA's shade plan fails, an
alternative plan for counteracting global warming with geoengineering
is to put sulfate aerosols into the stratosphere, either with big guns
or airplanes. These sulfate aerosols would scatter incoming sunlight,
cooling the planet. Estimate the total mass of sulfate aerosols that
would have to be put into the stratosphere to counteract global
warming. Give your answer in multiples of the mass of full 747
airplanes ($\approx 4 \times 10^5$~kg). You may assume that global
warming alters radiative balance by 4~W~m$^{-2}$, the solar constant
is 1365~W~m$^{-2}$, the mass scattering coefficient for sulfate
aerosols is about $\kappa \approx 4$~m$^2$~g$^{-1}$, the planetary
albedo is about 0.3, and the planetary radius is about 6,300~km. You
may also find the following expansion useful: $e^{-\tau} \approx
1-\tau$.

%% file: Midterms/midterm2_2019.tex
\subsection{Midterm 2, 2019}

\bigskip

The Starshot Breakthrough Initiative has the goal of sending
spacecraft to the nearest stellar system (Proxima Centauri) with a
travel time of 20 years or less within the next 50 years. The current
plan is to build very light spacecraft with large lightsails and to
accelerate them using lasers based on Earth (see the diagram
below). The lasers would accelerate the spacecraft via photon momentum
transfer. A large number of spacecraft would be launched because of
the dangers of interstellar dust particles and the potential failure
of instrumentation.

\begin{figure}[h!]
\begin{center}
  \includegraphics[width=0.8\textwidth]{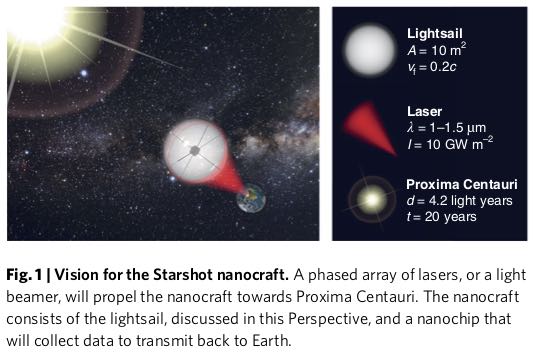}
\end{center}
\end{figure}

Recently a team of CalTech materials scientists and physicists
considered the practical engineering issues related to the Starshot
Breakthrough Initiative plan \cite{atwater2018materials}. In this
midterm we will be thinking through some of the results from their
paper. Here are some quantities that may be of interest to you:
Proxima Centauri is 4.2 light years from Earth. The speed of light is
c=3$\times$10$^8$~m~s$^{-1}$ and the Stefan-Boltzmann constant is
$\sigma$=6$\times$10$^{-8}$~W~m$^{-2}$~K$^{-4}$.  The spacecraft will
travel at a speed of approximately 20\% of the speed of light and it
will take approximately 150~s for the laser to accelerate them to this
speed. The laser will emit electromagnetic radiation in the wavelength
range of 1--1.5~$\mu$m with an intensity of I=10~GW~m$^{-2}$, where
1~GW=10$^9$~W. The average penetration depth of hydrogen and helium
atoms is about 1~mm at a speed of 0.2c. Interstellar dust particles
have a typical mass of about m=10$^{-16}$~kg. The light sail will have
an area of 10~m$^2$, be made out of a material with density of
5$\times$10$^3$~kg~m$^{-3}$, and have a total mass of 1 gram. The
diameter of an atom of the lightsail material is
2$\times$10$^{-10}$~m. The lightsail material has an absorption
coefficient of $\kappa$=0.5~cm$^{-1}$ and a melting temperature of
1000~K.

\clearpage

\noindent (1) Should the lightsail material have a high or low
reflectivity in the wavelength range of the laser? Why?

\vspace{1.7in}

\noindent (2) Should the lightsail material have a high or low
emissivity in the wavelength range of the laser? Why?

\vspace{1.7in}

\noindent (3) Should the lightsail have a high or low emissivity at
all other wavelengths than the wavelength range of the laser? Why?

\clearpage

\noindent (4) The Breakthrough Starshot laser will have a wavelength
of 1--1.5~$\mu$m. What frequency range does this correspond to?

\vspace{2in}

\noindent (5) Using the chart below, what part of the electromagnetic
spectrum will the Breakthrough Starshot laser be in?

\begin{figure}[h!]
\begin{center}
  \includegraphics[width=0.8\textwidth]{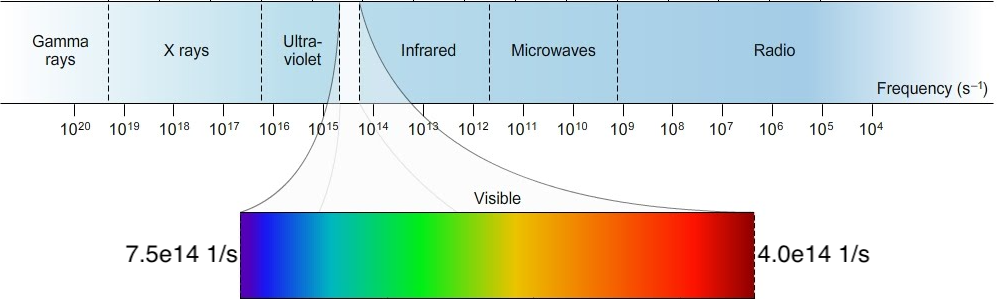}
\end{center}
\end{figure}

\clearpage

\noindent (6) Approximately how many atomic layers thick will the
lightsail be?

\vspace{2.2in}

\noindent (7) What fraction of the laser light will the lightsail absorb?

\vspace{2.2in}

\noindent (8) Assume that the lightsail has an emissivity of
$\epsilon$ at all wavelengths except for the very narrow range of the
laser. Write down a formula for the equilibrium temperature of the
lightsail when laser light is hitting it.

\clearpage

\noindent (9) Assume $\epsilon=0.1$. What is the equilibrium temperature of the lightsail? Would the lightsail melt?

\vspace{2.5in}

\noindent (10) Assume $\epsilon=0.001$. What is the equilibrium
temperature of the lightsail? Would the lightsail melt?

%% file: Finals/final_2017.tex
\subsection{Final, 2017}
\bigskip

Gliese 1214b (GJ1214b) is an extrasolar planet orbiting the star
Gliese 1214 that is located 42 light years ($4.0 \times 10^{17}$~m)
from the Sun \cite{charbonneau2009super}. GJ1214b is a transiting
planet, which means it passes between Gliese 1214 and us as it orbits
Gliese 1214. It has a size between Earth and Neptune, and is closer to
its star than Earth is.  GJ1214b is extremely exciting because it
allows us to test methods for observing the atmosphere of a relatively
small and not super hot planet. The hope is that we can refine our
techniques so that over the next twenty years as technology develops
we will be able to measure the atmospheres of more Earth-like planets,
and possibly detect signs of life. Using the material you have learned
in this course, you can now understand interesting aspects of the
atmosphere of GJ1214b! This final will walk you through this. Here is
some information about the star Gliese 1214 and the planet GJ1214b
that may be useful:

Gliese 1214 is a small, M4.5 star with an emission temperature of
about 3000~K, which compares to the Sun's value of 5800~K, and a
luminosity of 0.33\% of the Sun's ($L_\odot=3.8 \times 10^{26}$~W).
Gliese 1214 also has a mass of 16\% of the Sun's
($M_\odot=2.0 \times 10^{30}$~kg) and a radius of 21\% of the Sun's
($R_\odot=7.0 \times 10^{8}$~m). It is about 6 billion years old,
whereas the Sun is about 4.5 billion years old. The $\odot$ symbol
means the Sun's value.

GJ1214b has an orbital period of 1.5 Earth days and a mean orbital
distance of 0.014 AU (1 AU=$1.5 \times 10^{11}$~m is the distance
between the Earth and the Sun). The planet is tidally locked so that
it rotates once per orbit and always shows the same side to its star
(like the Moon does to Earth). Its mass is approximately 6.5 times
Earth's mass ($M_\oplus = 6.0 \times 10^{24}$~kg) and its radius is
approximately 2.65 times Earth's radius
($R_\oplus = 6.3 \times 10^6$~m), so that its surface gravity is
approximately 0.9 times Earth's surface gravity
($g_\oplus=9.8$~m~s$^{-2}$). The $\oplus$ symbol means Earth's value.

\clearpage

\noindent (1) If you were on a spaceship that traveled to the Gliese
1214 system and you looked out the window at the star, what color what
it be? Make a calculation to support your answer.

\clearpage

\noindent (2) Calculate the flux of stellar radiation GJ1214b receives
from its star in terms of multiples of the insolation Earth receives
($S_0=1360$~W~m$^{-2}$).

\vspace{3in}

\noindent (3) Assume that GJ1214b has an albedo of zero. Calculate
GJ1214b's emission temperature. You may use this emission temperature
later any time you need a temperature in a calculation.

\clearpage

In 2010 Jacob Bean, who is now a professor of astronomy at the
University of Chicago, published a famous paper measuring composition
of the atmosphere GJ1214b using a technique called transit
spectroscopy \cite{bean2010ground}. Transit spectroscopy involves
looking at the planet as the light from the star it is orbiting passes
through the planet's atmosphere on its way to us. The plot below shows
Jacob's transit measurements (black dots with error bars) of the
planetary radius as a function of wavelength compared to what would be
predicted by models incorporating various atmospheric
compositions. The wiggles in the models represent absorption by
atmospheric constituents, which change the apparent radius of the
planet.

\begin{figure}[h!]
\begin{center}
  \includegraphics[width=1.0\textwidth]{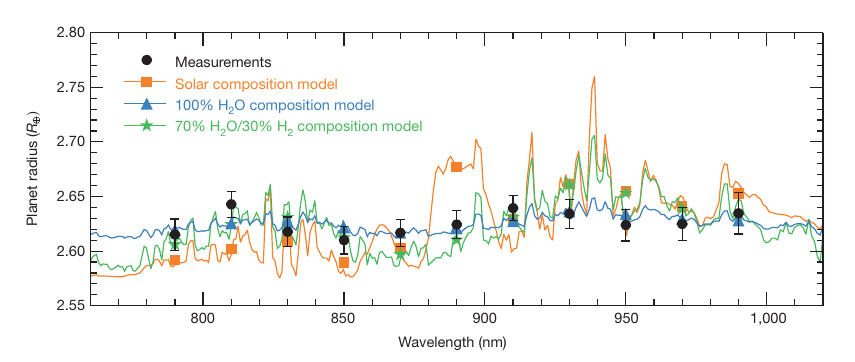} 
\end{center}
\end{figure}

\noindent (4) Suppose we are looking at a wavelength where GJ1214b's
atmosphere strongly absorbs the stellar light passing through it. Will
the planetary radius appear relatively large or relatively small at
this wavelength?

\clearpage

\noindent (5) Which model fits Jacob's measurements best?

\vspace{2.5in}

\noindent (6) Calculate the gas constant of GJ1214b's atmosphere
assuming the model that best fits the data. You can assume that the
``solar composition'' model is composed entirely of H$_2$. Remember
that the molecular mass of H is 1~g~mol$^{-1}$, the molecular mass
of O is 16~g~mol$^{-1}$, and the ideal gas constant is
8.3~J~mol$^{-1}$~K$^{-1}$.

\clearpage

\noindent (7) We can define a planetary Rossby number using the radius
as the length scale and the Coriolis parameter at the North Pole. The lowest
speed we can expect in GJ1214b's atmosphere is about
$U_{min}=1$~m~s$^{-1}$. Calculate the planetary Rossby number for
GJ1214b corresponding to a characteristic atmospheric speed of
$U_{min}$. Do you expect that geostrophic balance would prevail in
most of GJ1214b's atmosphere if the true characteristic atmospheric
speed is $U_{min}$?

\vspace{2.5in}

\noindent (8) The highest speed we can expect in GJ1214b's atmosphere
is the speed of sound, which we can calculate as $U_{max}=\sqrt{RT}$.
Calculate the planetary Rossby number for GJ1214b corresponding to a
characteristic atmospheric speed of $U_{max}$. Do you expect that
geostrophic balance would prevail in most of GJ1214b's atmosphere if
the true characteristic atmospheric speed is $U_{max}$?

\clearpage

\noindent (9) In Jacob's measurements the apparent radius of GJ1214b
at different wavelengths is given in terms of multiples of the radius
of Earth, $R_\oplus$. Estimate to one significant digit the size of
variations in planetary radius (in m) for the ``solar composition''
model and for the ``H$_2$O composition'' model from the figure.

\vspace{3in}

\noindent (10) The size of variations in the apparent radius of a
planet due to transit spectroscopy should be of the same order as the
scale height of the atmosphere. Estimate to one significant digit the
scale height for the ``solar composition'' model and for the ``H$_2$O
composition'' model. Compare your answers with the variations in the
apparent planetary radius that you found for these models.

%% file: Finals/final_2018.tex
\subsection{Final, 2018}
\bigskip

Hot Jupiters are gas giant exoplanets that orbit very close to their
host star. One such planet, WASP-33b, was discovered in
2010 \cite{cameron2010line} and is located 380 light years from planet
Earth. In this fun final we're going to demonstrate our knowledge of
the atmosphere by investigating WASP-33b.  WASP-33b orbits the star
WASP-33, which has a luminosity of $L=6.14L_{\odot}$, where
$L_{\odot}$ is the luminosity of the Sun, a mass of $M=1.50M_\odot$,
and a radius of $R=1.44R_\odot$. WASP-33b orbits WASP-33 at a distance
of 0.026~AU, where an AU is an astronomical unit, or the average
distance from the Sun to Earth, and is tidally locked, such that both
it's orbital and rotational periods are 1.22 days.  WASP-33b has a
radius that is 1.6 times Jupiter's radius, which is 71,500~km, or 11.2
times Earth's radius, and a mass of 2.1 times Jupiter's mass, which is
318 times Earth's mass. For this final, you may assume that the
typical speed of WASP-33b's atmosphere is 50~m~s$^{-1}$ and that it
has an albedo of zero. You may also need the Stefan-Boltzmann
constant, $\sigma = 5.67 \times 10^{-8}$~W~m$^{-2}$~K$^{-4}$. Finally,
it may be useful to know that Earth's rotation rate is $\Omega = 7.3
\times 10^{-5}$~rad~s$^{-1}$, that Earth's solar constant, the solar
radiative flux at Earth's distance from the Sun, is 1367~W~m$^{-2}$,
that Earth's average albedo is 0.31, that Earth's surface gravity is
9.81~m~s$^{-2}$, that Earth's mean density is 5.51~g~cm$^{-3}$, that
Earth's radius is 6,370~km, and that Earth's mass is $5.97 \times
10^{24}$~kg. You may also need to know that the surface gravity of a
planet or star scales like its mass divided by its surface area.

\clearpage

\noindent (1) We can use a planetary Rossby number to estimate whether
rotation, and the Coriolis force, is important for the atmospheric
circulation of a planet. For the planetary Rossby number, we use the
radius of the planet for the length scale, and we use the value of the
Coriolis parameter at the pole. Remember that $f=2\Omega \sin \phi$,
where $\phi$ is the latitude.  Estimate the planetary Rossby number of
WASP-33b.

\vspace{5in}

\noindent (2) Does rotation strongly affect the atmospheric dynamics
of WASP-33b?

\clearpage

\noindent Tad Komacek graduated from the Department of the Geophysical
Sciences in 2013, and recently defended his PhD thesis at the
University of Arizona. As part of his PhD thesis, he simulated the
atmospheric circulation of idealized planets like WASP-33b
\cite{komacek2016atmospheric}. The plots below show output from two of
these simulations assuming either high atmospheric friction or low
atmospheric friction.  The plots show velocity vectors and colormaps
of atmospheric temperature (yellow is hot and blue is cold). The
sub-stellar point (directly under the central star) is located at
(longitude,latitude)=(0,0). The anti-stellar point is located at
(longitude,latitude)=($\pm$180,0).

\begin{figure}[h!]
\begin{center}
  \includegraphics[width=\textwidth]{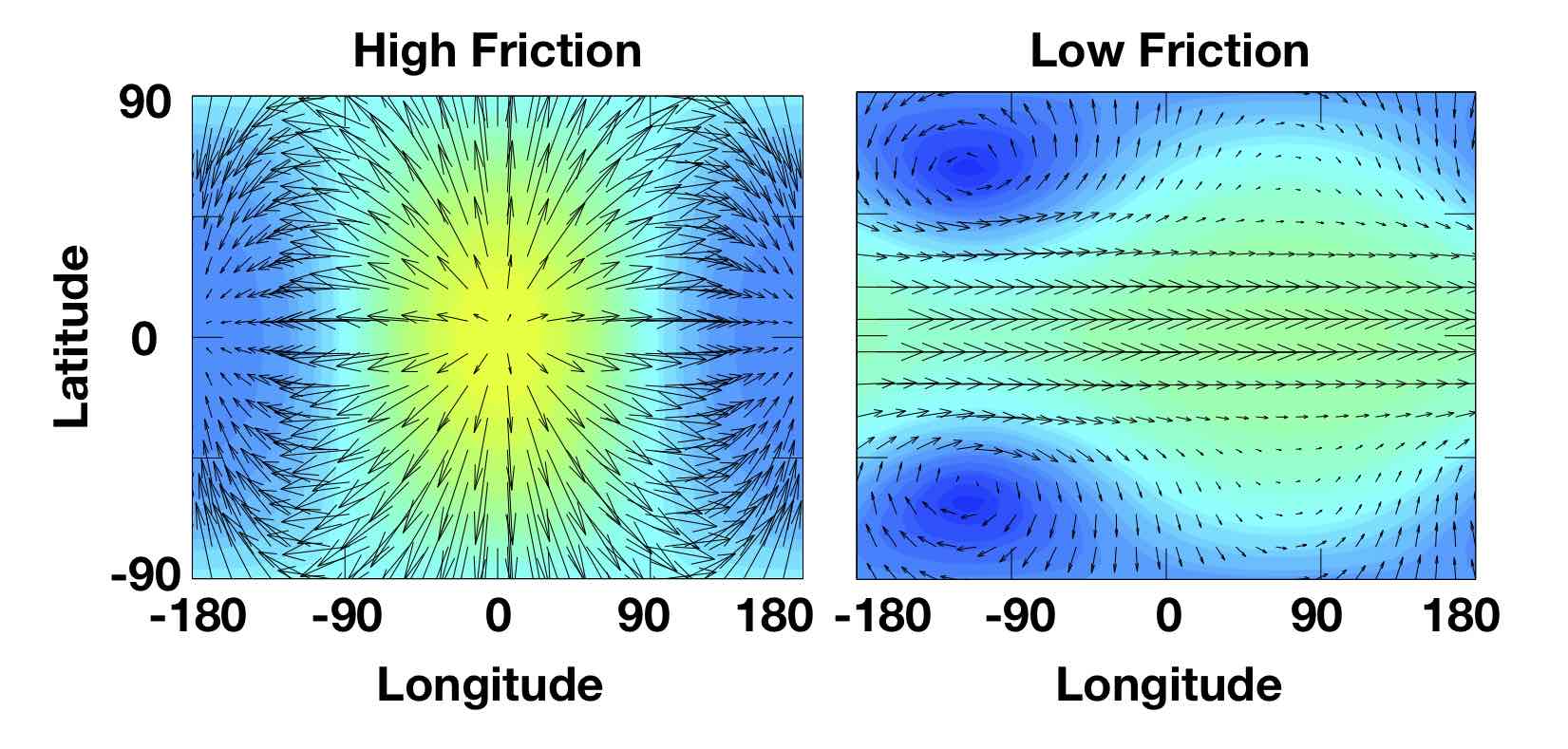}
\end{center}
\end{figure}

\noindent (3) Is the divergence of the flow in the high friction case
positive, negative, or zero at the substellar point?

\vspace{1.5in}

\noindent (4) Is the divergence of the flow in the high friction case
positive, negative, or zero at the anti-stellar point?

\clearpage

\noindent (5) The circulation in the low friction case has a feature
at the equator. What is this type of feature called?

\vspace{2in}

\noindent (6) There is a vortex in the low friction case located at
approximately \newline (longitude,latitude)=(-120,+60). Is the vorticity
associated with this vortex positive or negative?

\vspace{2in}

\noindent (7) Is the flow associated with the vortex at
(longitude,latitude)=(-120,+60) in the low friction case cyclonic or
anti-cyclonic?

\clearpage

\noindent (8) Estimate the emission temperature of WASP-33b, or the
temperature that it would have if it were a spherical blackbody with
an emissivity of one that received spatially uniform irradiation from
its star.

\clearpage

\noindent (9) WASP-33b is made mostly of H$_2$ gas. Estimate the scale
height of WASP-33b's atmosphere near the emission level.

\clearpage 

The figure below shows measurements and models of planetary radiative
emission normalized by stellar emission as a function of wavelength
for WASP-33b \cite{haynes2015spectroscopic}.

\begin{figure}[h!]
\begin{center}
  \includegraphics[width=0.8\textwidth]{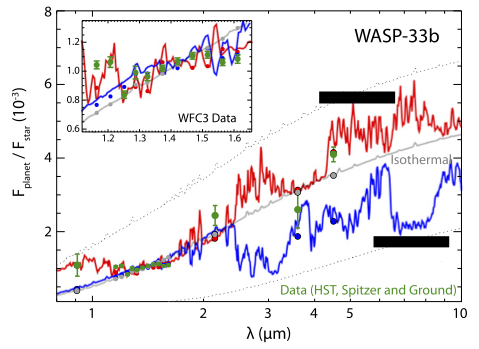}
\end{center}
\end{figure}

\noindent (10) The grey curve shows a modeled spectrum with an
isothermal temperature profile. Either the red or the blue curve shows
a modeled spectrum including a stratosphere, and the other one doesn't
have a stratosphere (the temperature just decreases as you go up). The
green dots are the data and the other colored dots are the models
averaged over the same bins as the data. Does the model that produced
the red or the blue curve include a stratosphere?

%% file: Finals/final_2019.tex
\subsection{Final, 2019}

\bigskip

\noindent This final concerns an effect on tropical cyclone motion
called ``$\beta$ drift.'' Consider the tropical cyclone represented
schematically by the diagram below. You can imagine that the diagram
represents winds in the eyewall of the tropical cyclone as we look
down on it from above, and that circulation extends outward from
it. The arrow in the upper left of the diagram shows the direction of
North. For this problem we will assume that there are no background
winds, which is to say that the only wind in the system is associated
with the tropical cyclone. We'll assume that absolute vorticity is
conserved, or that the thickness of the atmosphere in this region is
roughly constant. This is a reasonable approximation since tropical
cyclones mostly occur over oceans, where there are no mountains.

\begin{figure}[h!]
\begin{center}
  \includegraphics[width=0.8\textwidth]{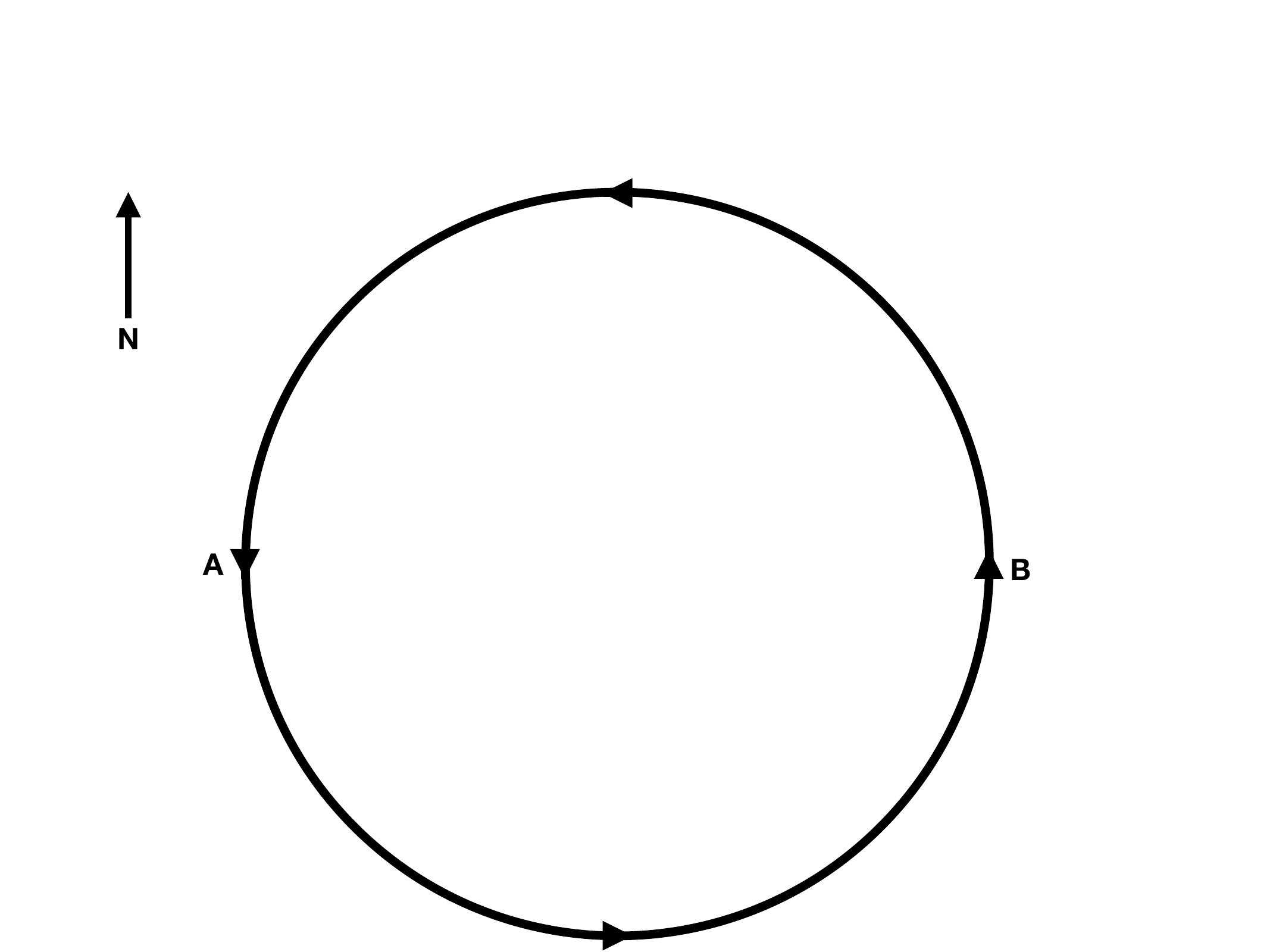}
\end{center}
\end{figure}

\bigskip

\noindent (1) Do the near-surface winds converge or diverge at the
center of the tropical cyclone?

\clearpage

\noindent (2) What is the sign of the divergence of the near-surface
winds  at the center of the tropical cyclone?

\vspace{1in}

\noindent (3) Do the winds near the tropopause converge or diverge at the
center of the tropical cyclone?

\vspace{1.1in}

\noindent (4) What is the sign of the divergence of the
winds near the tropopause at the center of the tropical cyclone?

\vspace{1.1in}

\noindent (5) What is the sign of the relative vorticity, $\zeta$,
associated with the tropical cyclone.

\vspace{1.1in}

\noindent (6) What hemisphere is the tropical cyclone in?

\vspace{1.1in}

\noindent (7) What is the sign of the planetary vorticity, $f$?

\clearpage

\noindent (8) At the point marked ``A,'' air is being advected
southward. Is the planetary vorticity of an air parcel at point A
increasing or decreasing?

\vspace{1in}

\noindent (9) Is the planetary vorticity of an air parcel at point B
increasing or decreasing?

\vspace{1in}

\noindent (10) Is the relative vorticity of an air parcel at point A
increasing or decreasing?

\vspace{1in}

\noindent (11) Is the relative vorticity of an air parcel at point B
increasing or decreasing?

\clearpage

\noindent (12, double points) On the diagram below, write $f\uparrow$,
$f\downarrow$, $\zeta\uparrow$, and $\zeta\downarrow$ in the
appropriate places next to points A and B. $f\uparrow$ indicates an
increase in $f$, for example.

\begin{figure}[h!]
\begin{center}
  \includegraphics[width=0.8\textwidth]{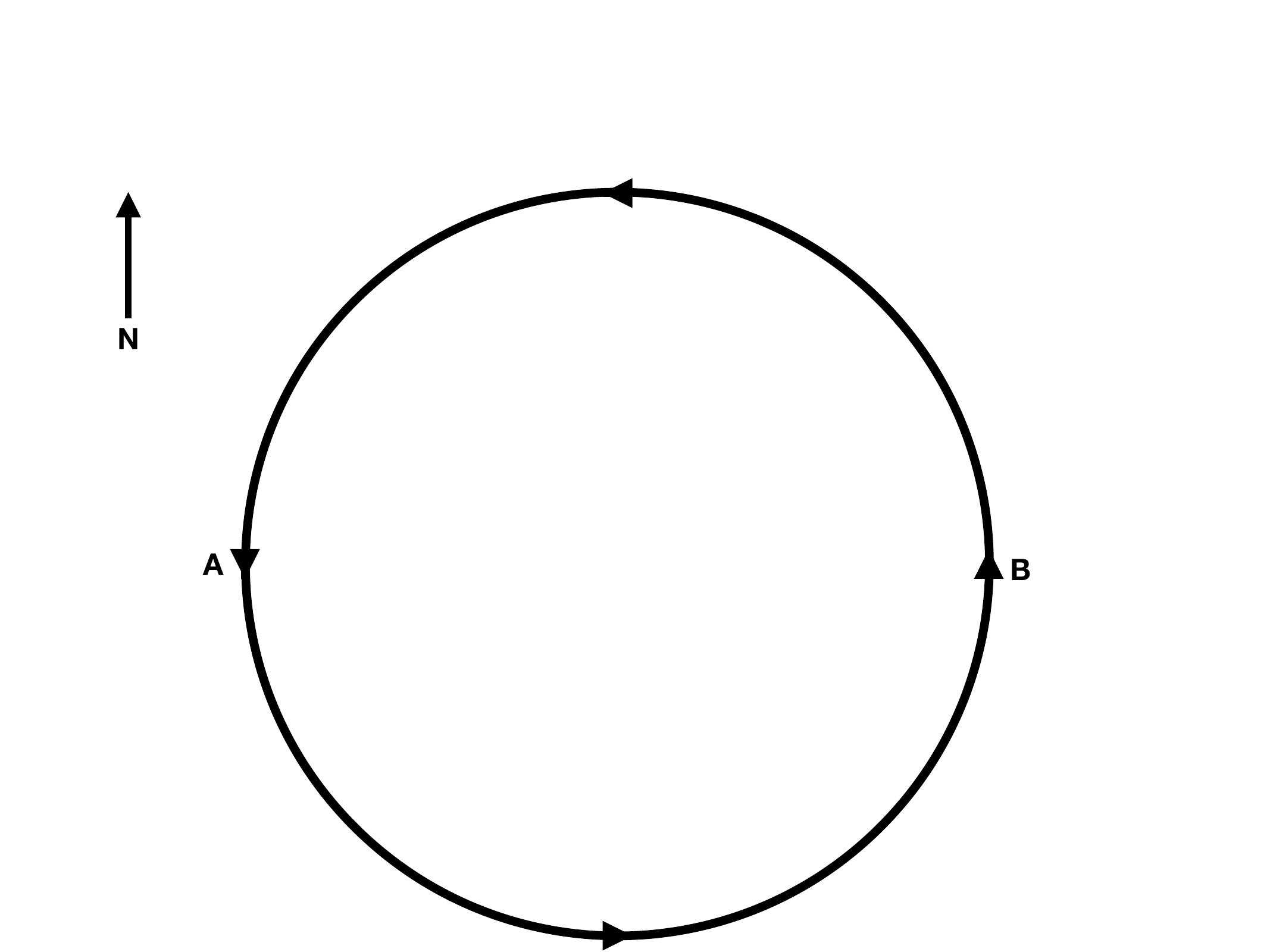}
\end{center}
\end{figure}

\noindent (13) The center of the tropical cyclone is a maximum or
minimum in relative vorticity. The changes in relative vorticity on
the west side (represented by point A) and the east side (represented
by point B) of the tropical cyclone change the relative vorticity
pattern so that the maximum or minimum is
advected horizontally. Notice that this occurs even though there are
no background winds. Based on the changes in relative vorticity we
have figured out so far, in which direction will the tropical cyclone
be advected?

\clearpage

\noindent (14) On the diagram below, draw an arrow through the center
of the tropical cyclone representing the direction of the advection
due to changes in relative vorticity that we have figured out so far
and label it $v_1$.

\begin{figure}[h!]
\begin{center}
  \includegraphics[width=0.8\textwidth]{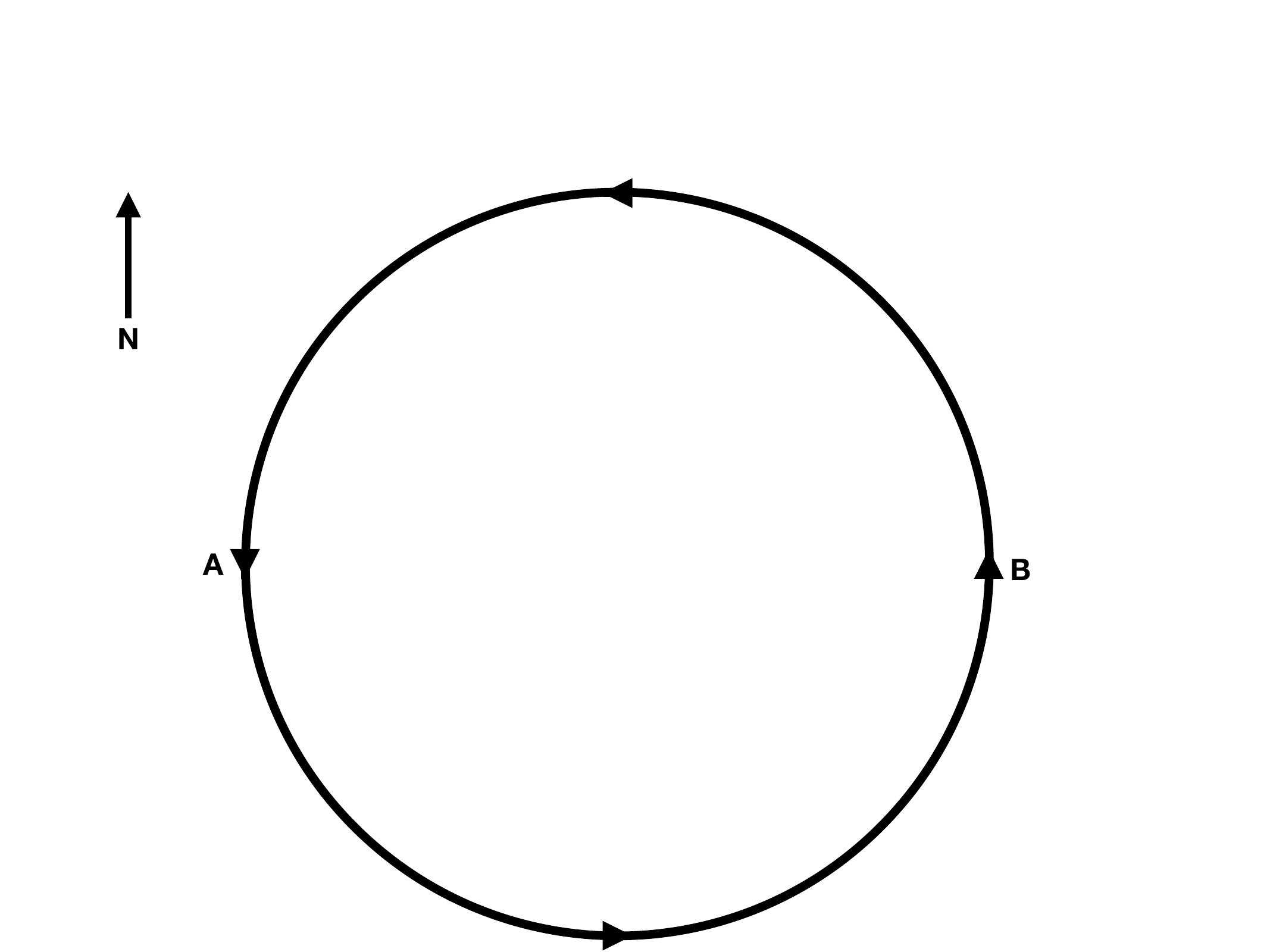}
\end{center}
\end{figure}

\noindent (15) As a result of the change in relative vorticity, will
the air at point A get additional clockwise or counterclockwise motion?

\vspace{1in}

\noindent (16) As a result of the change in relative vorticity, will
the air at point B get additional clockwise or counterclockwise motion?

\clearpage

\noindent (17) On the diagram below, draw circles around
points A and B with arrows to represent the additional vorticity that
they get simply from advection around the center of the tropical
cyclone.

\begin{figure}[h!]
\begin{center}
  \includegraphics[width=0.8\textwidth]{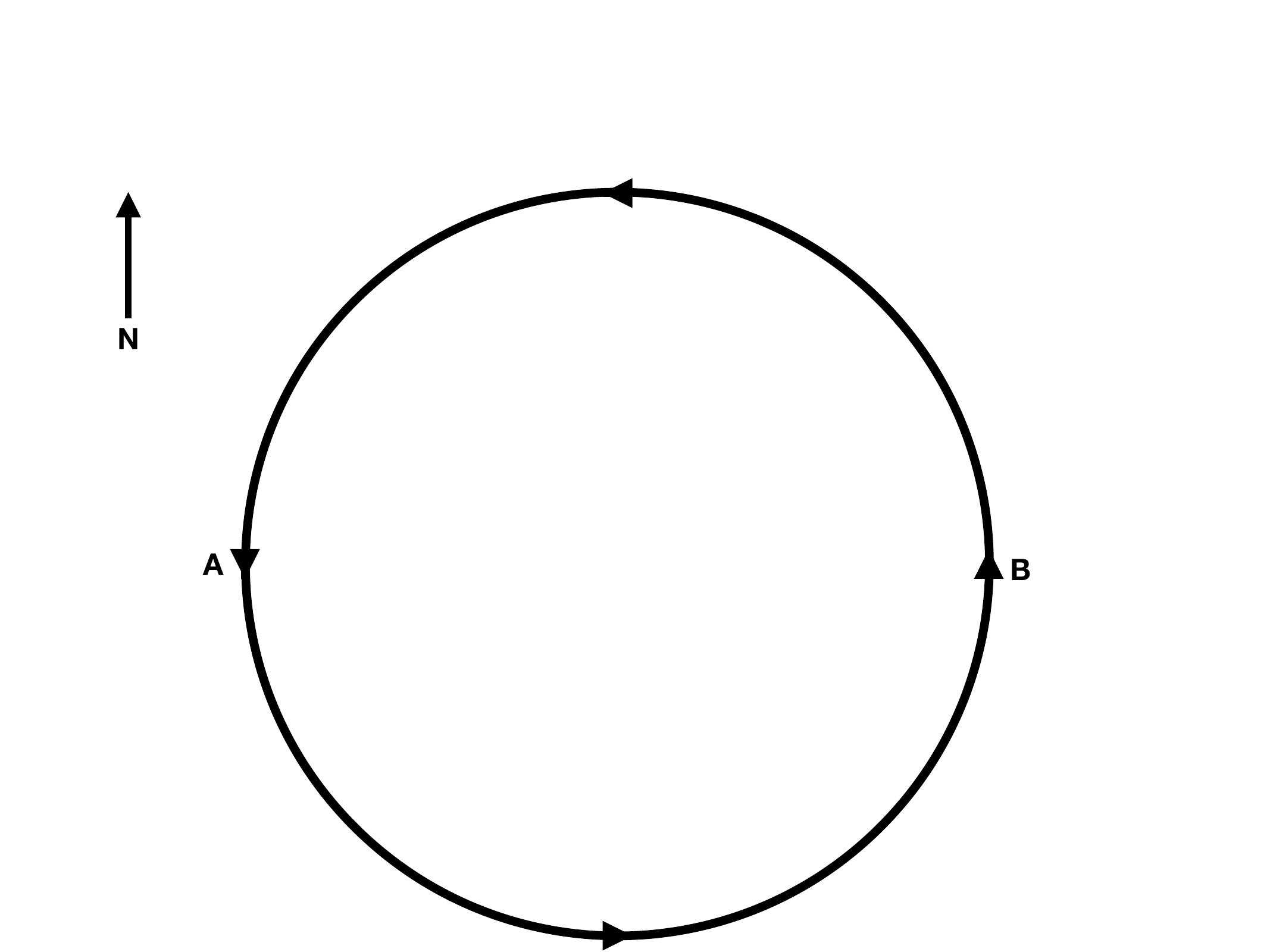}
\end{center}
\end{figure}

\clearpage

\noindent (18) These vorticity diagrams you have drawn should both be
pushing air at the center of the tropical cyclone in the same
direction. What direction is this?

\vspace{1in}

\noindent (19) On the diagram below, draw an arrow through the center
of the tropical cyclone representing the direction of advection
due to this second result of changes in relative vorticity and label
it $v_2$.

\begin{figure}[h!]
\begin{center}
  \includegraphics[width=0.8\textwidth]{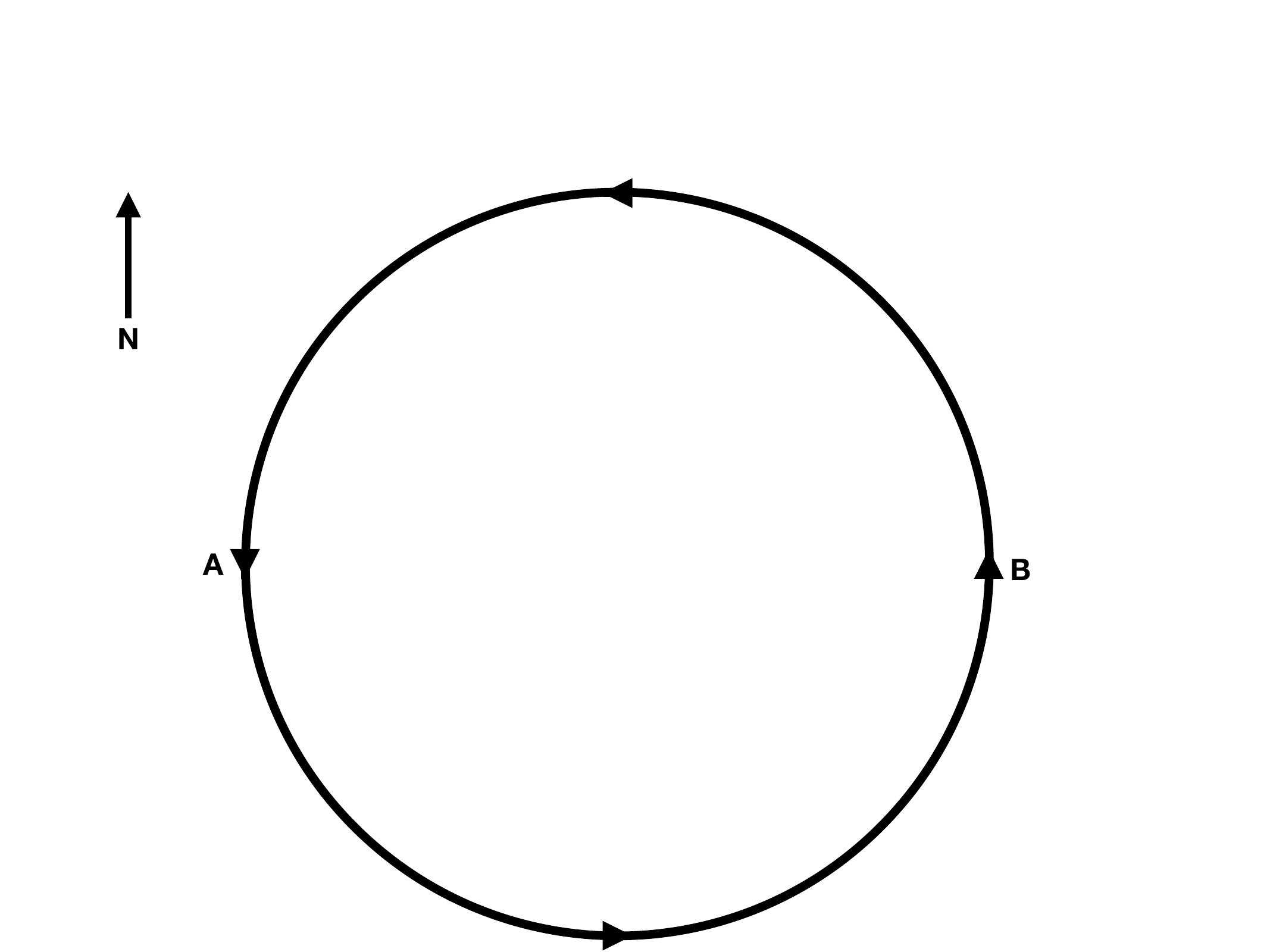}
\end{center}
\end{figure}

\clearpage
                                   
\noindent (20) On the diagram below, draw $v_1$, $v_2$, and
$v_T=v_1+v_2$. $v_T$ is the total advection of the tropical cyclone
simply due to its swirling motion.

\begin{figure}[h!]
\begin{center}
  \includegraphics[width=0.8\textwidth]{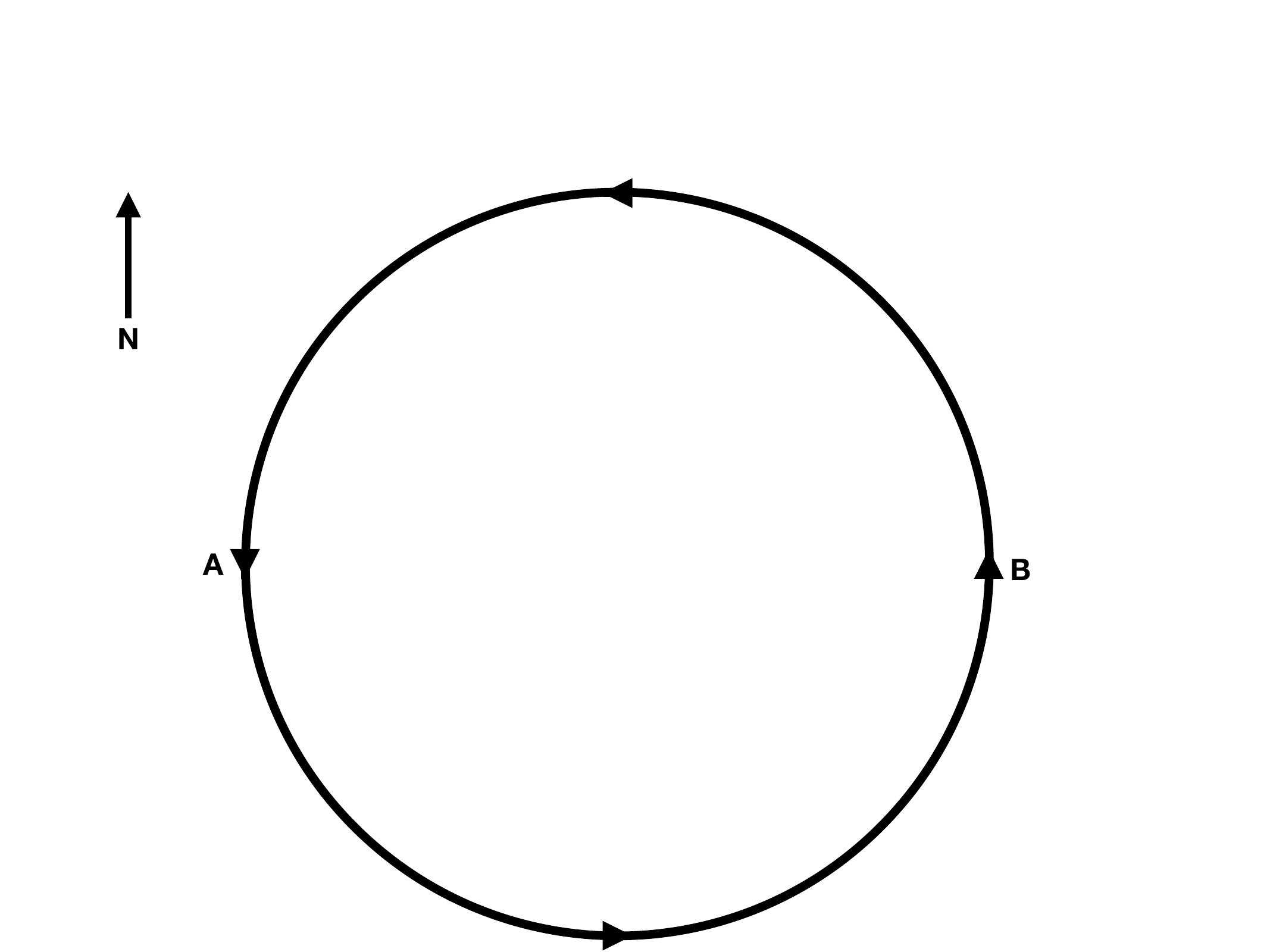}
\end{center}
\end{figure}

\vspace{1in}

\noindent (21) In which direction is $v_T$ pointing?

\clearpage

\noindent (22) If a tropical cyclone forms in the tropical Atlantic
ocean and there are no background winds, will it tend to move towards
the USA or away from the USA?

\vspace{2in}

\noindent (23) Would the drift we have worked out occur if the
planetary vorticity were constant?

\vspace{2in}

\noindent (24) The northward gradient of the planetary vorticity is
often referred to mathematically as $\beta$. Explain why the drift in
tropical cyclone we have worked out here is called ``$\beta$ drift.''

%% file: TheAtmosphere.bbl
\begin{thebibliography}{10}
\expandafter\ifx\csname url\endcsname\relax
  \def\url#1{\texttt{#1}}\fi
\expandafter\ifx\csname urlprefix\endcsname\relax\def\urlprefix{URL }\fi
\providecommand{\bibinfo}[2]{#2}
\providecommand{\eprint}[2][]{\url{#2}}

\bibitem{wallace2006atmospheric}
\bibinfo{author}{Wallace, J.~M.} \& \bibinfo{author}{Hobbs, P.~V.}
\newblock \emph{\bibinfo{title}{Atmospheric science: an introductory survey}}
  (\bibinfo{publisher}{Elsevier}, \bibinfo{year}{2006}).

\bibitem{SAGAN:1972p1233}
\bibinfo{author}{Sagan, C.} \& \bibinfo{author}{Mullen, G.}
\newblock \bibinfo{title}{Earth and {M}ars - evolution of atmospheres and
  surface temperatures}.
\newblock \emph{\bibinfo{journal}{Science}} \textbf{\bibinfo{volume}{177}},
  \bibinfo{pages}{52--56} (\bibinfo{year}{1972}).

\bibitem{seiff1996structure}
\bibinfo{author}{Seiff, A.} \emph{et~al.}
\newblock \bibinfo{title}{Structure of the atmosphere of {J}upiter: Galileo
  probe measurements}.
\newblock \emph{\bibinfo{journal}{Science}} \textbf{\bibinfo{volume}{272}},
  \bibinfo{pages}{844--845} (\bibinfo{year}{1996}).

\bibitem{manabe1967thermal}
\bibinfo{author}{Manabe, S.} \& \bibinfo{author}{Wetherald, R.~T.}
\newblock \bibinfo{title}{Thermal equilibrium of the atmosphere with a given
  distribution of relative humidity}.
\newblock \emph{\bibinfo{journal}{Journal of the Atmospheric Sciences}}
  \textbf{\bibinfo{volume}{24}}, \bibinfo{pages}{241--259}
  (\bibinfo{year}{1967}).

\bibitem{emanuel1986air}
\bibinfo{author}{Emanuel, K.~A.}
\newblock \bibinfo{title}{An air-sea interaction theory for tropical cyclones.
  {P}art i: Steady-state maintenance}.
\newblock \emph{\bibinfo{journal}{Journal of the Atmospheric Sciences}}
  \textbf{\bibinfo{volume}{43}}, \bibinfo{pages}{585--605}
  (\bibinfo{year}{1986}).

\bibitem{Sagan-1960}
\bibinfo{author}{Sagan, C.}
\newblock \emph{\bibinfo{title}{Physical Studies of Planets}}.
\newblock Ph.D. thesis, \bibinfo{school}{The University of Chicago}
  (\bibinfo{year}{1960}).
\newblock
  \bibinfo{note}{\href{https://drive.google.com/open?id=1d84ILZz1K7Gn_nFCOo7j08z5QUk5O7A0}{https://drive.google.com/open?id=1d84ILZz1K7Gn\_nFCOo7j08z5QUk5O7A0}}.

\bibitem{som2016earth}
\bibinfo{author}{Som, S.~M.} \emph{et~al.}
\newblock \bibinfo{title}{Earth's air pressure 2.7 billion years ago
  constrained to less than half of modern levels}.
\newblock \emph{\bibinfo{journal}{Nature Geoscience}}
  \textbf{\bibinfo{volume}{9}}, \bibinfo{pages}{448} (\bibinfo{year}{2016}).

\bibitem{fulchignoni2005situ}
\bibinfo{author}{Fulchignoni, M.} \emph{et~al.}
\newblock \bibinfo{title}{In situ measurements of the physical characteristics
  of {T}itan's environment}.
\newblock \emph{\bibinfo{journal}{Nature}} \textbf{\bibinfo{volume}{438}},
  \bibinfo{pages}{785} (\bibinfo{year}{2005}).

\bibitem{wolkenberg2010atmospheric}
\bibinfo{author}{Wolkenberg, P.}, \bibinfo{author}{Formisano, V.},
  \bibinfo{author}{Rinaldi, G.} \& \bibinfo{author}{Geminale, A.}
\newblock \bibinfo{title}{{The atmospheric temperatures over Olympus Mons on
  Mars: An atmospheric hot ring}}.
\newblock \emph{\bibinfo{journal}{Icarus}} \textbf{\bibinfo{volume}{207}},
  \bibinfo{pages}{110--123} (\bibinfo{year}{2010}).

\bibitem{zhai2017scalable}
\bibinfo{author}{Zhai, Y.} \emph{et~al.}
\newblock \bibinfo{title}{Scalable-manufactured randomized glass-polymer hybrid
  metamaterial for daytime radiative cooling}.
\newblock \emph{\bibinfo{journal}{Science}} \textbf{\bibinfo{volume}{355}},
  \bibinfo{pages}{1062--1066} (\bibinfo{year}{2017}).

\bibitem{atwater2018materials}
\bibinfo{author}{Atwater, H.~A.} \emph{et~al.}
\newblock \bibinfo{title}{{Materials challenges for the Starshot lightsail}}.
\newblock \emph{\bibinfo{journal}{Nature materials}} \bibinfo{pages}{1}
  (\bibinfo{year}{2018}).

\bibitem{charbonneau2009super}
\bibinfo{author}{Charbonneau, D.} \emph{et~al.}
\newblock \bibinfo{title}{A super-{E}arth transiting a nearby low-mass star}.
\newblock \emph{\bibinfo{journal}{Nature}} \textbf{\bibinfo{volume}{462}},
  \bibinfo{pages}{891} (\bibinfo{year}{2009}).

\bibitem{bean2010ground}
\bibinfo{author}{Bean, J.~L.}, \bibinfo{author}{Kempton, E. M.-R.} \&
  \bibinfo{author}{Homeier, D.}
\newblock \bibinfo{title}{{A ground-based transmission spectrum of the
  super-Earth exoplanet GJ 1214b}}.
\newblock \emph{\bibinfo{journal}{Nature}} \textbf{\bibinfo{volume}{468}},
  \bibinfo{pages}{669} (\bibinfo{year}{2010}).

\bibitem{cameron2010line}
\bibinfo{author}{Cameron, A.~C.} \emph{et~al.}
\newblock \bibinfo{title}{{Line-profile tomography of exoplanet transits--II. A
  gas-giant planet transiting a rapidly rotating A5 star}}.
\newblock \emph{\bibinfo{journal}{Monthly Notices of the Royal Astronomical
  Society}} \textbf{\bibinfo{volume}{407}}, \bibinfo{pages}{507--514}
  (\bibinfo{year}{2010}).

\bibitem{komacek2016atmospheric}
\bibinfo{author}{Komacek, T.~D.} \& \bibinfo{author}{Showman, A.~P.}
\newblock \bibinfo{title}{Atmospheric circulation of {H}ot {J}upiters:
  Dayside--nightside temperature differences}.
\newblock \emph{\bibinfo{journal}{The Astrophysical Journal}}
  \textbf{\bibinfo{volume}{821}}, \bibinfo{pages}{16} (\bibinfo{year}{2016}).

\bibitem{haynes2015spectroscopic}
\bibinfo{author}{Haynes, K.}, \bibinfo{author}{Mandell, A.~M.},
  \bibinfo{author}{Madhusudhan, N.}, \bibinfo{author}{Deming, D.} \&
  \bibinfo{author}{Knutson, H.}
\newblock \bibinfo{title}{{Spectroscopic evidence for a temperature inversion
  in the dayside atmosphere of hot Jupiter WASP-33b}}.
\newblock \emph{\bibinfo{journal}{The Astrophysical Journal}}
  \textbf{\bibinfo{volume}{806}}, \bibinfo{pages}{146} (\bibinfo{year}{2015}).

\end{thebibliography}
